\documentclass[longauth,traditabstract]{aa}  
\pdfoutput=1 
\usepackage{graphicx}

\usepackage{txfonts}
\usepackage{caption}
\usepackage{color, colortbl}
\usepackage{enumitem}

\usepackage{multirow}
\usepackage[normalem]{ulem} 
\usepackage{tabularx}

\DeclareCaptionLabelFormat{continued}{#1~#2 (Continued)}
\bibpunct{(}{)}{;}{a}{}{,}

\newcolumntype{Y}{>{\centering\arraybackslash}X}

\usepackage{multicol}

\begin{document} 

   \title{PHIBSS2: survey design and $z=0.5-0.8$ results\\\vspace{0.3cm} \Large Molecular gas reservoirs during the winding-down of star formation}

\titlerunning{Molecular gas reservoirs at $z=0.5-0.8$}
\authorrunning{Freundlich et al.}

\author{J.~Freundlich\inst{1,2,3}
        \and
        F.~Combes\inst{1,2}
        \and
        L. J.~Tacconi\inst{4}          
        \and
        R.~Genzel\inst{4,5,6}                       
        \and
        S.~Garcia-Burillo\inst{7}
        \and
        R.~Neri\inst{8}
        \and
        T.~Contini\inst{9}
        \and
        A.~Bolatto\inst{10}
        \and
        S.~Lilly\inst{11}
        \and
        P.~Salomé\inst{1}      
        \and
        I.~C.~Bicalho\inst{1}        
        \and
        J.~Boissier\inst{8}
        \and
        F.~Boone\inst{9}
        \and
        N.~Bouch\'e\inst{9}
        \and
        F.~Bournaud\inst{12}
        \and
        A.~Burkert\inst{13}
        \and
        M.~Carollo\inst{11}  
        \and
        M.~C.~Cooper\inst{14}             
        \and
        P.~Cox\inst{15}
        \and
        C.~Feruglio\inst{16}
        \and
        N. M.~F\"orster Schreiber\inst{6}
        \and
        S.~Juneau\inst{12}
        \and
        M.~Lippa\inst{6}
        \and
        D.~Lutz\inst{6}
        \and
        T.~Naab\inst{17}
        \and
        A.~Renzini\inst{18}
        \and
        A.~Saintonge\inst{19}
        \and
        A.~Sternberg\inst{20}
        \and
        F.~Walter\inst{21}
        \and
        B.~Weiner\inst{22}
        \and
        A.~Wei\ss\inst{23}
        \and
        S.~Wuyts\inst{24}
}

\institute{
        LERMA, Observatoire de Paris, PSL Research University, CNRS, Sorbonne Universit\'es, UPMC Univ. Paris 06, F-75014, Paris, France\\
        \email{jonathan.freundlich@obspm.fr}
        \and
        Coll\`ege de France, Paris, France
        \and
        Centre for Astrophysics and Planetary Science, Racah Institute of Physics, The Hebrew University, Jerusalem 91904 Israel
        \and
        Max-Planck-Institute f\"{u}r extraterrestrische Physik (MPE), Giessenbachstrasse 1, 85748 Garching, Germany
        \and
        Dept. of Physics, Le Conte Hall, University of California, CA 94720 Berkeley, USA
        \and
        Dept. of Astronomy, Campbell Hall, University of California, CA 94720 Berkeley, USA
        \and
        Observatorio Astron\'omico Nacional - OAN, Apartado 1143, 28800 Alcal\'a de Henares, Madrid, Spain
        \and
        IRAM, 300 rue de la Piscine, F-38406 St. Martin d'Heres, Grenoble, France  
        \and
        Institut de Recherche en Astrophysique et Plan\'etologie – IRAP, CNRS, Universit\'e de Toulouse, UPS-OMP, 14, avenue E. Belin,
        31400 Toulouse, France
        \and
        Dept. of Astronomy, University of Maryland, College Park, MD 20742-2421, USA 
        \and
        Institute for Astronomy, Department of Physics, ETH Zurich, CH-8093 Zurich, Switzerland
        \and
        CEA, IRFU, SAp, 91191 Gif-sur-Yvette, France
        \and
        Universit\"atsternwarte der Ludwig-Maximiliansuniversit\"at, Scheinerstrasse 1, 81679 M\"unchen, Germany,
        MPG-Fellow at MPE
        \and
        Dept. of Physics \& Astronomy, Frederick Reines Hall, University of California, Irvine, CA 92697, USA
        \and
        Institut d’Astrophysique de Paris, Sorbonne Universit\'e, CNRS, UMR 7095, 98 bis bd Arago, 7014 Paris, France
        \and
        Scuola Normale Superiore, Piazza dei Cavalieri 7, 56126 Pisa, Italy
        \and
        Max Planck Institut f\"ur Astrophysik, Karl Schwarzshildstrasse 1, D-85748 Garching, Germany
        \and
        INAF-Osservatorio Astronomico di Padova, Vicolo dell’Osservatorio 5, I-35122 Padova, Italy
        \and
        Department of Physics and Astronomy, University College London, Gower Street, London WC1E 6BT, UK
        \and
        School of Physics and Astronomy, Tel Aviv University, Tel Aviv 69978, Israel
        \and
        Max-Planck-Institut f\"ur Astronomie, K\"onigstuhl 17, D-69117 Heidelberg, Germany
        \and
        Steward Observatory, 933 N. Cherry Ave., University of Arizona, Tucson, AZ 85721-0065, USA      
        \and
        Max-Planck-Institut f\"ur Radioastronomie, Auf dem H\"ugel 69, D-53121 Bonn, Germany
        \and
        Department of Physics, University of Bath, Claverton Down, Bath, BA2 7AY, UK
}

\date{Received 1 November 2017 / Accepted 18 December 2018}


\abstract
{Following the success of the Plateau de Bure high-$z$ Blue Sequence Survey (PHIBSS), we present the PHIBSS2 legacy program, a survey of the molecular gas properties of star-forming galaxies on and around the star-formation main sequence (MS) at different redshifts using IRAM's NOrthern Extended Millimeter Array (NOEMA). This survey significantly extends the existing sample of star-forming galaxies with CO molecular gas measurements, probing the peak epoch of star formation ($z=1-1.6$) as well as its building-up ($z=2-3$) and winding-down ($z=0.5-0.8$) phases. The targets are drawn from the well-studied GOODS, COSMOS, and AEGIS cosmological deep fields and uniformly sample the MS in the stellar mass ($\rm M_{\star}$) -- star formation rate (SFR) plane with $\rm \log(M_{\star}/M_\odot) = 10-11.8$ and $\rm SFR = 3.5-500~M_\odot yr^{-1}$ without morphological selection, thus providing a statistically meaningful census of star-forming galaxies at different epochs. We describe the survey strategy and sample selection before focusing on the results obtained at redshift $z=0.5-0.8$, where we report 60 CO(2-1) detections out of 61 targets. We determine molecular gas masses between $2.10^9$ and $5.10^{10}~\rm M_\odot$ and separately obtain disc sizes and  bulge-to-total (B/T) luminosity ratios from HST I-band images. The median molecular gas-to-stellar mass ratio $\widetilde{\mu_{\rm gas}} = 0.28 \pm 0.04$, gas fraction $ \widetilde{f_{\rm gas}} = 0.22 \pm 0.02$, and depletion time $\widetilde{t_{\rm depl}} = 0.84 \pm 0.07~\rm Gyr$ as well as their dependence with stellar mass and offset from the MS follow 
published scaling relations  for a much larger sample of galaxies spanning a significantly wider range of redshifts, the cosmic evolution of the SFR being mainly driven by that of the molecular gas fraction. 
The galaxy-averaged molecular Kennicutt-Schmidt (KS) relation between molecular gas and SFR surface densities is strikingly linear, pointing towards similar star formation timescales within galaxies at any given epoch. In terms of morphology, 
the molecular gas content, the SFR, the disc stellar mass, and the disc molecular gas fraction do not seem to correlate with B/T and the stellar surface density, which suggests an ongoing supply of fresh molecular gas to compensate for the build-up of the bulge. Our measurements do not yield any significant variation of the depletion time with B/T and hence no strong evidence for morphological quenching within the scatter of the MS. 
}


\keywords{
        Galaxies: evolution -- Galaxies: high redshift -- Galaxies: star formation -- Galaxies: ISM -- ISM: molecules
}

\maketitle


\section{Introduction}
\label{section:introduction}

\subsection*{The main sequence of star formation}

Observed massive galaxies in the distant universe form stars at much higher rates than their local counterparts, with a peak epoch of star formation in the range $z=1 -3$ \citep{Noeske2007, Daddi2007, Reddy2009, Bouwens2010, Cucciati2012, Lilly2013, MadauDickinson2014}. 
%
At each epoch, there is a bimodality between red passive galaxies on one side and blue star-forming galaxies on the other, most of the latter lying on a relatively tight, almost linear relation between their stellar mass ($\rm M_{{\star}}$) and star formation rate (SFR), known as the star formation `main sequence'  \citep[MS;][]{Baldry2004, Brinchmann2004,Noeske2007, Elbaz2007, Elbaz2011, Daddi2007, Schiminovich2007, Damen2009, Santini2009, Rodighiero2010, Rodighiero2011, Peng2010,  Wuyts2011b, Sargent2012, Whitaker2012, Whitaker2014, Speagle2014, Renzini2015, Schreiber2015}. About 90\% of the cosmic star formation history since $z = 2.5$ took place near this MS.
The $\leq 0.3 ~\rm dex$  scatter of the MS, the rotating disc morphology of most galaxies that constitute it \citep{ForsterSchreiber2006, ForsterSchreiber2009, Genzel2006, Genzel2008, Stark2008, Daddi2010, Wuyts2011b} and 
the long star formation cycles inferred from the number of star-forming galaxies observed at $z=1-2$ \citep{Daddi2005, Daddi2007, Caputi2006} argue in favour of a relatively smooth mode of star formation. The large molecular gas reservoirs fueling star formation \citep{Erb2006, Daddi2010, Tacconi2010, Tacconi2013} are thought to be maintained by a continuous supply of fresh gas from the cosmic web and minor mergers \citep{Birnboim2003, Keres2005, Keres2009, Dekel2006, Cattaneo2006, Ocvirk2008, Dekel2009, Genel2010}.
Typical star-forming galaxies are expected to progress along the MS in a slowly evolving gas-regulated quasi equilibrium between inflows, outflows, and star formation \citep{Bouche2010, Dave2011a, Dave2012, Feldmann2013, Lilly2013,Dekel2013, Peng2014, DekelMandelker2014} until their star formation is quenched  when they enter denser environments or grow past the Schechter mass \citep[$\rm M_{\star} \sim 10^{10.8-11} M_\odot$; ][]{Conroy2009, Peng2010}, and then to rapidly transit down to the red sequence. 
Episodes of gas compaction, depletion, and replenishment could confine them within the scatter of the MS before the final quenching occurs \citep{Dekel2014, Zolotov2015, Tacchella2016a, Tacchella2016b}.
Quenching might be due to a combination of factors, including gas removal by winds driven by supernovae or active galactic nuclei (AGNs), gas streams from the cosmic web that stop penetrating galactic haloes above a critical halo mass, 
a sudden drop in the gas cooling, a change in morphology, and/or environmental effects. 

The equilibrium model predicts strong correlations between the specific SFR ($\rm sSFR = SFR/M_{star}$) and the gas fraction (or equivalently, the gas-to-stellar mass ratio $\rm \mu_{gas} = M_{gas}/M_{star}$, where $\rm M_{gas}$ is either the total or the molecular gaseous mass of the galaxy) as they evolve with redshift, while the gas compaction scenario further implies gradients of the central gas density, gas fraction, and depletion time ($t_{\rm  depl} = \rm M_{gas}/SFR$) across the MS. 
Measurements of these quantities at different redshifts thus provide crucial observational tests to understand the building-up and the winding-down phases of normal MS star-forming galaxies. The depletion time, which measures the star formation efficiency (SFE), has notably been suggested to decrease with redshift up to $z=1$ \citep{Combes2011, Combes2013}. 
The Kennicutt-Schmidt (KS) relation between the molecular gas and SFR surface densities, $\Sigma_{\rm SFR} \propto (\Sigma_{\rm gas})^N$, further characterises the SFE averaged over entire galaxies or subregions within them. It has been shown to be near linear on galactic and subgalactic scales for $\rm \Sigma_{\rm gas}>10~M_\odot pc^{-2}$, with an exponent $N=0.9-1.3$ and a scatter of $\pm 0.3-0.4\rm ~dex$ \citep{Schmidt1959, Kennicutt1998a, Bigiel2008, Bigiel2011, Leroy2008, Leroy2013, Daddi2010, Daddi2010b, Saintonge2011, Schruba2011, Kennicutt2012}, indicating relatively uniform molecular gas depletion times around $1-2\rm ~Gyr$.

\subsection*{The PHIBSS survey up to now}

The Plateau de Bure High-z Blue Sequence Survey (PHIBSS, PI: L. Tacconi \& F. Combes) carried out at the IRAM Plateau de Bure interferometer \citep[PdBI; ][]{Guilloteau1992, Cox2011} aimed at better understanding the winding-down of star formation within normal MS star-forming galaxies from the point of view of their molecular gas reservoirs.
It focused on the massive tail of the MS at $z=1.2$ and $2.2$, with $\rm log(M_{\star}/M_\odot) \geqslant 10.4$ and $\rm log(SFR/M_\odot yr^{-1}) \geqslant 1.5$, and comprised 52 CO (3-2) detections and 8 higher-resolution imaging observations with beam sizes in the range $0.3^{\prime\prime}-1^{\prime\prime}$. It uncovered large molecular gas reservoirs, with mean molecular gas fractions $f_{\rm gas} = \rm M_{gas}/(M_{gas} + M_{\star}) $ of 33\% at $z=1.2$ and 47\% at $z=2.2$ \citep{Tacconi2010, Tacconi2013}, when they only reach 7-10\% in local giant spirals \citep{Leroy2008, Saintonge2011}, showing that the cosmic evolution of the SFR is mainly due to the diminishing molecular gas content. 
\cite{Tacconi2013} further showed that $f_{\rm gas}$ decreases with $\rm M_\star$, which can be interpreted in terms of feedback models where the first generations of stars remove part of the gas or in terms of a mass dependence of the accretion effiency or the SFE, since $\mu_{\rm gas} = {\rm sSFR}\times t_{\rm depl}$ \citep[e.g. ][]{Bouche2010, Dave2011b}.
The PHIBSS also obtained a near linear KS relation lying in the continuity of low redshift measurements, albeit with a slightly lower mean depletion time \citep{Genzel2010, Tacconi2013}. 
The sub-arcsecond follow-up observations enabled to obtain good-quality rotation curves and resolved velocity dispersion maps, showing an increased turbulent support compared to low redshift which is compatible with models where cosmic streams feed the disc and trigger violent gravitational instabilities \citep[e.g. ][]{Dekel2009b}.
Resolved kinematics also enabled to separate smoothed ensembles of clumps due to their different velocities, and to obtain a resolved KS relation at sub-galactic scale for four galaxies of the sample \citep{Freundlich2013}. \cite{Genzel2013} further obtained a pixel by pixel KS relation for one typical $z=1.53$ massive star-forming galaxy from the PHIBSS sample.

\subsection*{The PHIBSS2 legacy program}

Built on the success of the PHIBSS, the IRAM PHIBSS2 legacy program (PIs: F. Combes, S. Garc\'ia-Burillo, R. Neri \& L. Tacconi) intends to extend these results to a wider range of redshifts and to better sample the $\rm M_{\star}$-SFR plane at each redshift. 
This four-year program was phased to optimise and exploit the NOrthern Extended Millimeter Array \citep[NOEMA; ][]{Schuster2014} capabilities as they came online at the PdBI, which enabled a significant statistical gain with the smaller integration times and the increased sensitivity. 
PHIBSS2 has measured mean molecular gas fractions and depletion times in different redshift bins and across the MS, with the aim of studying the connection between star formation and molecular gas reservoirs and its evolution with redshift. \cite{Genzel2015} and \cite{Tacconi2018} already use PHIBSS2 detections together with other measurements to quantify precisely how the depletion time and the gas fraction depend on redshift, stellar mass, and the offset from the MS reference line. 
They notably find that while the gas fraction decreases steeply with time, the depletion time slowly increases. They show how the gas fraction progressively increases above the MS and decreases below and how the depletion time follows the opposite trend without depending much on the stellar mass. 

In this paper, we present the PHIBSS2 strategy and its results at $z=0.5-0.8$. In this redshift bin, we report 60 \mbox{CO(2-1)} detections within a sample of 61 star-forming galaxies, hence constituting the first systematic census of the molecular gas in this redshift range. This sample bridges the gap between observations of the molecular gas in the nearby universe and at the peak epoch of star formation, probing the crucial period of the winding-down of star formation in the last 10 Gyr of the history of the universe. Until now,  paradoxically, the molecular gas content of galaxies in this redshift range has not been  studied as much as that of higher-redshift galaxies, partly because the CO line flux increase with frequency makes high CO rotational transitions more easily observable at higher redshifts \citep{Combes1999}.
We determine the molecular gas content, the sizes and morphology of these galaxies and describe the star-formation conditions within them in terms of gas fraction, depletion time, and surface densities. 
%
Section \ref{section:observations} presents the general PHIBSS2 strategy, the sample selection and its implementation with the IRAM Plateau de Bure interferometer and its NOEMA upgrade. 
Section \ref{section:results} reports the CO molecular gas measurements we obtain at $z=0.5-0.8$ and the results of our morphological study. In section \ref{section:discussion}, we interpret our results in terms of molecular gas fraction and depletion time, their dependence on morphology and the KS relation. 
We conclude in Section \ref{section:conclusion}.
%
Throughout this paper, we assume a flat $\Lambda$CDM universe with $\Omega_{\rm m} = 0.3$, $\Omega_\Lambda = 1-\Omega_{\rm m}$ and $H_0 = 70~\rm km.s^{-1}.Mpc^{-1}$.


\section{Observations}
\label{section:observations}

\begin{figure}
        \centering
        \includegraphics[width=0.98\hsize]{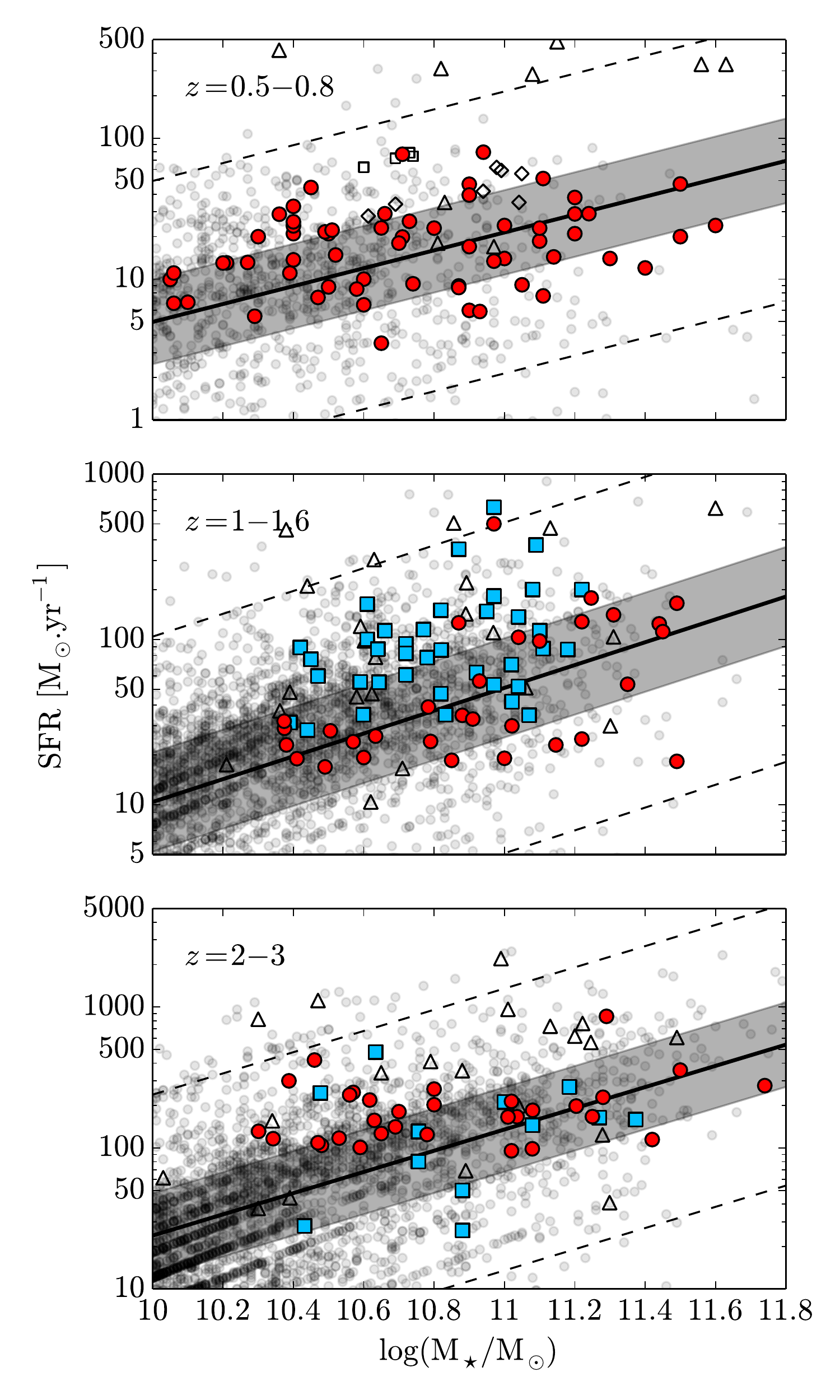}
        \vspace{-0.1cm}
        \caption{Location of the PHIBSS2 sample in the three redshift ranges $z=0.5-0.8$, $z=1-1.6$ and $z=2-3$ as a function of stellar mass $\rm M_\star$ and SFR. 
                The PHIBSS2 galaxies at $z=0.5-0.8$ presented in this paper and the targets in the other redshift ranges are all marked as plain red circles. 
                The PHIBSS sample \citep{Tacconi2010, Tacconi2013} is indicated as plain blue squares while other existing CO measurements in the same redshift ranges are displayed as open triangles, including ULIRGs at $z=0.5-0.8$ observed with the IRAM 30m telescope \citep{Combes2011, Combes2013}, near MS star-forming galaxies in the range $z=0.5-3$ \citep{Daddi2010, Magdis2012b, Magnelli2012}, above MS submillimeter galaxies in the range $z=1.2-3$ \citep{Greve2005, Tacconi2006, Tacconi2008, Bothwell2013} and lensed MS galaxies in the range $z=1.2-3$ observed with the IRAM PdBI \citep[][ and references therein]{Saintonge2013}, as well as additional sources between $z=1$ and $z=2.5$ observed with ALMA and NOEMA \citep[][Genzel et al. in prep.]{Silverman2015, Decarli2016}. 
                The upper panel also shows sources observed by \cite{Geach2011} at $z\sim 0.4$ as open diamonds and \cite{Bauermeister2013} at $z\sim 0.5$ as open squares. 
                The background data points correspond to 3D-HST galaxies in the AEGIS, GOODS-North, COSMOS and UDS fields \citep{Brammer2012, Momcheva2016}, while the solid line highlights the mean MS line from \cite{Speagle2014} in each redshift range. The shaded area corresponds to the $0.3~\rm dex$ scatter of the MS and the dashed line to $\rm \pm 1~dex$. 
        }
        \label{fig:sample}
\end{figure}

\subsection{Survey goals and strategy}

The PHIBSS2 four-year legacy program is designed as a comprehensive and systematic study of the CO molecular gas content of galaxies during the build-up ($z=2-3$), peak ($z=1-1.6$), and subsequent winding-down ($z=0.5-0.8$) phases of star formation in the universe. 
%
As shown in Fig.~\ref{fig:sample}, it targets more than 120 sources over three redshift bins sampling the $\rm M_{\star}$-SFR plane in the well-studied GOODS-North, COSMOS, and AEGIS cosmological fields on and around the MS. It significantly expands the first PHIBSS sample, which focused on galaxies at $z=1.2$ and $2.2$, doubles the number of CO measurements available at $z>2$, systematically probes the winding-down epoch at $z=0.5-0.8$ and includes measurements below the MS at $z=1-1.6$. 
The interconnected science goals of the PHIBSS2 survey are as follows.
\begin{enumerate}
        \item {Firstly, we aim to follow the evolution of the molecular gas fraction and the depletion time in normal MS star-forming galaxies at different epochs} to establish quantitatively whether the evolution of the cosmic SFR is mostly driven by the available molecular gas reservoirs.
        \\
        \item {Secondly, we aim to characterise the dependence of the gas fraction and the depletion time on the sSFR to compare the galaxy population on, above, and below the MS.} 
        Quantifying this dependence will allow us to investigate whether "out of equilibrium" systems above the MS are fundamentally different from MS galaxies, how the quenching of star formation occurs below it, and in future studies to estimate gas fractions and depletion times directly from SFR data. 
        \\
        \item {We futher intend to quantify the dependence of the gas fraction on the stellar mass to test feedback and quenching models.} PHIBSS2 indeed enables to confirm and quantify the decrease of the gas fraction with stellar mass uncovered by PHIBSS over a broader range of stellar masses $\rm \log(M_{\star}/M_\odot) = 10-11.6$ spanning from the stellar feedback regime to the quenching regime. 
        \\
        \item {Subsequently, we want to test the impact of AGNs, environment, and morphology on quenching using a purely mass-selected sample above the Schechter mass.} PHIBSS and PHIBSS2 together indeed provide a large enough sample of about 50 star-forming galaxies with $\rm \log(M_{\star}/M_\odot)>10.8$ to test how gas properties correlate with the presence of an AGN, environment, and morphological indicators such as the bulge-to-total ratio and the stellar mass surface density. 
        \\
        \item {We also plan to search for molecular outflows to test stellar and AGN feedback models.} While powerful galactic winds of ionized gas are found to be ubiquitous amongst star-forming galaxies at high-redshift \citep{Pettini2000, Weiner2009, Genzel2011, Newman2012, ForsterSchreiber2014}, detecting molecular gas outflows is still challenging and often limited to nearby quasars and ultra-luminous IR galaxies \citep{Feruglio2010, Sturm2011, Geach2014, Cicone2014}. With more than 180 spectra from PHIBSS and PHIBSS2, we will be able to use deep stacking techniques to detect molecular outflows both in the stellar and in the AGN feedback regimes.
        \\
        \item {We plan to determine the molecular gas distribution and kinematics from sub-arcsecond follow-ups of selected targets to establish spatially resolved KS relations, rotation curves, and velocity dispersion maps at different redshifts.} In addition to galaxy-averaged measurements, PHIBSS2 indeed includes spatially resolved molecular gas observations of selected targets with NOEMA and ALMA that can be compared to the stellar, SFR and ionized gas distributions from complementary observations. 
        \\
        \item {Lasty, we aim to probe the physical state of the gas from the CO line excitation at different $\rm M_\star$, SFR, and redshifts.} Although recent observations \citep{Ivison2011, Sharon2016, Dessauges-Zavadsky2017} indicate that the CO spectral energy distributions (SEDs) of star-forming galaxies at high redshift are similar to but slightly more excited than that of the Milky Way, measurements are still scarce. The combined PHIBSS and PHIBSS2 sample will constitute a benchmark to more systematically investigate the gas excitation at high redshift with additional CO transitions. 
        Follow-ups to probe dense gas tracers such as HCN will further help characterise the conditions ruling star formation. 
\end{enumerate}

The first three points are addressed in \cite{Genzel2015} and \cite{Tacconi2018}, whose scope extends beyond the PHIBSS and PHIBSS2 programs as these articles use a wealth of molecular gas data, combining both CO and dust measurements in order to yield quantitative scaling relations for the molecular gas fraction and the depletion time and to eliminate concerns about the CO to molecular gas mass conversion factor.
%
The PHIBSS2 program comprises approximately 1068 hours of on-source data gathered by the IRAM NOEMA interferometer over four years, and participates in current efforts to better understand star-forming galaxies and their evolution. It is indeed one element among different large interconnected surveys, including KMOS-3D \citep{Wisnioski2015}, Large Binocular Telescope (LBT) LUCI spectroscopic observations \citep{Wuyts2014}, SINS/zC-SINF \citep{ForsterSchreiber2006, ForsterSchreiber2009, ForsterSchreiber2014, Genzel2006, Genzel2011}, MUSE imaging and kinematics \citep{Contini2016}, Hubble Space Telescope (HST) imaging with CANDELS \citep{Grogin2011, Koekemoer2011}, and 3D-HST grism spectroscopy \citep{Brammer2012, Momcheva2016}; many of these surveys draw their samples from the same parent population.

\subsection{Sample selection}

The PHIBSS2 sample was drawn from large panchromatic imaging surveys with good spectroscopic redshifts and well-calibrated stellar masses, SFR, and HST morphologies, namely the North field of the Great Observatories Origins Deep Survey \citep[GOODS-N; R.A. = 12h36m, DEC. = 62$^\circ$14$^\prime$; ][]{Giavalisco2004}, the Cosmic Evolution Survey \citep[COSMOS; R.A. = 10h00m, DEC. = 02$^\circ$12$^\prime$; ][]{Scoville2007, Laigle2016}, and the All-Wavelength Extended Groth Strip International Survey \citep[AEGIS; R.A. = 14h17m, DEC. = 52$^\circ$30$^\prime$; ][]{Davis2007, Noeske2007}. These three fields constitute well-understood parent samples with excellent multi-band ancillary data from the X-ray to the radio, which enables us to quantitatively relate the PHIBSS2 sample to a much larger census of typical star-forming galaxies. 
Selected galaxies in the GOODS-N and AEGIS fields are part of the 3D-HST sample \citep{Brammer2012, Momcheva2016} while those in the COSMOS field were taken from the $z$COSMOS survey \citep{Lilly2007,Lilly2009}.
%
The stellar masses across the different parent samples are determined through SED modelling based on a Chabrier initial mass function (IMF). Such modelling uses standard assumptions on the star formation histories and the dust attenuation within these galaxies \citep{Erb2006, ForsterSchreiber2009, Wuyts2011a}. 
%
The SFRs are estimated following a combination of UV and IR luminosities to account for both unobscured and dust-embedded star formation according to Eq. 1 of \cite{Wuyts2011a}, which is based on \cite{Kennicutt1998b} and corrected for a Chabrier IMF. The total IR luminosity is extrapolated from Spitzer 24 $\mu$m assuming a single luminosity independent far-infrared (FIR) SED following \cite{Wuyts2008}. %
The typical systematic uncertainties on both the stellar mass and the SFR are conservatively estimated at about $0.2~\rm dex$ \citep[e.g.][]{Wuyts2011a,Whitaker2014, Roediger2015}.

The PHIBSS2 targets are chosen to have deep HST imagery, high-quality spectroscopic redshifts, as well as good rest-frame UV and Herschel PACS and/or 24 $\mu$m observations for the SFR estimate. 
We aimed at a homogeneous coverage of the MS and its scatter in the $\rm M_{\star}$-SFR plane, with $\rm \log(M_{\star}/M_\odot) \geq 10.1$ and $\rm SFR \geq 3.5~M_\odot yr^{-1}$ to assure a high probability of detection for reasonable on-source integration times. As shown in Fig.~\ref{fig:sample}, this coverage excludes the lower-mass end of the MS but fully covers the MS above the cuts in stellar mass and SFR.  
We applied no morphological selection and the relatively high masses surveyed ensure that the selected galaxies have metallicities close to  solar metallicity, which minimises the metallicity-induced variations of the CO-to-molecular-gas-mass conversion factor. 
We further selected galaxies with $\rm H\alpha$ and, when possible, H$\beta$ and [OIII] emission free from atmospheric line contamination for ionized gas kinematics and metallicity determinations. 
%
Considering Poisson errors and our past experience with PHIBSS, we required at least ten measurements in any given part of the parameter space, for example below $\rm \log(M_{\star}/M_\odot) = 10.4$ at $z=0.5-0.8$ or above the Schechter mass in each redshift bin, to establish well-determined average gas fraction and depletion time. 
As shown by \cite{Tacconi2018} in their Appendix, establishing the redshift, MS-offset, and stellar mass dependencies of the gas fraction and depletion time requires at least 40 sources in more than two redshift slices and a coverage of over 1 dex in both stellar mass and MS offset. 
Such constraints on the sample size further allow us to test the impact of AGNs and environment, as we can split the sample into mass-matched sub-samples with an expected AGN fraction of 25–50\% \citep{Mullaney2012,Juneau2013} while 20 to 30 \% of the targets display interacting satellites.

\subsection{Implementation with the NOEMA interferometer}

The CO observations were carried out between June 2014 and June 2017 with the IRAM Plateau de Bure millimeter interferometer and its NOEMA upgrade \citep{Guilloteau1992, Cox2011, Schuster2014}. The interferometer comprised six 15-m antennas at the beginning of the project and was upgraded to seven and eight antennas in September 2014 and April 2016, enabling us to reach higher sensitivities. Table~\ref{table:2} summarises the observations for the $z=0.5-0.8$ sample, including the interferometer's configurations, the total integration time $t_{\rm int}$ over all configurations included for a given galaxy, and the beam size. 
%
At this redshift, we observed the $^{12}$CO(2-1) rotational transition (rest-frame frequency 230.538 GHz), shifted into the 2-mm band, with the interferometer in compact `C' and `D' configurations. Given the integration times, these configurations yield beam sizes in the range $1^{\prime\prime}-5^{\prime\prime}$. 
Galaxies L14EG008, XA54, and XG55 were also observed in more extended configurations for higher-resolution follow-ups, which will be presented in a future paper. 
%
The integration time per target was initially determined from the expected CO flux estimated from the SFR assuming a linear KS relation, requiring a signal-to-noise ratio (S/N) of at least $4$, and later adapted in real time, galaxy per galaxy, during the observation campaign to ensure secure detections. One of the main goals of PHIBSS and PHIBSS2 is indeed to provide molecular gas estimates for a sample of star-forming galaxies covering different stellar masses and MS offsets in a way that is as unbiased as possible. 

Given the large number of observed hours, the weather conditions varied from excellent to very bad, with system temperatures ranging between 100 and 500 K depending on atmospheric conditions and season. We alternated source observations with bright quasar calibrators every 20 minutes to measure and remove the instrumental and atmospheric phase and amplitude fluctuations with time. The instrument response per frequency was further measured once per observational track on a strong quasar without spectral lines. The absolute flux scale was derived from secondary flux calibrators (MWC349 and LkH$\alpha$101), whose fluxes are regularly measured using Jupiter satellites or planets. We mostly used receivers of temperature $T_{\rm rec} \sim 35-70~\rm K$ in band 2, but also the latest NOEMA receivers with better $T_{\rm rec} \sim 30-50~\rm K$ in the last 1.5 years. The observations were carried out using dual polarization in the Single Side Band mode and we used the Widex backend correlator with 3.6-GHz coverage per polarization. 
The source integration times lie around 7 hours to achieve similar S/N under different weather conditions, except for the high-resolution follow-ups to be presented later, for which the integration times can reach about 30 hours.
The data were calibrated using the \texttt{CLIC} package of the IRAM \texttt{GILDAS} software and further analysed and mapped in its \texttt{MAPPING} environment. The spectra were analysed with the \texttt{CLASS} package within \texttt{GILDAS}.

As for the PHIBBS survey and the data compilation used in \cite{Tacconi2018}, the data from PHIBSS2 are to be made publicly available at the end of the reduction and interpretation procedure\footnote{\url{http://www.iram.fr/~phibss2/Data_Release.html}}.


\section{Results}
\label{section:results}

\begin{figure*}
        \centering
        \includegraphics[width=1\hsize]{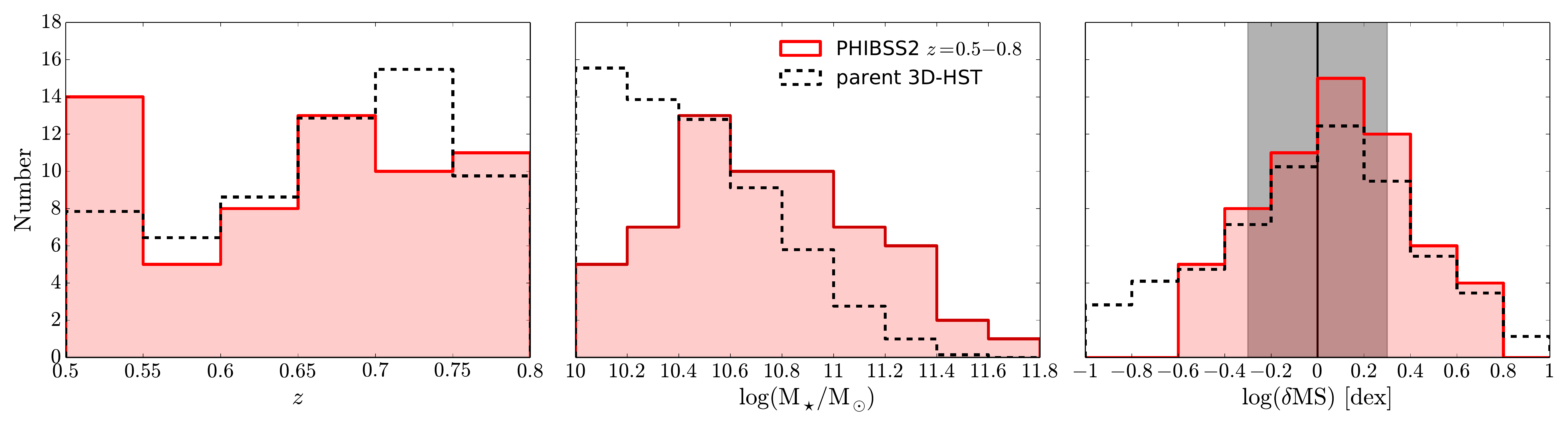}
        \caption{Distribution of redshift (left panel), stellar mass (middle panel), and offset from the MS (right panel) for the PHIBSS2 $z=0.5-0.8$ sample presented here, displayed in red. The offset from the MS is defined as $\delta {\rm MS} = {\rm sSFR/sSFR_{MS}}(z,{\rm M}_{\star})$, where ${\rm sSFR_{MS}}(z,{\rm M}_{\star})$ is the analytical prescription for the centre of the MS proposed in the compilation by \cite{Speagle2014}. In the right panel, the solid black line indicates $\delta {\rm MS} = 1$ and the grey shaded area shows the $\sim$$0.3~\rm dex$ scatter of the MS. While the PHIBSS2 $z=0.5-0.8$ sample is highlighted in red, the dashed line corresponds to the parent 3D-HST distribution at $z=0.5-0.8$ displayed in Fig.~\ref{fig:sample}, with $10 < \log({\rm M}_{\star}/{\rm M}_\odot) < 11.8$ and $|\log(\delta {\rm MS})| < 1$ (i.e. within $1~\rm dex$ of the MS line), normalized to match the same number of galaxies.  The dip at low redshift is probably due to either sky frequencies that made redshift determination more difficult or to cosmic variance. The mean and median stellar mass for the $z=0.5-0.8$ subsample is $\log({\rm M}_{\star}/{\rm M}_\odot) = 10.7$. 
        }
        \label{fig:histograms}
\end{figure*}

\begin{table*}
        \caption{The PHIBSS2 sample at $z=0.5-0.8$}
        \label{table:1}
        \centering
        \tiny

\begin{tabularx}{\textwidth}{l l l l c c Y Y Y Y Y}
        \hline\hline
        \noalign{\vskip 1mm} \# & ID & Field & Source$^a$  & R.A.$_{\rm optical}$ & DEC.$_{\rm optical}$ & $z_{\rm optical}$ & Morphology$^b$ & ${\rm M_{\star}}^c$  & SFR$^d$ & sSFR$^e$ \\
        & &  &       &              &              &             &            & ($\rm M_\odot$) & ($\rm M_\odot.yr^{-1}$) & ($\rm Gyr^{-1}$) \\
        \hline
        \noalign{\vskip 1mm} 1 & XA53 & COSMOS & 822872    & 10:02:02.09  & +02:09:37.40        & 0.7000           & DC           & 2.9E+11      & 47.3      & 0.16         \\
        2 & XC53 & COSMOS & 805007$^{\star}$  & 10:00:58.20      & +01:45:59.00        & 0.6227          & B           & 8.4E+10      & 47.1      & 0.56         \\
        3 & XD53 & COSMOS & 822965  & 10:01:58.73      & +02:15:34.20        & 0.7028          & DSb           & 8.9E+10      & 39.5      & 0.44         \\
        4 & XE53 & COSMOS & 811360  & 10:01:00.74      & +01:49:53.00        & 0.5297          & DSC           & 2.3E+10      & 25.5      & 1.13         \\
        5 & XF53 & COSMOS & 834187  & 09:58:33.86      & +02:19:50.90        & 0.5020          & DSbA           & 1.2E+11      & 18.6      & 0.16         \\
        6 & XG53 & COSMOS & 800405  & 10:02:16.78      & +01:37:25.00        & 0.6223          & DSAC           & 1.6E+11      & 21.0      & 0.13         \\
        7 & XH53 & COSMOS & 837919  & 10:01:09.67      & +02:30:00.70        & 0.7028          & DSAC           & 5.4E+10      & 18.2      & 0.34         \\
        8 & XI53 & COSMOS & 838956  & 10:00:24.70      & +02:29:12.10        & 0.7026          & B           & 2.9E+11      & 20.3      & 0.07         \\
        9 & XL53 & COSMOS & 824759  & 10:00:28.27      & +02:16:00.50        & 0.7506          & BR           & 1.7E+11      & 28.6      & 0.17         \\
        10 & XM53 & COSMOS & 810344  & 10:01:53.57      & +01:54:14.80        & 0.7007          & D           & 4.4E+11      & 23.9      & 0.05         \\
        11 & XN53 & COSMOS & 839268  & 10:00:11.16      & +02:35:41.60        & 0.6967          & DSb           & 1.1E+11      & 24.2      & 0.22         \\
        12 & XO53 & COSMOS & 828590  & 10:02:51.41      & +02:18:49.70        & 0.6077          & D           & 2.5E+11      & 11.7      & 0.05         \\
        13 & XQ53 & COSMOS & 838696  & 10:00:35.69      & +02:31:15.60        & 0.6793          & DC           & 8.3E+10      & 26.9      & 0.32         \\
        14 & XR53 & COSMOS & 816955  & 10:01:41.85      & +02:07:09.80        & 0.5165          & DSb           & 1.9E+11      & 14.5      & 0.08         \\
        15 & XT53 & COSMOS & 823380  & 10:01:39.31      & +02:17:25.80        & 0.7021          & DSA           & 1.1E+11      & 22.7      & 0.20         \\
        16 & XU53 & COSMOS & 831385  & 10:00:40.37      & +02:23:23.60        & 0.5172          & DSbA           & 1.9E+10      & 28.0      & 1.51         \\
        17 & XV53 & COSMOS & 850140  & 10:01:43.66      & +02:48:09.40        & 0.6248          & DA           & 6.3E+10      & 23.1      & 0.36         \\
        18 & XW53 & COSMOS & 824627$^{\star}$  & 10:00:35.52      & +02:16:34.30        & 0.7503          & DS           & 2.5E+10      & 13.7      & 0.54         \\
        19 & L14CO001 & COSMOS & 831870  & 10:00:18.91      & +02:18:10.10        & 0.5024          & DSA           & 1.5E+10      & 29.0      & 1.87         \\
        20 & L14CO004 & COSMOS & 831386  & 10:00:40.29      & +02:20:32.60        & 0.6885          & DC           & 2.8E+10      & 8.8      & 0.31         \\
        21 & L14CO007 & COSMOS & 838945  & 10:00:25.18      & +02:29:53.90        & 0.5015          & DC           & 5.1E+10      & 4.1      & 0.08         \\
        22 & L14CO008 & COSMOS & 820898  & 09:58:09.07      & +02:05:29.76        & 0.6081          & DSb           & 8.8E+10      & 13.9      & 0.16         \\
        23 & L14CO009 & COSMOS & 826687  & 09:58:56.45      & +02:08:06.72        & 0.6976          & DS           & 2.8E+10      & 21.4      & 0.76         \\
        24 & L14CO011 & COSMOS & 839183  & 10:00:14.30      & +02:30:47.16        & 0.6985          & DSbAC           & 2.6E+10      & 29.3      & 1.15         \\
        25 & L14CO012 & COSMOS & 838449  & 10:00:45.53      & +02:33:39.60        & 0.7007          & B           & 3.9E+10      & 10.0      & 0.25         \\
        26 & XA54 & AEGIS & 30084 (10098)  & 14:19:17.33      & +52:50:35.30        & 0.6590          & DS           & 1.3E+11      & 51.7      & 0.40         \\
        27 & XB54 & AEGIS & 17329 (5038)  & 14:19:37.26      & +52:51:03.40        & 0.6702          & DS           & 1.7E+11      & 29.1      & 0.17         \\
        28 & XC54 & AEGIS & 14885 (4097)  & 14:19:49.14      & +52:52:35.80        & 0.5093          & DSC           & 1.6E+11      & 37.9      & 0.24         \\
        29 & XD54 & AEGIS & 24556 (8538)  & 14:19:46.35      & +52:54:37.20        & 0.7541          & DSA           & 2.3E+10      & 28.9      & 1.26         \\
        30 & XE54 & AEGIS & 25608 (8310)  & 14:19:35.27      & +52:52:49.90        & 0.5090          & DS           & 2.5E+10      & 11.0      & 0.45         \\
        31 & XF54 & AEGIS & 32878 (11378)  & 14:19:41.70      & +52:55:41.30        & 0.7683          & DS           & 5.1E+10      & 19.9      & 0.39         \\
        32 & XG54 & AEGIS & 3654 (169)  & 14:20:13.43      & +52:54:05.90        & 0.6593          & DSbAC           & 1.4E+11      & 14.4      & 0.10         \\
        33 & XH54 & AEGIS & 30516 (10745)  & 14:19:45.42      & +52:55:51.00        & 0.7560          & DSA           & 1.9E+10      & 13.1      & 0.70         \\
        34 & L14EG006 & AEGIS & 23488 (7652)  & 14:18:45.52      & +52:43:24.10        & 0.5010          & DSbC           & 3.0E+10      & 7.4      & 0.25         \\
        35 & L14EG008 & AEGIS & 21351 (7021)  & 14:19:39.46      & +52:52:33.60        & 0.7315          & DS           & 8.7E+10      & 79.5      & 0.91         \\
        36 & L14EG009 & AEGIS & 31909 (11332)  & 14:20:04.88      & +52:59:38.84        & 0.7359          & DA           & 1.1E+10      & 9.9      & 0.89         \\
        37 & L14EG010 & AEGIS & 4004 (725)  & 14:20:22.80      & +52:55:56.28        & 0.6702          & B           & 5.5E+10      & 9.3      & 0.17         \\
        38 & L14EG011 & AEGIS & 6274  & 14:20:26.20      & +52:57:04.85        & 0.5705          & DSb           & 5.4E+10      & 25.7      & 0.48         \\
        39 & L14EG012 & AEGIS & 6449 (515)  & 14:19:52.95      & +52:51:11.06        & 0.5447          & DSb           & 1.1E+11      & 9.1      & 0.08         \\
        40 & L14EG014 & AEGIS & 9743  & 14:20:33.58      & +52:59:17.46        & 0.7099          & BC           & 8.5E+10      & 5.9      & 0.07         \\
        41 & L14EG015 & AEGIS & 26964  & 14:20:45.61      & +53:05:31.18        & 0.7369          & B           & 9.3E+10      & 13.4      & 0.14         \\
        42 & L14EG016 & AEGIS & 34302  & 14:18:28.90      & +52:43:05.28        & 0.6445          & DSb           & 4.0E+10      & 6.6      & 0.17         \\
        43 & XA55 & GOODS-N & 21285 (9335)$^{\star}$  & 12:36:59.92      & +62:14:50.00        & 0.7610          & DSA           & 2.8E+10      & 44.7      & 1.58         \\
        44 & XB55 & GOODS-N & 6666 (3091)$^{\dagger}$  & 12:36:08.13      & +62:10:35.90        & 0.6790          & B           & 4.5E+10      & 23.1      & 0.52         \\
        45 & XC55 & GOODS-N & 19725 (8738)  & 12:36:09.76      & +62:14:22.60        & 0.7800          & DS           & 4.6E+10      & 29.1      & 0.64         \\
        46 & XD55 & GOODS-N & 12097 (5385)  & 12:36:21.04      & +62:12:08.50        & 0.7790          & DRA           & 3.1E+10      & 21.6      & 0.70         \\
        47 & XE55 & GOODS-N & 19815 (8798)$^{\star}$  & 12:36:11.26      & +62:14:20.90        & 0.7720          & D           & 3.3E+10      & 14.8      & 0.45         \\
        48 & XF55 & GOODS-N & 7906 (3565)  & 12:35:55.43      & +62:10:56.80        & 0.6382          & DC           & 1.1E+10      & 11.1      & 0.96         \\
        49 & XG55 & GOODS-N & 19257 (8697)  & 12:37:02.93      & +62:14:23.60        & 0.5110          & DSA           & 3.8E+10      & 8.5      & 0.22         \\
        50 & XH55 & GOODS-N & 16987 (7668)  & 12:37:13.87      & +62:13:35.00        & 0.7784          & DS           & 1.6E+10      & 13.0      & 0.80         \\
        51 & XL55 & GOODS-N & 10134 (4568)  & 12:37:10.56      & +62:11:40.70        & 0.7880          & DSC           & 3.2E+10      & 22.2      & 0.69         \\
        52 & L14GN006 & GOODS-N & 30883 (12248)  & 12:36:34.41      & +62:17:50.50        & 0.6825          & DS           & 2.5E+10      & 23.8      & 0.95         \\
        53 & L14GN007 & GOODS-N & 939 (334)  & 12:36:32.38      & +62:07:34.10        & 0.5950          & DS           & 7.4E+10      & 8.9      & 0.12         \\
        54 & L14GN008 & GOODS-N & 11532 (5128)  & 12:36:07.83      & +62:12:00.60        & 0.5035          & DSb           & 1.9E+10      & 5.5      & 0.28         \\
        55 & L14GN018 & GOODS-N & 25413 (10807)  & 12:36:31.66      & +62:16:04.10        & 0.7837          & DA           & 2.5E+10      & 32.8      & 1.31         \\
        56 & L14GN021 & GOODS-N & 8738 (3875)  & 12:36:03.26      & +62:11:10.98        & 0.6380          & B           & 5.1E+10      & 76.9      & 1.50         \\
        57 & L14GN022 & GOODS-N & 11460 (5127)$^{\star}$  & 12:36:36.76      & +62:11:56.09        & 0.5561          & B           & 1.3E+10      & 6.8      & 0.54         \\
        58 & L14GN025 & GOODS-N & 36596 (14032)  & 12:37:13.99      & +62:20:36.60        & 0.5320          & BC           & 4.5E+10      & 3.5      & 0.08         \\
        59 & L14GN032 & GOODS-N & 21683 (9558)  & 12:37:16.32      & +62:15:12.30        & 0.5605          & BC           & 1.3E+11      & 7.6      & 0.06         \\
        60 & L14GN033 & GOODS-N & 1964 (918)  & 12:36:53.81      & +62:08:27.70        & 0.5609          & DSb           & 1.1E+10      & 6.7      & 0.59         \\
        61 & L14GN034 & GOODS-N & 33895  & 12:36:19.68      & +62:19:08.10        & 0.5200          & DCA           & 7.4E+10      & 8.7      & 0.12         \\
        \hline
\end{tabularx}
        
        \tablefoot{\\
                $^a$ The source numbers correspond to the $z$COSMOS nomenclature in the COSMOS field and to the 3D-HST v4.0 nomenclature in the other fields, with the 3D-HST v2.1 nomenclature indicated inside parentheses when applicable.\\
                $^b$ Morphology derived by eye from the HST I-band images presented in Appendix~\ref{appendix:HST}, with the following non-exclusive denominations: 
                D for disc-dominated; 
                B for bulge-dominated;  
                S for spiral; 
                Sb for barred spiral; 
                R for ring; 
                A for asymmetric or perturbed; 
                C for the presence of companions.\\
                $^c$ Stellar masses from SED fitting, assuming a Chabrier IMF, with assumed systematic uncertainties of $0.2$ dex. \\
                $^d$ Extinction-corrected $\rm SFR_{UV+IR}$ from UV continuum measurements and IR 24 $\mu$m  luminosities extrapolated with Herschel PACS calibrations to total IR luminosities, with assumed systematic uncertainties of $0.2$ dex. \\ 
                $^e$ $\rm sSFR=SFR/M_{\star}$.\\
                $^\star$ Marginal detections.\\ 
                $^\dagger$ Non-detection.
        }
\end{table*}

\subsection{The PHIBSS2 sample at $z=0.5-0.8$}

The PHIBSS2 $z=0.5-0.8$ sample of 61 near-MS star-forming galaxies which is the focus of this article is presented in Table~\ref{table:1} and Figs.~\ref{fig:sample}~and~\ref{fig:histograms}. Appendix \ref{appendix:HST} further presents HST I-band images of these galaxies, while Sect.~\ref{section:galfit} presents their sizes and bulge-to-total (B/T) luminosity ratios.
%
As can be seen in the HST I-band images and indicated in Table~\ref{table:1}, the  sample comprises 49 disc-dominated and 12 bulge-dominated galaxies (80\% and 20\% of the sample, respectively), 36 of them having clear spiral or ring features (59\%), which are highlighted in the residual maps. This repartition agrees very well with that found by \cite{Tacconi2013} for the PHIBSS sample and with larger HST imaging surveys \citep{Wuyts2011b}. A visual inspection also shows that 14 galaxies out of 61 (23\%) harbour bars despite the intermediate redshift (Table~\ref{table:1}). 
The relatively regular morphologies observed for most galaxies of the sample are compatible with them being isolated and not undergoing major mergers. 
%
Indeed, a visual inspection of the HST images indicates that only 4 galaxies out of 61 (7\%) have both asymmetries and companions, which is comparable to the fraction of mergers from other MS studies \citep[e.g.][]{Tacconi2013, Wisnioski2015}. 
Beyond the sample at $z=0.5-0.8$ presented here, higher-redshift PHIBSS2 samples at $z=1-1.6$ and $z=2-3$ will be presented in future articles.

\begin{table*}
        \caption{CO observations}
        \label{table:2}
        \centering
        \tiny
 
	\begin{tabularx}{\textwidth}{l l l l c c c Y Y Y Y Y Y Y}
        \hline\hline
        \noalign{\vskip 1mm} \# & ID &  Field & Source & Config.$^a$ & $t_{\rm int}$$^a$ & CO beam$^a$ & ${\Delta z}^b$ & $\Delta$R.A.$^b$ & $\Delta$DEC.$^b$ & ${s_{\rm peak}}^c$ & ${\sigma_{\rm 30}}^d$ & FWHM$^e$              & dFWHM$^e$             \\
        & &  &    &     & (hr)          &         &           & (arsec)      & (arcsec)     & (mJy)          & (mJy)             & ($\rm km.s^{-1}$) & ($\rm km.s^{-1}$) \\
        \hline
        \noalign{\vskip 1mm} 1 & XA53 & COSMOS & 822872 & D & 2.2 & $4.9^{\prime\prime} \times 3.9^{\prime\prime} $ & $-0.0018$ & 0.45 & 0.99 & 3.9 & 1.7 & 296 & 54\\
        2 & XC53& COSMOS & 805007$^{\star}$ & D & 11.8 & $4.4^{\prime\prime} \times 3.4^{\prime\prime} $ & $-0.0064$ & -0.90 & -2.05 & 0.8 & 1.5 & 670 & 380\\
        3 & XD53& COSMOS & 822965 & D & 4.3 & $4.2^{\prime\prime} \times 2.3^{\prime\prime} $ & $-0.0008$ & -0.45 & 0.22 & 3.5 & 1.2 & 414 & 58\\
        4 & XE53& COSMOS & 811360 & D & 1.7 & $5.4^{\prime\prime} \times 3.4^{\prime\prime} $ & $-0.0007$ & 0.15 & -0.75 & 5.4 & 2.1 & 233 & 45\\
        5 & XF53& COSMOS & 834187 & D & 0.6 & $4.7^{\prime\prime} \times 2.8^{\prime\prime} $ & $-0.0001$ & 0.45 & 0.73 & 3.2 & 1.9 & 524 & 97\\
        6 & XG53& COSMOS & 800405 & D & 5.3 & $3.0^{\prime\prime} \times 2.0^{\prime\prime} $ & $-0.0006$ & -0.45 & 1.23 & 1.9 & 0.8 & 474 & 84\\
        7 & XH53& COSMOS & 837919 & C & 9.4 & $2.7^{\prime\prime} \times 2.0^{\prime\prime} $ & $-0.0009$ & -0.15 & -0.14 & 2.1 & 0.9 & 112 & 30\\
        8 & XI53& COSMOS & 838956 & C & 10 & $2.6^{\prime\prime} \times 1.9^{\prime\prime} $ & $-0.0015$ & -2.55 & 0.19 & 3.4 & 1.2 & 110 & 29\\
        9 & XL53& COSMOS & 824759 & C & 3.3 & $3.0^{\prime\prime} \times 1.8^{\prime\prime} $ & $-0.0017$ & 0.45 & -0.61 & 1.8 & 1.0 & 318 & 77\\
        10 & XM53& COSMOS & 810344 & C & 3.9 & $2.8^{\prime\prime} \times 1.9^{\prime\prime} $ & $-0.0002$ & -0.30 & -0.15 & 1.4 & 0.7 & 713 & 107\\
        11 & XN53& COSMOS & 839268 & C & 5.4 & $4.6^{\prime\prime} \times 3.8^{\prime\prime} $ & $-0.0002$ & -0.30 & 0.17 & 2.6 & 1.2 & 210 & 53\\
        12 & XO53& COSMOS & 828590 & C & 6.3 & $3.4^{\prime\prime} \times 1.6^{\prime\prime} $ & $-0.0018$ & 0.15 & -0.68 & 1.2 & 0.5 & 161 & 51\\
        13 & XQ53& COSMOS & 838696 & D & 4.3 & $4.4^{\prime\prime} \times 3.8^{\prime\prime} $ & $-0.0012$ & 0.30 & 1.41 & 2.9 & 1.0 & 103 & 29\\
        14 & XR53& COSMOS & 816955 & D & 3.7 & $3.5^{\prime\prime} \times 2.3^{\prime\prime} $ & $-0.0003$ & -0.15 & -0.07 & 2.2 & 0.8 & 159 & 35\\
        15 & XT53& COSMOS & 823380 & D & 3.9 & $4.1^{\prime\prime} \times 2.6^{\prime\prime} $ & $-0.0009$ & 1.05 & -0.25 & 2.3 & 0.8 & 286 & 72\\
        16 & XU53& COSMOS & 831385 & D & 2.2 & $3.9^{\prime\prime} \times 2.0^{\prime\prime} $ & $-0.0008$ & -0.15 & -0.26 & 2.4 & 0.9 & 269 & 39\\
        17 & XV53& COSMOS & 850140 & C & 4.2 & $2.5^{\prime\prime} \times 1.8^{\prime\prime} $ & $-0.0012$ & -0.30 & -0.08 & 2.7 & 0.9 & 428 & 50\\
        18 & XW53& COSMOS & 824627$^{\star}$ & C & 9.3 & $3.1^{\prime\prime} \times 1.8^{\prime\prime} $ & $-0.0005$ & 0.30 & -1.63 & 3.3 & 1.2 & 110 & 26\\
        19 & L14CO001& COSMOS & 831870 & CD & 3.5 & $3.0^{\prime\prime} \times 2.2^{\prime\prime} $ & $-0.0003$ & -0.90 & 0.20 & 4.1 & 1.0 & 164 & 25\\
        20 & L14CO004& COSMOS & 831386 & CD & 14.2 & $2.8^{\prime\prime} \times 1.5^{\prime\prime} $ & $-0.0011$ & -0.30 & 0.13 & 0.6 & 0.4 & 475 & 215\\
        21 & L14CO007& COSMOS & 838945 & D & 8.3 & $5.2^{\prime\prime} \times 2.8^{\prime\prime} $ & $-0.0001$ & -0.45 & -0.80 & 1.1 & 0.6 & 402 & 94\\
        22 & L14CO008& COSMOS & 820898 & D & 7.5 & $4.2^{\prime\prime} \times 2.8^{\prime\prime} $ & $-0.0016$ & 0.15 & 0.13 & 1.3 & 0.7 & 501 & 125\\
        23 & L14CO009& COSMOS & 826687 & CD & 7 & $3.1^{\prime\prime} \times 1.4^{\prime\prime} $ & $-0.0001$ & 0.00 & -0.12 & 1.4 & 0.6 & 255 & 112\\
        24 & L14CO011& COSMOS & 839183 & CD & 5.6 & $4.4^{\prime\prime} \times 2.8^{\prime\prime} $ & $-0.0016$ & -0.15 & 0.72 & 2.5 & 0.6 & 273 & 35\\
        25 & L14CO012& COSMOS & 838449 & CD & 4.2 & $3.2^{\prime\prime} \times 1.5^{\prime\prime} $ & $-0.0011$ & 0.45 & 2.51 & 3.9 & 0.9 & 39 & 15\\
        26 & XA54& AEGIS & 30084 & ABD & 30.3 & $0.8^{\prime\prime} \times 0.7^{\prime\prime} $ & $-0.0005$ & 0.00 & -0.07 & 2.7 & 0.3 & 341 & 47\\
        27 & XB54& AEGIS & 17329 & D & 4.7 & $5.3^{\prime\prime} \times 2.9^{\prime\prime} $ & $-0.0006$ & 1.95 & -0.17 & 4.0 & 1.2 & 234 & 58\\
        28 & XC54& AEGIS & 14885 & D & 2.6 & $3.2^{\prime\prime} \times 2.6^{\prime\prime} $ & $+0.0000$ & -1.95 & 0.36 & 1.8 & 1.1 & 667 & 110\\
        29 & XD54& AEGIS & 24556 & D & 11.3 & $4.5^{\prime\prime} \times 3.0^{\prime\prime} $ & $-0.0008$ & 0.60 & 0.74 & 2.2 & 0.6 & 196 & 40\\
        30 & XE54& AEGIS & 25608 & D & 1.8 & $3.3^{\prime\prime} \times 2.7^{\prime\prime} $ & $-0.0004$ & 0.90 & -0.38 & 2.2 & 0.9 & 284 & 60\\
        31 & XF54& AEGIS & 32878 & D & 9 & $4.8^{\prime\prime} \times 4.3^{\prime\prime} $ & $+0.0000$ & 1.20 & -0.41 & 2.1 & 1.0 & 242 & 66\\
        32 & XG54& AEGIS & 3654 & D & 7 & $4.5^{\prime\prime} \times 3.7^{\prime\prime} $ & $-0.0003$ & 0.30 & -0.73 & 2.7 & 1.0 & 341 & 110\\
        33 & XH54& AEGIS & 30516 & C & 10.6 & $2.5^{\prime\prime} \times 2.3^{\prime\prime} $ & $-0.0003$ & -0.15 & 0.27 & 1.4 & 0.5 & 77 & 21\\
        34 & L14EG006& AEGIS & 23488 & D & 10.2 & $3.9^{\prime\prime} \times 2.8^{\prime\prime} $ & $-0.0005$ & -1.35 & -1.01 & 2.8 & 0.6 & 75 & 19\\
        35 & L14EG008& AEGIS & 21351 & AC & 27 & $2.6^{\prime\prime} \times 2.1^{\prime\prime} $ & $-0.0002$ & 0.00 & -0.20 & 4.2 & 0.8 & 257 & 29\\
        36 & L14EG009& AEGIS & 31909 & CD & 10 & $2.7^{\prime\prime} \times 1.8^{\prime\prime} $ & $-0.0005$ & -0.15 & -0.09 & 1.5 & 0.6 & 217 & 49\\
        37 & L14EG010& AEGIS & 4004 & CD & 9.4 & $2.3^{\prime\prime} \times 2.3^{\prime\prime} $ & $-0.0007$ & 4.20 & -0.18 & 4.8 & 0.6 & 31 & 8\\
        38 & L14EG011& AEGIS & 6274 & D & 6.8 & $3.2^{\prime\prime} \times 2.9^{\prime\prime} $ & $-0.0005$ & -0.45 & 0.13 & 1.9 & 0.7 & 387 & 48\\
        39 & L14EG012& AEGIS & 6449 & D & 4.8 & $3.2^{\prime\prime} \times 2.3^{\prime\prime} $ & $-0.0009$ & 1.20 & -0.53 & 2.6 & 0.8 & 103 & 43\\
        40 & L14EG014& AEGIS & 9743 & CD & 14.1 & $1.7^{\prime\prime} \times 1.7^{\prime\prime} $ & $-0.0002$ & -0.18 & 0.18 & 1.1 & 0.5 & 79 & 24\\
        41 & L14EG015& AEGIS & 26964 & CD & 16.9 & $1.9^{\prime\prime} \times 1.6^{\prime\prime} $ & $-0.0013$ & 2.86 & 1.40 & 1.1 & 0.5 & 261 & 72\\
        42 & L14EG016& AEGIS & 34302 & CD & 12.9 & $1.6^{\prime\prime} \times 1.3^{\prime\prime} $ & $-0.0011$ & -1.73 & -2.72 & 0.8 & 0.5 & 537 & 127\\
        43 & XA55& GOODS-N & 21285$^{\star}$ & D & 3.9 & $4.9^{\prime\prime} \times 3.1^{\prime\prime} $ & $-0.0001$ & -2.10 & -1.55 & 2.3 & 1.3 & 92 & 200\\
        44 & XB55& GOODS-N & 6666$^{\dagger}$ & D & 9.3 & $3.7^{\prime\prime} \times 2.8^{\prime\prime} $ &  &  &  &  & 0.9 &  & \\
        45 & XC55& GOODS-N & 19725 & D & 5.8 & $5.5^{\prime\prime} \times 3.4^{\prime\prime} $ & $-0.0002$ & -0.90 & -0.58 & 2.2 & 1.0 & 339 & 77\\
        46 & XD55& GOODS-N & 12097 & D & 9.5 & $4.5^{\prime\prime} \times 4.3^{\prime\prime} $ & $-0.0004$ & 1.05 & 0.63 & 1.8 & 1.0 & 229 & 79\\
        47 & XE55& GOODS-N & 19815$^{\star}$ & D & 12.7 & $4.6^{\prime\prime} \times 4.0^{\prime\prime} $ & $-0.0040$ & 2.25 & -0.93 & 1.2 & 0.8 & 200 & 61\\
        48 & XF55& GOODS-N & 7906 & D & 10.9 & $4.2^{\prime\prime} \times 3.5^{\prime\prime} $ & $-0.0008$ & -2.25 & -0.12 & 0.5 & 0.5 & 306 & 125\\
        49 & XG55& GOODS-N & 19257 & ABCD & 8.9 & $4.0^{\prime\prime} \times 3.7^{\prime\prime} $ & $-0.0003$ & 1.95 & -0.23 & 1.5 & 0.7 & 285 & 62\\
        50 & XH55& GOODS-N & 16987 & C & 9.4 & $2.5^{\prime\prime} \times 2.0^{\prime\prime} $ & $-0.0001$ & -0.30 & 0.18 & 1.1 & 0.5 & 288 & 51\\
        51 & XL55& GOODS-N & 10134 & D & 14.3 & $4.1^{\prime\prime} \times 3.5^{\prime\prime} $ & $+0.0000$ & 2.40 & -0.20 & 0.9 & 0.8 & 375 & 106\\
        52 & L14GN006& GOODS-N & 30883 & D & 2.6 & $3.4^{\prime\prime} \times 3.1^{\prime\prime} $ & $-0.0005$ & -0.15 & 0.25 & 2.3 & 1.0 & 383 & 85\\
        53 & L14GN007& GOODS-N & 939 & D & 4.2 & $2.8^{\prime\prime} \times 2.3^{\prime\prime} $ & $-0.0001$ & 0.75 & 0.79 & 1.6 & 0.6 & 371 & 76\\
        54 & L14GN008& GOODS-N & 11532 & D & 14.3 & $3.2^{\prime\prime} \times 2.8^{\prime\prime} $ & $-0.0004$ & 3.45 & 1.37 & 0.7 & 0.5 & 399 & 161\\
        55 & L14GN018& GOODS-N & 25413 & CD & 10.3 & $2.6^{\prime\prime} \times 2.1^{\prime\prime} $ & $-0.0005$ & -0.30 & 0.04 & 1.9 & 0.6 & 168 & 41\\
        56 & L14GN021& GOODS-N & 8738 & CD & 4.6 & $2.0^{\prime\prime} \times 1.4^{\prime\prime} $ & $-0.0002$ & -0.60 & 0.00 & 3.1 & 0.8 & 338 & 34\\
        57 & L14GN022& GOODS-N & 11460$^{\star}$ & CD & 16.9 & $1.5^{\prime\prime} \times 1.4^{\prime\prime} $ & $+0.0000$ & 0.30 & -0.63 & 1.1 & 0.4 & 97 & 28\\
        58 & L14GN025& GOODS-N & 36596 & CD & 10.5 & $1.9^{\prime\prime} \times 1.2^{\prime\prime} $ & $-0.0008$ & -0.75 & 0.21 & 1.0 & 0.5 & 264 & 95\\
        59 & L14GN032& GOODS-N & 21683 & CD & 22.5 & $3.4^{\prime\prime} \times 2.3^{\prime\prime} $ & $-0.0004$ & -7.96 & -1.98 & 2.3 & 1.1 & 60 & 61\\
        60 & L14GN033& GOODS-N & 1964 & D & 15.9 & $2.5^{\prime\prime} \times 2.1^{\prime\prime} $ & $-0.0003$ & 3.78 & 2.08 & 1.0 & 0.4 & 100 & 60\\
        61 & L14GN034& GOODS-N & 33895 & D & 7.2 & $2.8^{\prime\prime} \times 2.1^{\prime\prime} $ & $-0.0004$ & -0.32 & 0.05 & 1.6 & 0.8 & 504 & 90\\
        \hline
\end{tabularx}

        \tablefoot{\\
        		$^a$ Configuration of the interferometer, total on-source observation time, and the resulting CO beam size.\\
                $^b$ Redshift and position offsets of the detected CO emission: $\Delta z = z_{\rm CO} - z_{\rm optical}$, with $z_{\rm CO}$ from a Gaussian fit to the CO line, $\rm {\Delta}R.A. = R.A._{CO} - R.A._{optical}$, and $\rm {\Delta}DEC.= DEC._{CO} - DEC._{optical}$. With typical redshift errors $\sigma_z = 0.003 \times (1+z)$ 
                \citep{Momcheva2016}, i.e. on average 0.005 in our sample at $z=0.5-0.8$, all our detections except for XC53 and XE55 are within $\pm 0.4 \sigma_z$.\\
                $^c$ Peak CO luminosity from a single Gaussian fit to the spatially averaged spectrum.\\
                $^d$ Experimental RMS noise  per $30~\rm km.s^{-1}$ wide channel.\\
                $^e$ Full width at half maximum (FWHM) from the single Gaussian fit and its uncertainty dFWHM.
                \\
                $^\star$ Marginal detection.\\
                $^\dagger$ Non-detection.
        }
\end{table*}

\subsection{CO fluxes}

Tables~\ref{table:2} and \ref{table:3} summarise the PHIBSS2 CO(2-1) line observations at $z=0.5-0.8$. 
In Table~\ref{table:3}, we display the CO(2-1) line fluxes $\rm F(CO)$, which correspond to the mean value of three different estimates of the line intensity weighted by their respective uncertainties: (i) directly obtained from the spatially integrated spectrum over the velocity window centred on the peak emission that maximizes the flux; (ii) derived from a Gaussian fit to the spatially integrated spectrum, whose peak luminosity and FWHM are indicated in Table~\ref{table:2}; and (iii) determined from the velocity-integrated line map  (using the \texttt{GO FLUX} tool of \texttt{GILDAS}). 
The spatially averaged spectra and the Gaussian fits are displayed in Appendix~\ref{appendix:HST} along with the HST I-band images. Amongst them, eight galaxies clearly display double-horned profiles which are characteristic of thin rotating discs. 
The flux uncertainty $\rm dF(CO)$ is estimated from the RMS noise integrated over the full line width, leading to an integrated flux S/N of $\rm S/N = F(CO)/dF(CO)$. 
The dispersion between the three estimates of the line intensity is in most cases well below the quoted uncertainty $\rm dF(CO)$. In the few cases where there were significant discrepancies between the three estimates, we selected the most reliable one.
As shown in Table~\ref{table:3}, we detected the CO(2-1) line emission in 60 of the 61 galaxies of the $z=0.5-0.8$ sample, which corresponds to a detection rate of 97\%. Amongst these detections, 5 have low S/N and display a slight offset from the HST I-band image or an ambiguous spatially integrated spectrum: they are indicated as marginal in the table. 

For the non-detected galaxy XB55 (\texttt{GN-6666}), we evaluate the RMS noise on the integrated line flux at $\sigma_{\rm F} =0.12 \rm ~mJy.km.s^{-1}$ using 
\begin{equation}
\sigma_{\rm F} =  \sigma_{30} \sqrt{\Delta V \times 30\rm ~km.s^{-1}},
\end{equation}
where $\sigma_{30}$ is the noise per $\rm 30~km.s^{-1}$ -wide channel and $\Delta V$ the velocity width over which the signal is integrated when searching for the line, and quote $\rm F_{upper} =3\sigma_F$ as the upper limit, which corresponds to the flux that should have been detected at $\rm S/N > 3$ with a 50\% probability \citep[e.g.][]{Masci2011}. 

\begin{table*}
        \caption{Integrated CO line flux and derived quantities}
        \label{table:3}
        \centering
        \tiny

\begin{tabularx}{\textwidth}{l l l l Y Y Y c c Y Y Y}
        \hline\hline
        \noalign{\vskip 1mm} \# & ID & Field & Source & F(CO)$^a$                & dF(CO)$^a$           & S/N$^b$ & ${\rm L_{CO(2-1)}}^c$           & ${\rm M_{\rm gas}}^d$   & $\mu_{\rm gas}^e$& ${f_{\rm gas}}^f$ & ${t_{\rm depl}}^g$\\
        & & & & ($\rm Jy.km.s^{-1}$) & ($\rm Jy.km.s^{-1}$) &     & ($\rm K.km.s^{-1}.pc^{2}$) & ($\rm M_\odot$) &             &  & (Gyr)     \\
        \hline
        \noalign{\vskip 1mm} 1 & XA53 & COSMOS & 822872 & 1.45 & 0.46 & 3.2 & 9.4E+09 & 4.6E+10 & 0.16 & 0.14 & 1.0 \\
        2 & XC53 & COSMOS & 805007$^{\star}$ & 0.20 & 0.08 & 2.5 & 1.0E+09 & 5.0E+09 & 0.06 & 0.06 & 0.1\\
        3 & XD53 & COSMOS & 822965 & 1.00 & 0.24 & 4.2 & 6.6E+09 & 3.3E+10 & 0.37 & 0.27 & 0.8\\
        4 & XE53 & COSMOS & 811360 & 1.29 & 0.45 & 2.9 & 4.7E+09 & 2.6E+10 & 1.13 & 0.53 & 1.0\\
        5 & XF53 & COSMOS & 834187 & 1.71 & 0.46 & 3.7 & 5.6E+09 & 2.8E+10 & 0.24 & 0.19 & 1.5\\
        6 & XG53 & COSMOS & 800405 & 0.98 & 0.25 & 4.0 & 5.0E+09 & 2.5E+10 & 0.15 & 0.13 & 1.2\\
        7 & XH53 & COSMOS & 837919 & 0.26 & 0.10 & 2.6 & 1.7E+09 & 8.8E+09 & 0.16 & 0.14 & 0.5\\
        8 & XI53 & COSMOS & 838956 & 0.37 & 0.10 & 3.5 & 2.4E+09 & 1.2E+10 & 0.04 & 0.04 & 0.6\\
        9 & XL53 & COSMOS & 824759 & 0.71 & 0.19 & 3.8 & 5.3E+09 & 2.6E+10 & 0.16 & 0.14 & 0.9\\
        10 & XM53 & COSMOS & 810344 & 0.83 & 0.20 & 4.1 & 5.4E+09 & 2.7E+10 & 0.06 & 0.06 & 1.1\\
        11 & XN53 & COSMOS & 839268 & 0.58 & 0.17 & 3.5 & 3.7E+09 & 1.9E+10 & 0.17 & 0.14 & 0.8\\
        12 & XO53 & COSMOS & 828590 & 0.67 & 0.13 & 5.2 & 3.3E+09 & 1.6E+10 & 0.06 & 0.06 & 1.4\\
        13 & XQ53 & COSMOS & 838696 & 0.51 & 0.19 & 2.7 & 3.1E+09 & 1.6E+10 & 0.19 & 0.16 & 0.6\\
        14 & XR53 & COSMOS & 816955 & 0.50 & 0.11 & 4.5 & 1.8E+09 & 8.6E+09 & 0.04 & 0.04 & 0.6\\
        15 & XT53 & COSMOS & 823380 & 0.67 & 0.18 & 3.7 & 4.4E+09 & 2.2E+10 & 0.19 & 0.16 & 1.0\\
        16 & XU53 & COSMOS & 831385 & 0.82 & 0.15 & 5.7 & 2.9E+09 & 1.6E+10 & 0.85 & 0.46 & 0.6\\
        17 & XV53 & COSMOS & 850140 & 1.26 & 0.28 & 4.5 & 6.5E+09 & 3.3E+10 & 0.52 & 0.34 & 1.4\\
        18 & XW53 & COSMOS & 824627$^{\star}$ & 0.17 & 0.08 & 2.2 & 1.3E+09 & 7.3E+09 & 0.29 & 0.22 & 0.5\\
        19 & L14CO001 & COSMOS & 831870 & 0.90 & 0.20 & 4.5 & 3.0E+09 & 1.7E+10 & 1.08 & 0.52 & 0.6\\
        20 & L14CO004 & COSMOS & 831386 & 0.23 & 0.07 & 3.4 & 1.4E+09 & 8.0E+09 & 0.28 & 0.22 & 0.9\\
        21 & L14CO007 & COSMOS & 838945 & 0.42 & 0.12 & 3.4 & 1.4E+09 & 6.9E+09 & 0.14 & 0.12 & 1.7\\
        22 & L14CO008 & COSMOS & 820898 & 0.84 & 0.17 & 4.9 & 4.1E+09 & 2.0E+10 & 0.23 & 0.19 & 1.5\\
        23 & L14CO009 & COSMOS & 826687 & 0.41 & 0.10 & 3.9 & 2.7E+09 & 1.5E+10 & 0.52 & 0.34 & 0.7\\
        24 & L14CO011 & COSMOS & 839183 & 0.70 & 0.10 & 7.4 & 4.5E+09 & 2.5E+10 & 0.99 & 0.50 & 0.9\\
        25 & L14CO012 & COSMOS & 838449 & 0.40 & 0.11 & 3.7 & 2.6E+09 & 1.4E+10 & 0.36 & 0.26 & 1.4\\
        26 & XA54 & AEGIS & 30084 & 1.11 & 0.10 & 11.1 & 6.4E+09 & 3.1E+10 & 0.24 & 0.20 & 0.6\\
        27 & XB54 & AEGIS & 17329 & 1.00 & 0.25 & 4.1 & 6.0E+09 & 2.9E+10 & 0.17 & 0.14 & 1.0\\
        28 & XC54 & AEGIS & 14885 & 1.17 & 0.27 & 4.4 & 4.0E+09 & 1.9E+10 & 0.12 & 0.11 & 0.5\\
        29 & XD54 & AEGIS & 24556 & 0.50 & 0.08 & 6.1 & 3.8E+09 & 2.2E+10 & 0.96 & 0.49 & 0.8\\
        30 & XE54 & AEGIS & 25608 & 0.66 & 0.25 & 2.7 & 2.2E+09 & 1.2E+10 & 0.49 & 0.33 & 1.1\\
        31 & XF54 & AEGIS & 32878 & 0.40 & 0.11 & 3.5 & 3.2E+09 & 1.7E+10 & 0.32 & 0.24 & 0.8\\
        32 & XG54 & AEGIS & 3654 & 0.87 & 0.17 & 5.0 & 5.0E+09 & 2.5E+10 & 0.18 & 0.15 & 1.7\\
        33 & XH54 & AEGIS & 30516 & 0.11 & 0.03 & 3.4 & 8.4E+08 & 5.0E+09 & 0.27 & 0.21 & 0.4\\
        34 & L14EG006 & AEGIS & 23488 & 0.36 & 0.08 & 4.3 & 1.2E+09 & 6.2E+09 & 0.21 & 0.17 & 0.8\\
        35 & L14EG008 & AEGIS & 21351 & 1.16 & 0.10 & 12.0 & 8.3E+09 & 4.2E+10 & 0.48 & 0.32 & 0.5\\
        36 & L14EG009 & AEGIS & 31909 & 0.43 & 0.13 & 3.4 & 3.1E+09 & 2.0E+10 & 1.79 & 0.64 & 2.0\\
        37 & L14EG010 & AEGIS & 4004 & 0.21 & 0.06 & 3.7 & 1.3E+09 & 6.4E+09 & 0.12 & 0.10 & 0.7\\
        38 & L14EG011 & AEGIS & 6274 & 0.85 & 0.18 & 4.7 & 3.7E+09 & 1.9E+10 & 0.34 & 0.26 & 0.7\\
        39 & L14EG012 & AEGIS & 6449 & 0.32 & 0.10 & 3.3 & 1.2E+09 & 6.1E+09 & 0.05 & 0.05 & 0.7\\
        40 & L14EG014 & AEGIS & 9743 & 0.14 & 0.08 & 1.8 & 9.4E+08 & 4.7E+09 & 0.06 & 0.05 & 0.8\\
        41 & L14EG015 & AEGIS & 26964 & 0.14 & 0.04 & 3.5 & 1.0E+09 & 5.1E+09 & 0.05 & 0.05 & 0.4\\
        42 & L14EG016 & AEGIS & 34302 & 0.25 & 0.08 & 3.1 & 1.4E+09 & 7.2E+09 & 0.18 & 0.15 & 1.1\\
        43 & XA55 & GOODS-N & 21285$^{\star}$ & 0.38 & 0.12 & 3.2 & 2.9E+09 & 1.6E+10 & 0.58 & 0.37 & 0.4\\
        44 & XB55 & GOODS-N & 6666$^{\dagger}$ & $<$ 0.36 &  &  & $<$2.2E+09 & $<$1.2E+10 & $<$0.26 & $<$ 0.21 & $<$ 0.5\\
        45 & XC55 & GOODS-N & 19725 & 0.70 & 0.14 & 5.0 & 5.7E+09 & 3.0E+10 & 0.66 & 0.40 & 1.0\\
        46 & XD55 & GOODS-N & 12097 & 0.40 & 0.12 & 3.4 & 3.2E+09 & 1.8E+10 & 0.59 & 0.37 & 0.8\\
        47 & XE55 & GOODS-N & 19815$^{\star}$ & 0.27 & 0.11 & 2.5 & 2.1E+09 & 1.2E+10 & 0.35 & 0.26 & 0.8\\
        48 & XF55 & GOODS-N & 7906 & 0.23 & 0.09 & 2.6 & 1.2E+09 & 7.6E+09 & 0.66 & 0.40 & 0.7\\
        49 & XG55 & GOODS-N & 19257 & 0.68 & 0.17 & 3.9 & 2.3E+09 & 1.2E+10 & 0.31 & 0.24 & 1.4\\
        50 & XH55 & GOODS-N & 16987 & 0.25 & 0.07 & 3.4 & 2.0E+09 & 1.2E+10 & 0.76 & 0.43 & 1.0\\
        51 & XL55 & GOODS-N & 10134 & 0.46 & 0.18 & 2.6 & 3.8E+09 & 2.1E+10 & 0.66 & 0.40 & 1.0\\
        52 & L14GN006 & GOODS-N & 30883 & 0.82 & 0.20 & 4.2 & 5.1E+09 & 2.9E+10 & 1.14 & 0.53 & 1.2\\
        53 & L14GN007 & GOODS-N & 939 & 0.78 & 0.17 & 4.5 & 3.7E+09 & 1.8E+10 & 0.25 & 0.20 & 2.0\\
        54 & L14GN008 & GOODS-N & 11532 & 0.44 & 0.12 & 3.6 & 1.5E+09 & 8.0E+09 & 0.41 & 0.29 & 1.5\\
        55 & L14GN018 & GOODS-N & 25413 & 0.30 & 0.07 & 4.2 & 2.5E+09 & 1.4E+10 & 0.56 & 0.36 & 0.4\\
        56 & L14GN021 & GOODS-N & 8738 & 1.28 & 0.21 & 6.1 & 6.9E+09 & 3.6E+10 & 0.69 & 0.41 & 0.5\\
        57 & L14GN022 & GOODS-N & 11460$^{\star}$ & 0.09 & 0.04 & 2.3 & 3.7E+08 & 2.2E+09 & 0.17 & 0.15 & 0.3\\
        58 & L14GN025 & GOODS-N & 36596 & 0.21 & 0.06 & 3.5 & 7.8E+08 & 4.0E+09 & 0.09 & 0.08 & 1.1\\
        59 & L14GN032 & GOODS-N & 21683 & 0.17 & 0.08 & 2.1 & 7.0E+08 & 3.4E+09 & 0.03 & 0.03 & 0.5\\
        60 & L14GN033 & GOODS-N & 1964 & 0.19 & 0.06 & 3.0 & 7.9E+08 & 4.7E+09 & 0.41 & 0.29 & 0.7\\
        61 & L14GN034 & GOODS-N & 33895 & 1.90 & 0.31 & 6.2 & 6.7E+09 & 3.3E+10 & 0.45 & 0.31 & 3.8\\
        \hline
\end{tabularx}

        \tablefoot{\\
                $^a$ CO(2-1) integrated line flux and its uncertainty.\\
                $^b$ $\rm S/N = F(CO)/dF(CO)$.\\
                $^c$ Integrated CO(2-1) line luminosity as derived from Eq.~\ref{eq:luminosity}.\\
                $^d$ Molecular gas mass, corrected by a factor $1.36$ for interstellar Helium, using a Galactic CO-H$_2$ conversion factor $\alpha_{\rm CO}=4.36 ~\rm M_\odot /(K.km.s^{-1}.pc^{2})$ and a CO(2-1)/CO(1-0) line $r_{21} = 0.77$. The systematic uncertainties are evaluated at $\pm 50\%$.\\
                $^e$ $\mu_{\rm gas} = \rm M_{gas}/M_{\star}$.\\
                $^f$ $f_{\rm gas}=\rm M_{gas}/(M_{gas}+M_{\star}) = \mu_{gas}/(1+\mu_{gas})$.\\
                $^g$ $t_{\rm depl} = \rm M_{gas}/SFR$.\\
                $^\star$ Marginal detection.\\
                $^\dagger$ Non-detection.
        }
\end{table*}

\subsection{Molecular gas masses}
\label{section:Mgas}

Although the CO molecule only represents a small fraction of the total molecular gas mass and its lower rotational lines are almost always optically thick, observations of giant molecular clouds (GMCs) in the Milky Way have shown that the integrated line flux of its rotational lines could be used as a quantitative tracer of the molecular gas mass \citep{Dickman1986, Solomon1987, Combes1991, Young1991, Solomon1991,Dame2001, Bolatto2013}.
%
The intrinsic CO luminosity associated to any region of flux can be expressed as 
\begin{equation}
\label{eq:luminosity}
\left(\frac{\mathrm{L}^\prime_{\mathrm{CO}}}{\mathrm{K.km.s}^{-1}\mathrm{.pc}^2}\right) = \frac{3.25 \times 10^7}{1+z}~\left(\frac{\rm F(CO)}{\mathrm{Jy.km.s}^{-1}}\right)~ \left(\frac{\nu_{\rm rest}}{\mathrm{GHz}}\right)^{-2}~ \left(\frac{\mathrm{D}_\mathrm{L}}{\mathrm{Mpc}}\right)^2,
\end{equation}
where $\rm F(CO)$ is the  velocity integrated flux, $\nu_{\rm rest}$ the rest-frame frequency -- $\rm 230.538~GHz$ in the case of CO(2-1) --, and $\rm D_L$ the luminosity distance \citep{Solomon1997}. 
Considering a certain $J \rightarrow J-1$ CO line, the total molecular gas mass including a 36\% correction to account for interstellar helium is then estimated as
\begin{equation}
{\rm M_{gas}} = \alpha_{\rm CO} {\rm L}_{\rm CO(J\rightarrow J-1)}^\prime / r_{\rm J1}
,\end{equation}
where $\alpha_{\rm CO}$ is the CO(1-0) luminosity-to-molecular-gas-mass conversion factor and $r_{J1}={\rm L_{CO(J\rightarrow J-1)}/L_{CO(1-0)}}$ the corresponding line ratio. 

The $\alpha_{\rm CO}$ conversion factor a priori depends on the average cloud density, the Rayleigh-Jeans brightness temperature of the CO transition, and the metallicity \citep{Strong2004, Leroy2011,Genzel2012, Papadopoulos2012b, Bolatto2013, Sandstrom2013}. 
In the Milky Way, its dense star-forming clumps, nearby MS star-forming galaxies, and low-metallicity galaxies, estimates of the conversion factor based on virial masses, optically thin tracers of the column density, and diffuse gamma-ray emission stemming from the interaction between cosmic rays and interstellar medium (ISM) protons seem to converge towards a relatively uniform value $\alpha_G = 4.36 \pm 0.9 \rm~ M_\odot/(K.km.s^{-1}.pc^{2})$ including helium \citep{Strong1996,Dame2001,Grenier2005, Bolatto2008,Abdo2010,Schinnerer2010, Leroy2011,Bolatto2013}. 
As the CO emission in the $z=0.5-0.8$ galaxies studied in this paper is likely to originate from virialized GMCs with mean densities of the same order of magnitude as their lower-redshift counterparts \citep{Daddi2008, Daddi2010, Dannerbauer2009} and similar dust temperatures \citep{Magnelli2009, Hwang2010, Elbaz2011}, their conversion factor should be relatively close to the `Galactic' conversion factor $\alpha_G$.
%
But since the CO conversion factor increases with decreasing metallicity as the CO molecule gets more photo-dissociated \citep{Wolfire2010,Bolatto2013}, we do account for its metallicity dependence. 
From the different metallicity corrections proposed in the literature, we adopt the geometric mean of the recipes by \cite{Bolatto2013} and \cite{Genzel2012} as adopted by \cite{Genzel2015} and \cite{Tacconi2018}: 
\begin{equation}
\alpha_{\rm CO} = \alpha_G \sqrt{0.67 \times \exp(0.36 \times 10^{8.67-\log{Z}}) \times 10^{-1.27\times (\log{Z}-8.67)}}
,\end{equation}
where $\log{Z} = 12+\log({\rm O/H})$ is the metallicity on the \cite{Pettini2004} scale estimated from the mass--metallicity relation 
\begin{equation}
\log{Z} = 8.74 - 0.087 \times (\log({\rm M_{\star}})-b)^2
,\end{equation}
with $b = 10.4 + 4.46 \times \log(1+z)-1.78 \times (\log(1+z))^2$ \citep[][and references therein]{Genzel2015}. This metallicity correction leads to a mean $\alpha_{\rm CO} = 4.0 \pm 0.3 \rm~ M_\odot/(K.km.s^{-1}.pc^{2})$ within the $z=0.5-0.8$ sample. 

The $r_{21}$ line ratio converts the observed CO(2-1) luminosity into the CO(1-0) luminosity for which the $\alpha_{\rm CO}$ conversion factor is calibrated. 
While a thermally excited transition in the Rayleigh-Jeans domain with $r_{21} = 1$ has often been assumed to derive molecular gas masses \citep{Combes2011, Combes2013, Bauermeister2013, Tacconi2013}, the CO(2-1) line could both be sub-thermally excited and require a Planck-correction, leading to $r_{21}<1$. 
In particular, 
\cite{Leroy2009} obtain values of $r_{21}$ between 0.6 and 1 within a sample of 18 nearby galaxies, with a typical value $r_{21}\sim 0.8$, 
while 
\cite{Dannerbauer2009} and \cite{Aravena2010} obtain $r_{21}\sim 0.85$ at $z\sim 1.5$,
and \cite{Papadopoulos2012a} around $r_{21}=0.91$ for a large sample of luminous and ultra-luminous IR galaxies in the local universe. \cite{Bothwell2013} further measure $r_{21}\sim 0.84$ within a sample of 40 luminous sub-millimitre galaxies in the range $z=1-4$ while 
\cite{Daddi2015} find an average $r_{21} = 0.76$ from a sample of four galaxies at $z=1.5$. 
In the following we assume $r_{21} = 0.77$, as also assumed by \cite{Genzel2015} and \cite{Tacconi2018}. 

\begin{figure*}
        \centering
        \includegraphics[width=1\hsize]{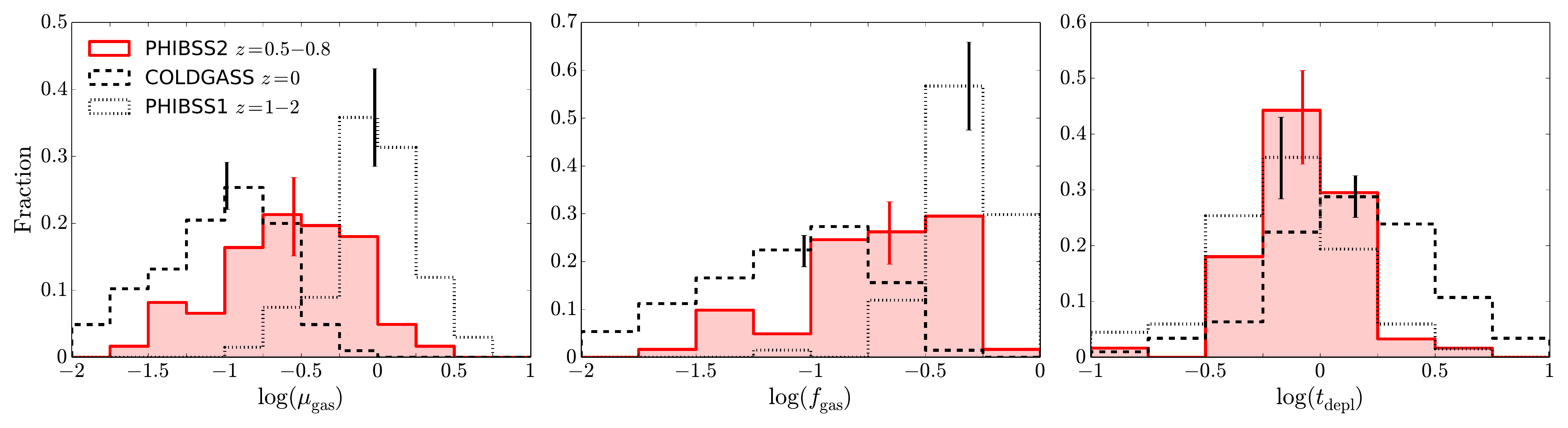}
        \caption{Distributions of the molecular-gas-to-stellar-mass ratio $\mu_{\rm gas}$, the gas fraction $f_{\rm gas}$ , and the depletion time $t_{\rm depl}$ for the PHIBSS2 $z=0.5-0.8$ sample, whose medians are $\widetilde{\mu_{\rm gas}} = 0.28 \pm 0.04$, $ \widetilde{f_{\rm gas}} = 0.22 \pm 0.02,$ and $\widetilde{t_{\rm depl}} = 0.84\pm 0.07~\rm Gyr$ with respective dispersions of $0.42$, $0.33,$ and $0.23~\rm dex$. 
                The corresponding distributions of the total, non-mass-matched $z\sim 0$ COLDGASS survey \citep[dashed lines;][]{Saintonge2011, Saintonge2011b}, and of the $z=1.2-2.2$ PHIBSS survey \citep[dotted lines;][]{Tacconi2010, Tacconi2013} are indicated for comparison. The medians are respectively $0.10$, $0.09,$ and $1.42~\rm Gyr$ for the COLDGASS survey and $0.96$, $0.49,$ and $0.67~\rm Gyr$ for the PHIBSS survey. Typical Poisson errors are shown at the positions of the medians for the different surveys. 
        }
        \label{fig:hist-tdepl-fgas}
\end{figure*}

The resulting values of the intrinsic CO(2-1) luminosity $\rm L_{CO(2-1)}$ and the molecular gas mass $\rm M_{gas}$ as well as the corresponding gas-to-stellar-mass ratio \mbox{$\mu_{\rm gas} = \rm M_{gas}/M_{\star}$}, gas fraction \mbox{$f_{\rm gas}=\rm M_{gas}/(M_{gas}+M_{\star}) = \mu_{\rm gas}/(1+\mu_{\rm gas}),$} and depletion time \mbox{$t_{\rm depl} = \rm M_{gas}/SFR$} are displayed in Table~\ref{table:3}. 
The relative uncertainty on the CO(2-1) line flux $\rm dF(CO)/F(CO)=1/(S/ N)$ is 30\% on average, and as high as about 50\%, which is transferred to the intrinsic CO luminosity $\rm L_{CO(2-1)}$. Considering 
the 30\% uncertainty on the Galactic conversion factor $\alpha_G$ \citep{Bolatto2013}, 
the systematic difference up to 20\% between the metallicity corrections of \cite{Bolatto2013} and \cite{Genzel2012} in the metallicity range of the $z=0.5-0.8$ sample that reflects the scatter in the $\alpha_{\rm CO}$-metallicity relation, and the more negligible 12\% uncertainty on the $r_{21}$ line ratio from \cite{Daddi2015} leads to a systematic uncertainty of at least 50\% on the final molecular gas masses.
Figure~\ref{fig:hist-tdepl-fgas} shows the distributions of $\mu_{\rm gas}$, $f_{\rm gas}$, and $t_{\rm depl}$, comparing them with those obtained at $z=0$ from the COLDGASS survey \citep{Saintonge2011, Saintonge2011b} and at $z=1-2$ with the first PHIBSS program \citep{Tacconi2010, Tacconi2013}. 
The gas-to-stellar-mass ratios $\mu_{\rm gas}$ range from 0.03, close to the detection limit, to 1.8, with a median of $\widetilde{\mu_{\rm gas}} = 0.28 \pm 0.04$.
The ranges for $f_{\rm gas}$ and $t_{\rm depl}$ are $0.03-0.64$ and $0.11-3.82~\rm Gyr$, with median values of  $ \widetilde{f_{\rm gas}} = 0.22 \pm 0.02$ and $\widetilde{ t_{\rm depl}} = 0.84 \pm 0.07~\rm Gyr$, respectively.
These values are intermediary between their low- and high-redshift counterparts, fitting well with a significantly increasing gas fraction and a slightly decreasing depletion time with redshift.
%
In fact, they are in excellent agreement with the values expected from the scaling relations obtained by \cite{Tacconi2018} within their comprehensive sample of about 1400 CO and dust molecular gas measurements between $z=0$ and $z=4.6$. Indeed, applying the scaling relations on the MS ($\delta \rm MS =0$) at the median redshift $z=0.67$ and $\log(\rm M_{\star}/M_\odot) = 10.7$ of the sample yields 0.27, 0.21, and 0.90 for the gas-to-stellar-mass ratio, the gas fraction, and the depletion time. 
Since the molecular gas content of galaxies increases strongly with redshift \citep{Daddi2010b,Tacconi2010,Tacconi2013,Tacconi2018,Genzel2015,Lagos2015} while their atomic gas content varies much more slowly \citep[e.g.][]{Bauermeister2010} with three times more $\rm HI$ mass at $z=0$ \citep{Saintonge2011}, molecular gas is expected to dominate above $z\sim 0.4$ and in particular in our $z=0.5-0.8$ sample. The molecular gas-to-stellar ratio $\mu_{\rm gas}$ and gas fraction $f_{\rm gas}$ thus approximately probe the total gas fractions.

Noting from Figs.~\ref{fig:sample}~and~\ref{fig:histograms} that the mass distribution of the PHIBSS2 $z=0.5-0.8$ sample differ from its CANDELS/3D-HST parent distribution, we further derive mass-matched median values for the molecular gas-to-stellar mass ratio, gas fraction, and depletion time. Following \cite{Catinella2010} and \cite{Saintonge2011}, we place galaxies in stellar mass bins of 0.2 dex width as in the central panel of Fig.~\ref{fig:histograms} and assign as weight the ratio between the number of galaxies in the CANDELS/3D-HST parent sample within 1 dex of the MS line and that in the PHIBSS2 sample at $z=0.5-0.8$ in each of these stellar-mass bins. Limiting ourselves to stellar masses above $\log(\rm M_\star/M_\odot)\geq 10.4$ to avoid being affected by the sparsely populated bins below this value, the resulting mass-weighted medians are $\overline{\mu_{\rm gas}}=0.30 \pm 0.04$,  $\overline{f_{\rm gas}} = 0.23 \pm 0.02$ and $\overline{t_{\rm depl}} =0.84 \pm 0.08\rm ~Gyr$, which correspond well to the values expected from \cite{Tacconi2018} at the median stellar mass and redshift of the CANDELS/3D-HST parent sample with $\rm 10.4 < \log(M_{\star}/M_\odot) < 11.8$ and $|\log(\delta {\rm MS})| < 1$ (respectively, 0.29, 0.22, and 0.89).  However, we leave the detailed study of the influence of mass selection on the \cite{Tacconi2018} scaling relations and in particular on their zero points to future works.

\begin{figure*}
        \centering
        \includegraphics[width=1\hsize]{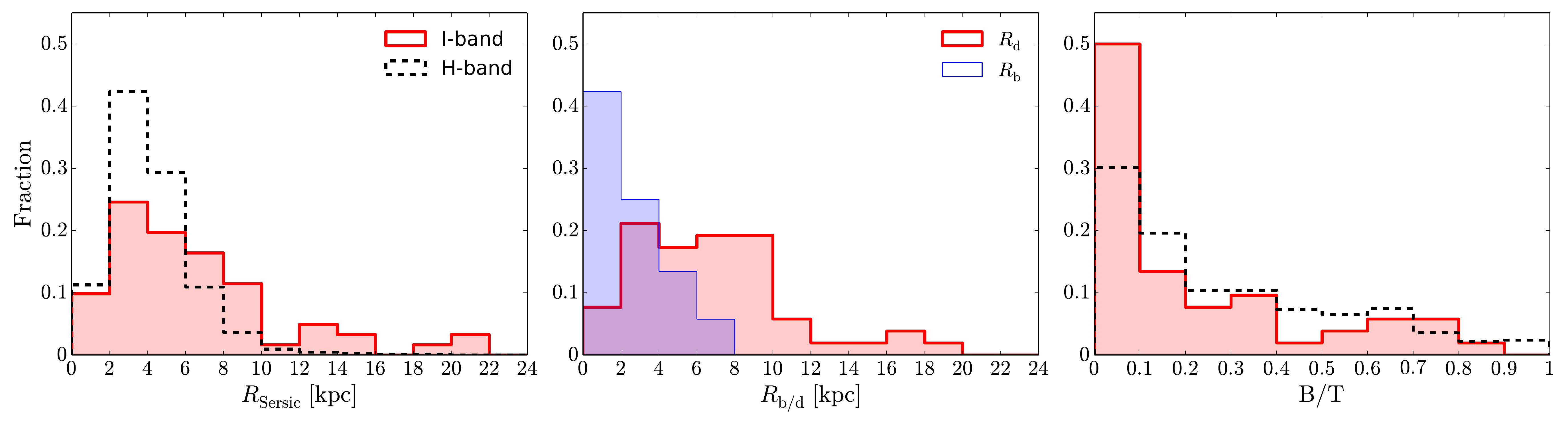}
        \caption{Distributions of the single-Sérsic, disc, and bulge half-light radii together with the bulge-to-total luminosity ratio of the PHIBSS2 $z=0.5-0.8$ sample determined with \texttt{galfit} from their HST/ACS I-band images. In the middle and right panels, we only consider galaxies with $\rm B/T\neq 1$. 
                The black dashed lines show the corresponding parent CANDELS/3D-HST H-band distributions at $z=0.5-0.8$ with $\rm 10 < \log(M_{\star}/M_\odot) < 11.8$ and $|\log(\delta {\rm MS})| < 1$  \citep{Vanderwel2012, Lang2014}.  
        }
        \label{fig:hist_galfit}
\end{figure*}

\subsection{Size and morphology}
\label{section:galfit}

The radial distribution of the star-forming molecular gas in most nearby galaxies follows an exponential profile reminiscent of the stellar disc \citep[e.g. ][]{Young1982a, Scoville1983, Young2000} while bulges are mostly made of old stars \citep[e.g. ][]{Wyse1997, Zoccali2003,Freeman2008}. 
To address the influence of morphology on star formation and to separate the contribution of disc and bulge, we not only determine the total half-light radius of the galaxies of the PHIBSS2 $z=0.5-0.8$ sample from single Sérsic fits but also decompose them as two-component bulge disc systems with the 2D morphology fitting code \texttt{galfit} \citep{Peng2002,Galfit2010}. 
The fits are carried out on publicly available high-resolution ($0.03$ arcsec per pixel) HST Advanced Camera for Survey (ACS) images in the F814W I-band. This band is optimal for our study as it is available for all the galaxies of the sample and probes the blue young stellar population at $z=0.5-0.8$ while avoiding the rest-frame UV light from very young stars. 
The disc is described by an exponential profile \citep{Freeman1970}
\begin{equation}
I(R) = I_{\rm d} e^{-R/R_{\rm e}}, 
\end{equation}
where $I_{\rm d}$ is the disc central density and $R_{\rm e}$ its scale length, which is proportional to the disc half-light radius $R_{\rm d} = 1.67835 R_{\rm e}$. 
The bulge is described by a \cite{Sersic1968} profile:
\begin{equation}
\label{eq:inclination}
I(R) = I_{\rm b} e^{-b_n (R/R_{\rm b})^{1/n}}
,\end{equation}
assuming a classic de Vaucouleurs Sérsic index $n=4$ as in \cite{Bruce2012}, \cite{Lang2014}, and \cite{Contini2016}, $I_{\rm b}$ being the bulge central density and $R_{\rm b}$ its half-light radius. The parameter $b_n$ depends on the Sérsic index $n$ and is derived from $\Gamma (2n) = 2\gamma(2n,b_n)$, where $\Gamma$ is the gamma function and $\gamma$ the lower incomplete gamma function. %
The centres of both components are left free but within 2 pixels of each other and their position angles are constrained to be equal. We impose the bulge not to be more elongated than the disc. Following \cite{Lang2014}, we also consider pure disc and pure $n=4$ bulge models. 
The \texttt{galfit} fits are carried out with a point spread function (PSF) obtained by averaging the 3D-HST \citep{Skelton2014, Momcheva2016} PSF in the three fields of interest (AEGIS, COSMOS and GOODS-N) and a uniform weight map motivated by the nearly uniform HST weight maps around the $z=0.5-0.8$ sources but which does not account for potential Poisson errors in the high-flux regions. 
We refer to \cite{Haussler2007}, \cite{Bruce2012}, \cite{Lang2014}, and \cite{Contini2016} for the influence of the PSF, the weight map and the background subtraction on \texttt{galfit} models. 
Neighbouring nearby galaxies or satellites are fitted simultaneously with a single Sérsic model to account for their luminosity distribution. 

One of the main difficulties when carrying out multi-component fits with \texttt{galfit} is to avoid being trapped in a local $\chi^2$ minimum depending on the initial guess instead of the global minimum \citep[e.g. ][]{Haussler2007, Galfit2010, Bruce2012}. 
Following \cite{Lang2014}, we first build ten initial guesses from the least degenerate single Sérsic fits using an empirical analysis of noise-free two-component models: (i) we generate a grid of ideal, noise-free two-component bulge disc models with different B/T and  $\rm R_{\rm b}/R_{\rm d}$; (ii) we obtain their global Sérsic indices and half-light radii with \texttt{galfit} using no PSF and a weight map corresponding to ideal Poisson noise; (iii) we estimate the B/T associated to a series of ten values of $\rm R_{\rm b}/R_{\rm d}$ for each galaxy from its Sérsic index using the result of the previous step; and (iv) we determine the corresponding bulge and disc to be used in the ten initial guesses for that galaxy. 
Noting that not all single Sérsic fits yield physical half-light radii, we also build initial guesses from the pure bulge and pure disc fits with different values of B/T and $\rm R_{\rm b}/R_{\rm d}$ (10 values of $\rm R_{\rm b}/R_{\rm d}$ between 0.1 and 1, $\rm B/T=0.1$ and $0.5$ when using the pure disc model, $\rm B/T=0.5$ and $0.9$ when using the pure bulge model), leading to a total of 52 \texttt{galfit} runs per galaxy for the bulge disc model, including the pure-bulge and pure-disc models. 
The best-fit model is that with the lowest reduced $\chi^2$, casting away models where the bulge is implausibly small ($R_{\rm b}<0.1$ pixel) or larger than the disc ($R_{\rm b}>R_{\rm d}$); except for L14GN025 (\texttt{GN4-36596}) where we release this latter condition as the low surface brightness of the stellar halo in which the disc is embedded makes all models with $R_{\rm b}<R_{\rm d}$ unsatisfactory. 
We note that as in \cite{Lang2014}, two-component decompositions are preferred over single Sérsic fits for about two thirds of the sample, namely for 37 galaxies out of 61 (61\%). 
The best-fit models are displayed with the I-band images in Appendix~\ref{appendix:HST}.

\begin{table*}
        \caption{I-band morphology and sizes }
        \label{table:4}
        \centering
        \tiny
        
        \begin{tabularx}{\textwidth}{l l l l Y Y Y Y Y Y}
                \hline\hline
                \noalign{\vskip 1mm}
                \# & ID & Field & Source & ${R_{\rm Sersic}}^a$ & $n_{\rm Sersic}^a$ & $q_{\rm Sersic}^a$ & ${R_{\rm d}}^b$ & ${R_{\rm b}}^b$ & ${\rm B/T}^c$ \\
                &    &       &        & (kpc)         &     &          & (kpc)     & (kpc)     &           \\
                \hline
                \noalign{\vskip 1mm} 1 & XA53 & COSMOS & 822872 & 7.70 & 2.17 & 0.62 &  6.82 & 6.65 & 0.45 \\
                2 & XC53 & COSMOS & 805007$^{\star}$ & 2.71 & 4.00 & 0.77 & 3.50 & 1.24 & 0.61 \\
                3 & XD53 & COSMOS & 822965 & 7.20 & 1.53 & 0.92 & 10.13 & 2.34 & 0.16 \\
                4 & XE53 & COSMOS & 811360 & 5.35 & 1.02$^{+}$ & 0.83 & 5.61 & 0.46 & 0.02 \\
                5 & XF53 & COSMOS & 834187 & 15.37 & 4.00 & 0.57 & 8.59 & 4.36 & 0.37 \\
                6 & XG53 & COSMOS & 800405 & 4.21 & 3.86$^{+}$ & 0.82 & - & 4.49 & 1.00 \\
                7 & XH53 & COSMOS & 837919 & 4.36 & 3.34 & 0.79 & 3.58 & 3.51 & 0.70 \\
                8 & XI53 & COSMOS & 838956 & 4.66 & 3.10$^{+}$ & 0.77 & - & 7.30 & 1.00 \\
                9 & XL53 & COSMOS & 824759 & 2.76 & 2.95$^{+}$ & 0.73 & - & 4.35 & 1.00 \\
                10 & XM53 & COSMOS & 810344 & 4.81 & 1.86 & 0.43 & 4.44 & 3.85 & 0.29 \\
                11 & XN53 & COSMOS & 839268 & 7.02 & 4.00 & 0.79 & - & 6.42 & 1.00 \\
                12 & XO53 & COSMOS & 828590 & 3.31 & 2.32$^{+}$ & 0.51 & 5.66 & 5.21 & 0.81 \\
                13 & XQ53 & COSMOS & 838696 & 32.42 & 4.00 & 0.45 & 16.41 & 7.75 & 0.38 \\
                14 & XR53 & COSMOS & 816955 & 18.95 & 4.00 & 0.73 & 15.75 & 4.78 & 0.30 \\
                15 & XT53 & COSMOS & 823380 & 7.21 & 2.73$^{+}$ & 0.66 & 6.07 & 5.93 & 0.54 \\
                16 & XU53 & COSMOS & 831385 & 5.84 & 0.90 & 0.47 & 6.36 & 0.61 & 0.02 \\
                17 & XV53 & COSMOS & 850140 & 3.93 & 1.19 & 0.41 & 4.42 & 2.04 & 0.13 \\
                18 & XW53 & COSMOS & 824627$^{\star}$ & 4.61 & 1.19 & 0.86 & 4.55 & 0.34 & 0.02 \\
                19 & L14CO001 & COSMOS & 831870 & 2.84 & 1.20 & 0.92 & 2.97 & 0.47 & 0.03 \\
                20 & L14CO004 & COSMOS & 831386 & 3.51 & 1.53 & 0.54 & 3.66 & 2.06 & 0.19 \\
                21 & L14CO007 & COSMOS & 838945 & 15.13 & 3.37 & 0.39 & 10.30 & 7.55 & 0.52 \\
                22 & L14CO008 & COSMOS & 820898 & 7.05 & 2.17 & 0.62 & 6.08 & 4.46 & 0.29 \\
                23 & L14CO009 & COSMOS & 826687 & 7.28 & 1.24$^{+}$ & 0.56 & 6.53 & - & 0.00 \\
                24 & L14CO011 & COSMOS & 839183 & 6.16 & 0.59 & 0.71 & 8.00 & 0.93 & 0.03 \\
                25 & L14CO012 & COSMOS & 838449 & 1.71 & 1.74 & 0.83 & 1.73 & 1.72 & 0.39 \\
                26 & XA54 & AEGIS & 30084 & 9.80 & 2.17 & 0.89 & 6.56 & 0.64 & 0.05 \\
                27 & XB54 & AEGIS & 17329 & 20.54 & 4.00 & 0.97 & 18.15 & 1.79 & 0.10 \\
                28 & XC54 & AEGIS & 14885 & 12.76 & 0.32$^{+}$ & 0.17 & 17.47 & 0.53 & 0.01 \\
                29 & XD54 & AEGIS & 24556 & 3.75 & 0.54$^{+}$ & 0.81 & 4.62 & - & 0.00 \\
                30 & XE54 & AEGIS & 25608 & 8.30 & 0.29$^{+}$ & 0.26 & 9.72 & - & 0.00 \\
                31 & XF54 & AEGIS & 32878 & 9.92 & 2.19 & 0.72 & 7.15 & 1.86 & 0.10 \\
                32 & XG54 & AEGIS & 3654 & 38.20 & 4.00$^{+}$ & 0.57 & 13.47 & 3.38 & 0.17 \\
                33 & XH54 & AEGIS & 30516 & 4.95 & 0.78 & 0.78 & 5.74 & 0.27 & 0.02 \\
                34 & L14EG006 & AEGIS & 23488 & 9.85 & 1.84 & 0.60 & 8.20 & 2.81 & 0.09 \\
                35 & L14EG008 & AEGIS & 21351 & 21.89 & 4.00 & 0.70 & 8.38 & 0.88 & 0.09 \\
                36 & L14EG009 & AEGIS & 31909 & 2.81 & 0.46$^{+}$ & 0.78 & 3.42 & - & 0.00 \\
                37 & L14EG010 & AEGIS & 4004 & 1.77 & 2.84$^{+}$ & 0.85 & - & 2.80 & 1.00 \\
                38 & L14EG011 & AEGIS & 6274 & 11.57 & 1.86 & 0.57 & 9.23 & 2.76 & 0.07 \\
                39 & L14EG012 & AEGIS & 6449 & 13.49 & 4.00 & 0.61 & 10.53 & 3.23 & 0.37 \\
                40 & L14EG014 & AEGIS & 9743 & 3.92 & 1.68$^{+}$ & 0.93 & 3.21 & 2.72 & 0.14 \\
                41 & L14EG015 & AEGIS & 26964 & 2.10 & 2.60$^{+}$ & 0.82 & - & 3.80 & 1.00 \\
                42 & L14EG016 & AEGIS & 34302 & 4.61 & 2.23 & 0.77 & 3.74 & 3.72 & 0.36 \\
                43 & XA55 & GOODS-N & 21285$^{\star}$ & 3.84 & 0.33$^{+}$ & 0.62 & 5.54 & - & 0.00 \\
                44 & XB55 & GOODS-N & 6666$^{\dagger}$ & 1.69 & 2.60$^{+}$ & 0.87 & - & 2.85 & 1.00 \\
                45 & XC55 & GOODS-N & 19725 & 3.11 & 3.00$^{+}$ & 0.98 & 3.77 & 3.16 & 0.77 \\
                46 & XD55 & GOODS-N & 12097 & 2.55 & 0.35$^{+}$ & 0.59 & 3.55 & - & 0.00 \\
                47 & XE55 & GOODS-N & 19815$^{\star}$ & 6.16 & 0.85 & 0.22 & 7.10 & 0.57 & 0.04 \\
                48 & XF55 & GOODS-N & 7906 & 7.16 & 0.54$^{+}$ & 0.17 & 8.43 & - & 0.00 \\
                49 & XG55 & GOODS-N & 19257 & 4.32 & 2.71$^{+}$ & 0.76 & 3.80 & 3.78 & 0.61 \\
                50 & XH55 & GOODS-N & 16987 & 5.59 & 0.48$^{+}$ & 0.78 & 7.18 & 0.45 & 0.01 \\
                51 & XL55 & GOODS-N & 10134 & 9.46 & 1.53 & 0.66 & 8.19 & 1.52 & 0.04 \\
                52 & L14GN006 & GOODS-N & 30883 & 4.51 & 1.08 & 0.29 & 4.94 & 4.71 & 0.15 \\
                53 & L14GN007 & GOODS-N & 939 & 6.55 & 3.03$^{+}$ & 0.69 & - & 11.43 & 1.00 \\
                54 & L14GN008 & GOODS-N & 11532 & 8.38 & 1.86 & 0.85 & 6.57 & 0.73 & 0.04 \\
                55 & L14GN018 & GOODS-N & 25413 & 3.51 & 1.00 & 0.81 & 3.75 & 0.11 & 0.01 \\
                56 & L14GN021 & GOODS-N & 8738 & 0.93 & 2.14 & 0.91 & 1.31 & 1.07 & 0.71 \\
                57 & L14GN022 & GOODS-N & 11460$^{\star}$ & 0.91 & 1.63 & 0.96 & 0.88 & 0.86 & 0.27 \\
                58 & L14GN025 & GOODS-N & 36596 & 0.97 & 1.72 & 0.52 & 0.67 & 5.44 & 0.64 \\
                59 & L14GN032 & GOODS-N & 21683 & 2.32 & 2.79$^{+}$ & 0.89 & - & 4.10 & 1.00 \\
                60 & L14GN033 & GOODS-N & 1964 & 13.50 & 4.00 & 0.96 & 8.21 & 0.91 & 0.12 \\
                61 & L14GN034 & GOODS-N & 33895 & 8.34 & 1.28 & 0.44 & 8.48 & 1.37 & 0.04 \\
                \hline
        \end{tabularx}
        
        \tablefoot{\\
                $^a$ Half-light radius, Sérsic index and axis ratio of the single Sérsic fits in the F814W I-band images. Sérsic indices followed by a cross ($+$) correspond to cases where the single-Sérsic fit is better than the two-component fit (39\% of the sample).\\
                $^b$ Half-light radii of the $n=1$ disc and $n=4$ bulge components of the two-component fits in the I-band images.\\
                $^c$ Bulge-to-total luminosity ratios defined from the two-component fits using Eq.~\ref{eq:BT}.\\
                $^\star$ Marginal detection;\\
                $^\dagger$ Non-detection.
        }
\end{table*}

Table~\ref{fig:hist_galfit} displays the results of both the single Sérsic and the two-component fits, with the half-light radius $R_{\rm Sersic}$, Sérsic index $n_{\rm Sersic}$, axis ratio $q_{\rm Sersic}$ resulting from the single Sérsic fits, the disc and bulge half-light radii $R_{\rm d}$ and $R_{\rm b}$ of the two-component fits, as well as the corresponding bulge-to-total-luminosity ratio B/T for the PHIBSS2 sample at $z=0.5-0.8$. 
The Sérsic index $n_{\rm Sersic}$ is constrained to be between 0.2 and 4, and we note that $n_{\rm Sersic}=4$ often coincides with relatively large values of the half-light radius $R_{\rm Sersic}$ that are usually corrected with the two-component fits. 
The  bulge-to-total-luminosity ratio is estimated from the two-component model as
\begin{equation}
\label{eq:BT}
{\rm B/T} = \frac{F_{\rm b}}{F_{\rm b}+F_{\rm d}} = \frac{1}{1+10^{(m_{\rm b}-m_{\rm d})/2.5}},
\end{equation}
where $F_{\rm b}$ and $F_{\rm d}$ are the total fluxes associated to the bulge and disc components, respectively, and $m_{\rm b}$, $m_{\rm d}$ are the associated magnitudes. 
%
Assuming that the uncertainties on the single-Sérsic fit parameters only depend on the S/N, we transpose the results of \cite{Vanderwel2012} to band I and conservatively evaluate the uncertainties on $R_{\rm Sersic}$, $n_{\rm Sersic}$ and $q_{\rm Sersic}$ at about 20\% given the magnitude of the sources. We note that \cite{Bruce2012} find that the background subtraction induces errors of about 5\% for $n_{\rm Sersic}$ and 10\% for $R_{\rm Sersic}$, to which we should add the uncertainties introduced by the PSF choice and those intrinsic to \texttt{galfit} such as the choice of the weight matrix. 
From \cite{Lang2014}, we infer a 0.05 uncertainty on B/T.
Figure~\ref{fig:hist_galfit} shows the distributions of the I-band $R_{\rm Sersic}$, $R_{\rm d}$, $R_{\rm b}$, and $\rm B/T$ within the sample. The parent  CANDELS/3D-HST measurements in the H-band \citep{Vanderwel2012,Lang2014} are shown for comparison in the case of $R_{\rm Sersic}$ and B/T. As most $\rm B/T=1$ cases correspond to galaxies harbouring clear spiral features that are not well accounted for by the cylindrically symmetric models adopted here, we do not show them in the middle and right panels. 
More generally, asymmetries and structures such as spiral arms, rings, and bars may introduce biases in the B/T measurements. Accounting for such features would require a case-by-case study that is beyond the scope of this article; here we favour a systematic approach to the determination of $\rm B/T$ compatible with large datasets, in line with other surveys \citep[e.g. ][]{Vanderwel2012,Lang2014,Contini2016}. 
Compared to the H-band measurements, the I-band half-light radii are more spread out with a higher median value while half of the sample are found to be discs with very faint or non-existent bulges ($\rm B/T < 0.1$). These trends relate to the lower characteristic wavelength of the I-band, which traces younger stars and hence highlights the disc relative to the H-band images.


\section{Discussion}
\label{section:discussion}

\subsection{Molecular gas fraction and depletion time}
\label{section:scaling}

\begin{figure*}
        \centering
        \includegraphics[height=0.3\hsize,trim={1.3cm 0 2.7cm 0},clip]{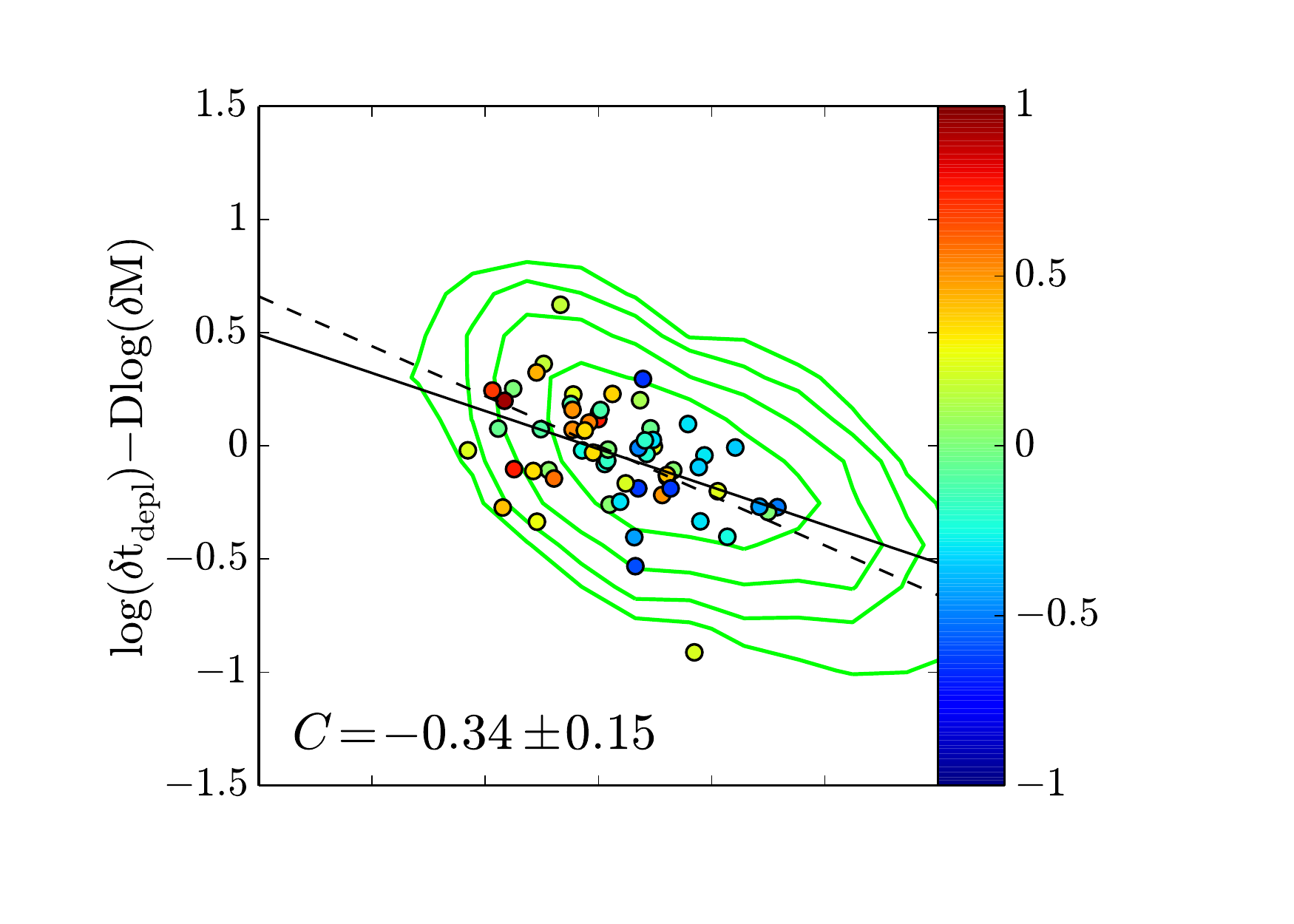}  
        \hfill
        \includegraphics[height=0.3\hsize,trim={1.3cm 0 2.7cm 0},clip]{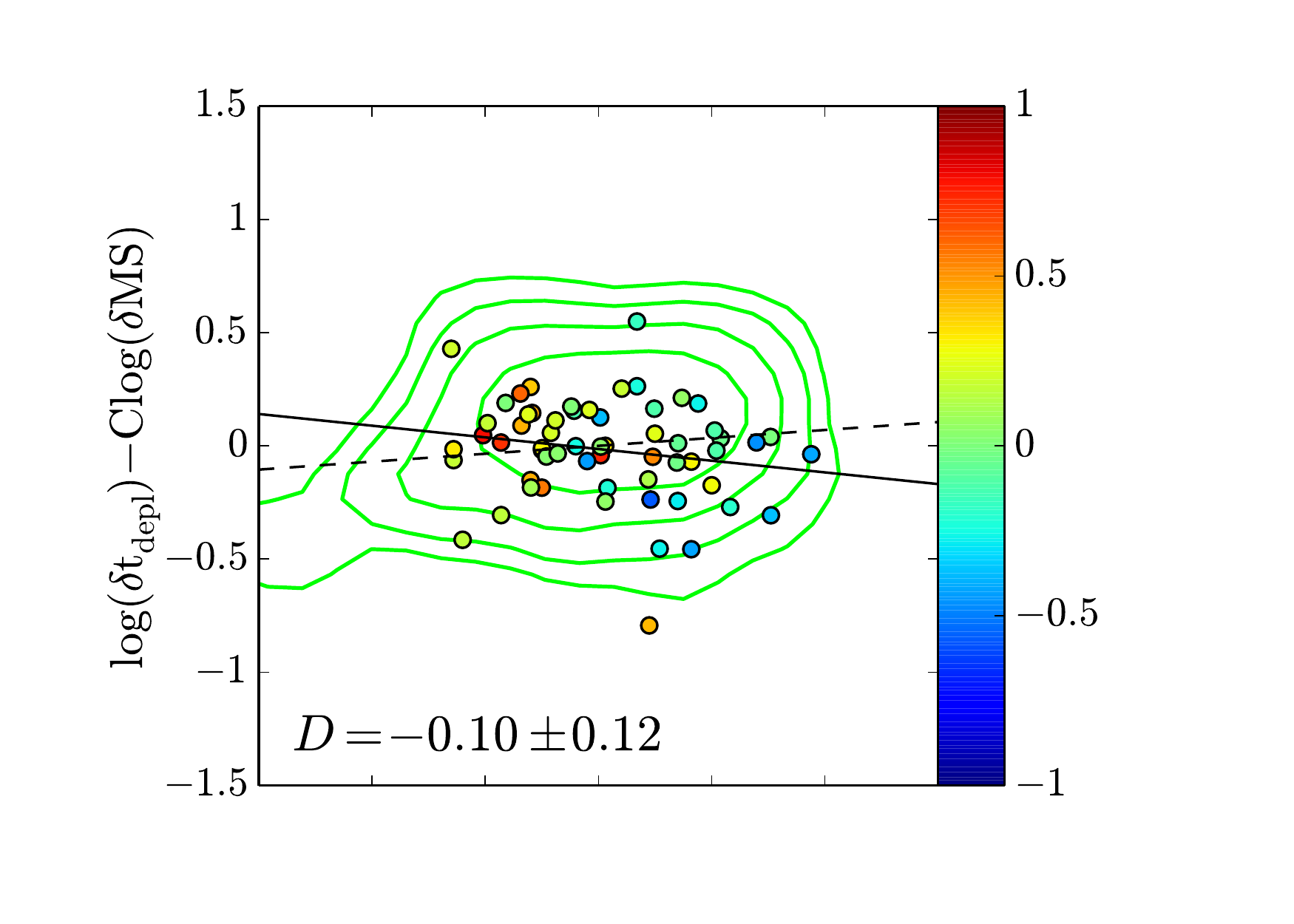}  
        \hfill
        \includegraphics[height=0.3\hsize,trim={1.3cm 0 3.1cm 0},clip]{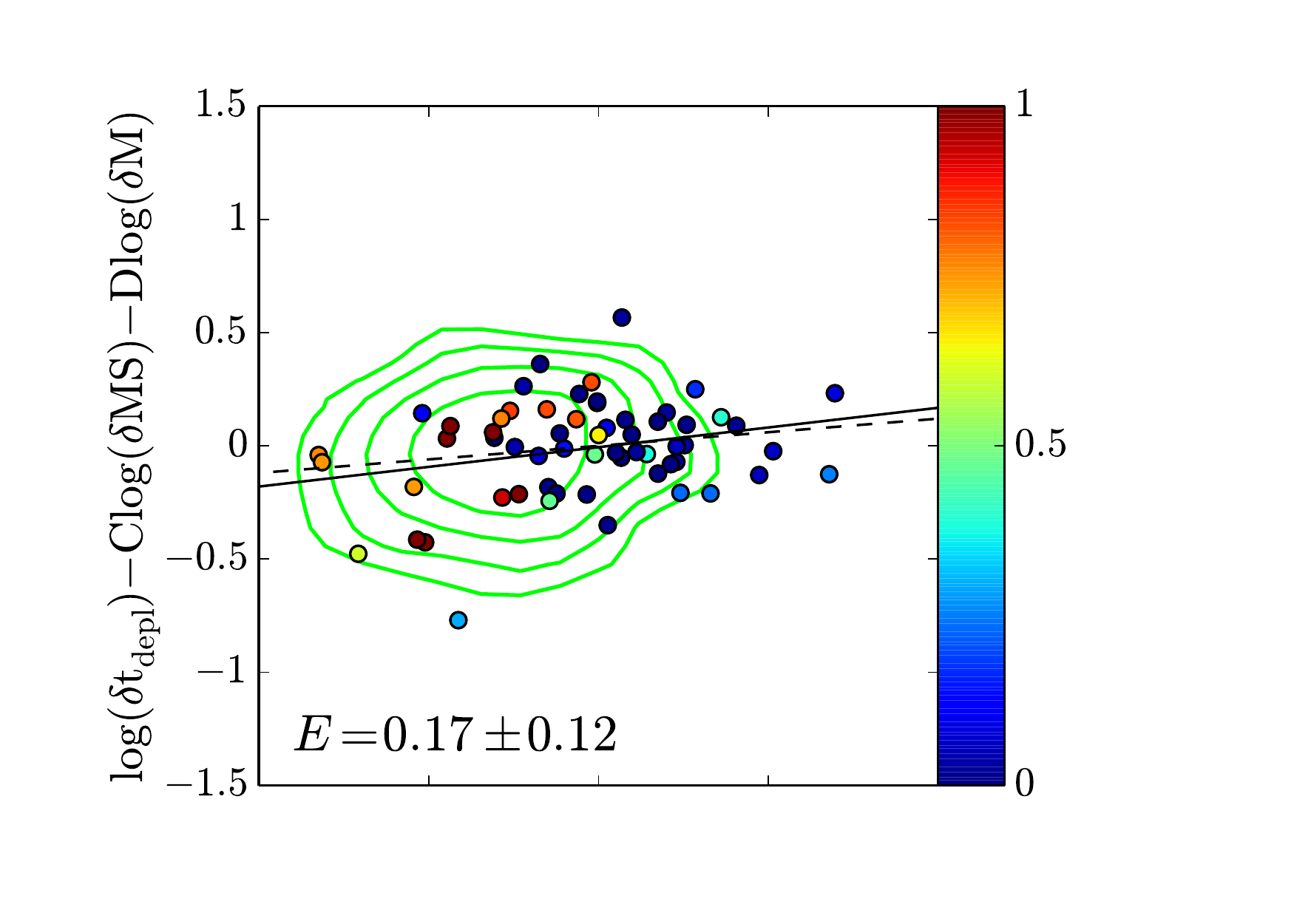}  
        \\\vspace{-0.65cm}
        \includegraphics[height=0.3\hsize,trim={1.3cm 0 2.7cm 0},clip]{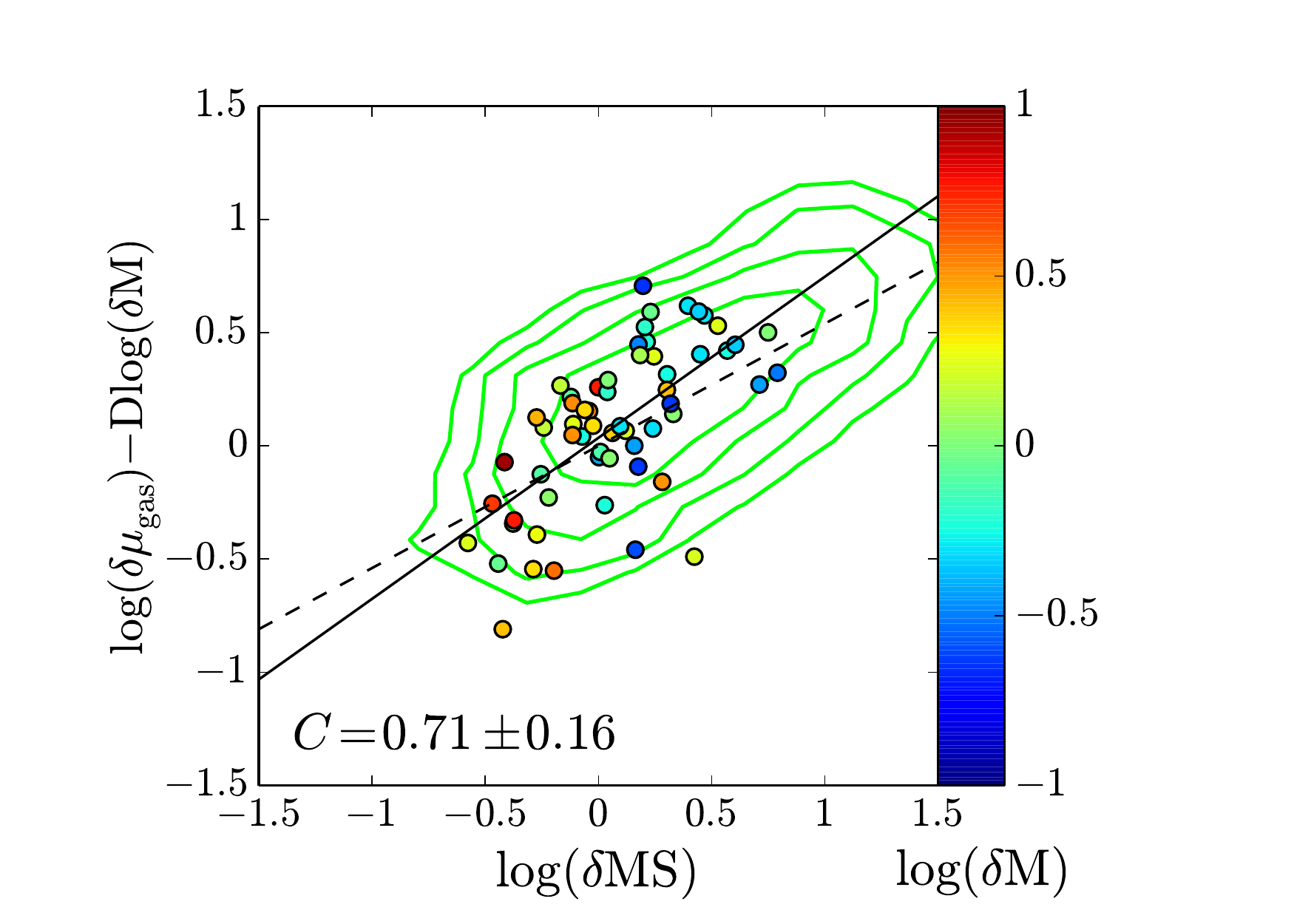}     
        \hfill
        \includegraphics[height=0.3\hsize,trim={1.3cm 0 2.7cm 0},clip]{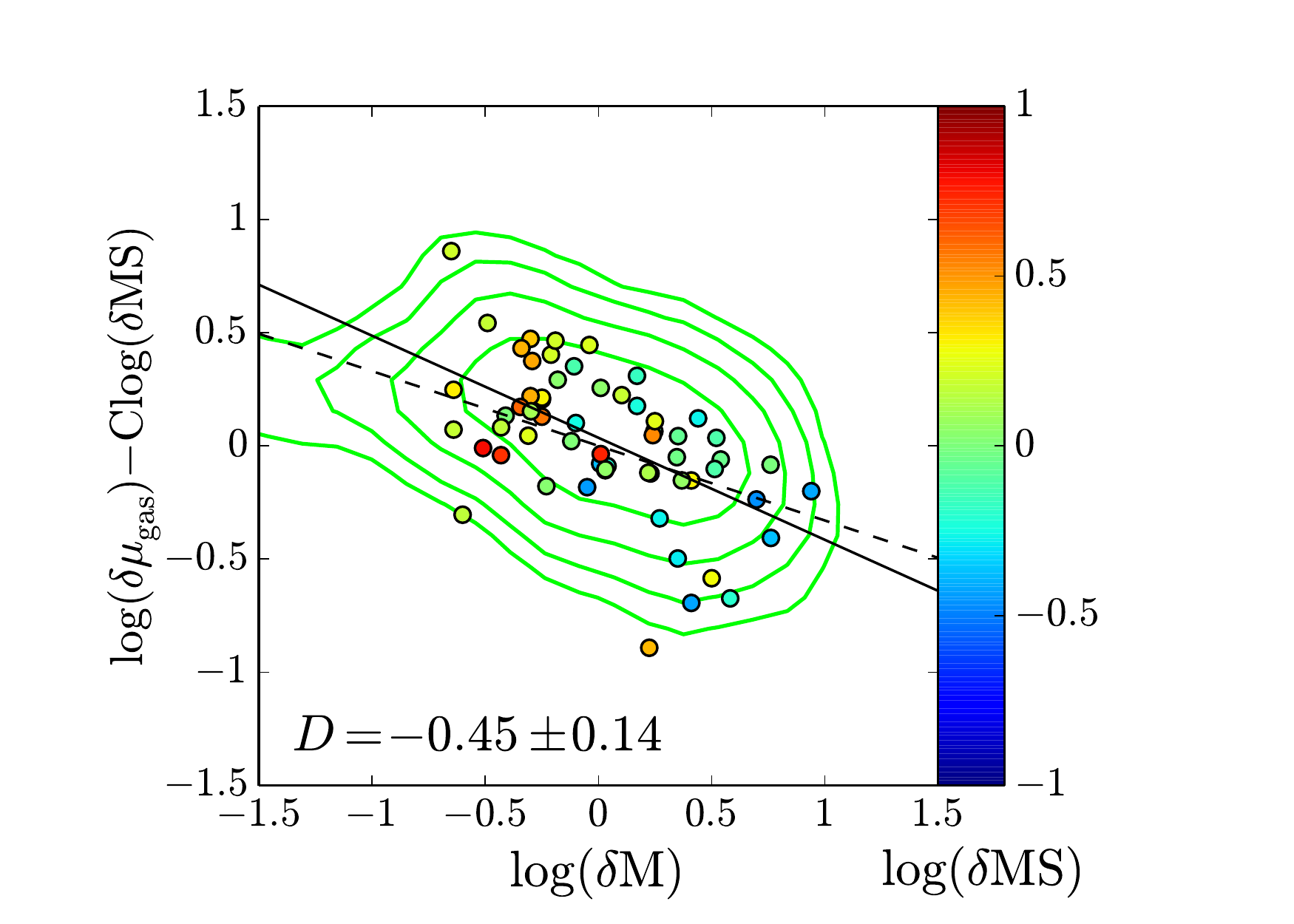}     
        \hfill
        \includegraphics[height=0.3\hsize,trim={1.3cm 0 3.1cm 0},clip]{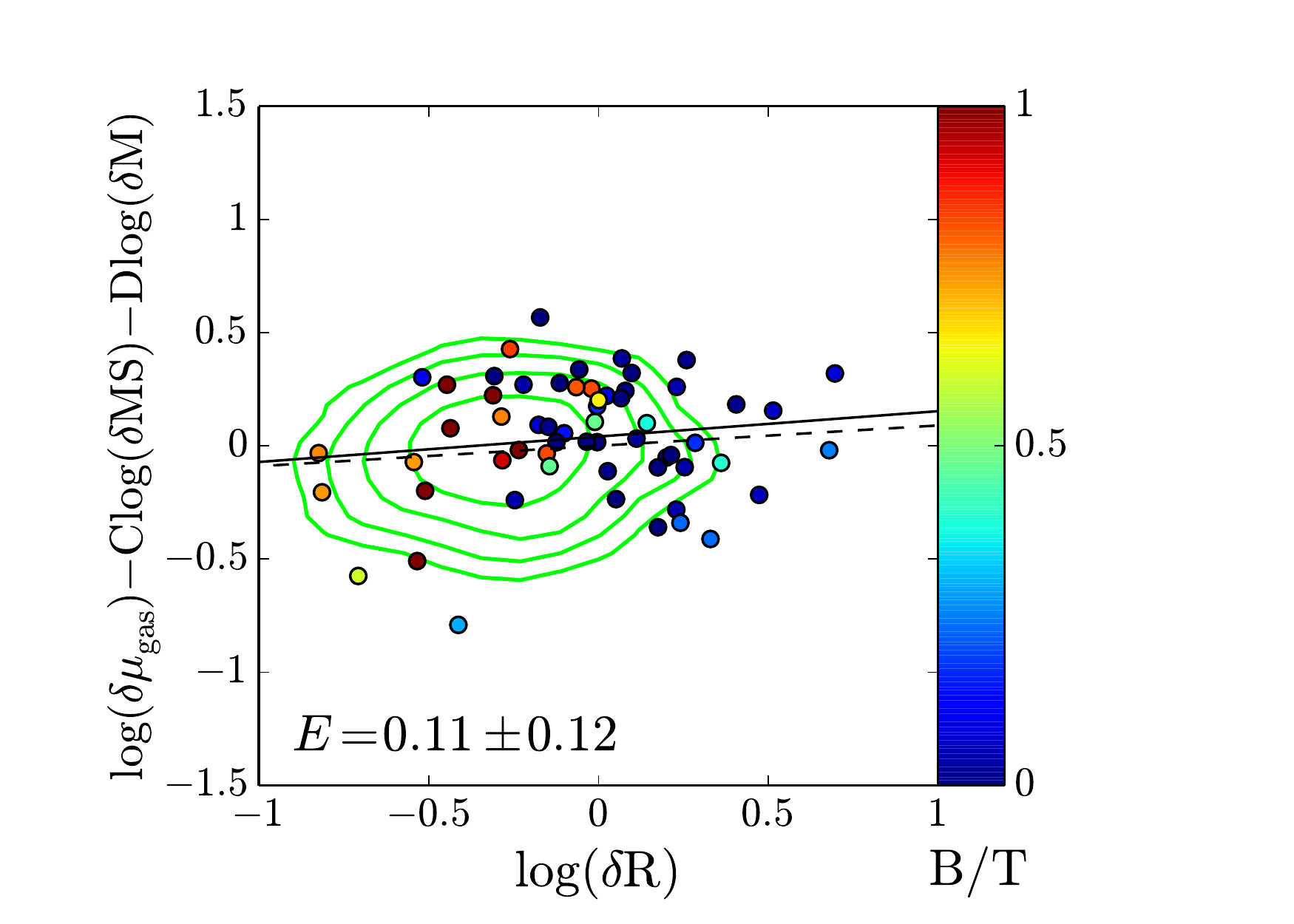}       
        \caption{Dependence of the residual molecular depletion time and gas-to-stellar-mass ratio after subtraction of the redshift dependence on the distance from the MS, stellar mass, and disc size within the PHIBSS2 $z=0.5-0.8$ sample, compared to the dependences derived from the scaling relations obtained by \cite{Tacconi2018}. 
                We assume that the different variables can be separated as in Eqs.~\ref{eq:logy-scaling} and \ref{eq:Dy}, and plot each dependency independently. 
                The black solid lines and the values indicated on the plots refer to the best fits for the $z=0.5-0.8$ subsample, while the green contours and the dashed lines correspond to the comprehensive data set studied by \cite{Tacconi2018} and its fitting formulae ($C=-0.44\pm0.04$, $D=+0.07\pm 0.05$ and $E=+0.12\pm0.12$ for $t_{\rm depl}$ and $C=+0.54\pm0.03$, $D=-0.32\pm 0.03$ and $E=+0.09\pm0.09$ for $\mu_{\rm gas}$). 
        }
        \label{fig:scaling}
\end{figure*}

The PHIBSS2 legacy program provides the largest sample to-date of CO molecular gas measurements at intermediate redshift with its 60 CO(2-1) detections in the range $z=0.5-0.8$. 
As shown in Sect.~\ref{section:Mgas}, the median molecular gas masses, gas-to-stellar-mass ratios, gas fractions, and depletion times obtained for the PHIBSS2 $z=0.5-0.8$ sample are in excellent agreement with the scaling relations established by \cite{Genzel2015} and \cite{Tacconi2018} on a much larger sample of about 1400 sources in the range $z=0-4.5$ including both CO and dust observations. 
These relations, which characterise the dependence of the molecular gas-to-stellar-mass ratio and the depletion time on redshift, stellar mass, MS offset, and galaxy size, constitute the main contribution of the PHIBSS2 program aimed at understanding star formation processes on the MS across cosmic time. 
The molecular gas-to-stellar-mass ratio $\mu_{\rm gas}$ and the depletion time $t_{\rm depl}$ are written as power-law functions of redshift, stellar mass, distance from the MS, and galaxy size such that their logarithms yield
\begin{equation}
\label{eq:logy-scaling}
\log(y) =A + B  \log(1+z)+C  \log({\rm \delta{MS}}) + D  \log({\rm \delta{M}}) + E  \log({\rm \delta{R}}),
\end{equation}
where $A$, $B$, $C,$ and $D$ are constants determined from the observations, ${\rm \delta{\rm MS}} = {\rm sSFR/sSFR}({\rm MS}, z, {\rm M_\star})$ with ${\rm sSFR} ({\rm MS}, z,{\rm M_{\star}})$ the mean sSFR on the MS, $\rm \delta{M} = M_{\star}/5.10^{10}M_\odot$ and $\delta{\rm R} = R_{\rm Sersic}/ R ({\rm MS},z,{\rm M_\star})$ with ${R} ({\rm MS},z,{\rm M_\star})$ the mean half-light radius on the MS, for example from \cite{Vanderwel2014}: ${R} ({\rm MS},z,{\rm M_\star}) = 8.9 \times (1+z)^{-0.75} ({\rm M_\star/M_\odot})^{0.23}~\rm kpc$. 
In particular, while part of the scatter of the MS is due to the stochasticity of the cosmic accretion, the  individual histories and environment, morphology, and fundamental physical quantities of star-forming galaxies, such as their molecular gas fraction and depletion time, vary progressively with $\rm \delta{MS}$. This variation has already been highlighted in different studies \citep[e.g.][]{Schiminovich2007, Wuyts2011b, Saintonge2011b, Saintonge2012, Magdis2012, Huang2014, Genzel2015, Tacchella2016a}, but \cite{Genzel2015} and \cite{Tacconi2018} quantify it precisely through the coefficient~$C$. 
\cite{Tacconi2018} further add a non-linearity in the redshift evolution of the molecular gas fraction to follow the observations more closely, namely considering a redshift evolution of the form $B \left(\log(1+z)-F\right)^\beta$ instead of the linear trend of Eq.~\ref{eq:logy-scaling}, where $F$ and $\beta$ are additional constants. 
We define the residual
\begin{equation}
\label{eq:Dy}
\log(\delta{y})= \log(y)-A-B \times \left(\log(1+z)-F\right)^\beta
,\end{equation}
when the redshift dependence is subtracted from the original quantity $\log(y)$. 
We determine $\delta t_{\rm depl}$ and $\rm \delta \mu_{\rm gas}$ for the galaxies of the PHIBSS2 $z=0.5-0.8$ sample, subtracting the redshift dependence obtained by \cite{Tacconi2018}, and study their dependence on $\rm \delta MS$, $\rm \delta M,$ and $\rm \delta R$.
As can be seen in Fig.~\ref{fig:scaling}, these dependences are in good agreement with the scaling relations of \cite{Tacconi2018}, although with much bigger uncertainties due to our more limited sample. 
Coefficients $C$ and $D$ were obtained through simultaneous linear fits of the redshift-subtracted quantities $\log(\delta{y})$ as a function of $\rm \log(\delta MS)$ and $\rm \log(\delta M)$ while $E$ results from a single linear fit of the residual ($\rm \log(\delta{y})-C\log(\delta MS)-D\log(\delta M)$) as a function of $\rm \log(\delta R)$. The uncertainties are evaluated by assuming a 0.3 dex uncertainty on $\mu_{\rm gas}$ and $t_{\rm depl}$ and 0.2 dex uncertainties on $\rm \delta {\rm MS}$, $\rm \delta {\rm M,}$ and $\rm \delta \rm R$.
This illustrates Appendix A of \cite{Tacconi2018}, which shows from model data sets driven by the actual data that the MS offset and stellar mass dependences of the molecular-gas-to-stellar-mass ratio and depletion time can be recovered from data sets with $\rm N\gtrsim 40$ sources as long as the coverage in $\delta \rm MS$ and $\delta \rm M$ exceeds 1 dex -- which is the case here.

\subsection{Kennicutt-Schmidt relation}
\label{section:KS}

\begin{figure}
        \centering
        \includegraphics[height=1\hsize,trim={0.8cm 1.4cm 3.5cm 2.5cm},clip]{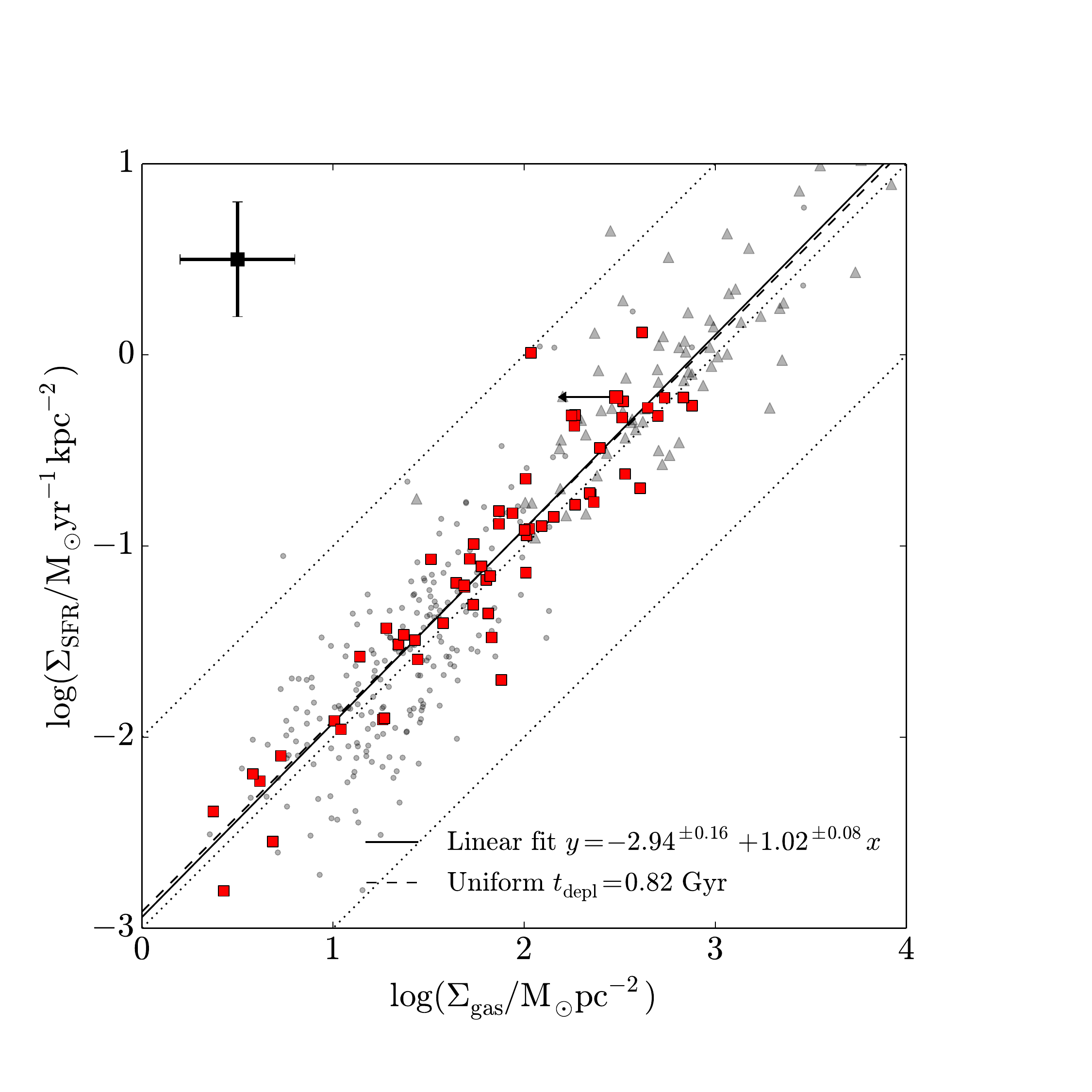}
        \caption{Kennicutt-Schmidt relation for the galaxies of the PHIBSS2 sample at $z=0.5-0.8$. The dotted diagonal lines correspond to constant depletion times of 0.1, 1, and 10 Gyr from top to bottom and the 0.3-dex errors assumed to assess the uncertainties are displayed at the upper left. The observed PHIBSS2 $z=0.4-0.8$ data points are indicated by squares and the upper limit by an arrow. 
                The underlying grey points correspond to COLDGASS data \citep{Saintonge2011,Saintonge2012} and the grey triangles to PHIBSS \citep{Tacconi2013}. 
                The black solid line corresponds to a linear least-square fits to the data points; the dashed lines to a uniform depletion time corresponding to the best-fitting value on the KS diagram.
                The Pearson correlation coefficient is 0.94 while the standard deviation from the linear fit is 0.24 dex. 
        }
        \label{fig:ks}
\end{figure}

In Fig.~\ref{fig:ks}, we plot the KS relation  between SFR and molecular gas mass surface densities $\Sigma_{\rm SFR} = 0.5~ {\rm SFR}/\pi R_{\rm Sersic}^2$ and $\Sigma_{\rm gas} = 0.5~ {\rm M_{gas}}/\pi R_{\rm Sersic}^2$ within the $z=0.5-0.8$ PHIBSS2 sample. 
A linear least-square fit to the data yields a KS exponent $N=1.02 \pm 0.08$ assuming 0.3-dex uncertainties in both SFR and molecular gas surface densities. This strikingly linear relation with Pearson correlation coefficient $r=0.94$ corresponds to a uniform depletion time of $t_{\rm depl} = 0.82\rm ~ Gyr$ in line with the \cite{Tacconi2018} scaling relations, the residual scatter being 0.24 dex. 
We also tested definitions of $\Sigma_{\rm SFR}$ and $\Sigma_{\rm gas}$ using the disc radius $R_{\rm d}$ instead of $R_{\rm Sersic}$, yielding similar results.
%
The main contributor to the 0.24 dex scatter may be the different evolutionary stages of the molecular clouds within galaxies and their dynamical environment \citep[e.g.][]{Lada2010, Lombardi2010, Onodera2010, Schruba2010,  Murray2010, Murray2011, Zamora2012, Zamora2014, Meidt2013, Davies2014, Kruijssen2014, Utomo2015,Kruijssen2018}. 
Alternatively, regions within a single galaxy could have different star-formation efficiencies \citep[e.g.][]{Freundlich2013, Cibinel2017} and the conversion factors used to determine the molecular gas mass and the SFR may also vary from region to region or between galaxies \citep[e.g.][]{Israel1997, Bolatto2013}. 
Furthermore, observations in the Milky Way and nearby galaxies reveal that the properties of the molecular gas in GMCs vary considerably from the disc to the central region of a galaxy, in particular in the presence of strong bars \citep[e.g. ][]{Oka2001, Regan2001, Jogee2005, Shetty2012, Kruijssen2013, Colombo2014, Leroy2015, Freeman2017}. This variety, which is also expected for the galaxies of the PHIBSS2 sample, is likely to contribute to the scatter in the KS relation since $\Sigma_{\rm gas}$  and $\Sigma_{\rm SFR}$ do not always probe the same regions within a galaxy. 

\begin{figure}
        \centering
        \includegraphics[height=0.52\hsize,trim={1.25cm 0 4.5cm 0},clip]{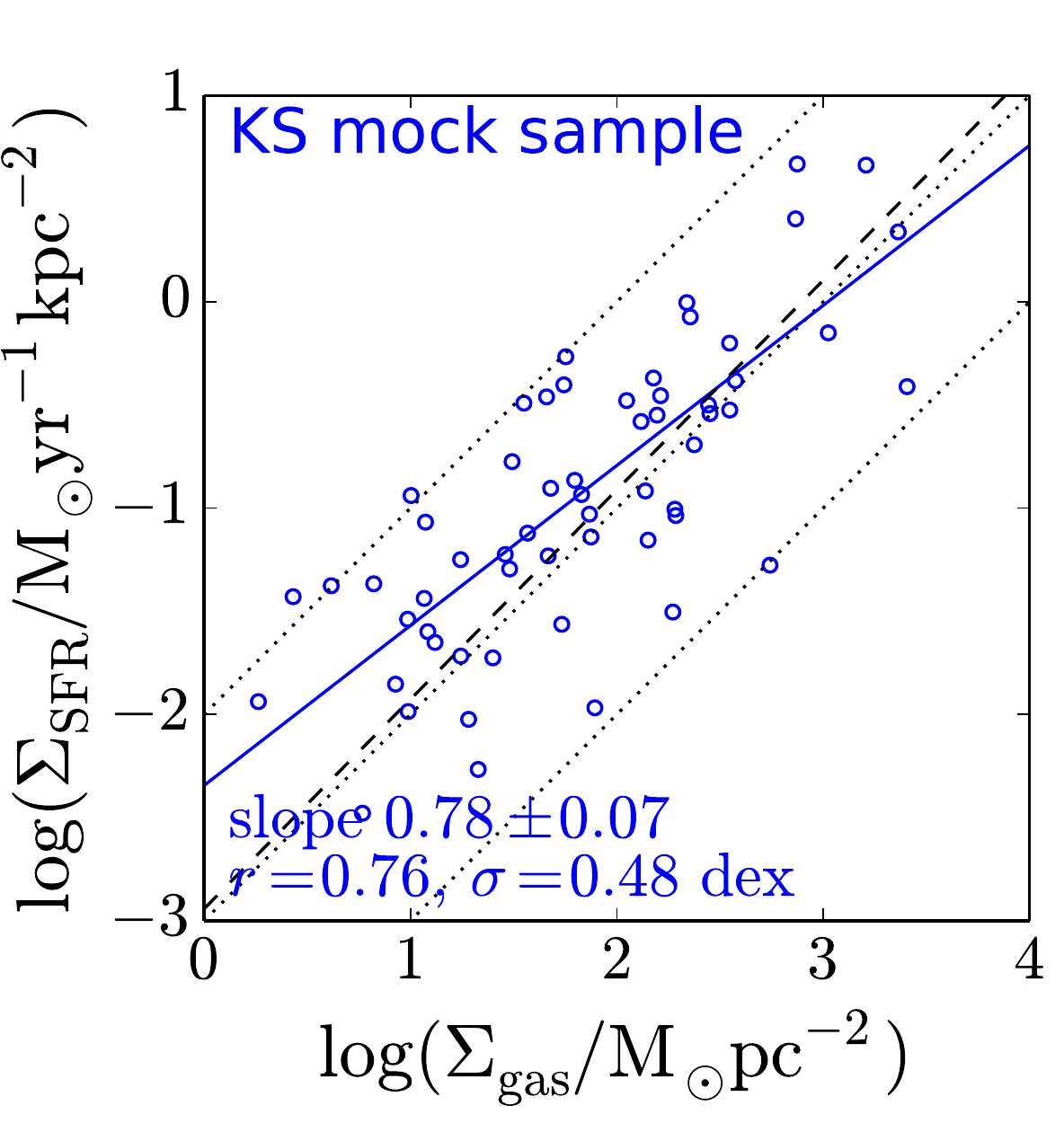}
        \hfill
        \includegraphics[height=0.52\hsize,trim={1.25cm 0 4.5cm 0},clip]{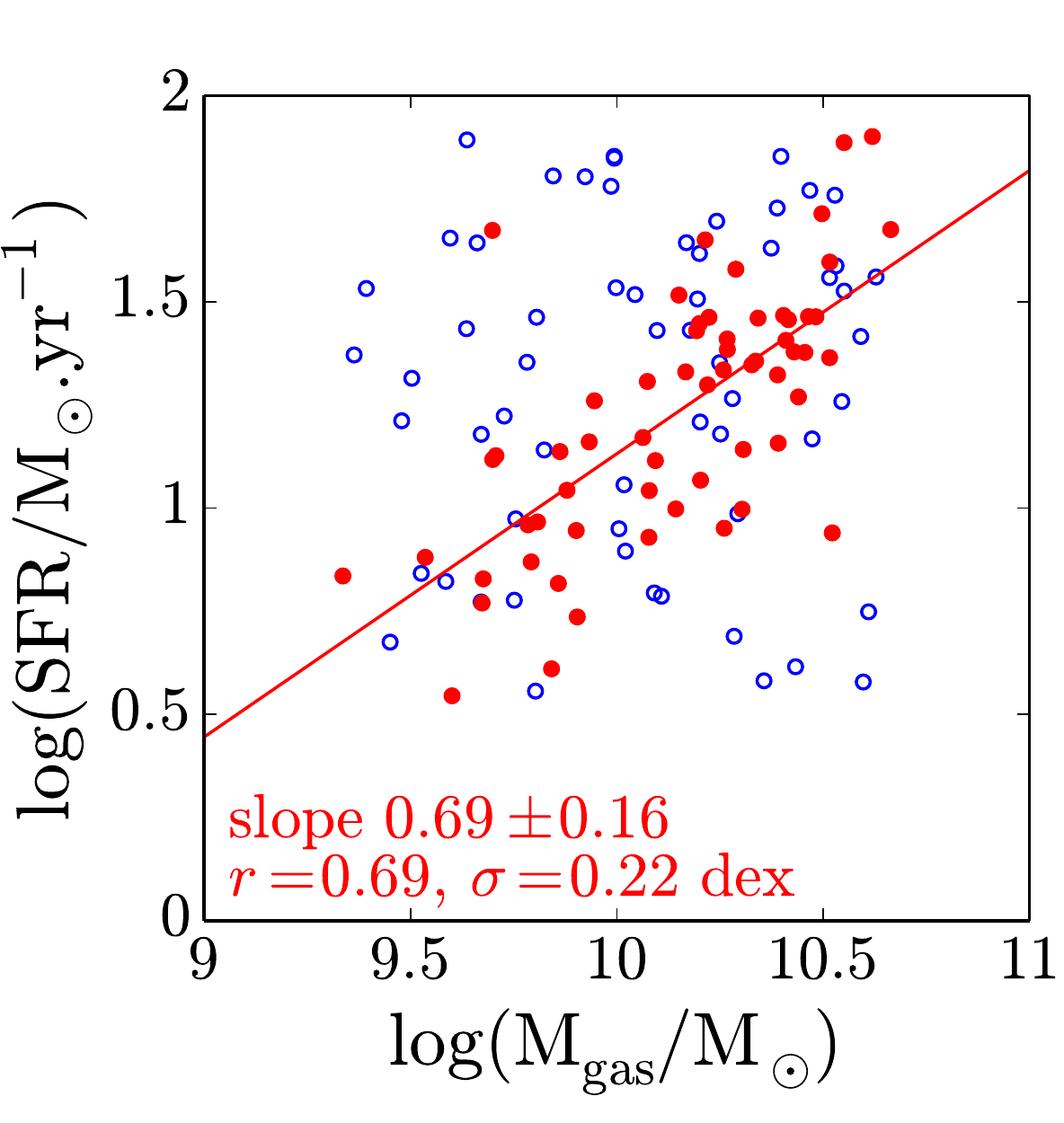}
        \caption{
                \textit{Left: } Kennicutt-Schmidt relation for a mock sample where the molecular gas mass and the SFR are not correlated and their logarithms uniformly distributed between the extrema of the PHIBSS2 $z=0.5-0.8$ sample, assuming a Gaussian distribution for $\log(R_{\rm Sersic}/\rm kpc)$ with mean and standard deviation corresponding to those of the PHIBSS2 sample.
                The solid blue line corresponds to a fit to the mock data, the dashed line recalls the best fit from Fig.~\ref{fig:ks}, and the dotted lines indicate uniform depletion times of 0.1, 1, and 10 Gyr from top to bottom. 
                \textit{Right: } Correlation between the SFR and the molecular gas mass of the PHIBSS2 $z=0.5-0.8$ together with the corresponding distribution of the non-correlated mock sample. The PHIBSS2 sample and the resulting fit are indicated in red, while the mock sample is displayed as open blue circles.
        }
        \label{fig:KS-fake}
\end{figure}

In the absence of separate size estimates for the SFR and molecular gas distributions, we use the half-light radius obtained from a single Sérsic fit $R_{\rm Sersic}$ to estimate both surface densities, which are consequently not independent from each other. To ensure that the striking correlation between $\Sigma_{\rm SFR}$ and $\Sigma_{\rm gas}$ from Fig.~\ref{fig:ks} does not stem from the dependence between the two variables, we study the KS relation that would be obtained for non-correlated uniform distributions of $\rm \log(M_{\rm gas}/M_\odot)$ and $\log(\rm SFR/M_\odot.yr^{-1})$ between the extrema of the PHIBSS2 $z=0.5-0.8$ sample. Figure~\ref{fig:KS-fake} shows that such distributions would yield a much greater scatter of about 0.50 dex and a less linear relation. 
While the figure shows the result for one such distribution, we confirmed the trend both in slope and scatter by reproducing the experiment 1000 times, obtaining slopes, Pearson correlation coefficients, and scatters of $0.6 \pm 0.1$, $0.62 \pm 0.08,$ and $0.48\pm 0.04,$ respectively.
We also note that sticking to the molecular gas mass and the SFR, which are independent variables unlike their surface densities, yields a clear correlation with a slope of $0.81 \pm 0.14$, a Pearson coefficient of 0.77, and a residual standard deviation 0.18 dex: the striking KS relation we obtain in Fig.~\ref{fig:ks} does not result from an artificial correlation induced by the dependency of both surface densities on galaxy size. 
We do however notice from Fig.~\ref{fig:KS-fake} that part of the correlation in the KS diagram is due to our selection of MS galaxies excluding starbursts and quenched galaxies and the fact that both surface densities are not independent variables. These issues are relevant for most KS studies and are not specific to the study presented here.

\subsection{Star formation and morphology}
\label{section:morphology}

\begin{figure*}
        \centering
        \includegraphics[height=0.24\hsize,trim={1.cm .cm 4.85cm 0.7cm},clip]{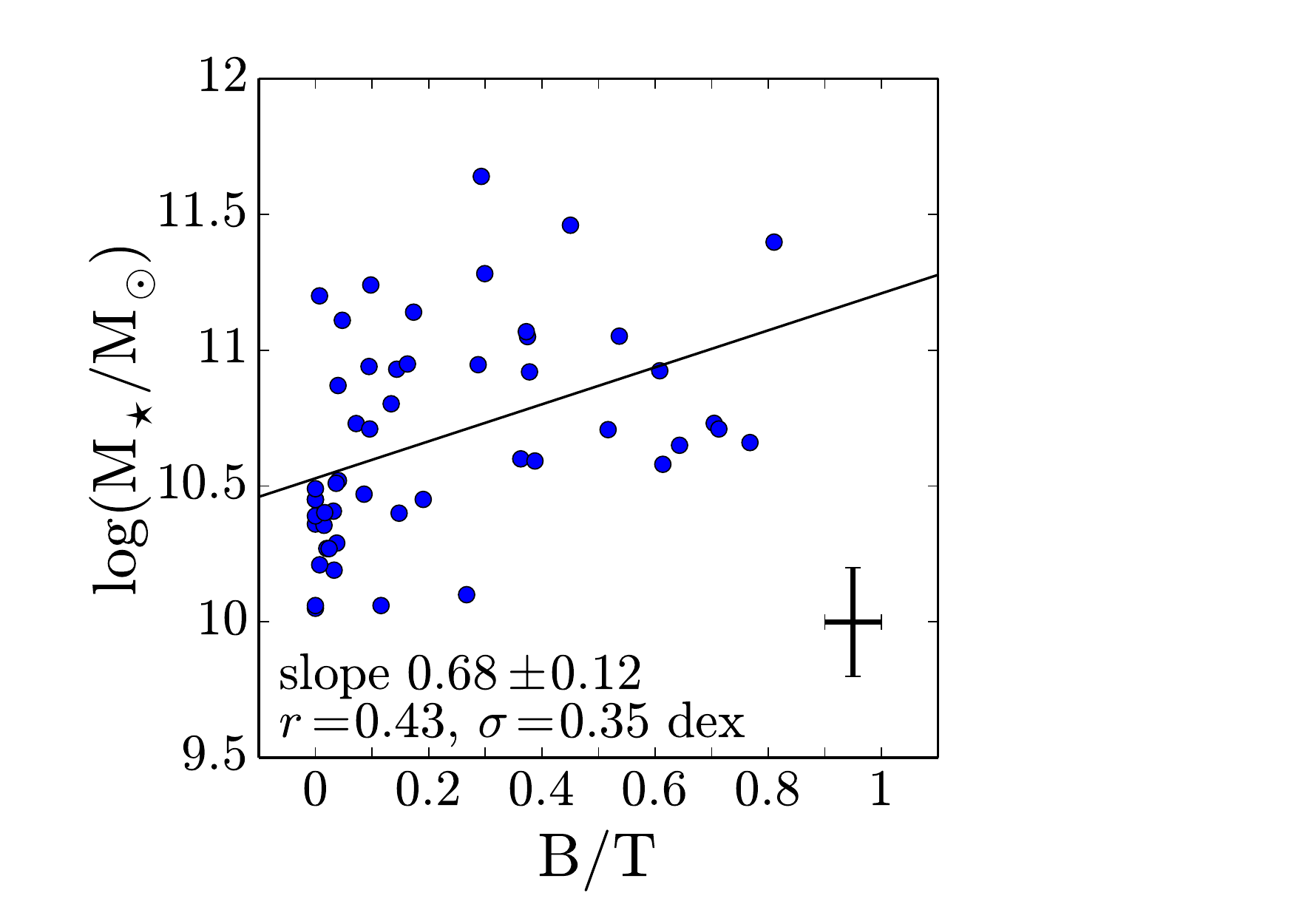}
        \hfill  
        \includegraphics[height=0.24\hsize,trim={1.cm .cm 4.85cm 0.7cm},clip]{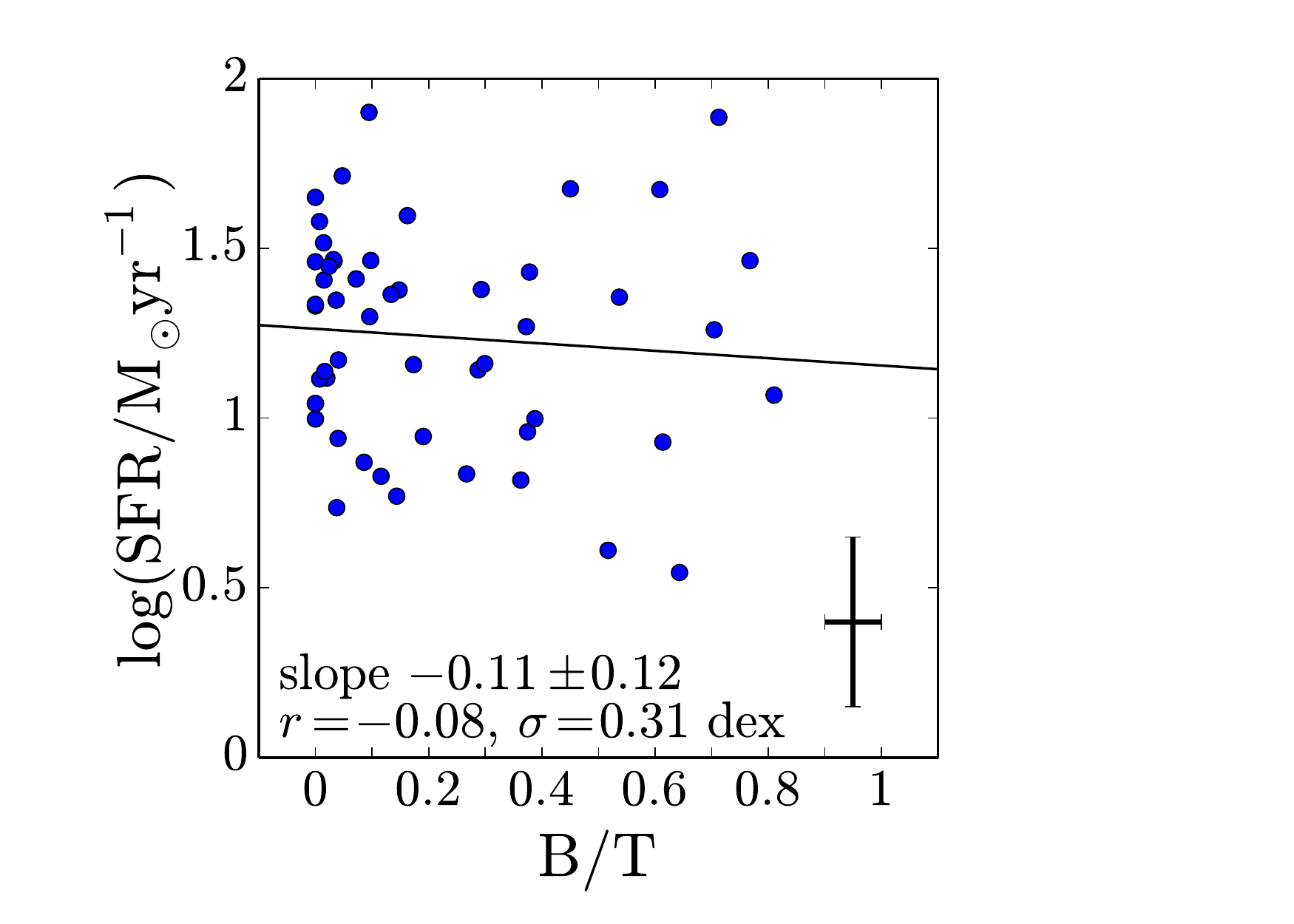}    
        \hfill   
        \includegraphics[height=0.24\hsize,trim={1.cm .cm 4.85cm 0.7cm},clip]{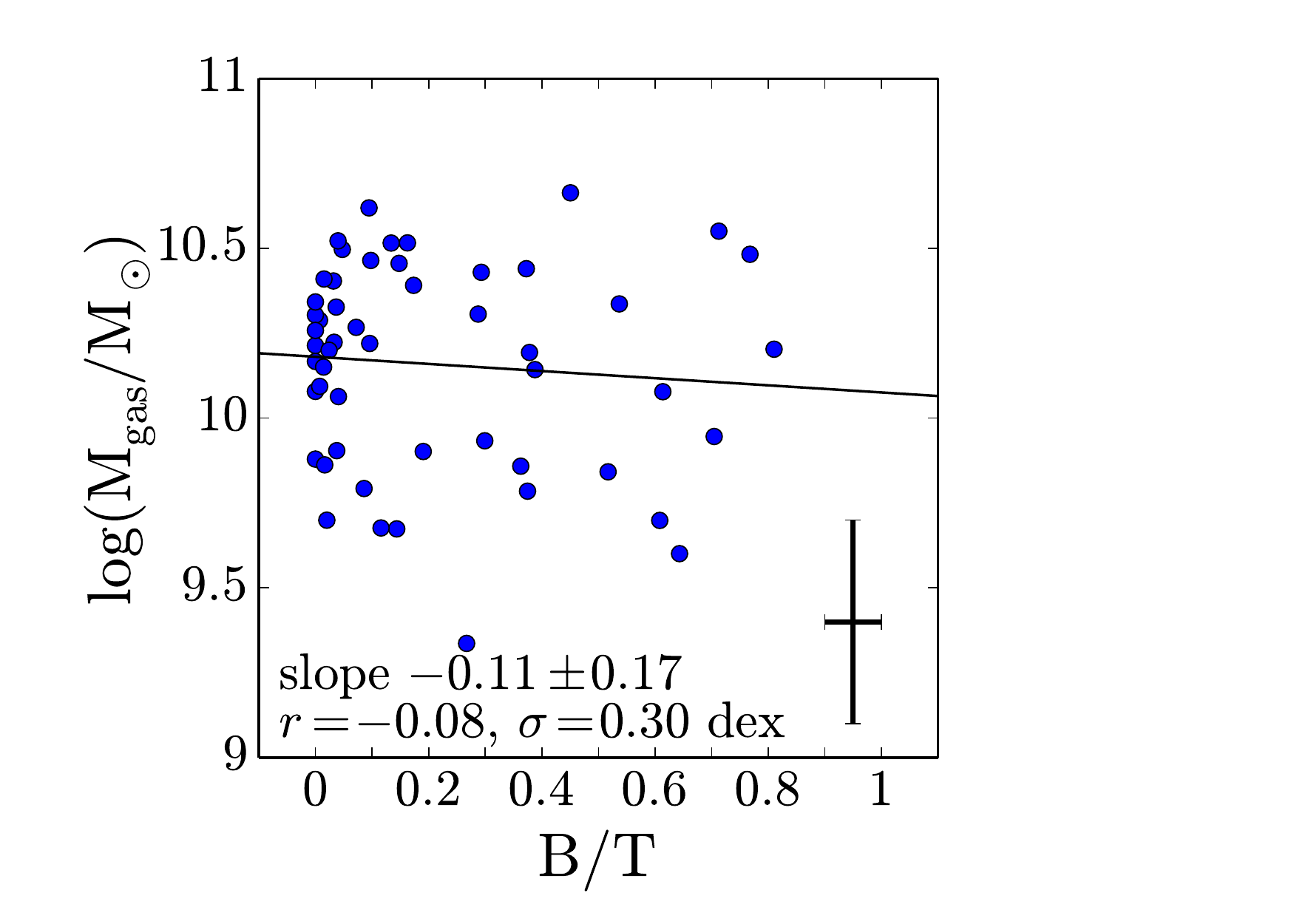}
        \hfill 
        \includegraphics[height=0.24\hsize,trim={1.cm .cm 4.85cm 0.7cm},clip]{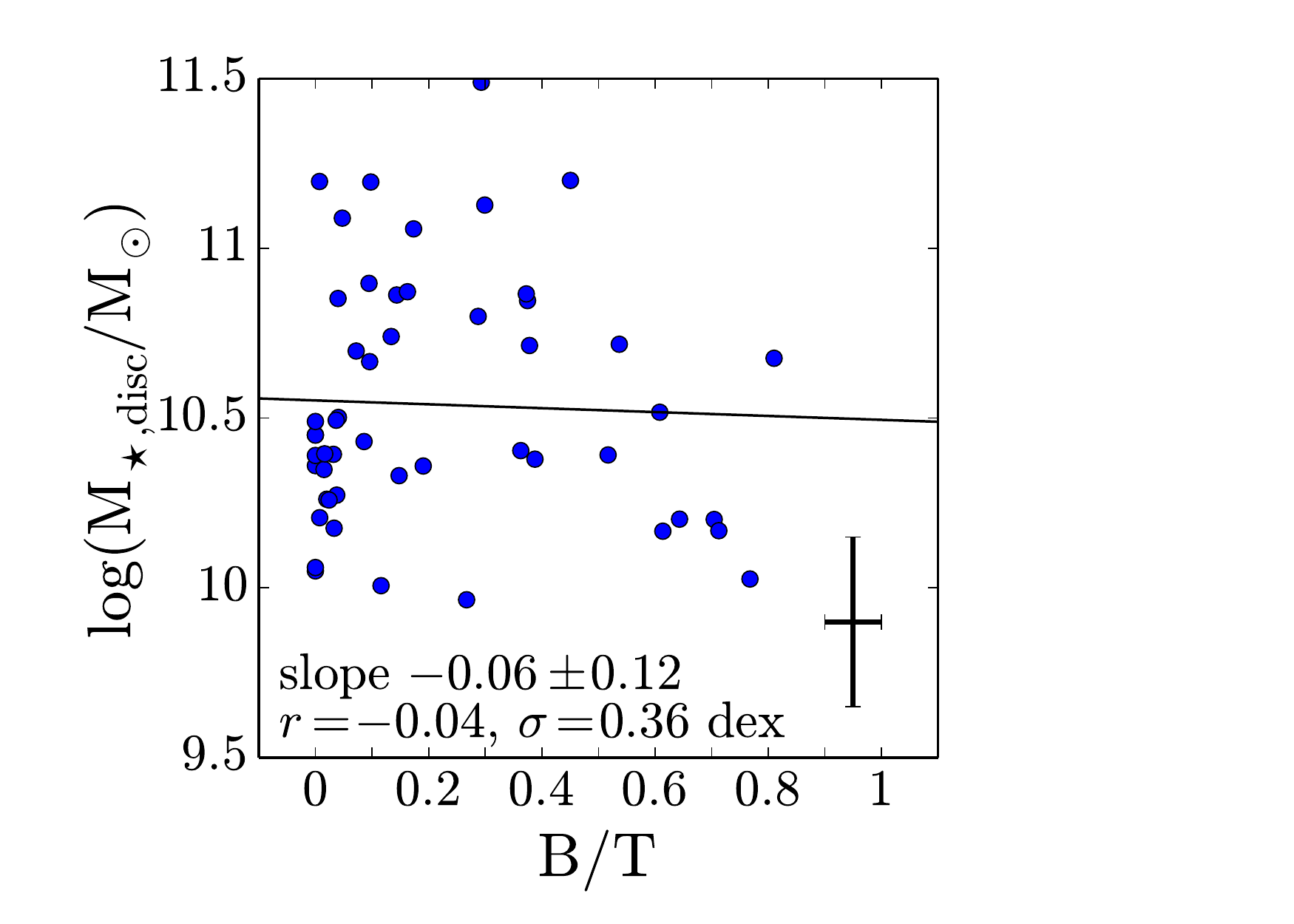}
        \\                 
        \includegraphics[height=0.24\hsize,trim={1.cm .cm 4.85cm 0.7cm},clip]{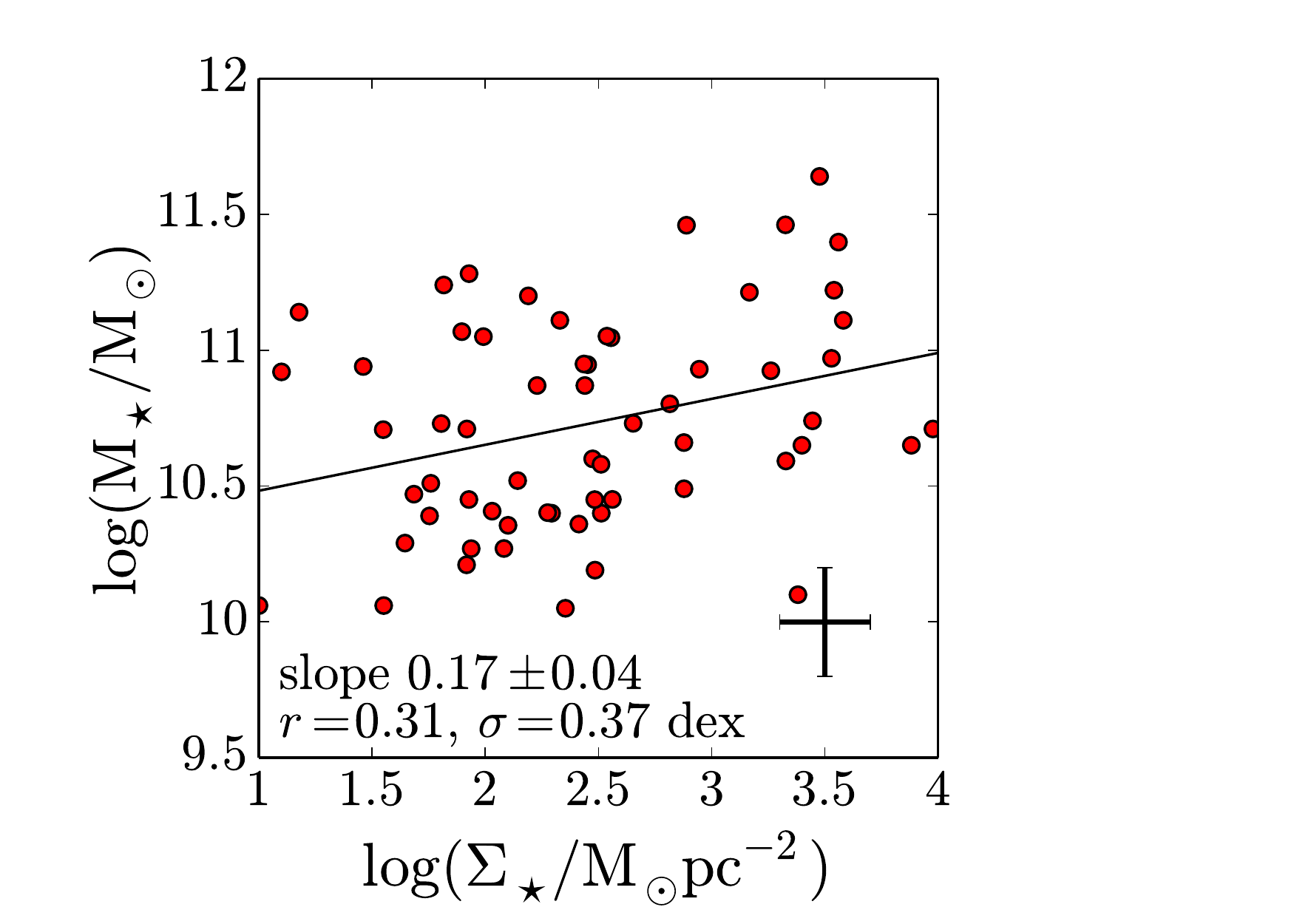}
        \hfill  
        \includegraphics[height=0.24\hsize,trim={1.cm .cm 4.85cm 0.7cm},clip]{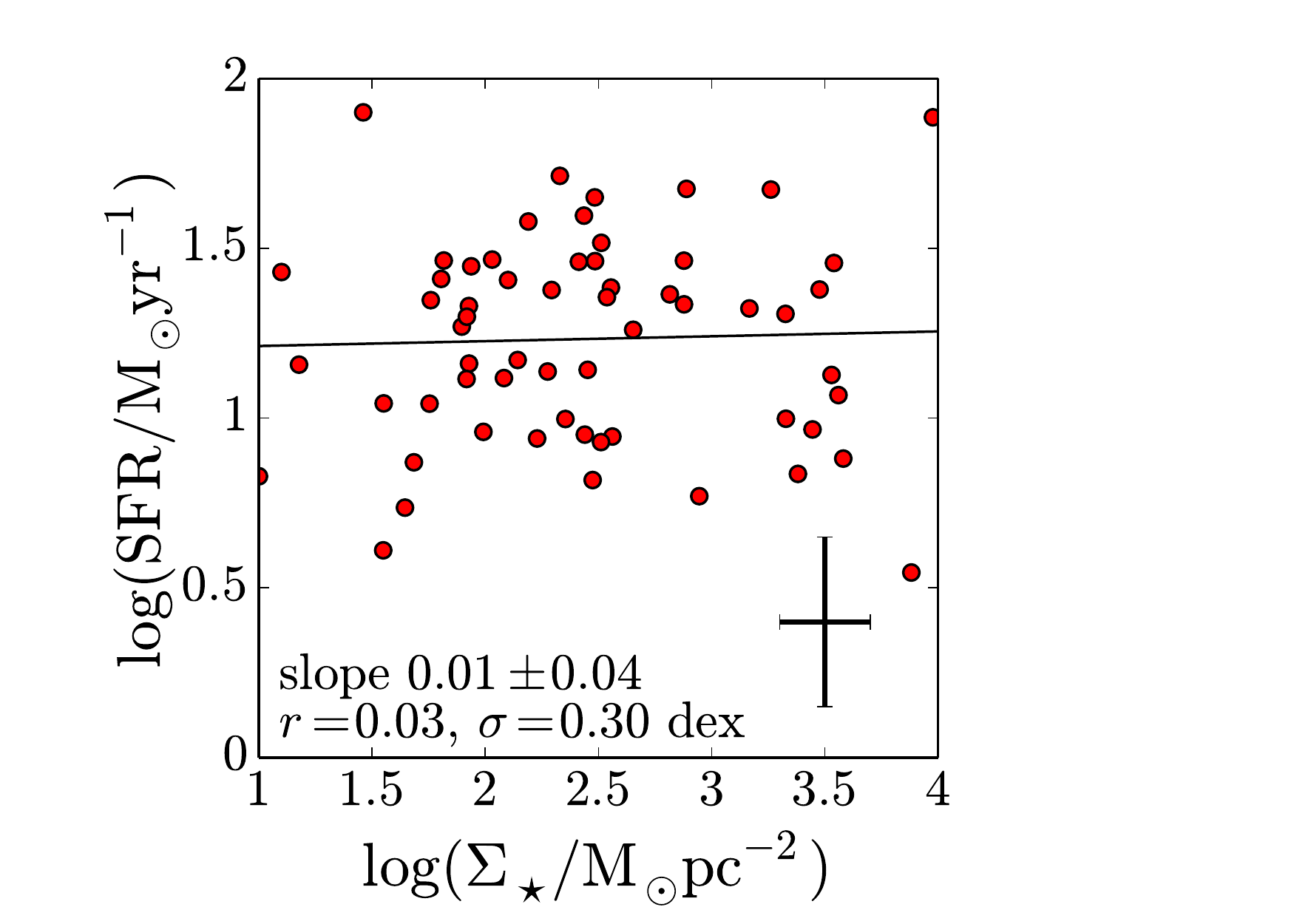} 
        \hfill   
        \includegraphics[height=0.24\hsize,trim={1.cm .cm 4.85cm 0.7cm},clip]{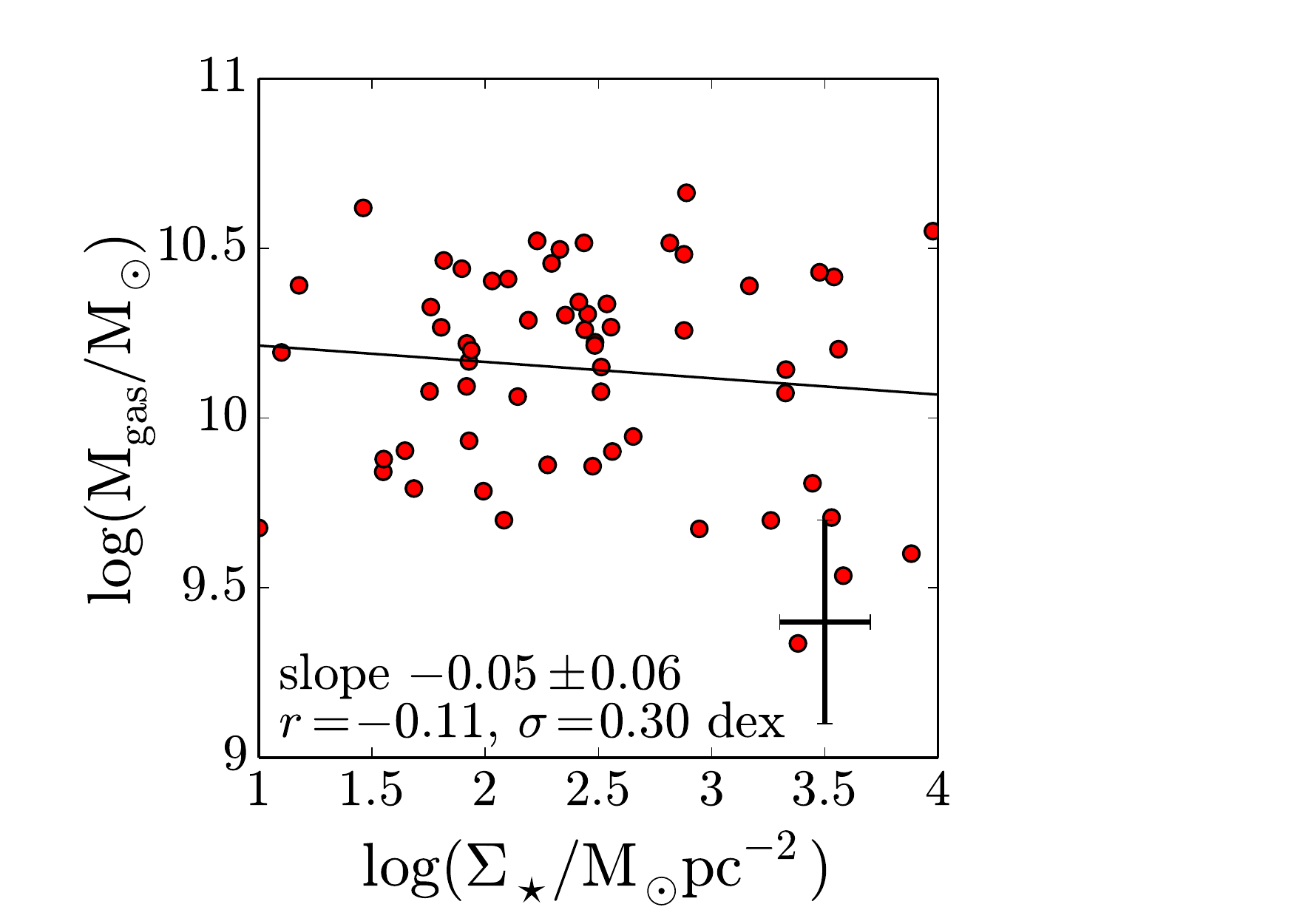}
        \hfill 
        \includegraphics[height=0.24\hsize,trim={1.cm .cm 4.85cm 0.7cm},clip]{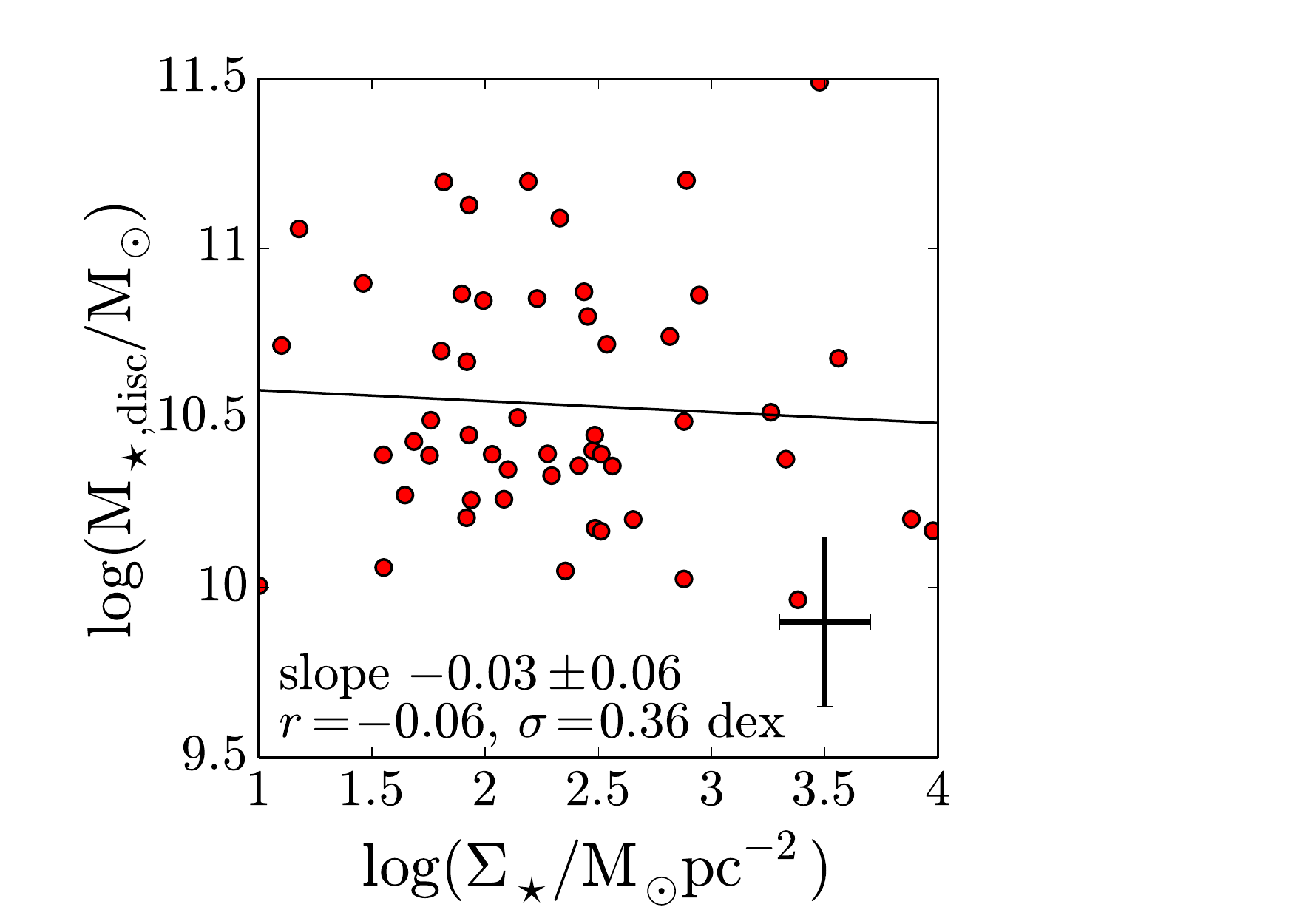}
        \\  
        \caption{Dependence of different galaxy parameters on the bulge-to-total ratio $\rm B/T$ within the PHIBSS2 sample at $z=0.5-0.8 $.
                In each panel, we carry out a linear least-square fit shown as a black solid line and indicate its slope, assuming $0.3\rm$-dex errors on the different quantities, as well as the Pearson correlation coefficient ($r$) and the scatter of the residuals of the best-fitting linear relation ($\sigma$). 
                Assumed errors are indicated in the lower-right corners. 
        }
        \label{fig:BT}
\end{figure*}

To address how morphology affects star formation within the PHIBSS2 $z=0.5-0.8$ sample, we investigate how global physical parameters such as the stellar mass $\rm M_\star$, the molecular gas mass $\rm M_{gas}$, and the SFR, as well as derived quantities like the sSFR, the molecular gas depletion time $t_{\rm depl}$, and the gas-to-stellar mass ratio $\rm \mu_{gas}$, depend on the bulge-to-total ratio B/T and the total stellar surface density $\Sigma_{\star} = 0.5 M_{\star}/\pi R_{\rm Sersic}^2$. 
\cite{Shi2011} notably show from a large sample of galaxies at different redshifts that the depletion time is a decreasing function of $\Sigma_{\star}$, which can be understood both in terms of the stellar contribution to the gravitational potential in which stars form and in terms of disc hydrostatic pressure acting on star-forming regions. A high disc pressure indeed enhances the production of $\rm H_2$ molecular gas from $\rm HI$ atomic gas and hence contributes to balance stellar feedback \citep{Blitz2004,Schaye2008, Shi2011}.  
%
To quantify the correlations between the different parameters with $\rm B/T$ and $\Sigma_{\star}$, we carry out linear least-square fits with errors on both axes, determine the Pearson correlation coefficient $r$ between them, and indicate the scatter $\sigma$ of the residuals. 
$\rm B/T$ and $\Sigma_{\star}$ are themselves correlated, with a Pearson correlation coefficient of $r=0.67$. 
As mentioned in Sect.~\ref{section:galfit}, most fits with $\rm B/T=1$ correspond to cases where the inner structure of the galaxy includes spiral arms not well accounted for; we therefore exclude these points from the correlations. 
As shown in Fig.~\ref{fig:BT}, we find that while the stellar mass increases with B/T and $\Sigma_{\star}$ (respectively with $r = 0.43$ and $0.31$), the SFR and the molecular gas mass do not seem to depend on these parameters (with $\lvert r \lvert \lesssim 0.10$). Derived quantities are consistent with these trends: the sSFR and $\rm \mu_{gas}$ decrease both with B/T and $\Sigma_{\star}$ (respectively with $r=-0.45$ and $-0.49$ with B/T, $-0.26$ and $-0.38$ with $\Sigma_{\star}$) while $t_{\rm depl}$ displays no correlation with B/T ($\lvert r \lvert < 0.05$) and a very weak negative correlation of slope $-0.06$ with $\Sigma_{\star}$ ($r=-0.17$). 
%
We also introduce 
\begin{equation}
\rm M_{\rm \star,disc}= (1-{\rm B/T}) \times M_{\star}
,\end{equation}
the stellar mass within the disc, which does not correlate with $\rm B/T$ or $\Sigma_{\star}$ ($\lvert r \lvert < 0.10$).
%
As can be seen in Fig.~\ref{fig:images}, the goodness of the best-fit \texttt{galfit} model varies from one galaxy to another, which could affect the correlations with $\rm B/T$ and $\Sigma_{\star}$. To test how the goodness-of-fit affects these  correlations, we also determine Pearson correlation coefficients weighted by the reduced $\chi^2$ of the best-fit models. We find no significant deviation from the trends indicated above (namely, the weighted correlation coefficients with $\rm B/T$ and $\Sigma_{\star}$ are respectively $0.37$ and $0.29$ for $\log(\rm M_\star/M_\odot)$, $-0.11$ and $0.11$ for $\log(\rm SFR/M_\odot yr^{-1})$, $-0.15$ and $-0.04$ for $\log(\rm M_{\rm gas}/M_\odot)$, $-0.08$ and $-0.12$ for $\log(\rm M_{\star,\rm disc}/M_\odot)$), 
advocating relatively robust correlations. 
%
The decrease of the sSFR with B/T, which was also observed at low-redshift by \cite{Saintonge2012}, can be either interpreted as a decreasing sSFR with bulge growth, or as a consequence of the fact that B/T traces the fraction of stars that formed early and are now part of the bulge while the sSFR traces on the contrary the fraction of stars that formed recently. 
%
Similarly, the total molecular-gas-to-stellar-mass ratio and the corresponding gas fraction decrease with B/T and $\Sigma_{\rm \star}$ in accordance to low- and high-redshift measurements where morphology is probed by the concentration parameter and the Sérsic index \citep{Saintonge2011,Papovich2015}. 
Assuming an evolutionary sequence from small to high B/T for the same objects, the fact that the molecular gas mass does not vary with B/T and $\Sigma_{\rm \star}$ suggests an ongoing supply of fresh molecular gas while the stellar bulge assembles, which could stem from mergers, infall from the cosmic web through streams of cold gas penetrating inside the hot circumgalactic medium \citep{Dekel2009}, or from efficient transformation from atomic to molecular gas owing to the pressure increase \citep{Blitz2004}. 
%
This continuing supply of gas would be reflected on the disc gas-to-stellar-mass ratio $\mu_{\rm gas,disc} = \rm  M_{\rm gas}/M_{\rm \star,disc}$, which neither correlates with B/T nor  $\Sigma_{\rm \star}$ ($\lvert r \lvert < 0.05$). 
Without invoking an evolutionary sequence, the absence of correlation for the molecular gas mass, the SFR, and hence the depletion time with B/T might indicate relatively uniform star formation processes in a given redshift bin, irrespective of the past history of star formation traced by B/T. 

Contrarily to morphological quenching scenarios \citep{Martig2009} and observations in the nearby universe \citep{Saintonge2011b}, we do not observe any variation of the depletion time with B/T or $\Sigma_{\rm \star}$  ($\lvert r \lvert < 0.05$). 
%
This variation is expected to be more pronounced when the gas fraction drops below 20\% \citep{Martig2009, Gobat2017}, so we separately searched for it in galaxies with $f_{\rm gas} < 20\%$ but did not observe any significant variation of the depletion time with B/T or $\Sigma_{\rm \star}$. 
%
Part of this negative result may come from the fact that \cite{Gobat2017} rely on dust observations encompassing both molecular and atomic gas, while we only have access to the molecular gas. When B/T increases, more gas may remain atomic instead of molecular as the bulge stabilises the disc against gravitational collapse and fragmentation. This means that the total depletion time including both molecular and atomic components may increase without any increase of the molecular gas depletion time. 
More importantly, \cite{Gobat2017} focus on quenched red and dead galaxies well below the MS while PHIBSS2 galaxies are precisely on and around the MS. By selecting star-forming galaxies, we may exclude those with high depletion times. Furthermore, although our sample does include bulge-dominated galaxies, those are still on the MS and are therefore potentially atypical with relatively high SFR. For example, they could have become bulge-dominated from mergers or violent disc instabilities recently, both processes also being able to trigger star formation. Morphological quenching and compaction events take time to settle down \citep{Gobat2017,Dekel2017} and may therefore not be observed in the recent bulge-dominated galaxies of our sample.


\section{Conclusion}
\label{section:conclusion}

This paper presents the strategy and the $z=0.5-0.8$ results of the PHIBSS2 survey, a four-year legacy program with the IRAM NOEMA interferometer designed to investigate early galaxy evolution from the perspective of the molecular gas reservoirs. This survey builds upon the successful PHIBSS program \citep{Tacconi2010,Tacconi2013}, which uncovered high gas fractions near the peak epoch of star formation and showed that the cosmic evolution of the SFR was mainly driven by the molecular gas reservoirs. 
The PHIBSS and PHIBSS2 surveys probe a representative sample of star-forming MS galaxies drawn from well-studied parent catalogues in the COSMOS, AEGIS, and GOODS-North cosmological deep fields.
%
While PHIBSS focused on galaxies at $z=1.2$ and $2.2$ on and above the MS, PHIBSS2 significantly enlarges the sample by probing the build-up epoch at $z>2$, the winding-down of star formation at $z<0.8,$ and galaxies below the MS at $z=1-1.6$. It aims at homogeneous coverage of the MS in the $M_{\rm \star}$-SFR plane without morphological selection (Figs.~\ref{fig:sample} and \ref{fig:histograms}). 
With a total of more than 120 sources, PHIBSS2 significantly adds to the number of molecular gas observations above $z=0.5$. Together with PHIBSS, it thus provides a benchmark sample of near MS galaxies at different redshifts with molecular gas measurements, which can be used for further CO and dust continuum follow-ups at high resolution with ALMA and NOEMA as well as for other complementary observations. 

In this paper we present the CO(2-1) molecular gas observations obtained at $z=0.5-0.8$ as part of the PHIBSS2 survey, reporting 60 detections from a sample of 61 galaxies (Tables~\ref{table:1}~and~\ref{table:2}).
We determine the molecular gas masses, gas fractions, gas-to-stellar-mass ratios, and depletion times of these galaxies and carry out single Sérsic and two-component bulge disc fits with the 2D morphology fitting code \texttt{galfit} to obtain the half-light radii of the galaxies and their bulge and disc components as well as their bulge-to-total-luminosity ratios and molecular gas mass and SFR surface densities. 
%
The molecular gas-to-stellar-mass ratio, gas fraction, and depletion time, respectively, yield values in the ranges $0.03-1.79$, $0.03-0.64,$ and $0.11-3.82\rm~Gyr$ with medians $\widetilde{\mu_{\rm gas}} = 0.28 \pm 0.04$, $ \widetilde{f_{\rm gas}} = 0.22 \pm 0.02,$ and $\widetilde{t_{\rm depl}} = 0.84\pm 0.07~\rm Gyr$ (Table~\ref{table:3} and Fig.~\ref{fig:hist-tdepl-fgas}). These values are consistent with the observed increase of the gas fraction and slight decrease of the depletion time with redshift \citep{Tacconi2013, Genzel2015,Tacconi2018}. 
%
They are indeed in excellent agreement with the scaling relations of the depletion time and the gas fraction as a function of stellar mass, offset from the MS, and galaxy size established by \cite{Tacconi2018} within a much more comprehensive sample of about 1400 galaxies between $z=0$ and $z=4.6$ (Fig.~\ref{fig:scaling}).
%
We show that the Kennicutt-Schmidt relation between molecular gas and SFR surface densities within our sample is strikingly linear (Figs.~\ref{fig:ks}~and~\ref{fig:KS-fake}), which argues in favour of uniform star-formation timescales within galaxies at any given epoch.
%
In terms of morphology, we study the dependence of different global parameters including the depletion time and the molecular gas fraction on the bulge-to-total ratio B/T and the stellar surface density $\Sigma_{\rm \star}$ (Table~\ref{table:4} and Fig.~\ref{fig:BT}). In particular, the total molecular gas mass, the SFR, the disc stellar mass, and the disc molecular gas fraction do not seem to depend on either B/T or  $\Sigma_{\rm \star}$. 
This either suggests an ongoing supply of fresh gas to the disc while the stellar bulge assembles or that star formation proceeds irrespectively of the past history of star formation traced by B/T. 
We find no strong evidence for morphological quenching, which we would expect to manifest as a dependence of the molecular gas depletion time on B/T and $\Sigma_{\rm \star}$. Our sample, however, only focuses on star-forming galaxies within the scatter of the MS; probing morphological quenching might require including galaxies well below it.
Therefore, the analysis presented here should not be interpreted as evidence against morphological quenching in general.


\begin{acknowledgements}
        The authors acknowledge the particularly careful reading of the anonymous referee, which greatly contributed to improving this article.
        J.~F. would like to thank Beno\^it Epinat and Marc Huertas Company for their advice using Galfit, and Avishai Dekel, Sandy Faber, Joel Primack, Sandro Tacchella, Sharon Lapiner and Casey Papovich for interesting discussions and insights. 
        The observations presented here were carried out at the IRAM Plateau de Bure/NOEMA interferometer and we are grateful to the astronomers on duty and telescope operators for the quality of the data. 
        This study makes use of data from AEGIS, a multi-wavelength sky survey conducted with the Chandra, GALEX, Hubble, Keck, CFHT, MMT, Subaru, Palomar, Spitzer, VLA, and other telescopes and supported in part by the NSF, NASA, and the STFC.
        It is also based on data products from observations made with ESO Telescopes at the La Silla Paranal Observatory under ESO programme ID 179.A-2005 and on
        data products produced by TERAPIX and the Cambridge Astronomy Survey
        Unit on behalf of the UltraVISTA consortium.
This study further uses observations made with the Spitzer Space Telescope, which is operated by the Jet Propulsion Laboratory, California Institute of Technology under a contract with NASA. 
        Finally, this work was based in part on observations made with the NASA/ESA Hubble Space Telescope, and obtained from the Hubble Legacy Archive, which is a collaboration between the Space Telescope Science Institute (STScI/NASA), the Space Telescope European Coordinating Facility (ST-ECF/ESA), and the Canadian Astronomy Data Centre (CADC/NRC/CSA).
\end{acknowledgements}

\bibliographystyle{aa} 
\bibliography{freundlich_phibss2} 

\begin{appendix}

\section{HST images and NOEMA CO spectra} 
\label{appendix:HST}

Fig.~A.1 presents the HST/ACS F814W I-band images, the best-fit two-component bulge disc models obtained with the method outlined in Sect. \ref{section:galfit} and the corresponding residuals, the radial density profiles, and the NOEMA CO(2-1) molecular gas spatially averaged line spectra for the 61 galaxies of the PHIBSS2 $z=0.5-0.8$ sample.

\newcommand{\captiontext}{
        HST/ACS F814W I-band image, best-fit two-component bulge disc model, residuals, averaged radial profile and CO spectrum for the different galaxies of the PHIBSS2 $z=0.5-0.8$ sample. For each galaxy, the HST image, the model, the residuals, and the radial profile have the same arbitrary log-scale units.
        The red and dashed blue ellipses respectively denote the disc and bulge half-light radii, the dashed lines the disc axes, and the scale at the bottom left corresponds to 10 kpc. In the light profiles, the solid red line is the averaged profile from the HST image while the dashed blue line that of the model. 
        Green ellipses correspond to neighbouring satellites or companions that were simultaneously fitted as single Sérsic light distributions. 
        The CO spectra display a Gaussian fit to the data.
}

\begin{figure*}
        \flushleft
        \includegraphics[height=0.175\textwidth,clip,trim=0 0 4cm 0]{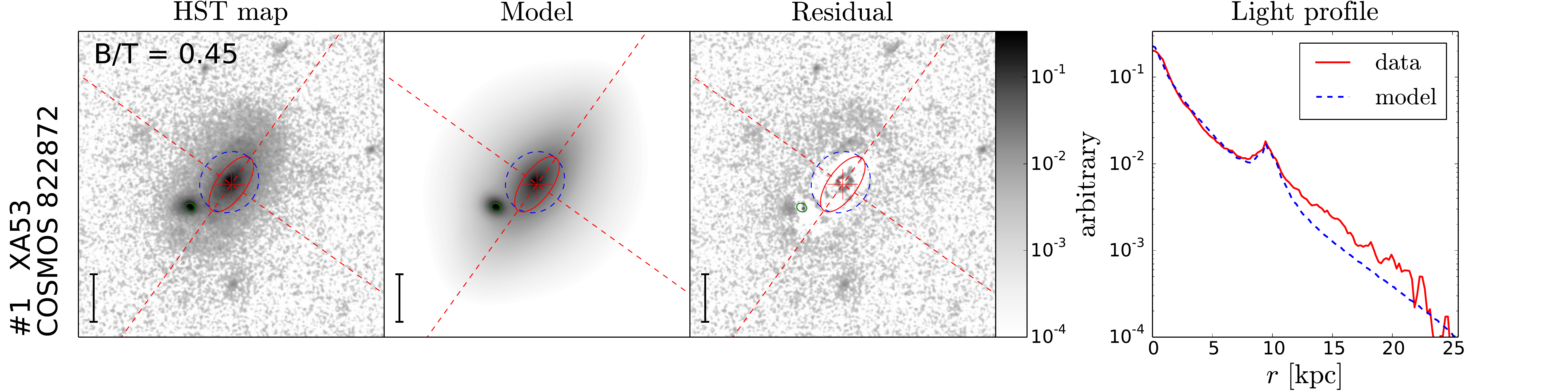}
        \hfill\includegraphics[height=0.175\textwidth,clip,trim=0 0 32cm 0]{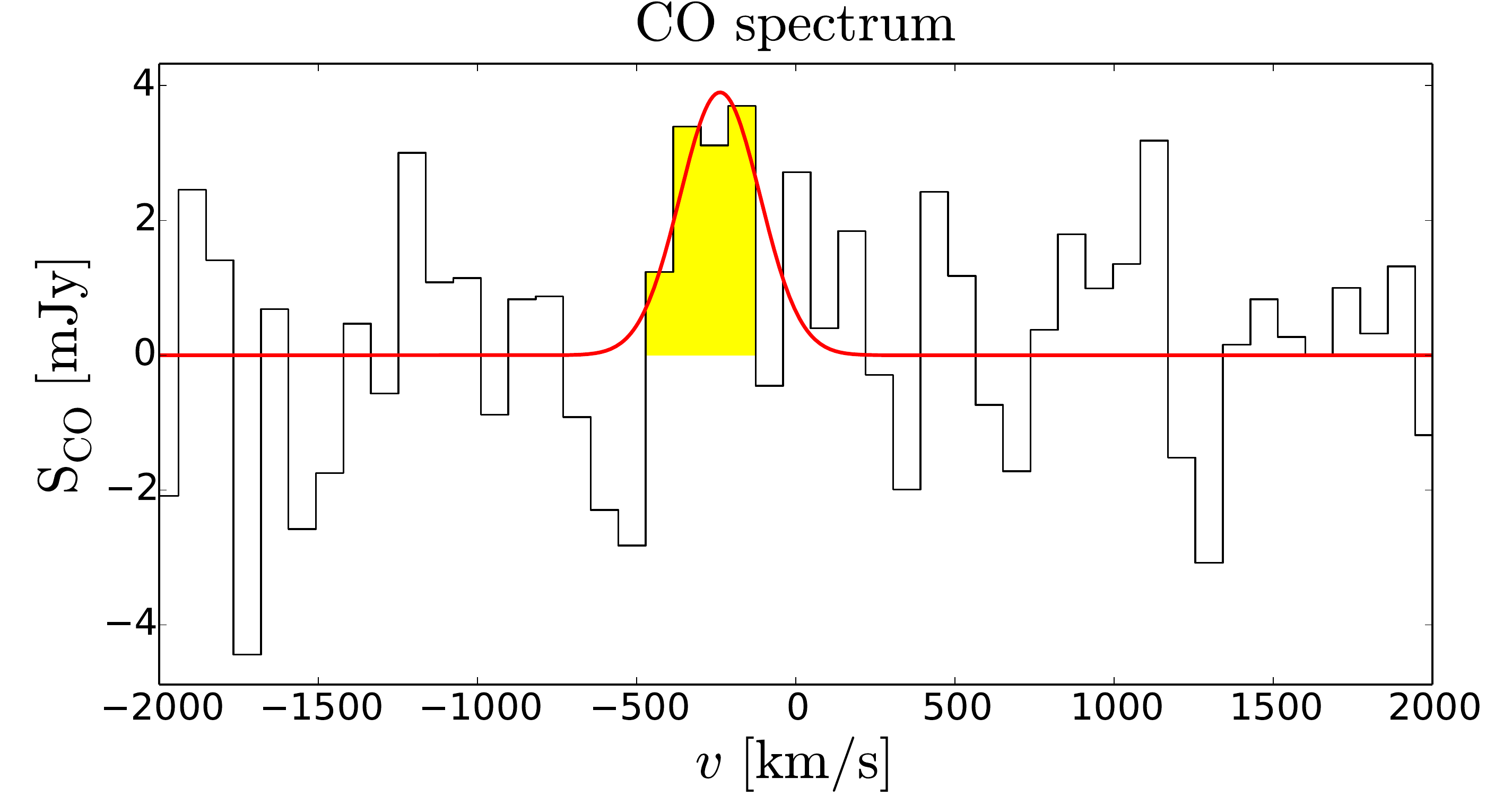}
        \\
        \includegraphics[height=0.1623\textwidth,clip,clip,trim=0 0 4cm 1.1cm]{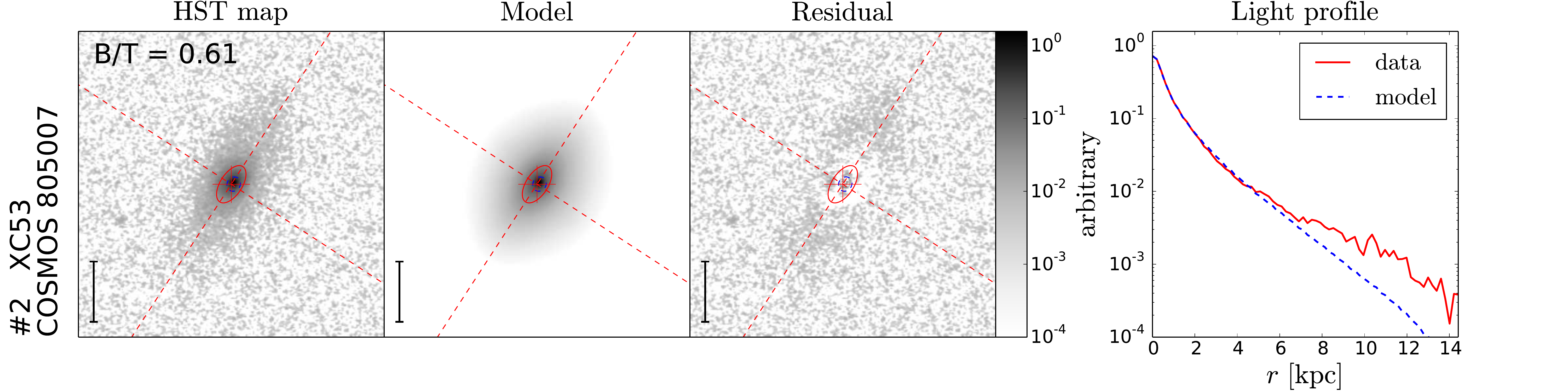}
        \hfill\includegraphics[height=0.1623\textwidth,clip,trim=0 0 32cm 1.1cm]{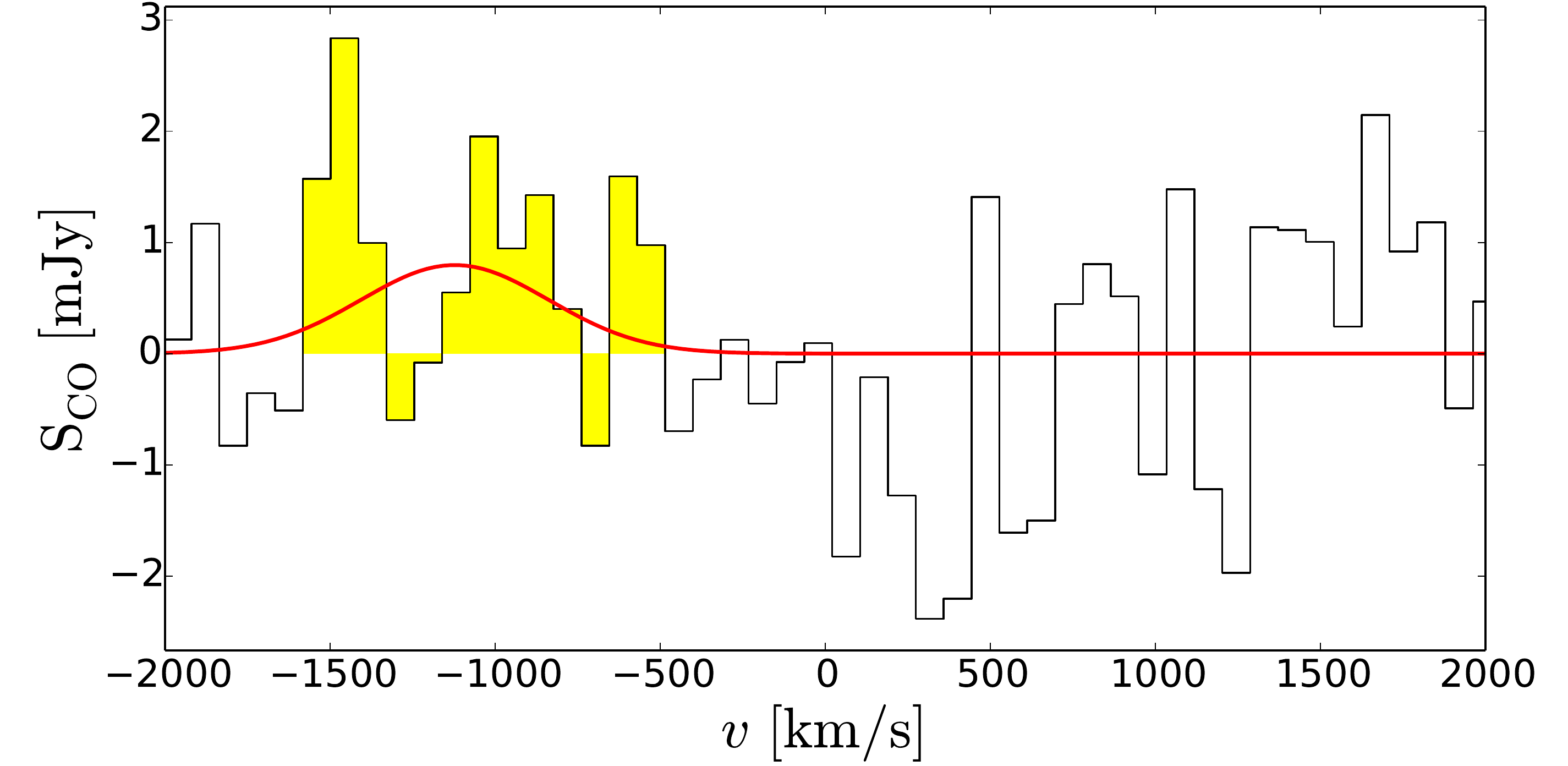}
        \\
        \includegraphics[height=0.1623\textwidth,clip,clip,trim=0 0 4cm 1.1cm]{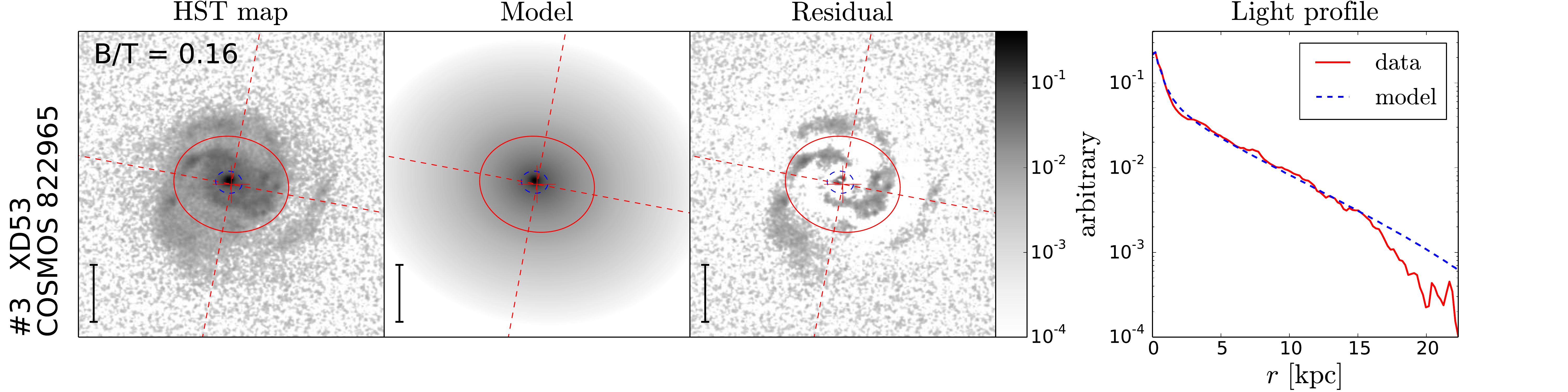}
        \hfill\includegraphics[height=0.1623\textwidth,clip,trim=0 0 32cm 1.1cm]{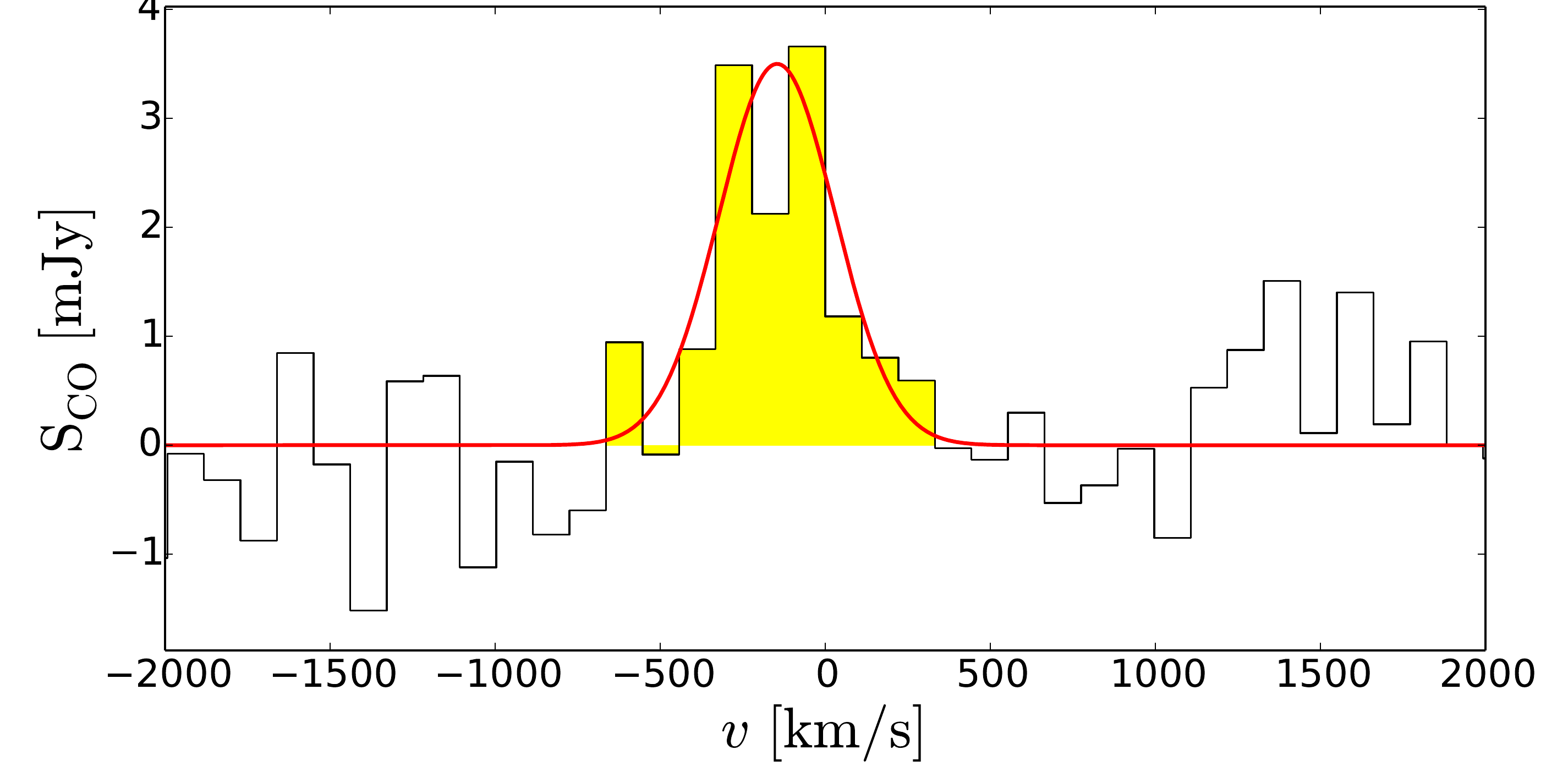}
        \\
        \includegraphics[height=0.1623\textwidth,clip,clip,trim=0 0 4cm 1.1cm]{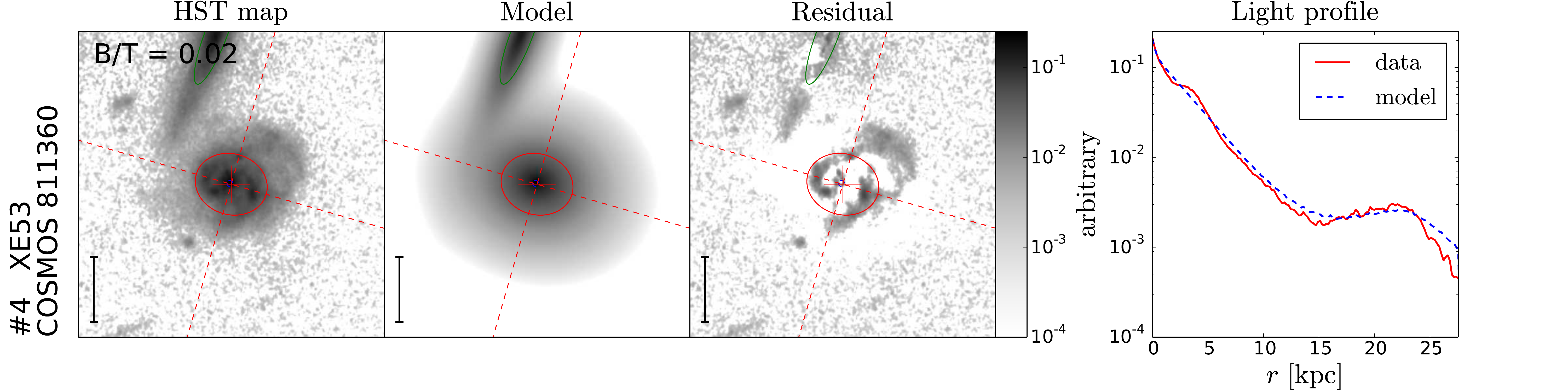}
        \hfill\includegraphics[height=0.1623\textwidth,clip,trim=0 0 32cm 1.1cm]{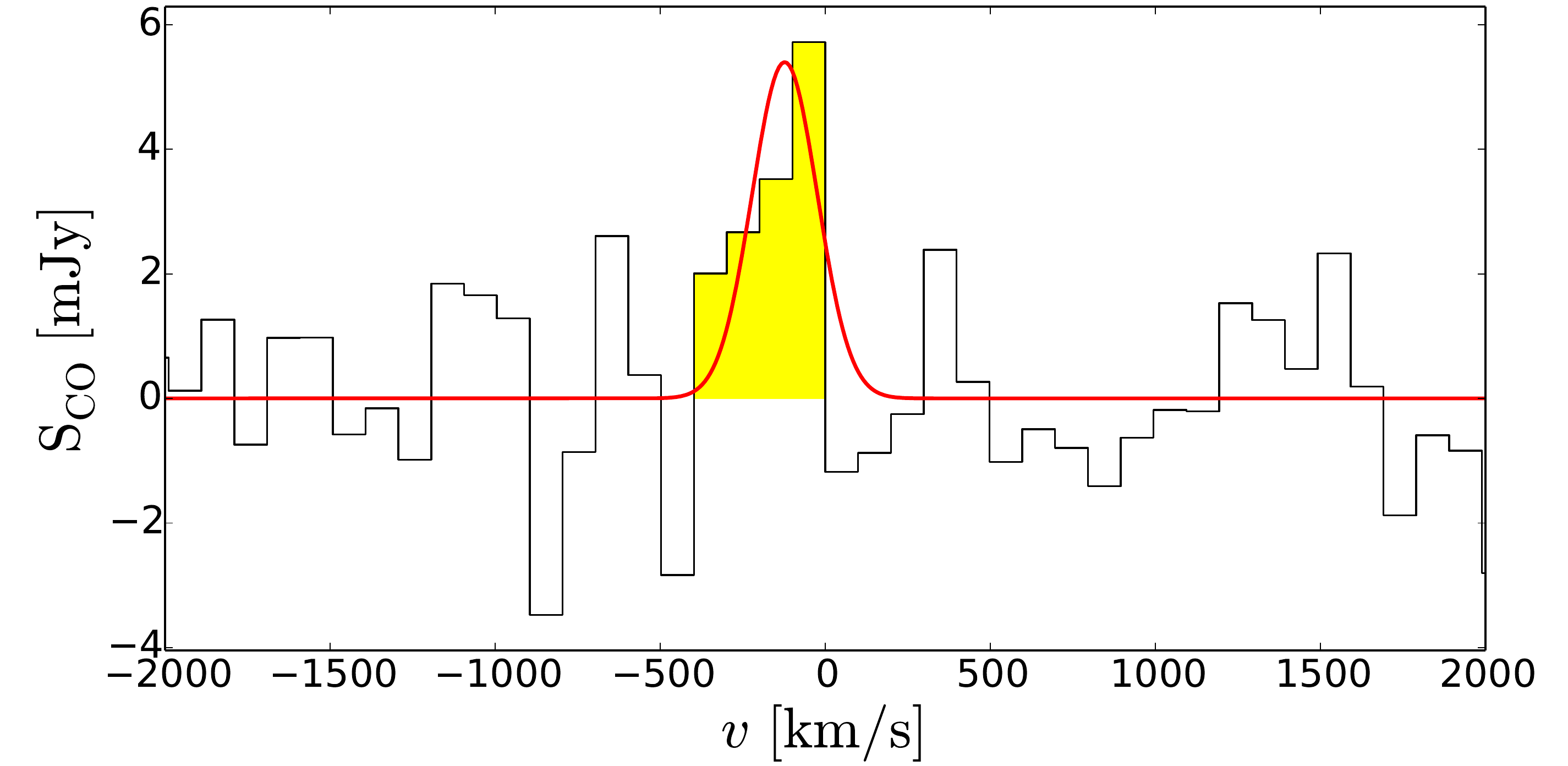}
        \\
        \includegraphics[height=0.1623\textwidth,clip,clip,trim=0 0 4cm 1.1cm]{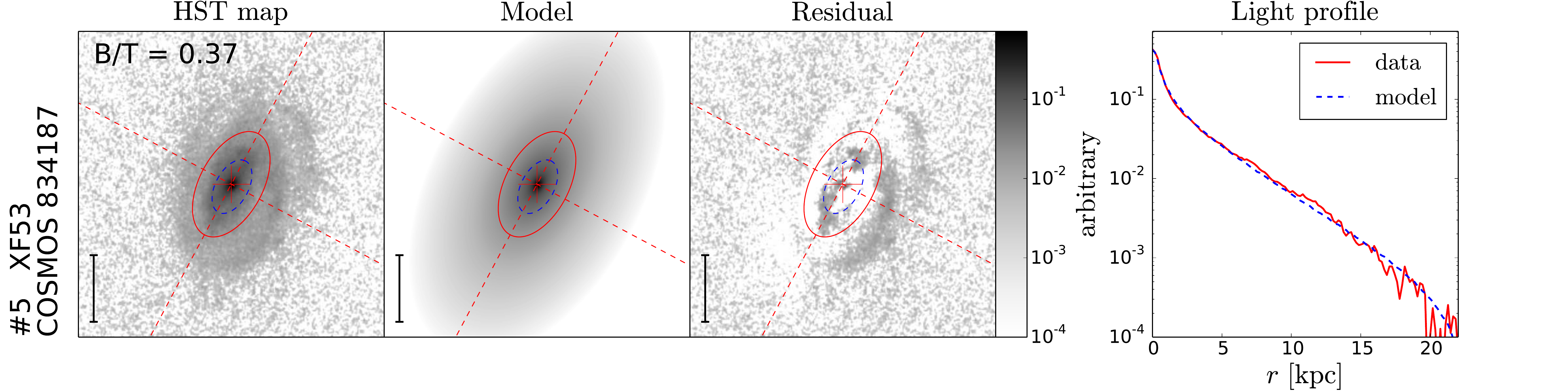}
        \hfill\includegraphics[height=0.1623\textwidth,clip,trim=0 0 32cm 1.1cm]{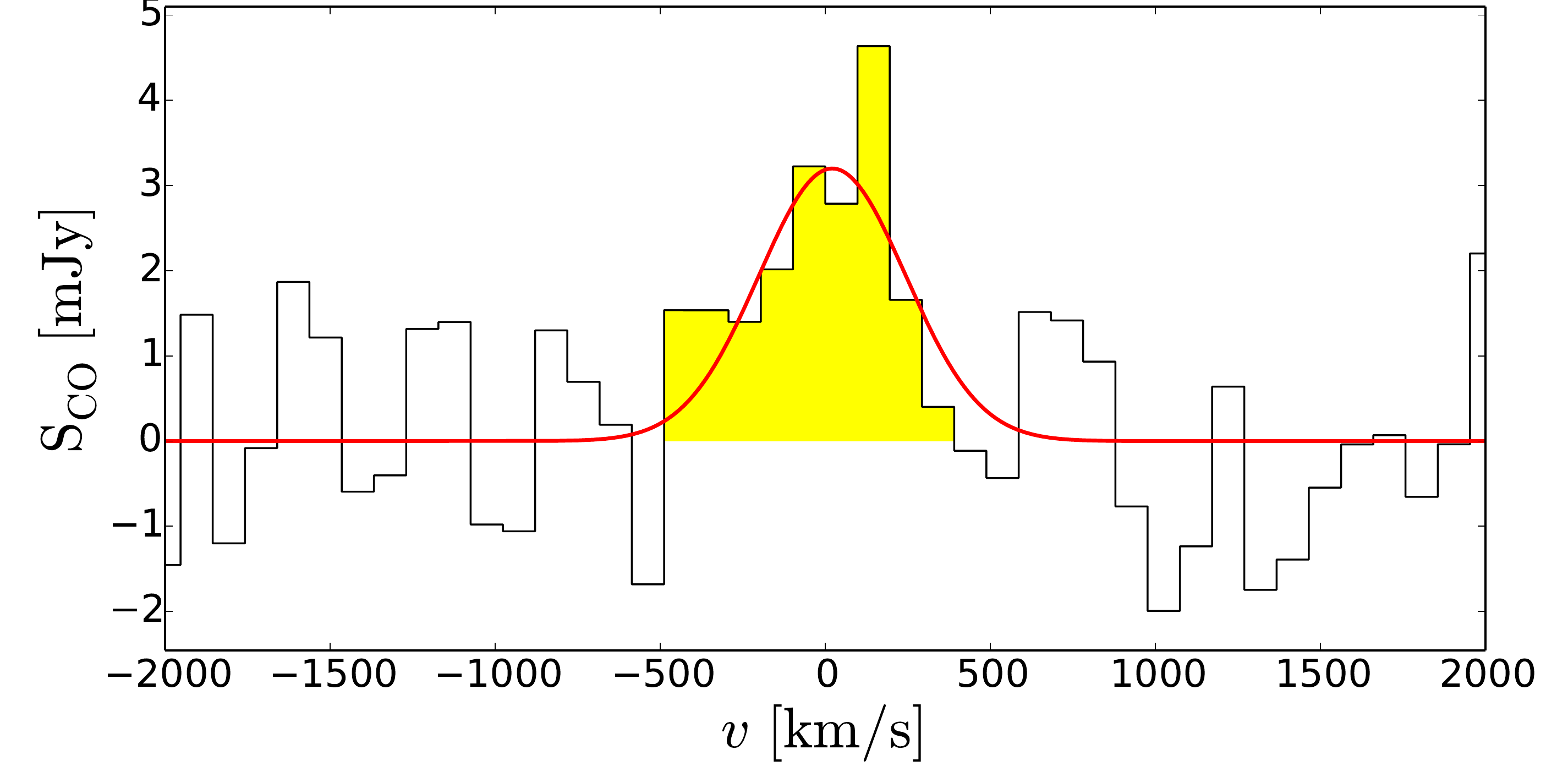}
        \\
        \includegraphics[height=0.1623\textwidth,clip,clip,trim=0 0 4cm 1.1cm]{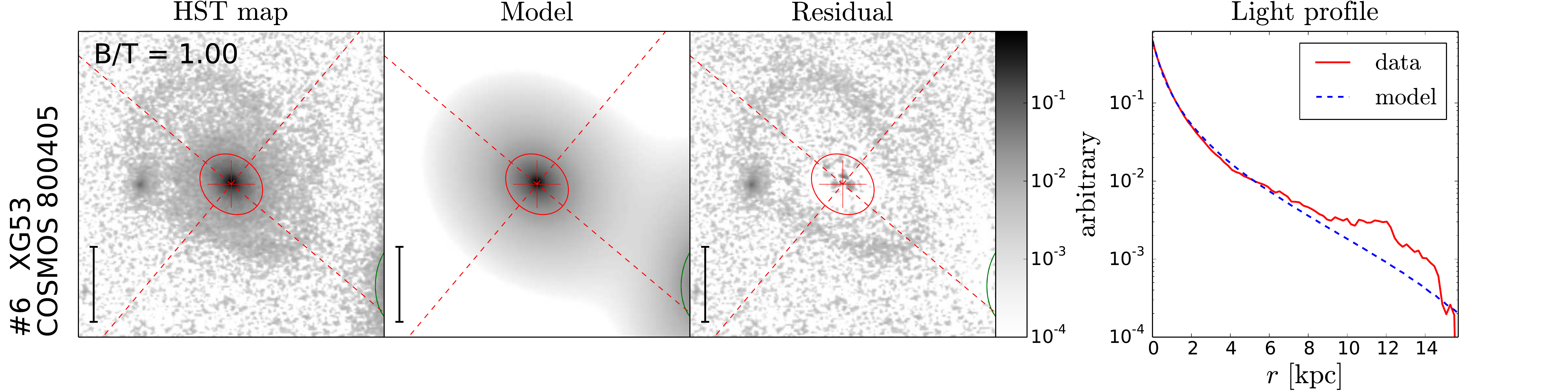}
        \hfill\includegraphics[height=0.1623\textwidth,clip,trim=0 0 32cm 1.1cm]{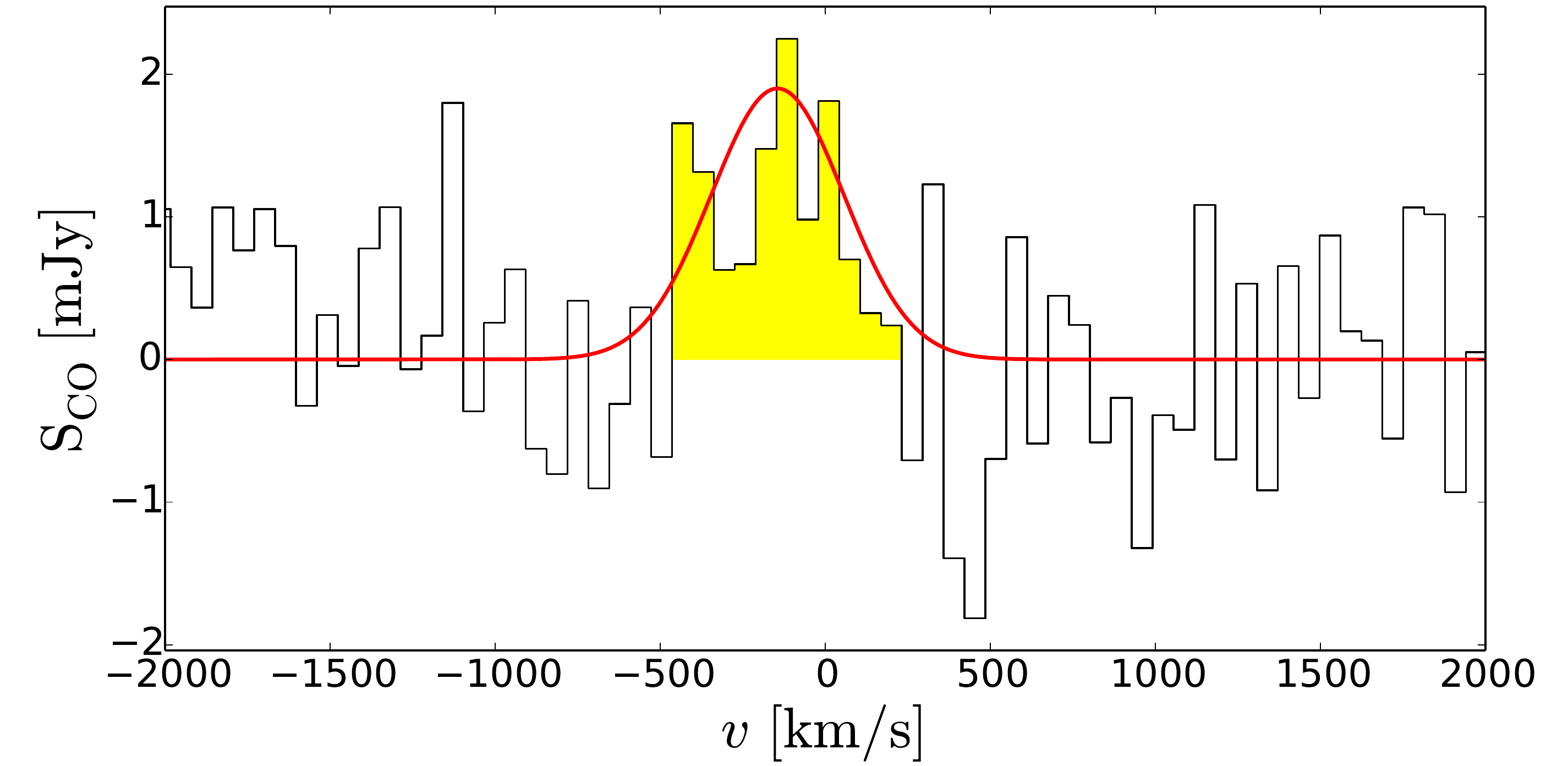}
        \\
        \includegraphics[height=0.1623\textwidth,clip,clip,trim=0 0 4cm 1.1cm]{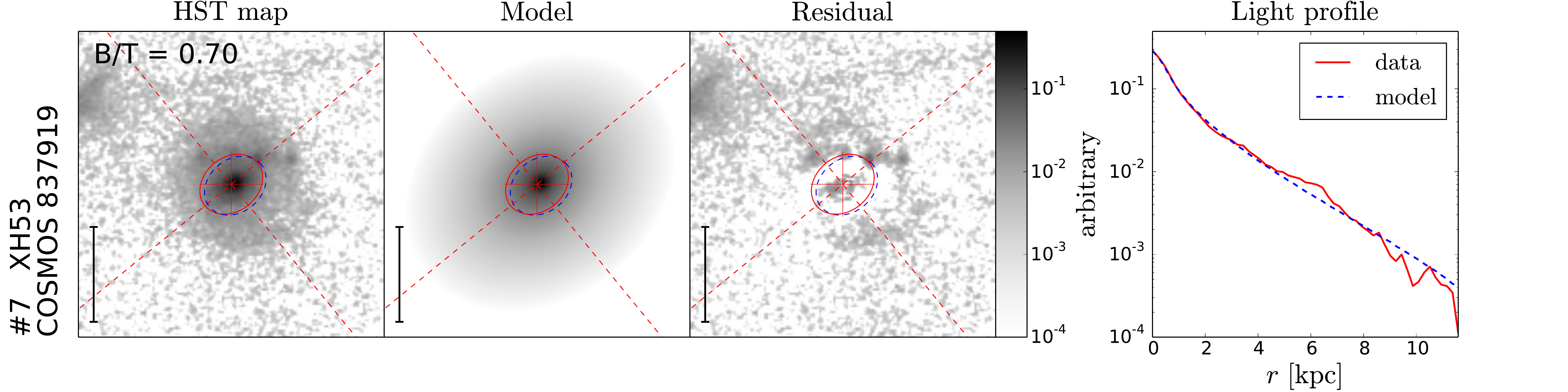}
        \hfill\includegraphics[height=0.1623\textwidth,clip,trim=0 0 32cm 1.1cm]{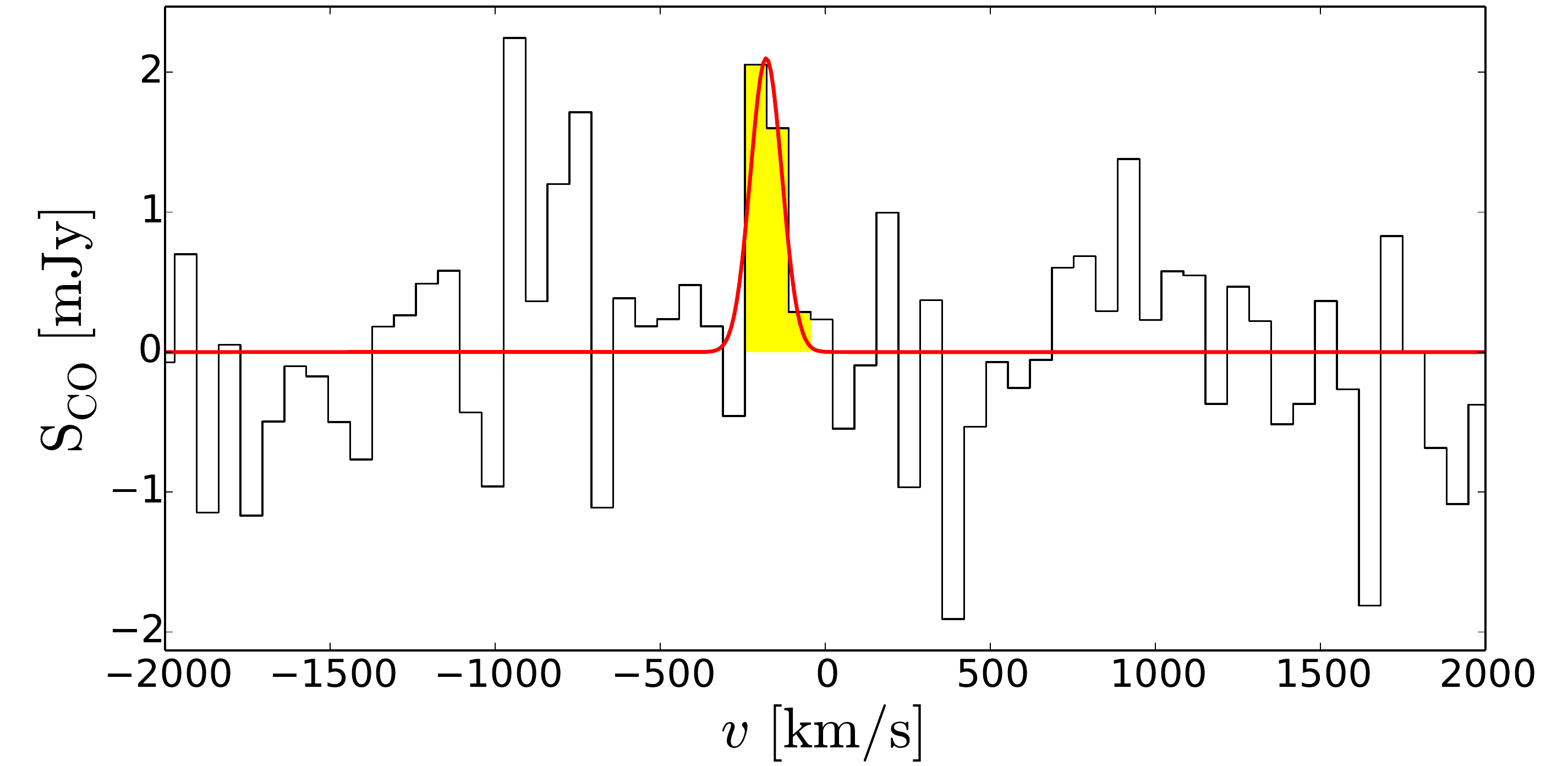}
        \\
        \caption{\captiontext}
        \label{fig:images}
\end{figure*}

\begin{figure*}[h!]
        \ContinuedFloat
        \flushleft
        \includegraphics[height=0.175\textwidth,clip,trim=0 0 4cm 0]{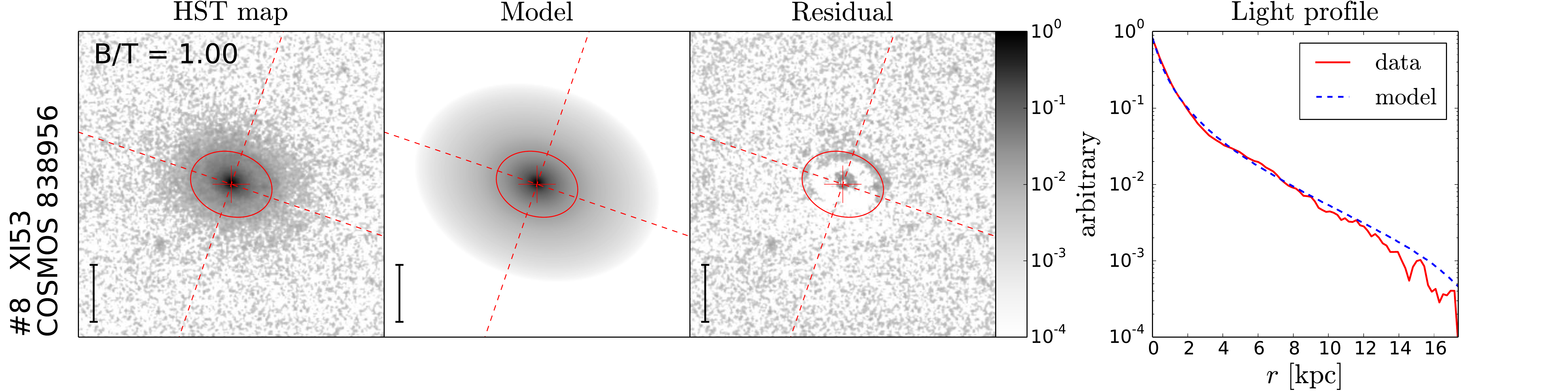}
        \hfill\includegraphics[height=0.175\textwidth,clip,trim=0 0 32cm 0]{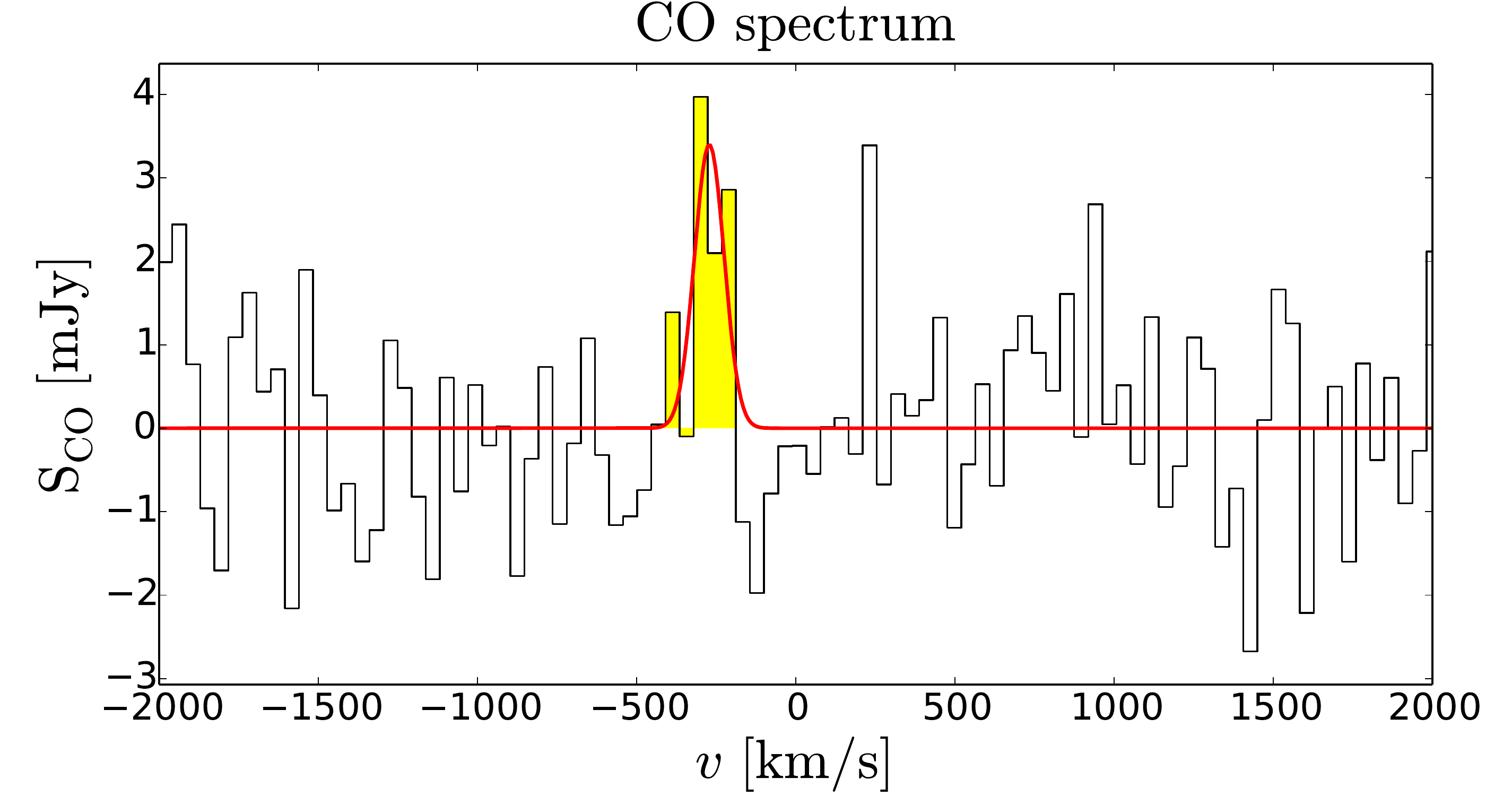}
        \\
        \includegraphics[height=0.1623\textwidth,clip,clip,trim=0 0 4cm 1.1cm]{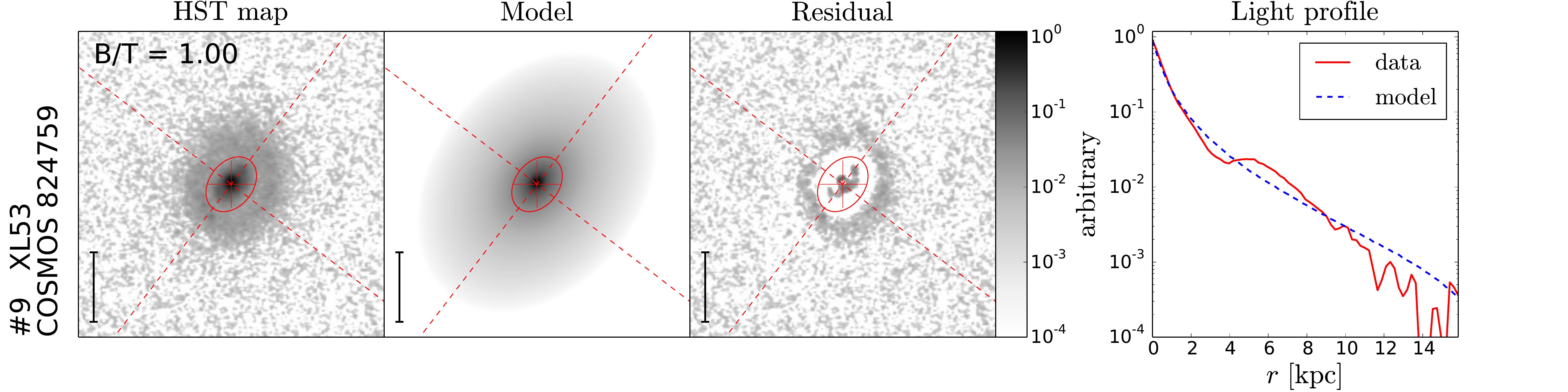}
        \hfill\includegraphics[height=0.1623\textwidth,clip,trim=0 0 32cm 1.1cm]{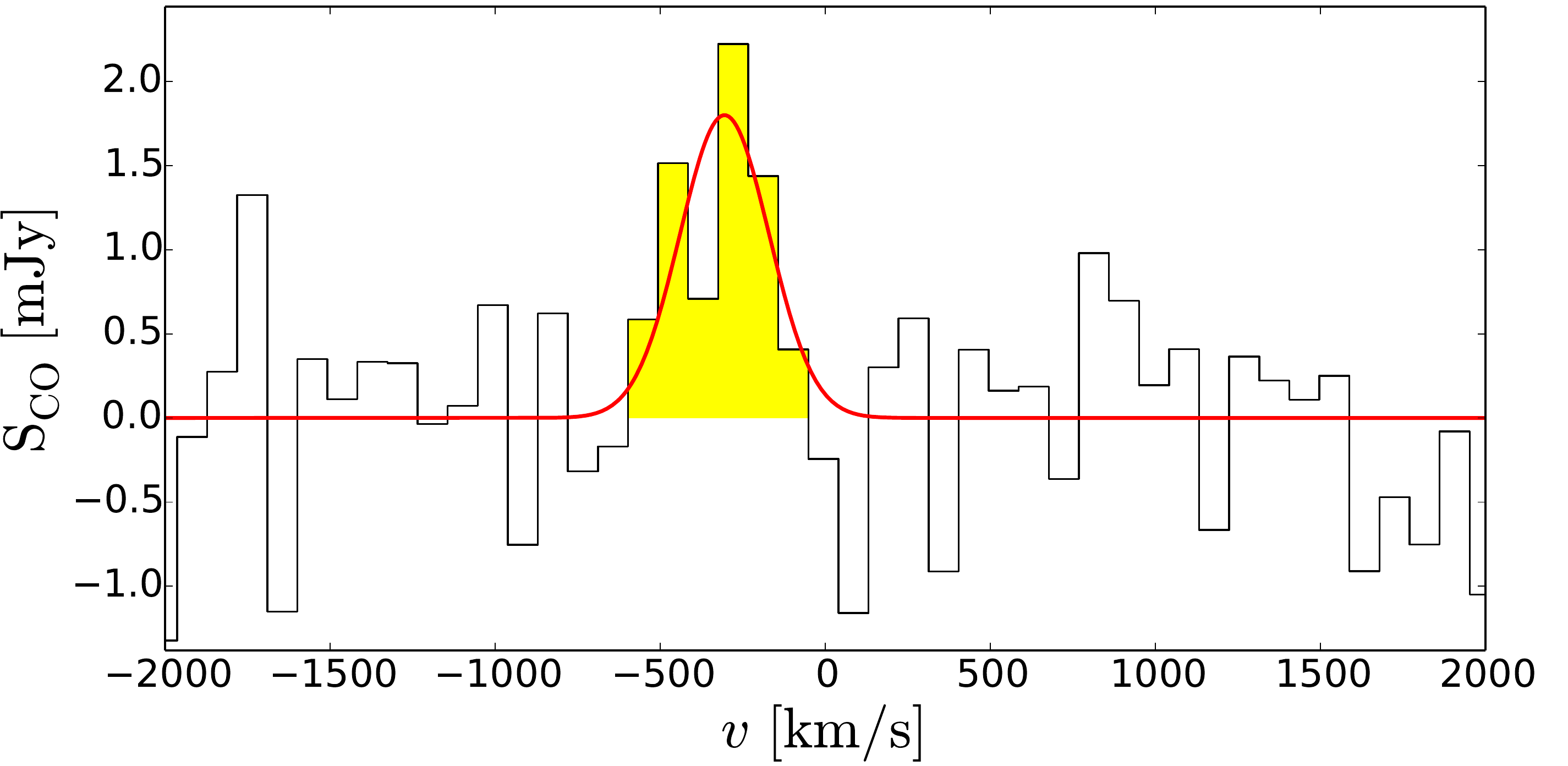}
        \\
        \includegraphics[height=0.1623\textwidth,clip,clip,trim=0 0 4cm 1.1cm]{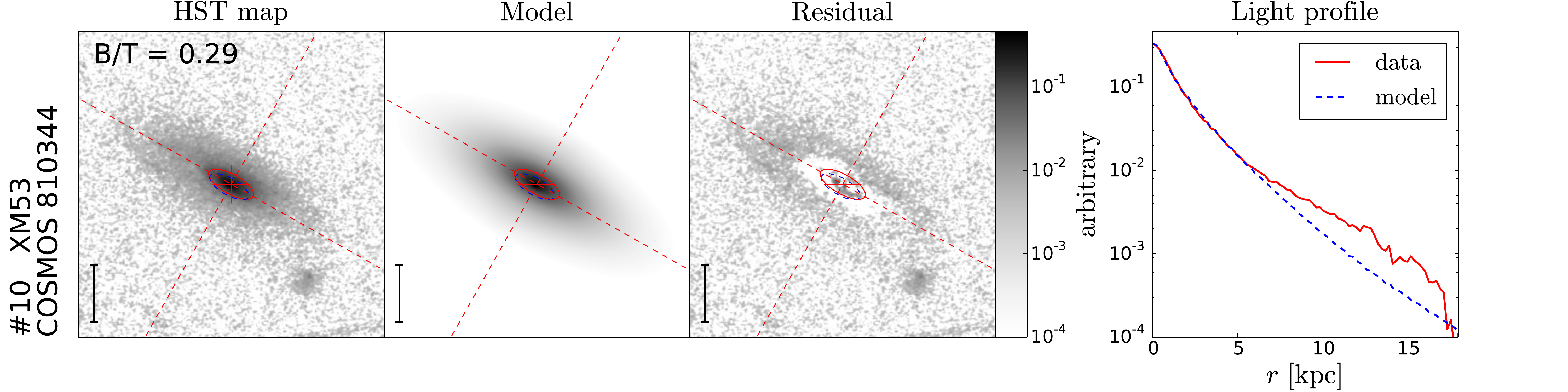}
        \hfill\includegraphics[height=0.1623\textwidth,clip,trim=0 0 32cm 1.1cm]{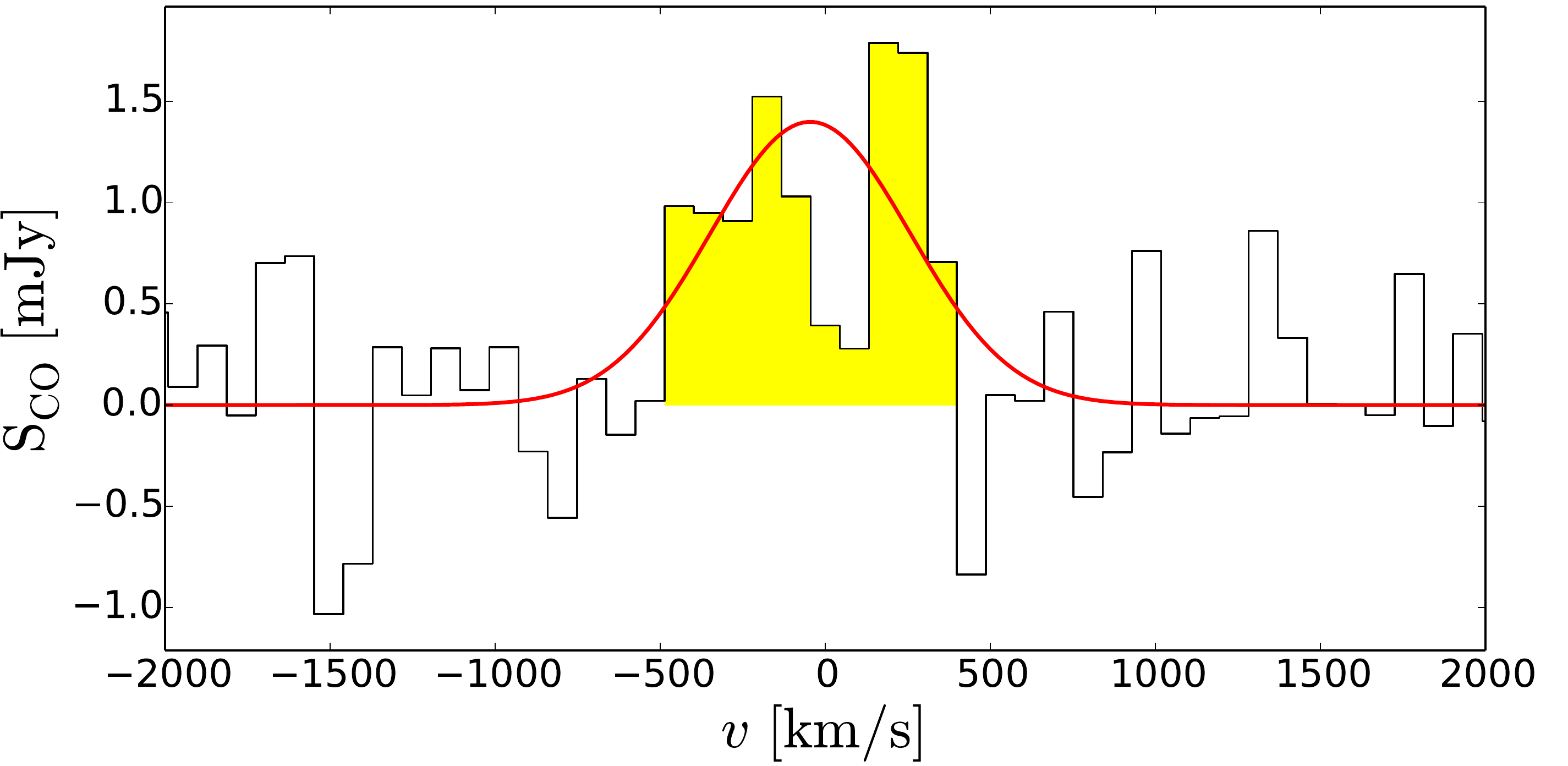}
        \\
        \includegraphics[height=0.1623\textwidth,clip,clip,trim=0 0 4cm 1.1cm]{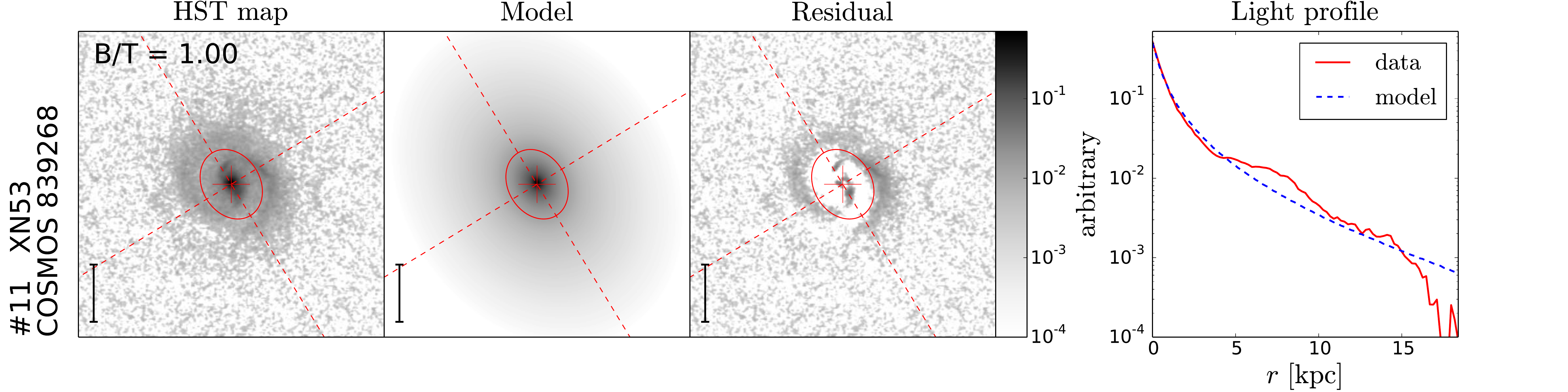}
        \hfill\includegraphics[height=0.1623\textwidth,clip,trim=0 0 32cm 1.1cm]{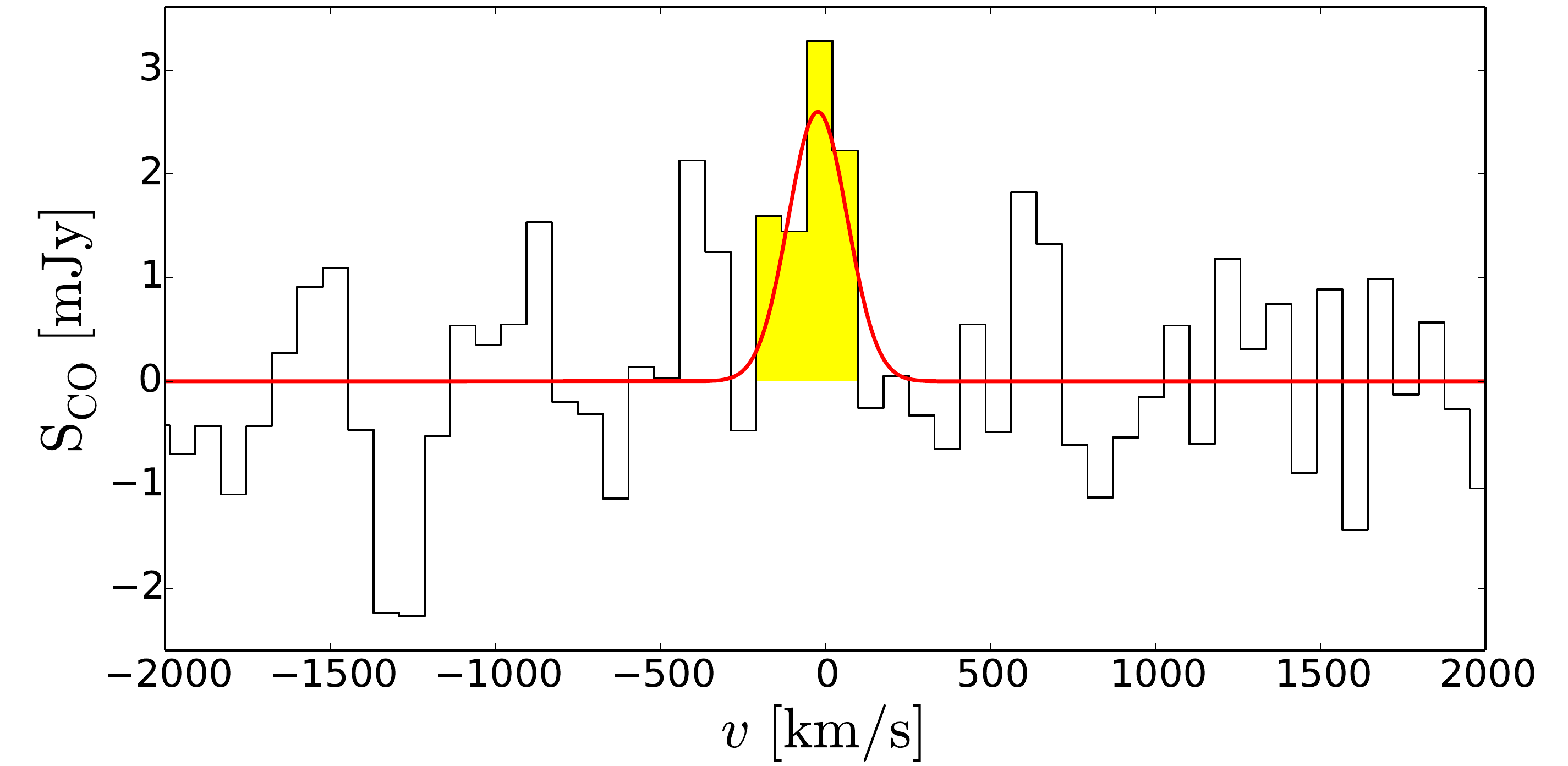}
        \\
        \includegraphics[height=0.1623\textwidth,clip,clip,trim=0 0 4cm 1.1cm]{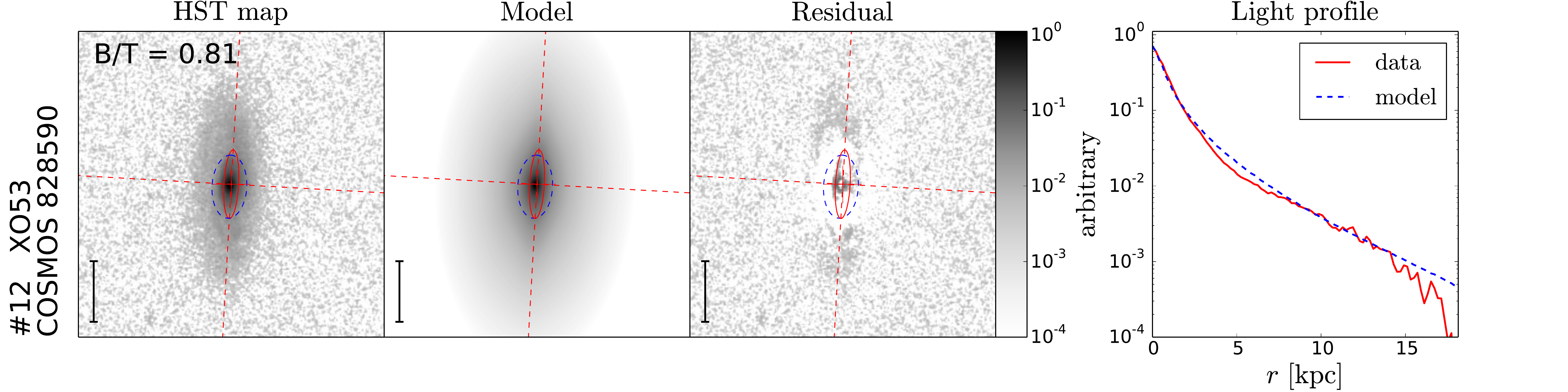}
        \hfill\includegraphics[height=0.1623\textwidth,clip,trim=0 0 32cm 1.1cm]{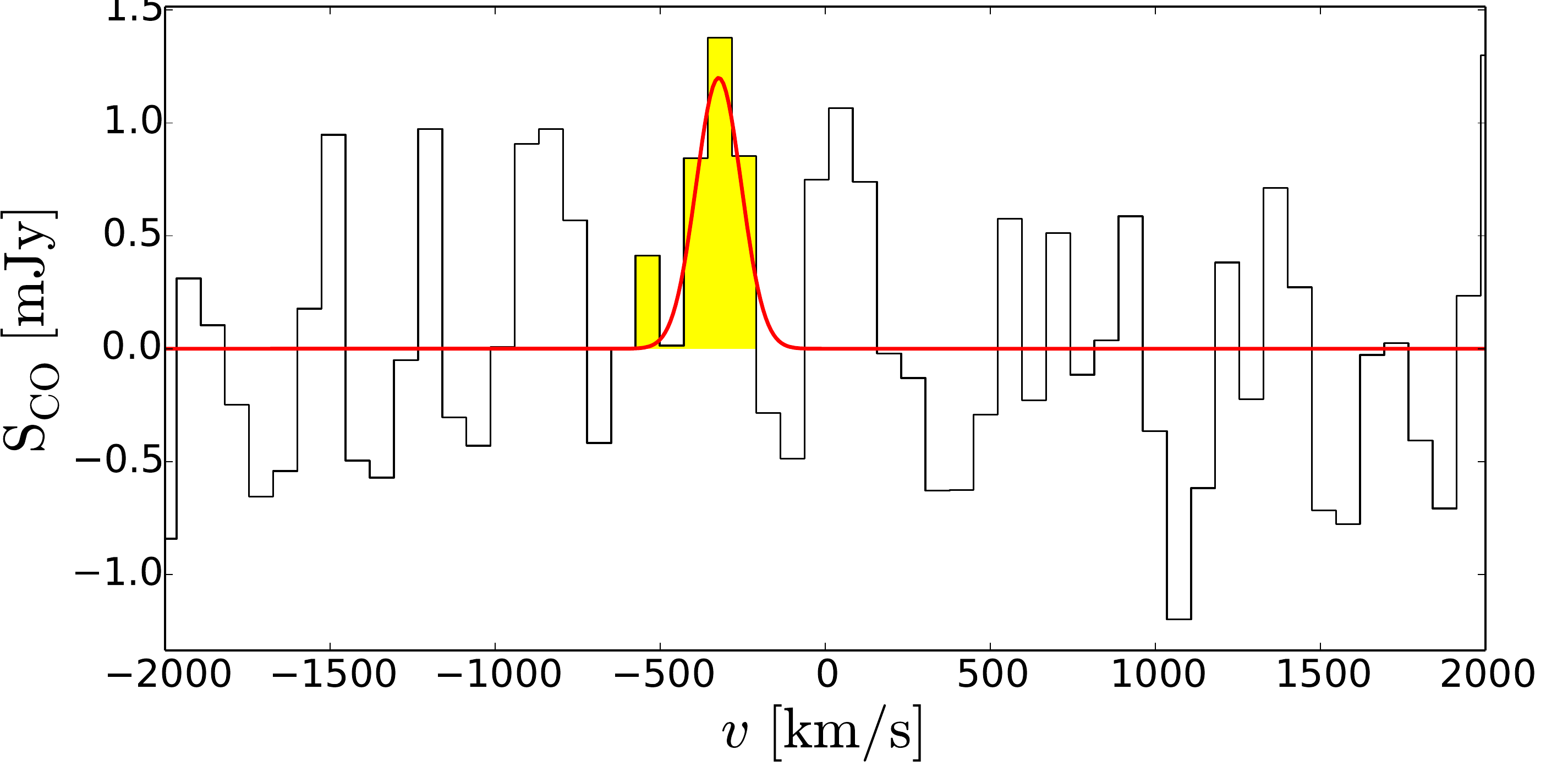}
        \\
        \includegraphics[height=0.1623\textwidth,clip,clip,trim=0 0 4cm 1.1cm]{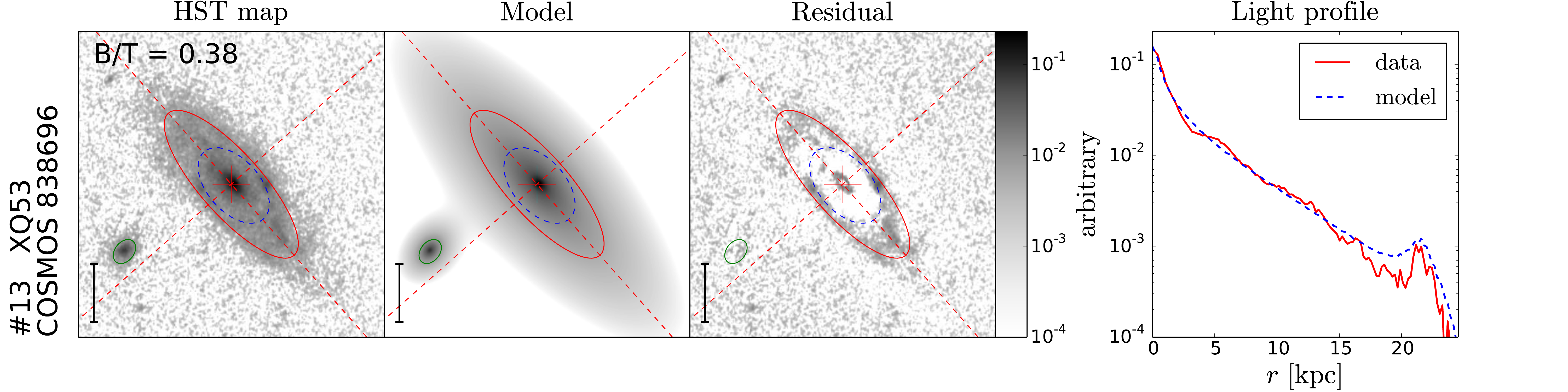}
        \hfill\includegraphics[height=0.1623\textwidth,clip,trim=0 0 32cm 1.1cm]{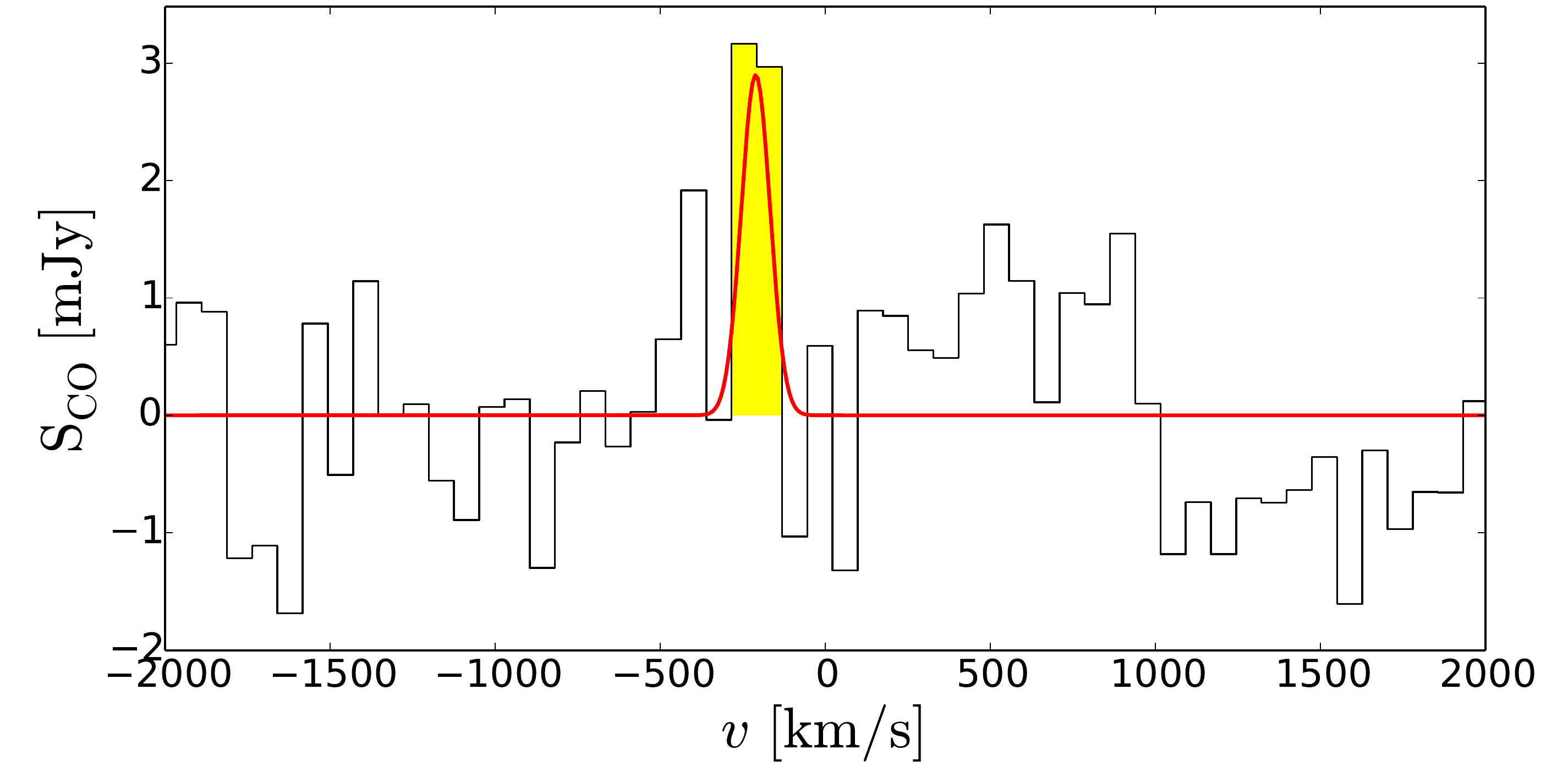}
        \\
        \includegraphics[height=0.1623\textwidth,clip,clip,trim=0 0 4cm 1.1cm]{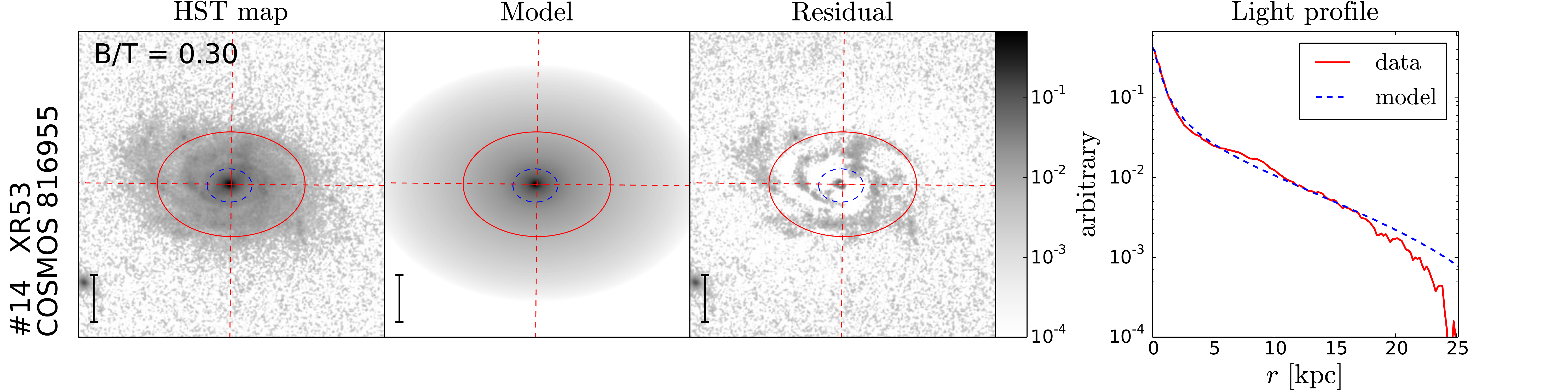}
        \hfill\includegraphics[height=0.1623\textwidth,clip,trim=0 0 32cm 1.1cm]{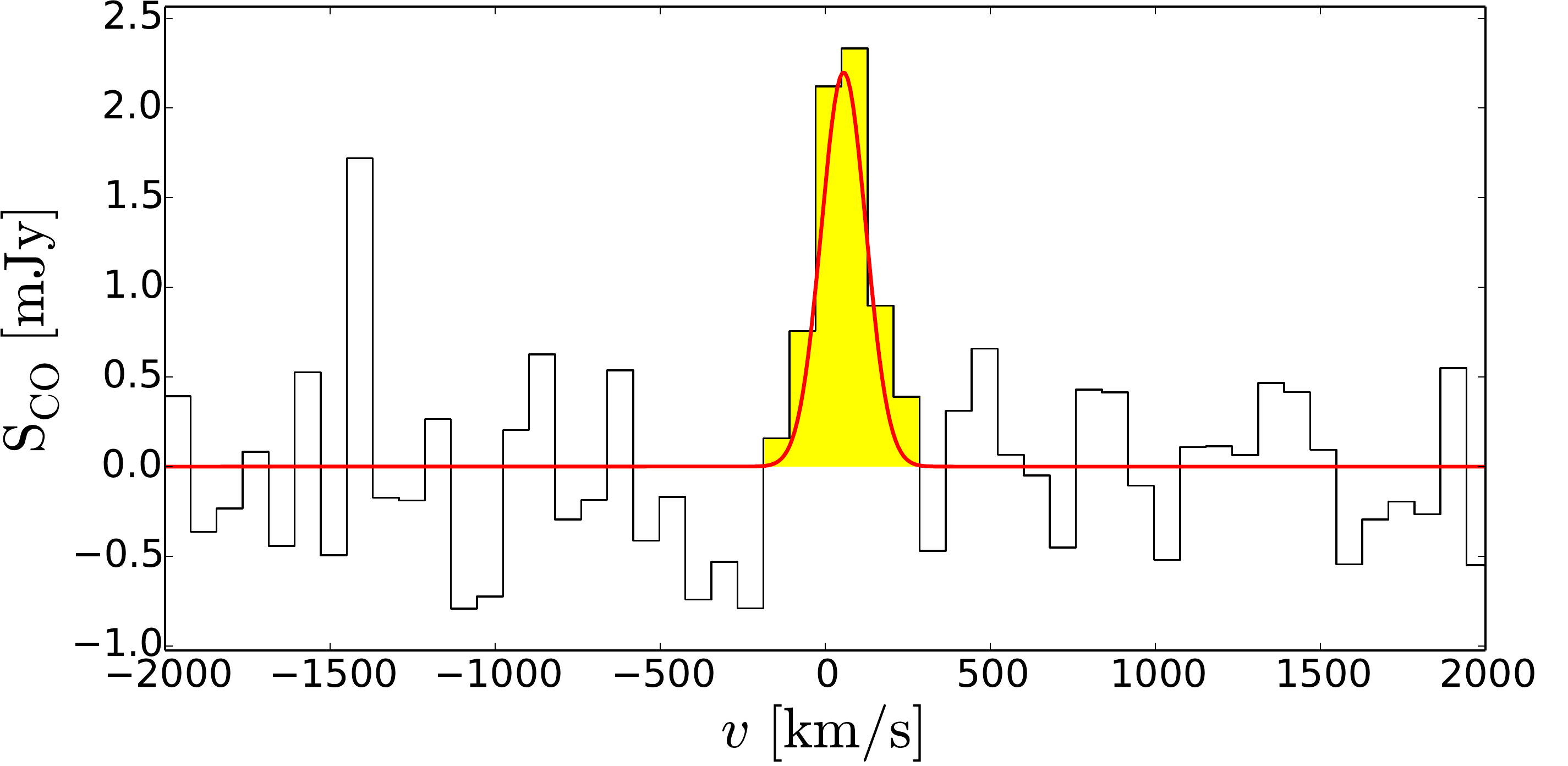}
        \\
\end{figure*}

\begin{figure*}[h!]
        \ContinuedFloat
        \flushleft
        \includegraphics[height=0.175\textwidth,clip,trim=0 0 4cm 0]{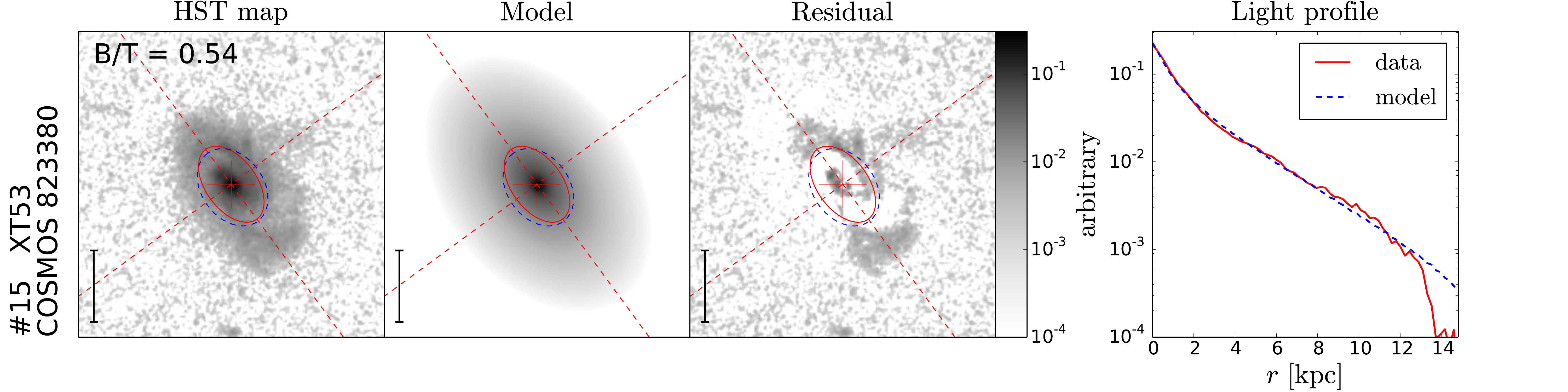}
        \hfill\includegraphics[height=0.175\textwidth,clip,trim=0 0 32cm 0]{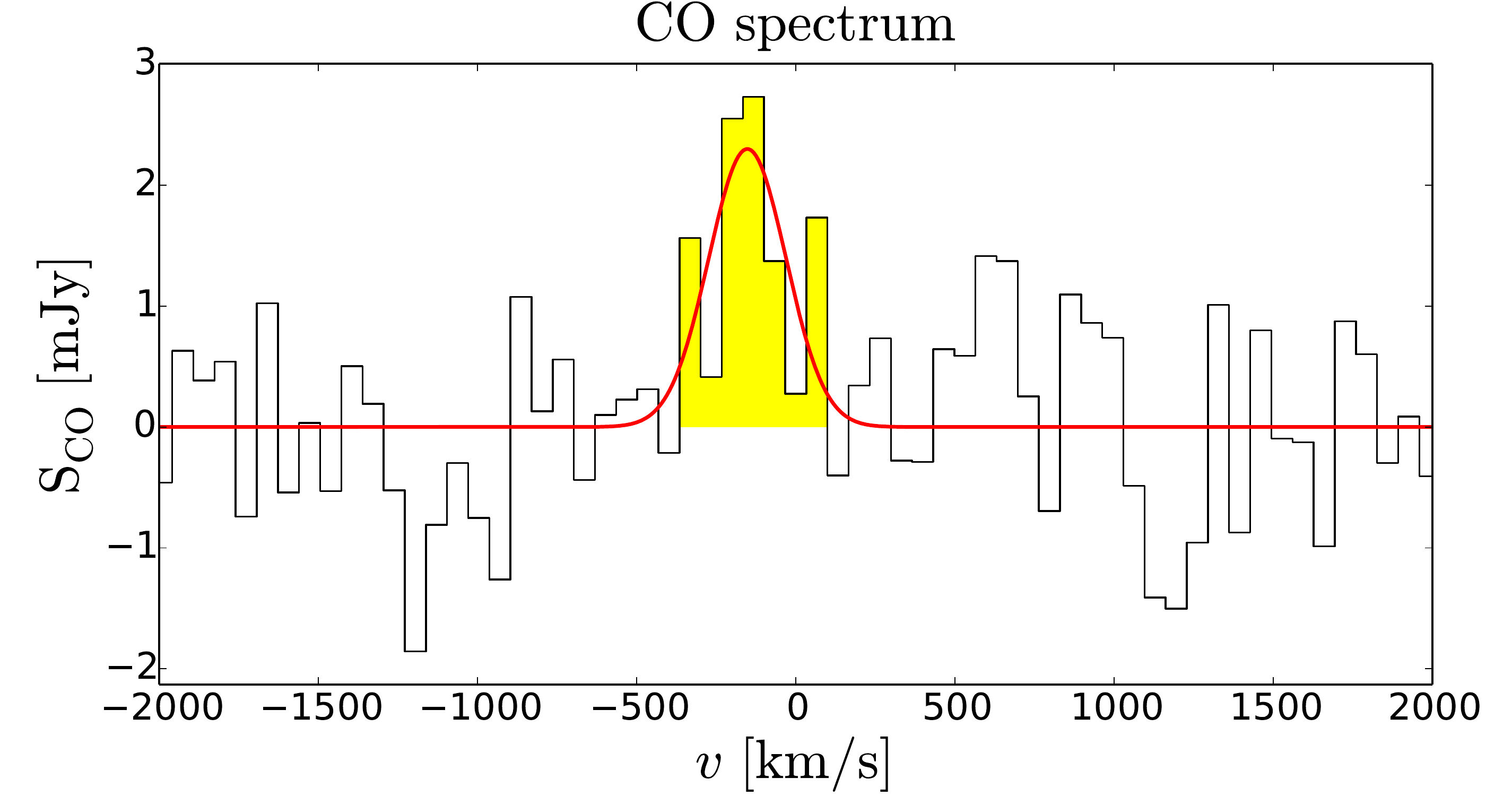}
        \\
        \includegraphics[height=0.1623\textwidth,clip,clip,trim=0 0 4cm 1.1cm]{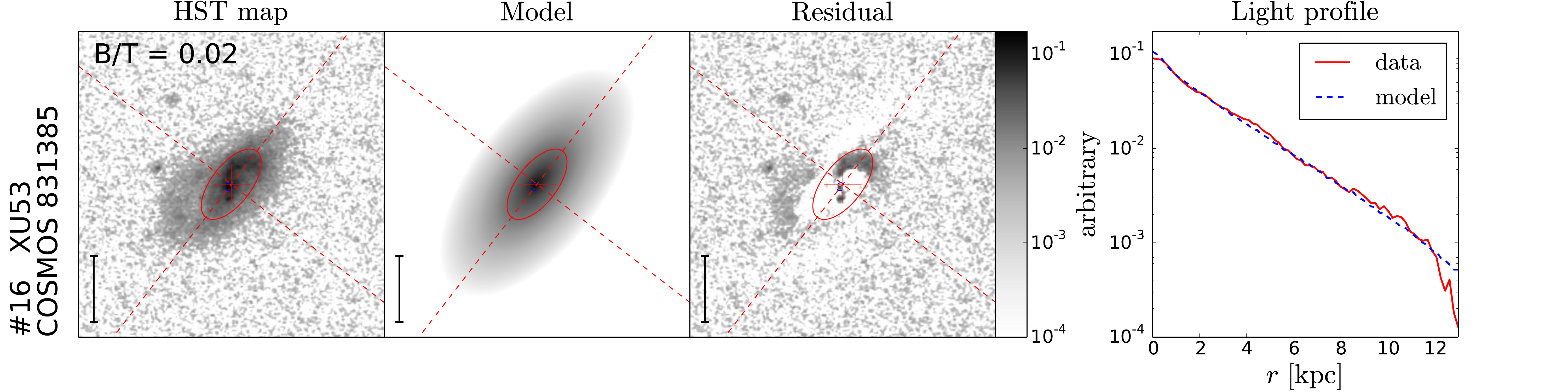}
        \hfill\includegraphics[height=0.1623\textwidth,clip,trim=0 0 32cm 1.1cm]{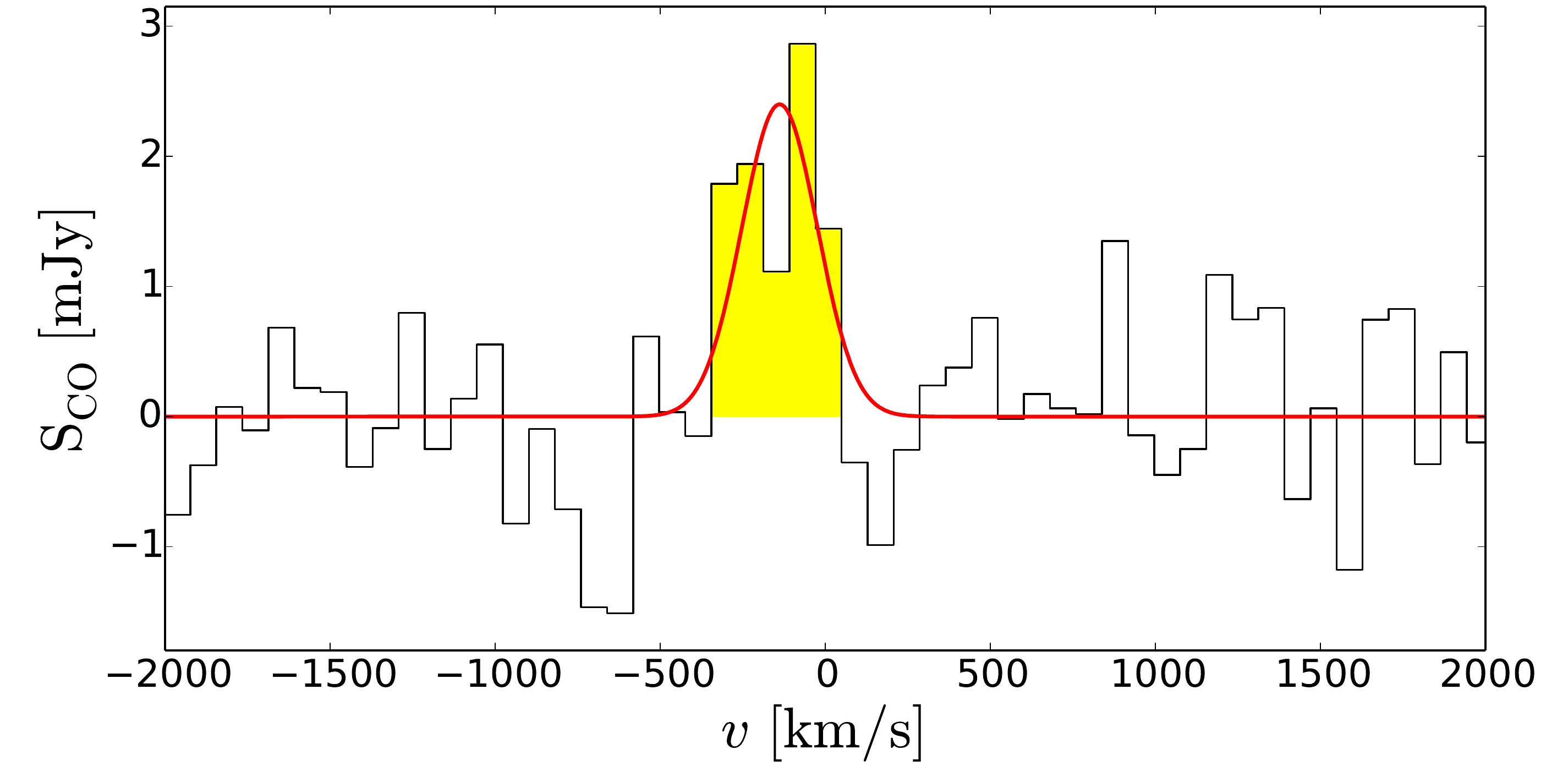}
        \\
        \includegraphics[height=0.1623\textwidth,clip,clip,trim=0 0 4cm 1.1cm]{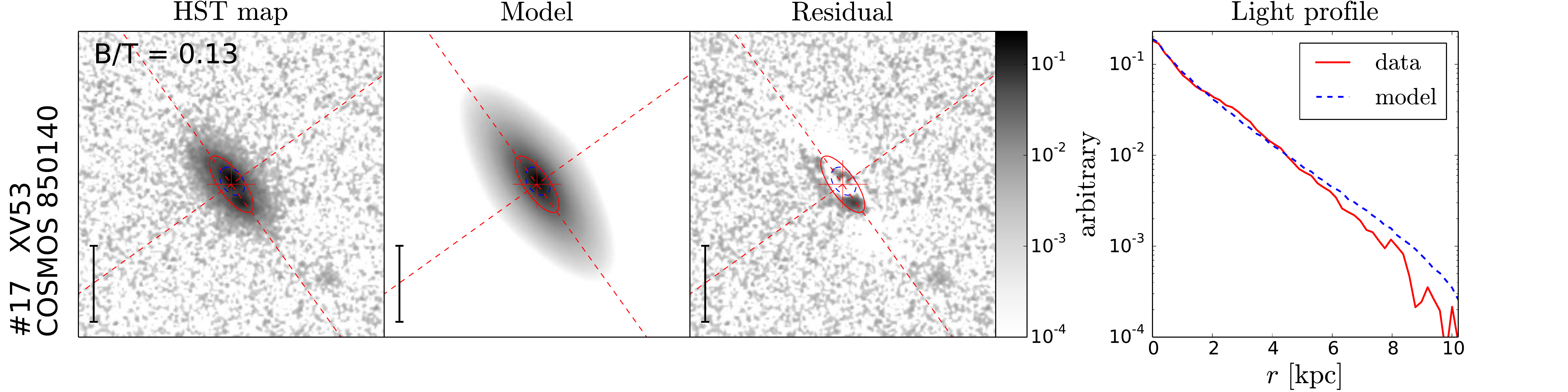}
        \hfill\includegraphics[height=0.1623\textwidth,clip,trim=0 0 32cm 1.1cm]{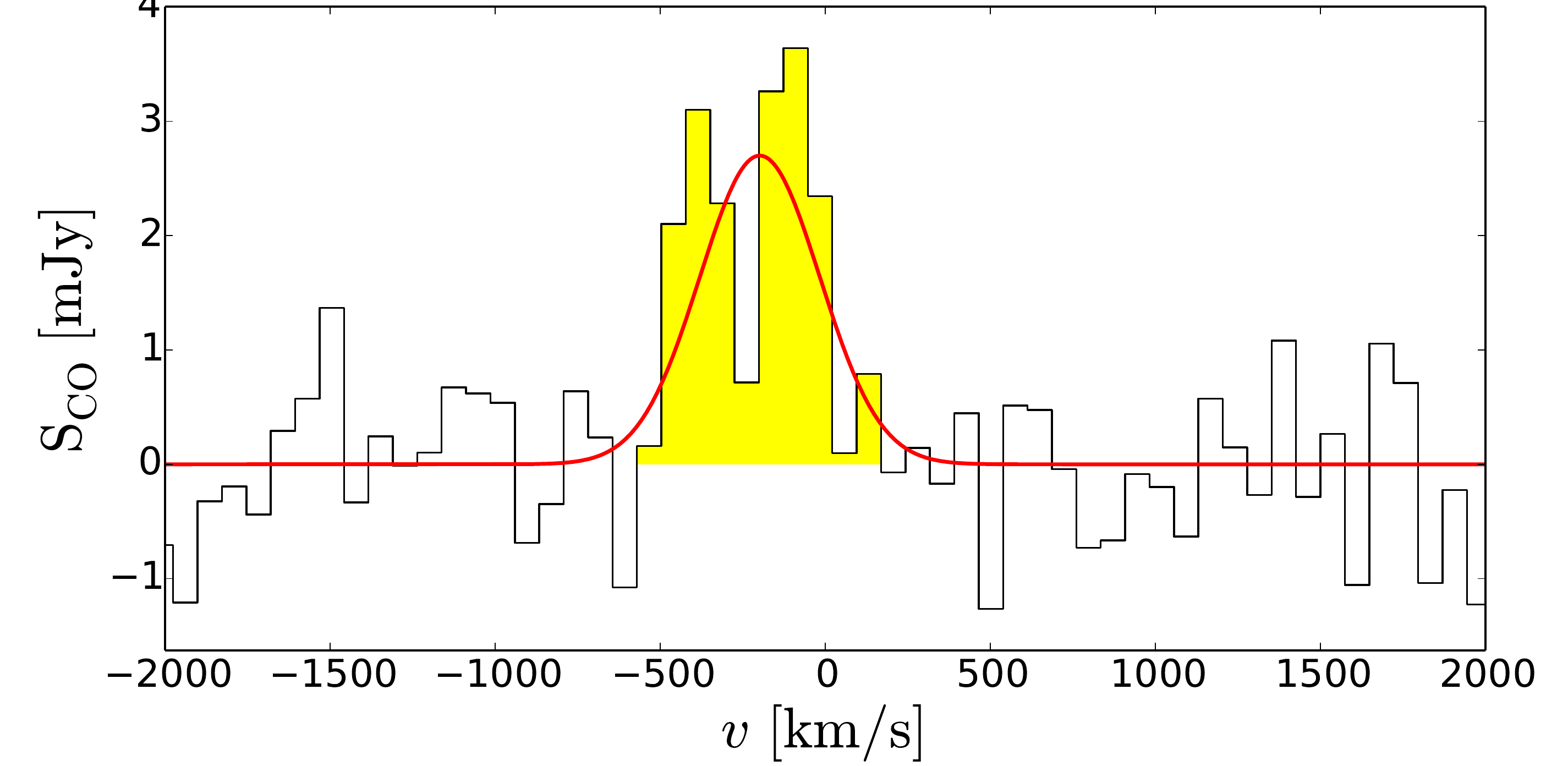}
        \\
        \includegraphics[height=0.1623\textwidth,clip,clip,trim=0 0 4cm 1.1cm]{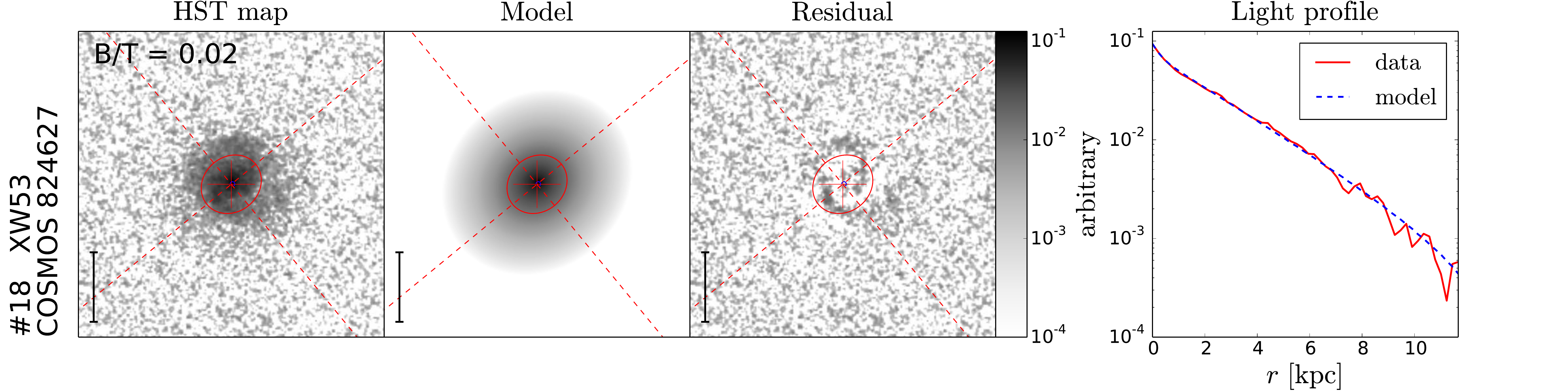}
        \hfill\includegraphics[height=0.1623\textwidth,clip,trim=0 0 32cm 1.1cm]{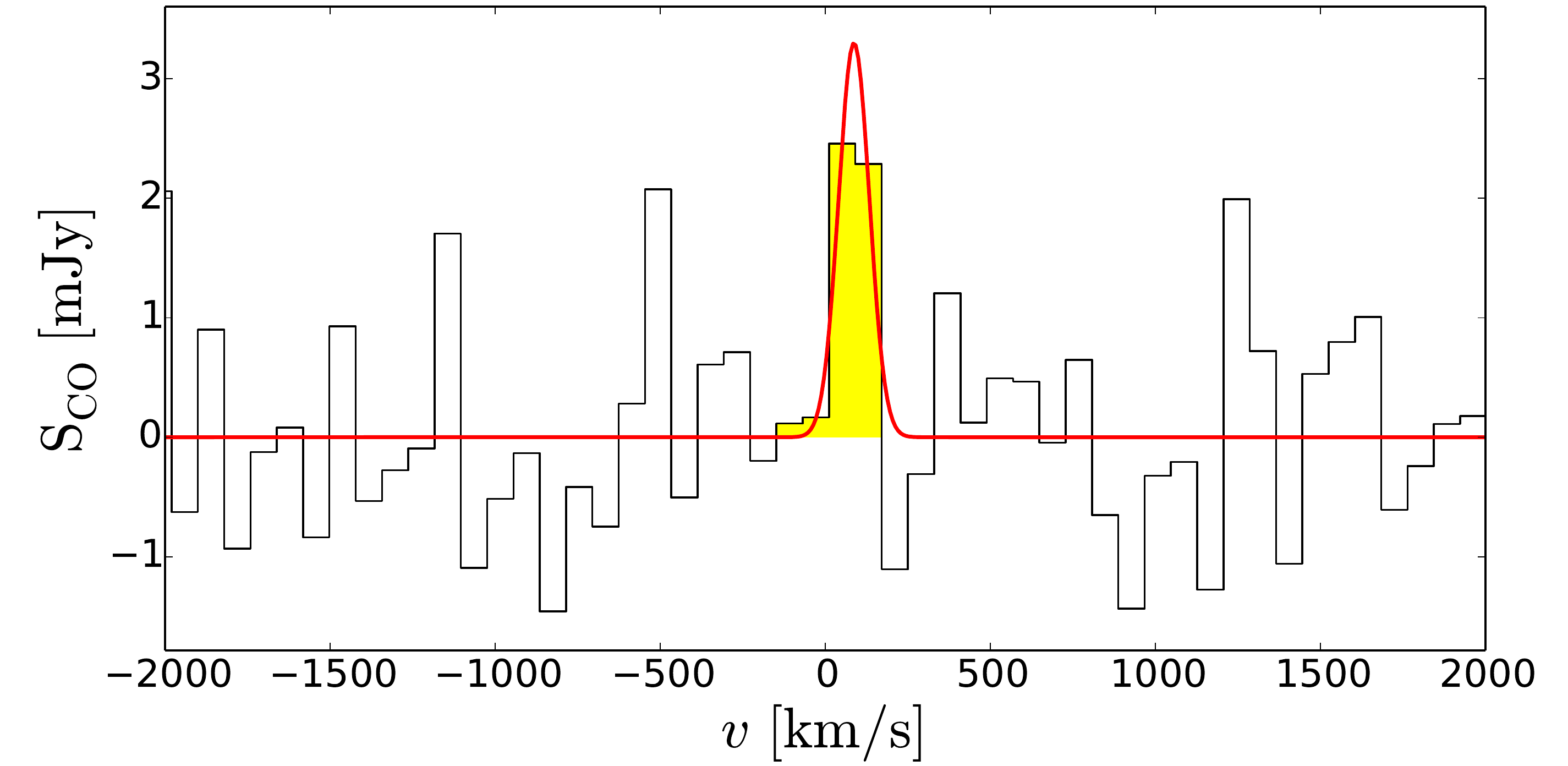}
        \\
        \includegraphics[height=0.1623\textwidth,clip,clip,trim=0 0 4cm 1.1cm]{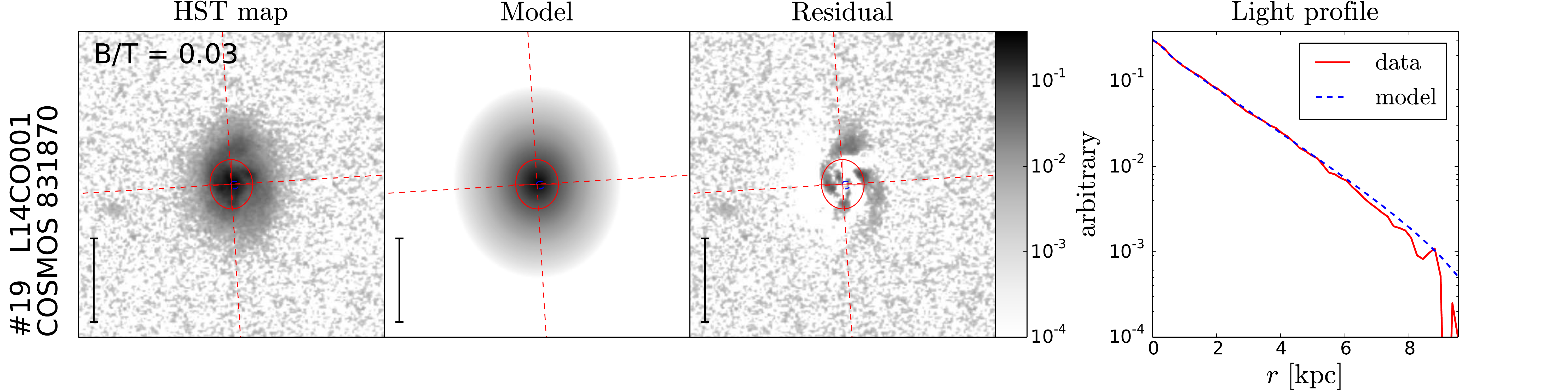}
        \hfill\includegraphics[height=0.1623\textwidth,clip,trim=0 0 32cm 1.1cm]{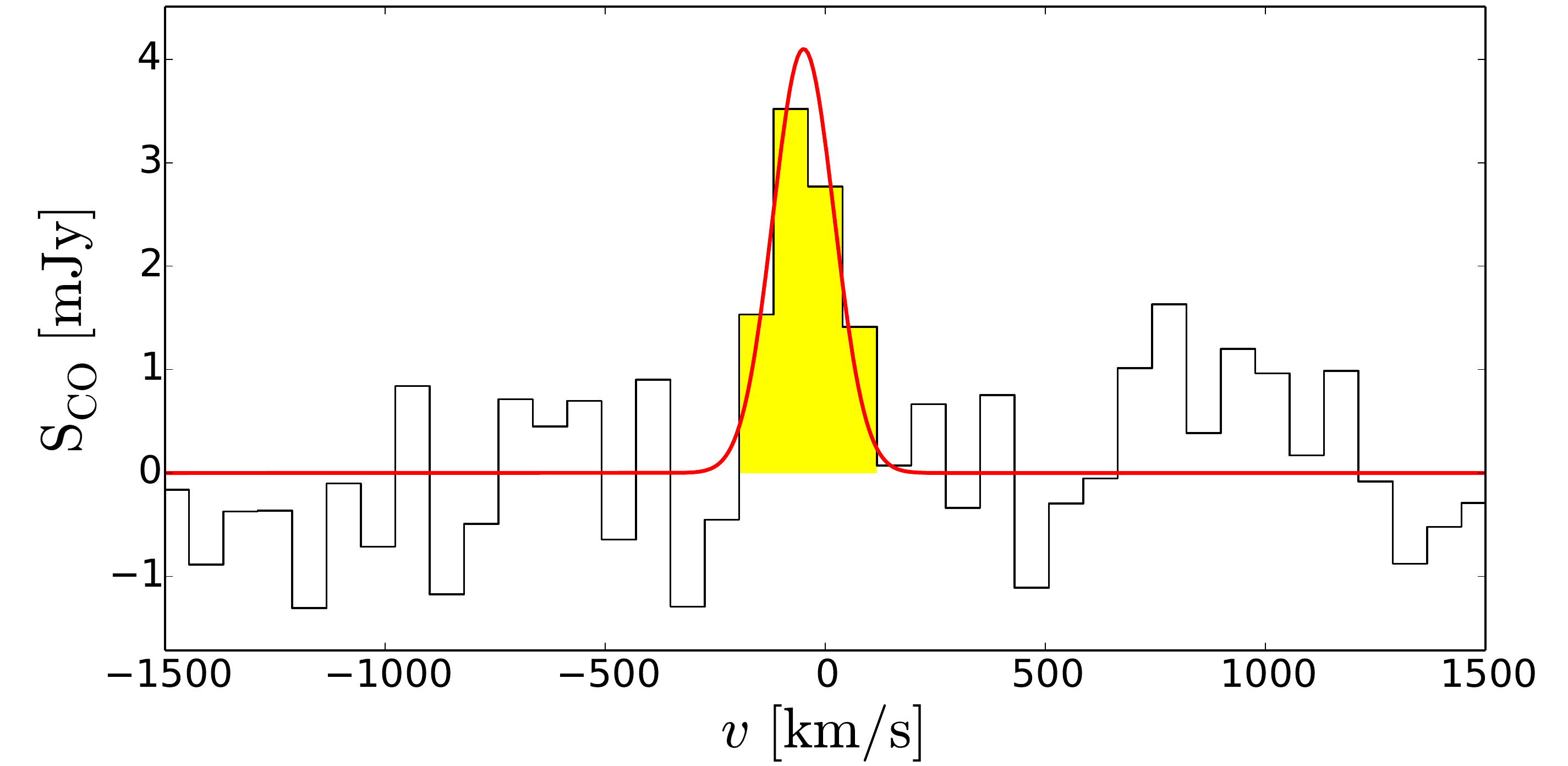}
        \\
        \includegraphics[height=0.1623\textwidth,clip,clip,trim=0 0 4cm 1.1cm]{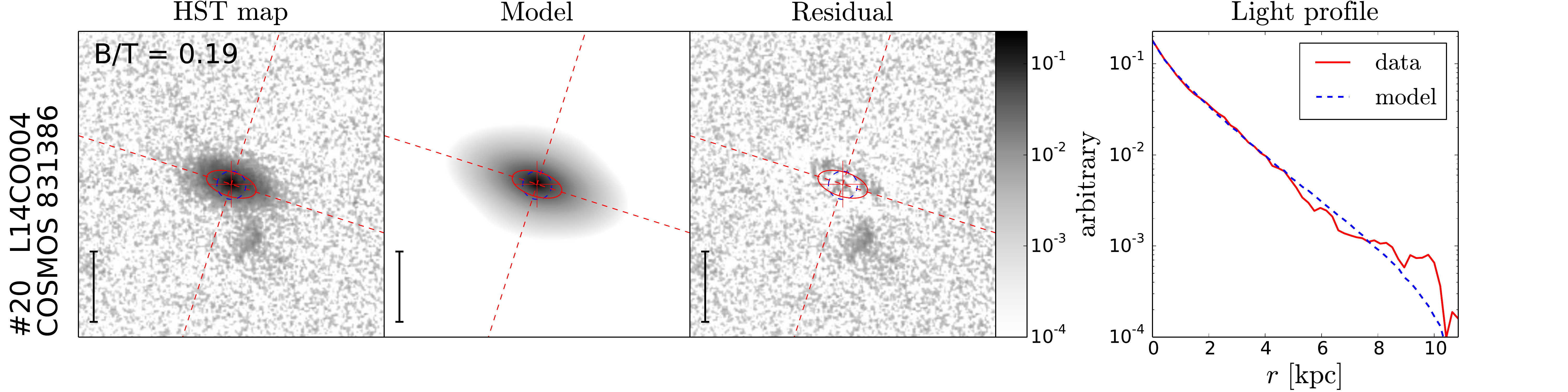}
        \hfill\includegraphics[height=0.1623\textwidth,clip,trim=0 0 32cm 1.1cm]{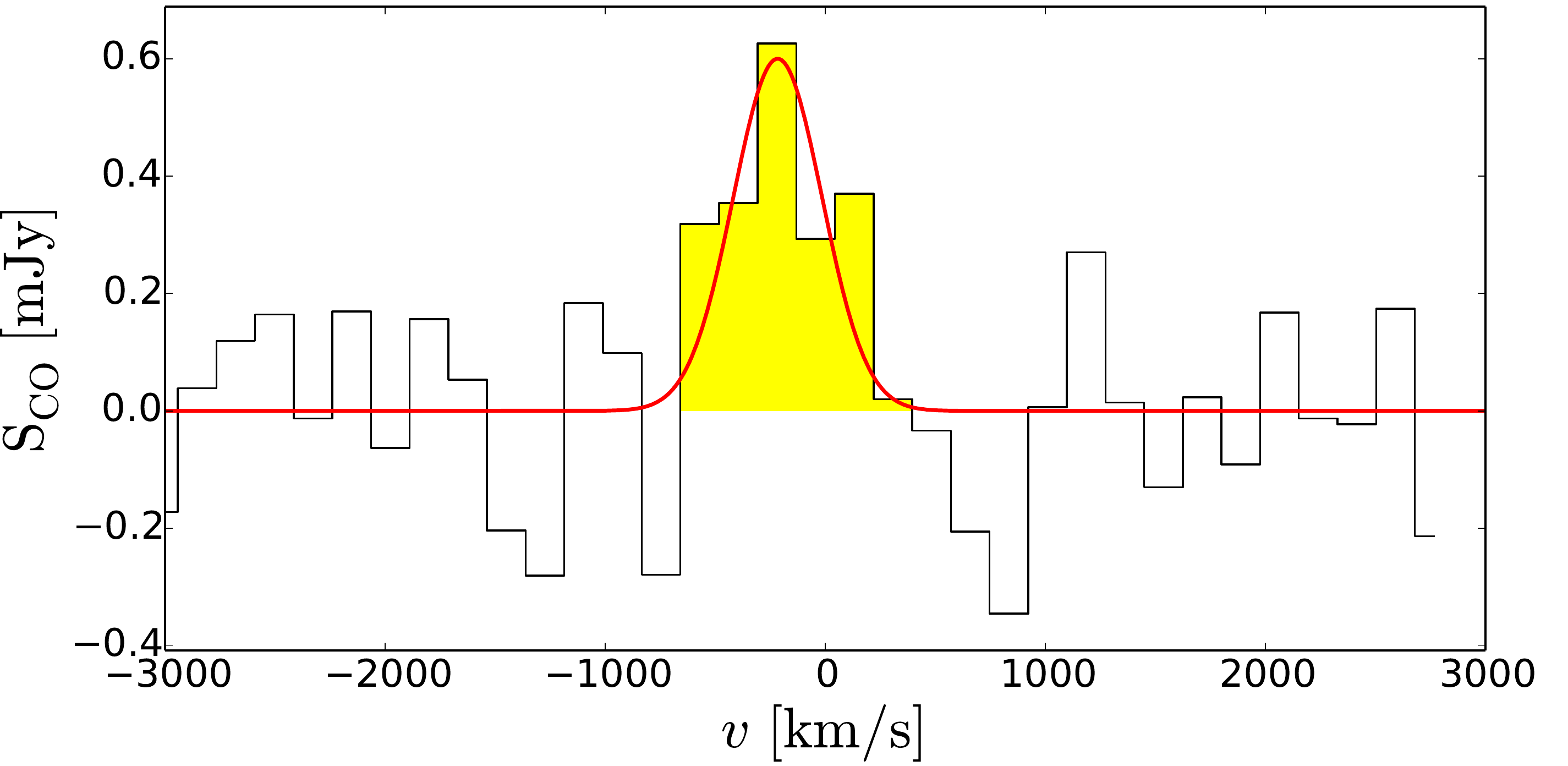}
        \\
        \includegraphics[height=0.1623\textwidth,clip,clip,trim=0 0 4cm 1.1cm]{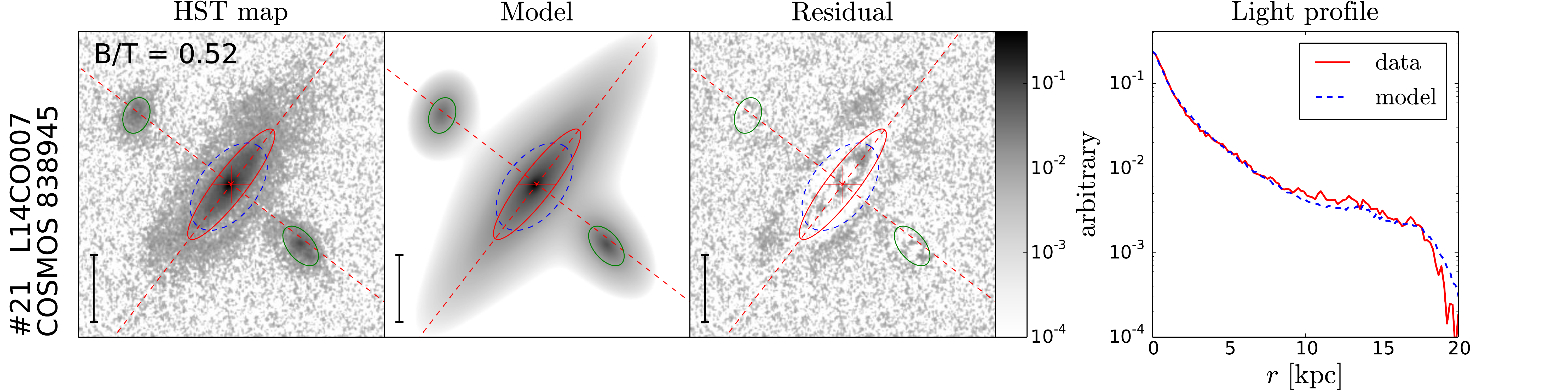}
        \hfill\includegraphics[height=0.1623\textwidth,clip,trim=0 0 32cm 1.1cm]{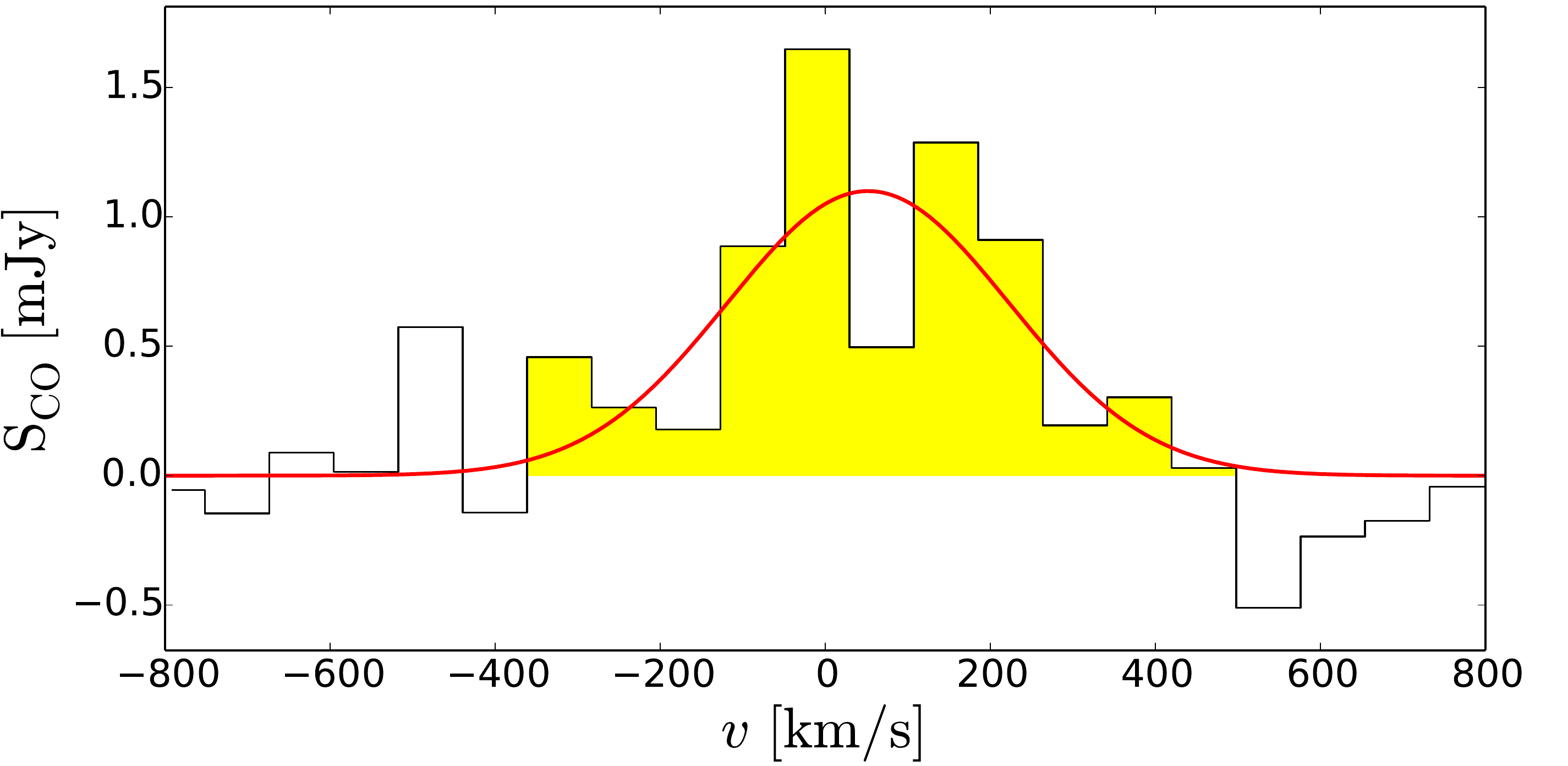}
        \\
\end{figure*}

\begin{figure*}[h!]
        \ContinuedFloat
        \flushleft
        \includegraphics[height=0.175\textwidth,clip,trim=0 0 4cm 0]{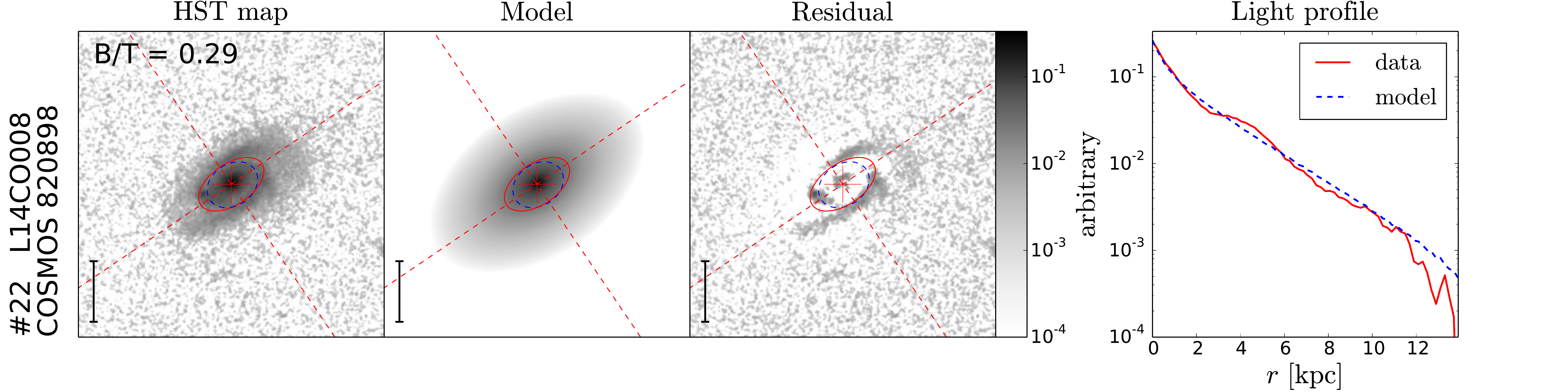}
        \hfill\includegraphics[height=0.175\textwidth,clip,trim=0 0 32cm 0]{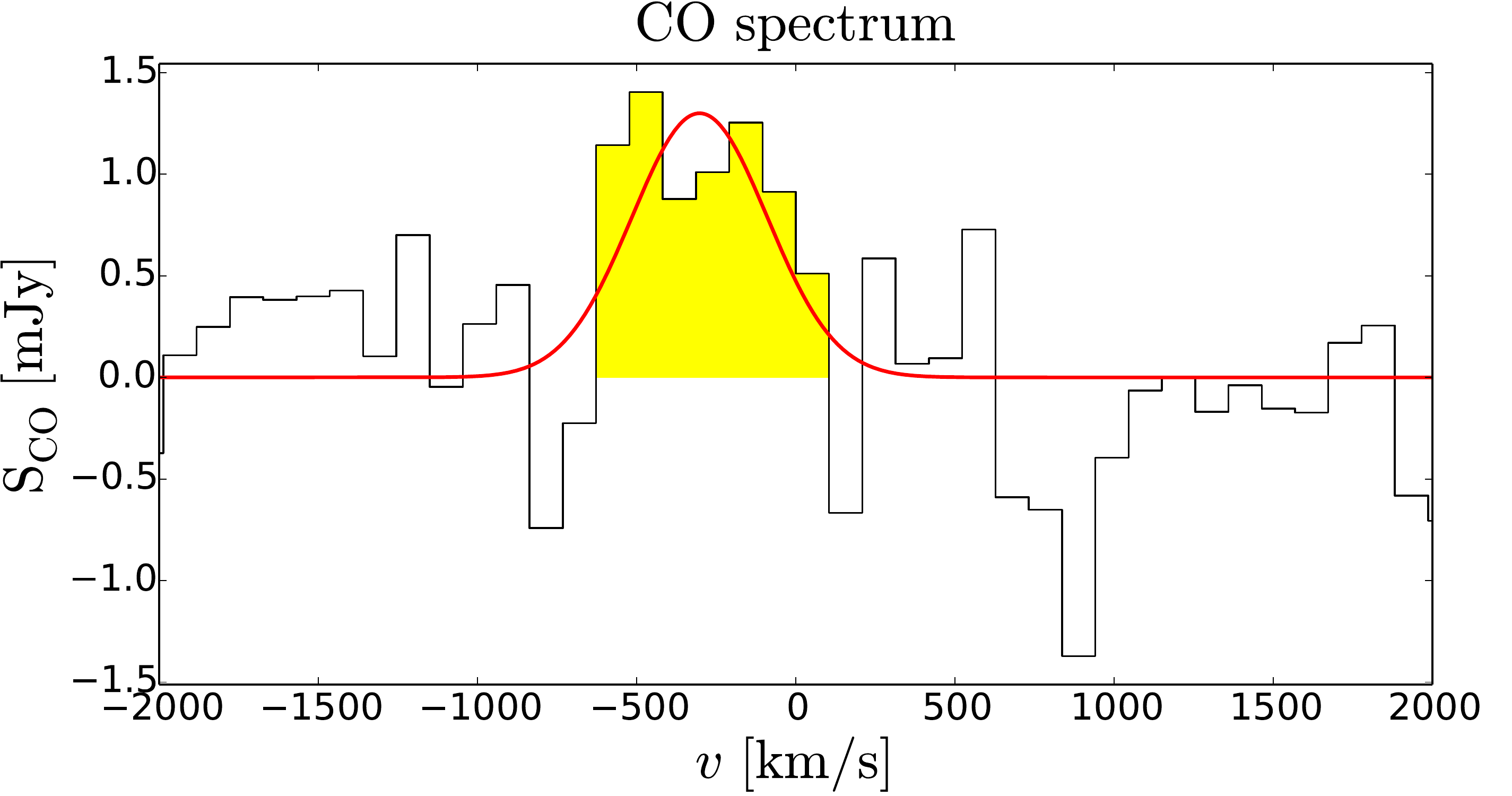}
        \\
        \includegraphics[height=0.1623\textwidth,clip,clip,trim=0 0 4cm 1.1cm]{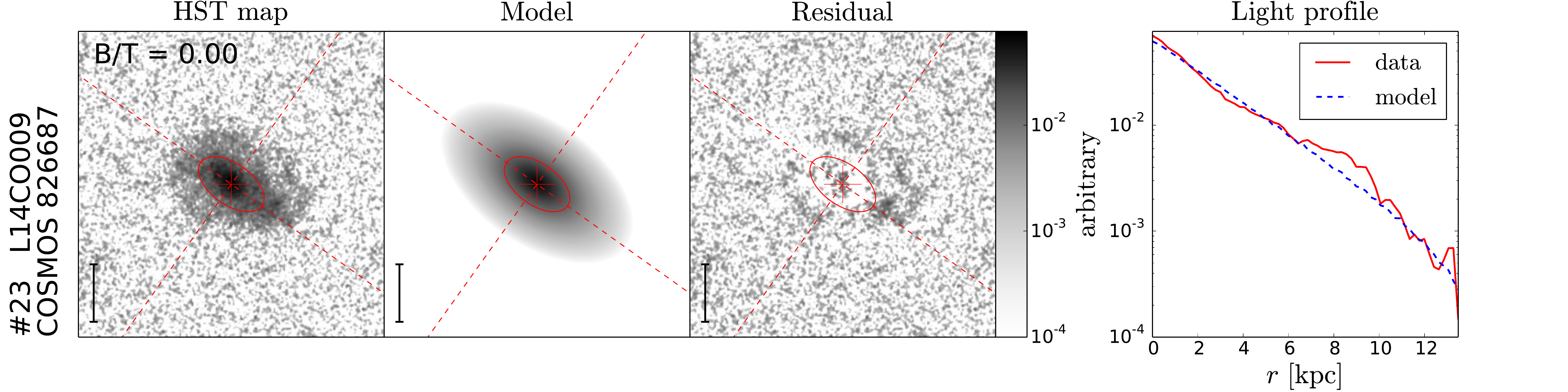}
        \hfill\includegraphics[height=0.1623\textwidth,clip,trim=0 0 32cm 1.1cm]{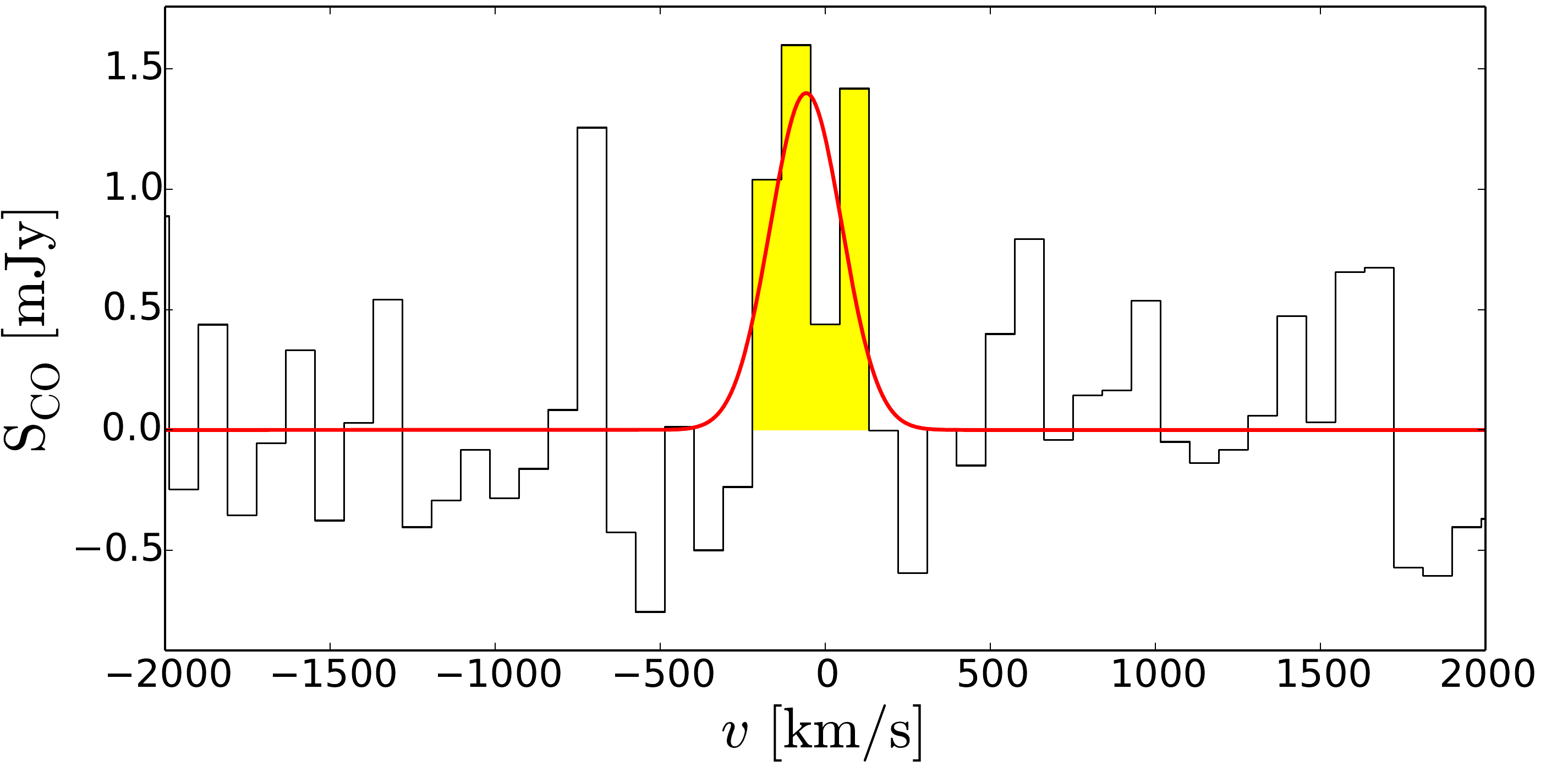}
        \\
        \includegraphics[height=0.1623\textwidth,clip,clip,trim=0 0 4cm 1.1cm]{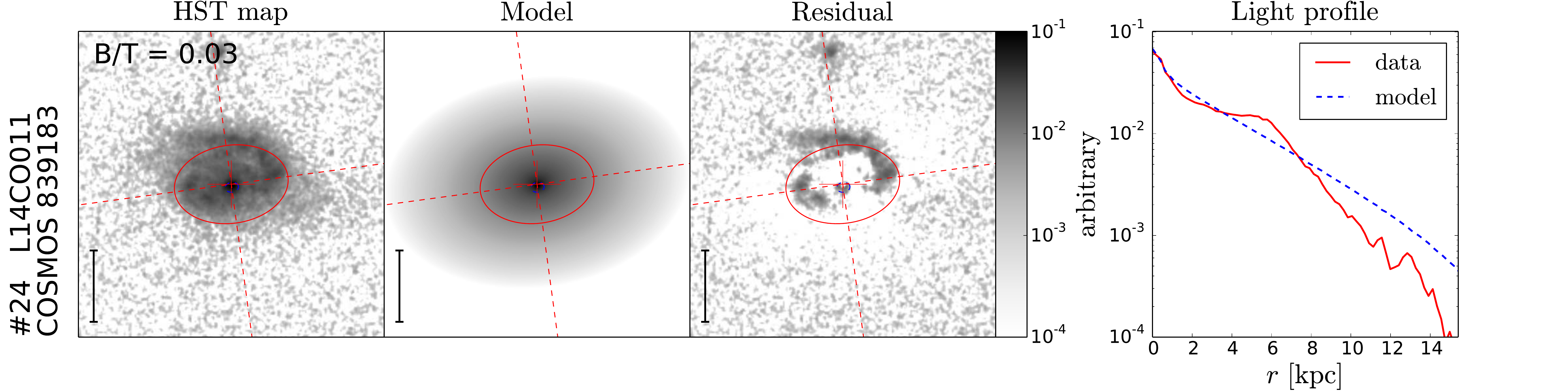}
        \hfill\includegraphics[height=0.1623\textwidth,clip,trim=0 0 32cm 1.1cm]{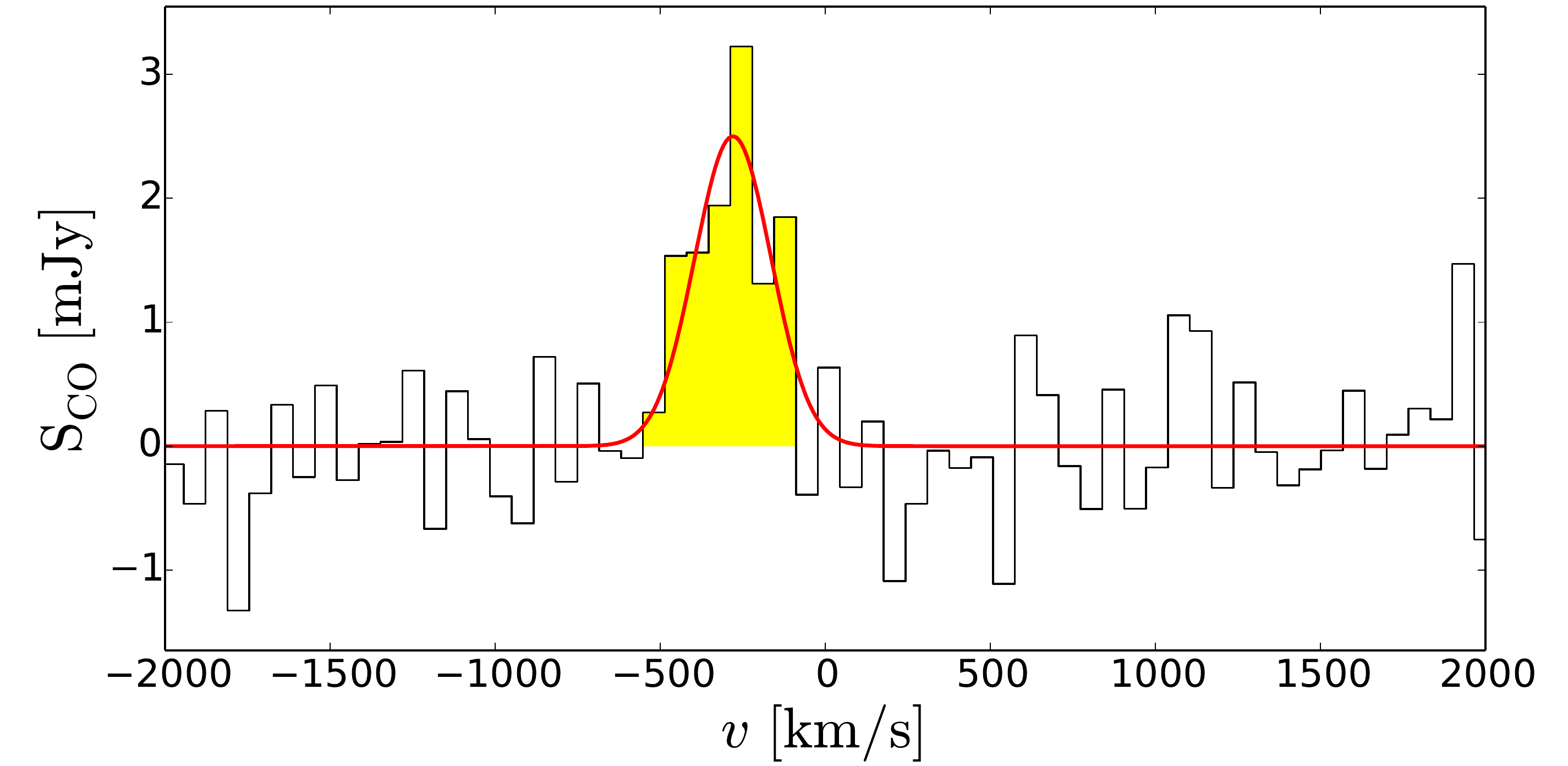}
        \\
        \includegraphics[height=0.1623\textwidth,clip,clip,trim=0 0 4cm 1.1cm]{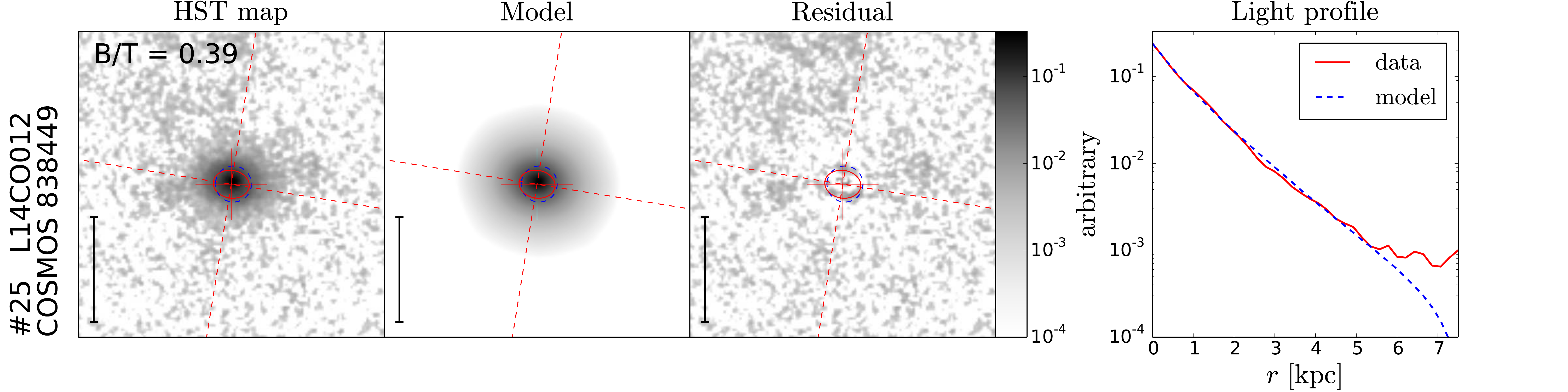}
        \hfill\includegraphics[height=0.1623\textwidth,clip,trim=0 0 32cm 1.1cm]{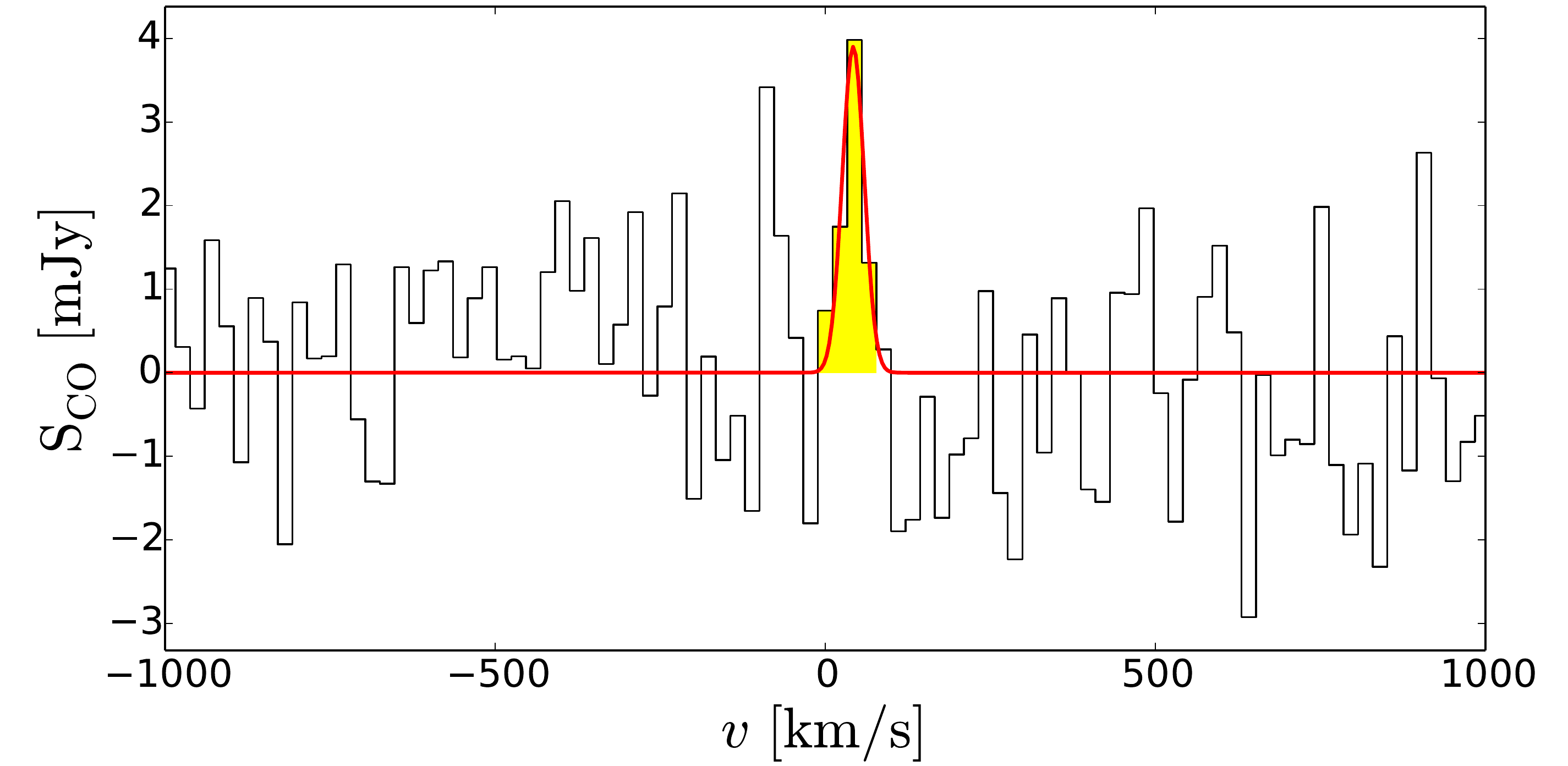}
        \\
        \includegraphics[height=0.1623\textwidth,clip,clip,trim=0 0 4cm 1.1cm]{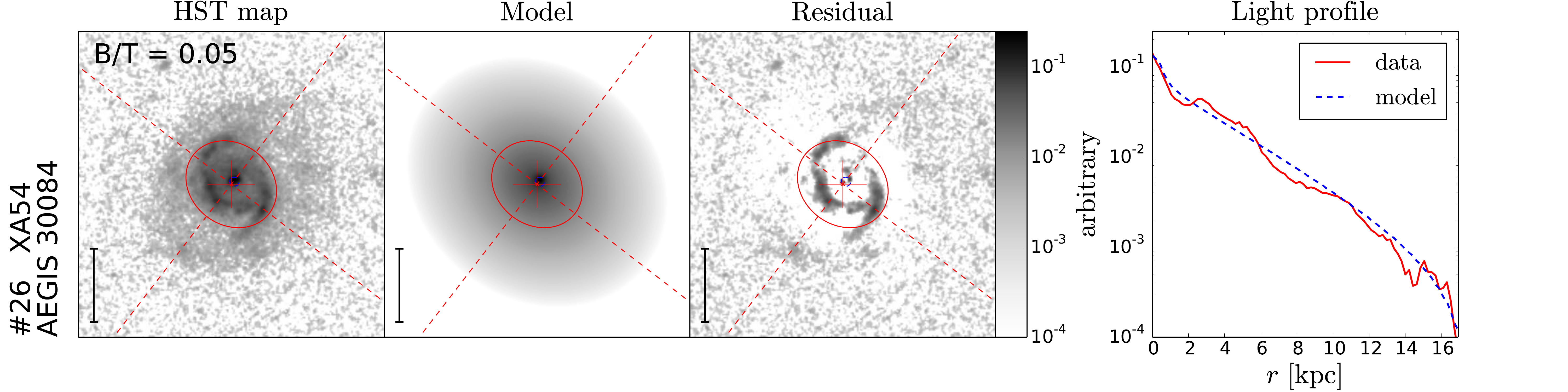}
        \hfill\includegraphics[height=0.1623\textwidth,clip,trim=0 0 32cm 1.1cm]{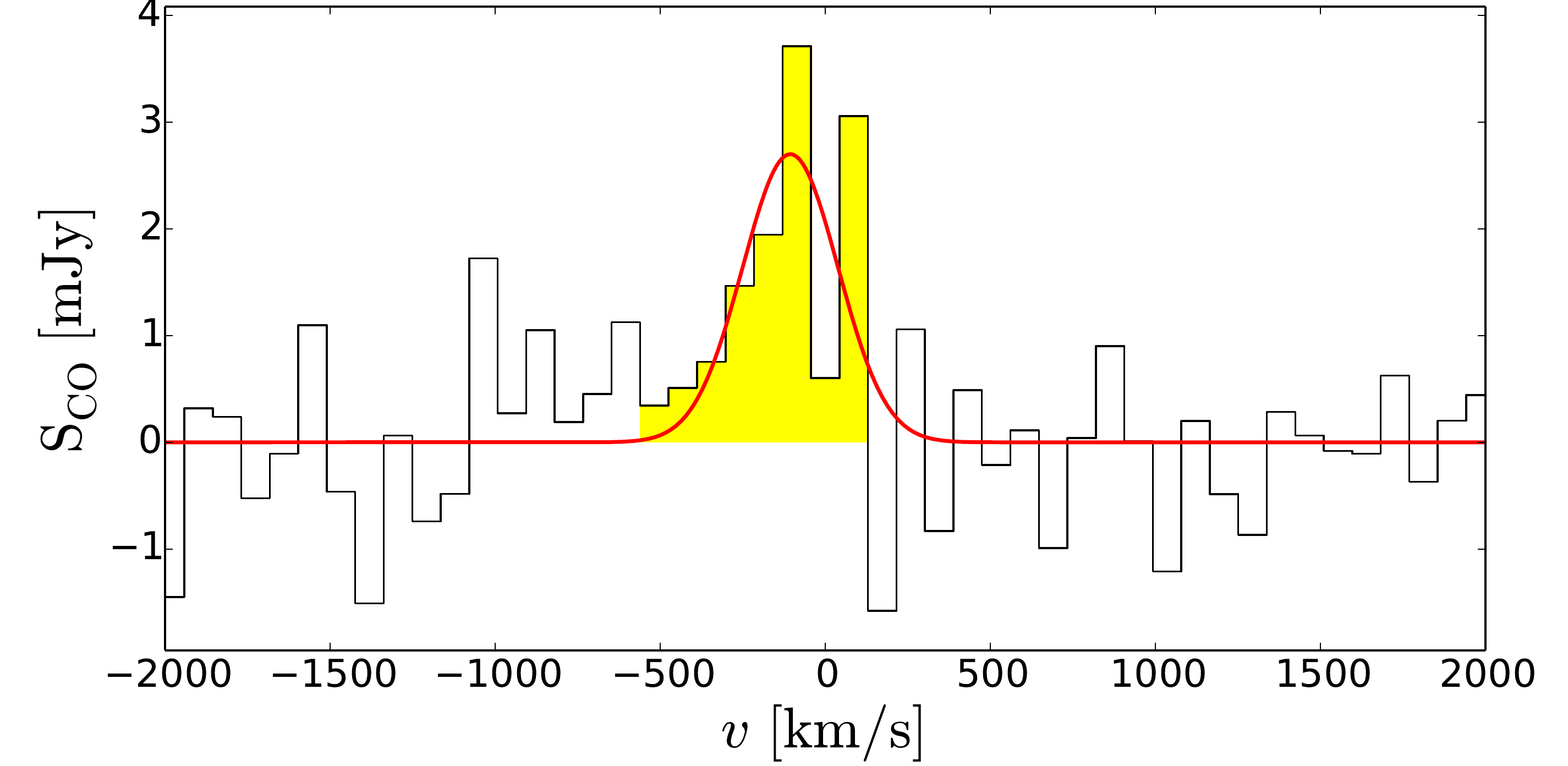}
        \\
        \includegraphics[height=0.1623\textwidth,clip,clip,trim=0 0 4cm 1.1cm]{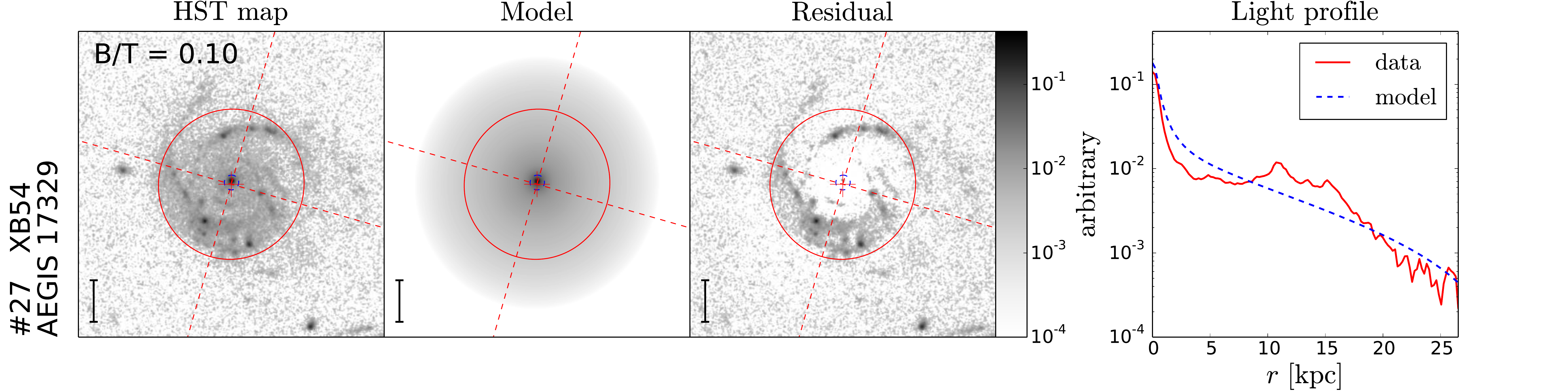}
        \hfill\includegraphics[height=0.1623\textwidth,clip,trim=0 0 32cm 1.1cm]{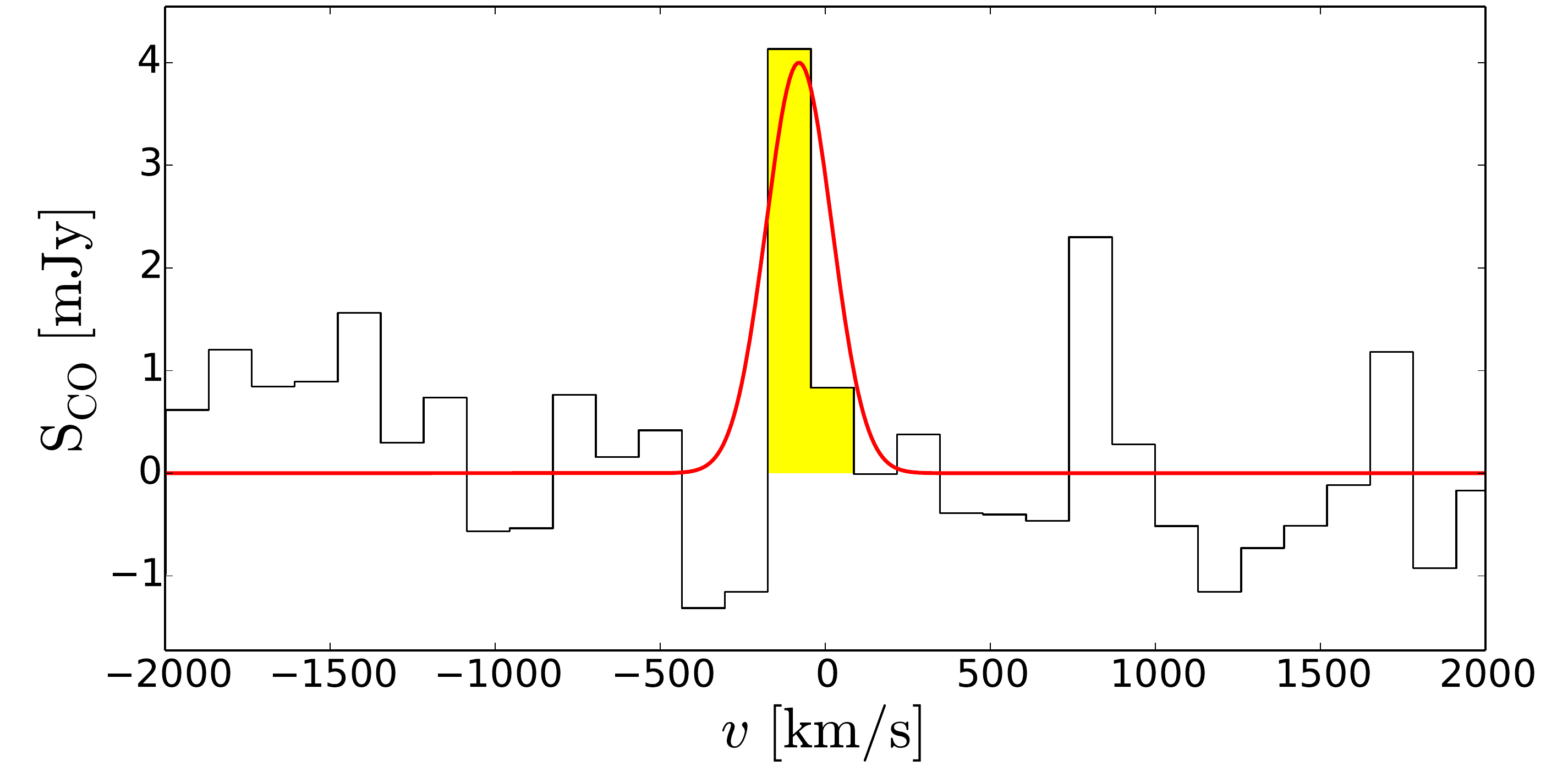}
        \\
        \includegraphics[height=0.1623\textwidth,clip,clip,trim=0 0 4cm 1.1cm]{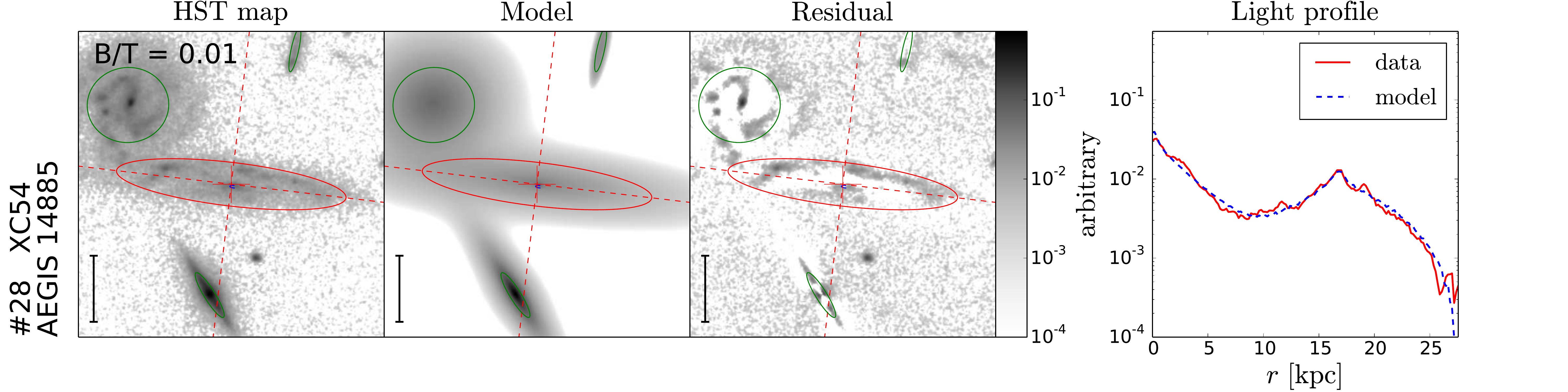}
        \hfill\includegraphics[height=0.1623\textwidth,clip,trim=0 0 32cm 1.1cm]{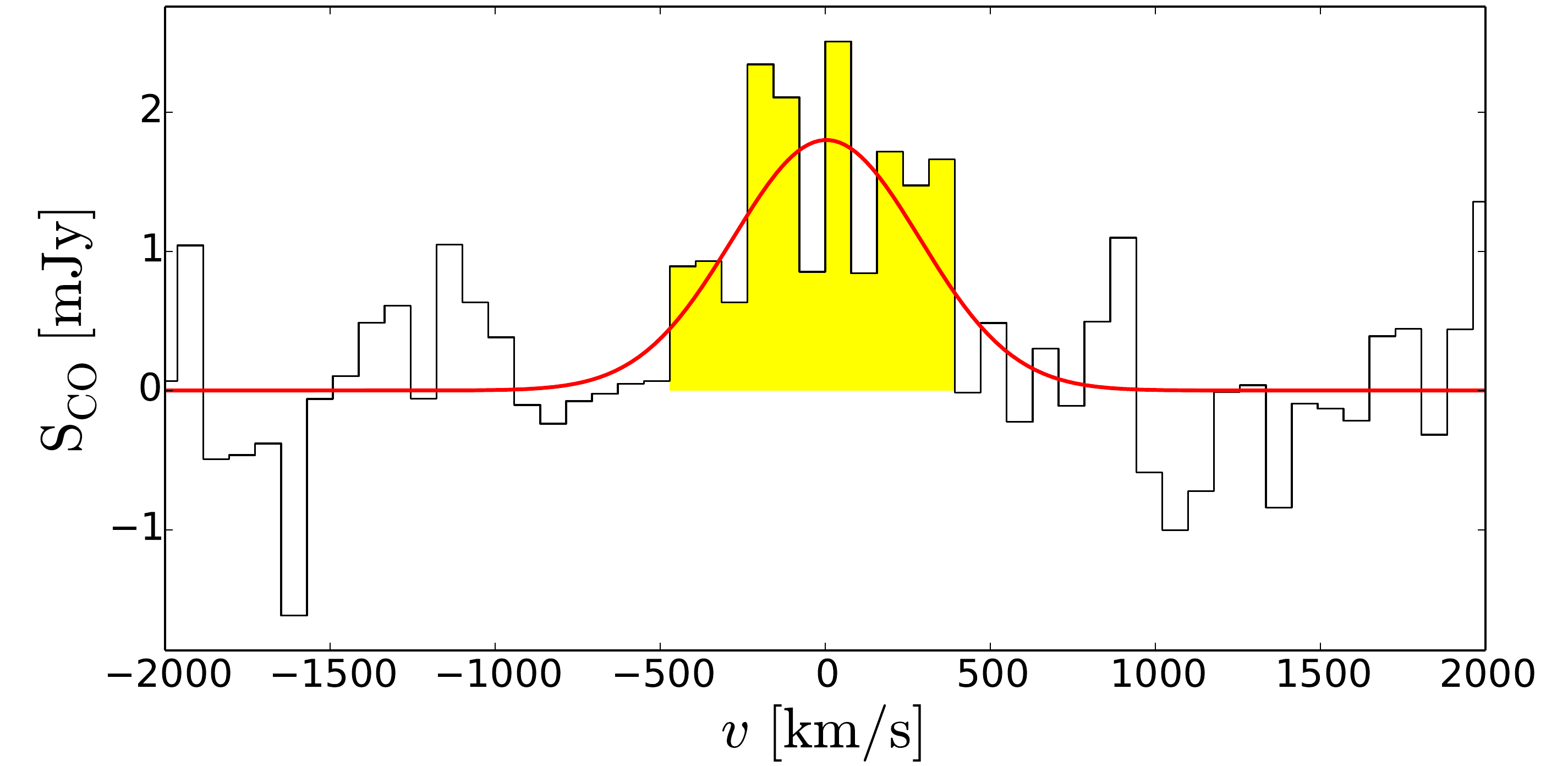}
        \\
\end{figure*}

\begin{figure*}[h!]
        \ContinuedFloat
        \flushleft
        \includegraphics[height=0.175\textwidth,clip,trim=0 0 4cm 0]{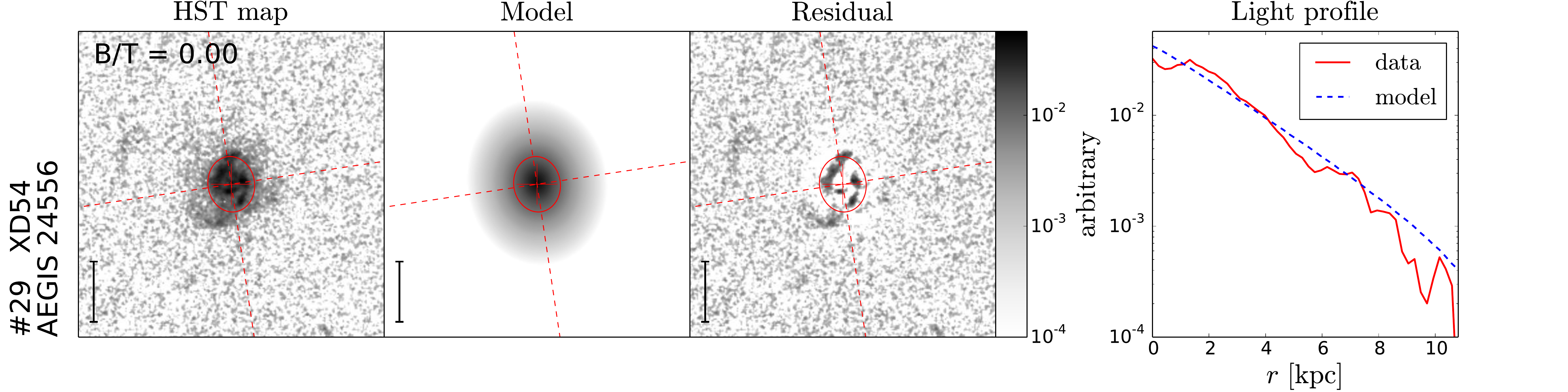}
        \hfill\includegraphics[height=0.175\textwidth,clip,trim=0 0 32cm 0]{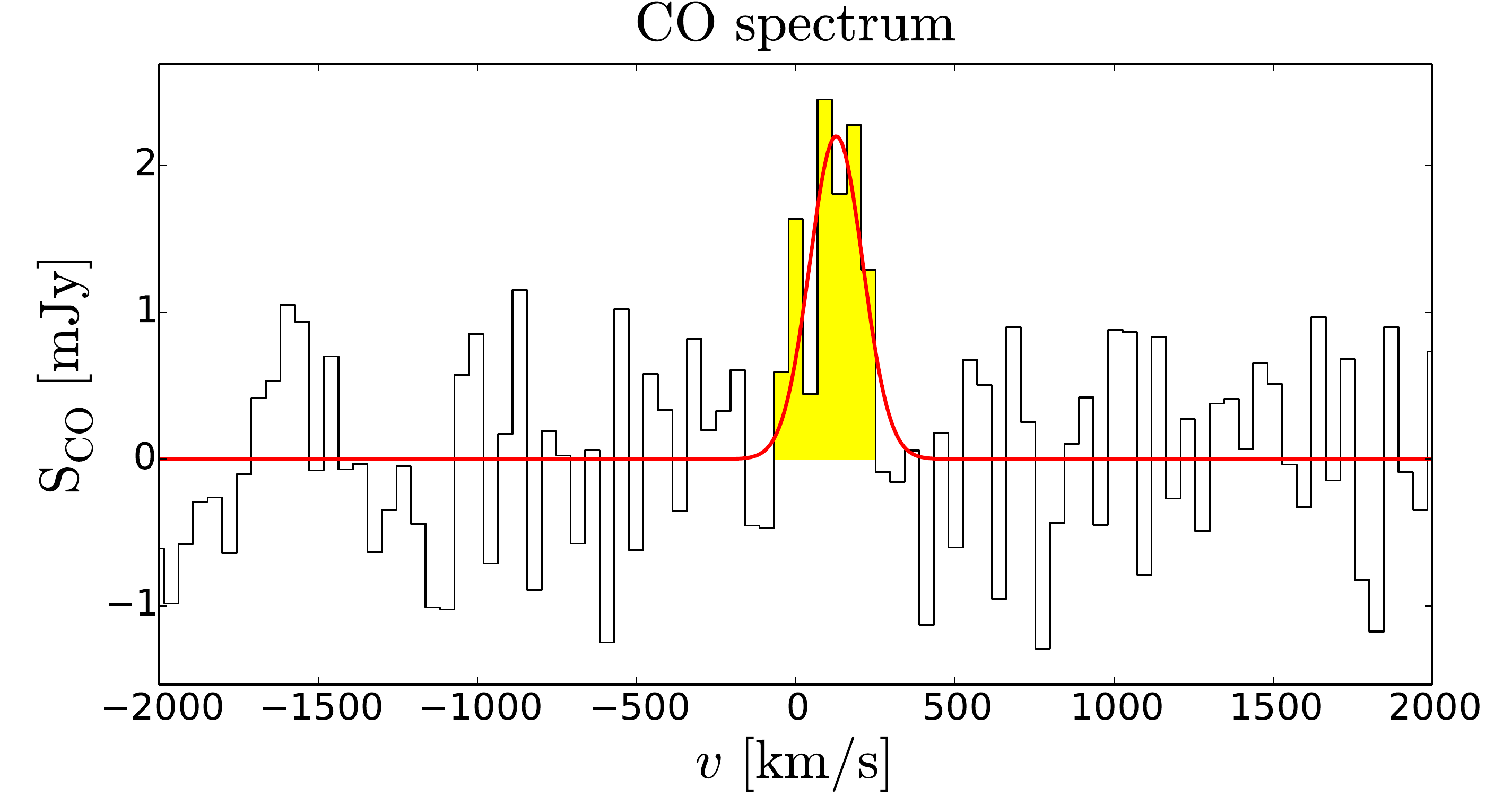}
        \\
        \includegraphics[height=0.1623\textwidth,clip,clip,trim=0 0 4cm 1.1cm]{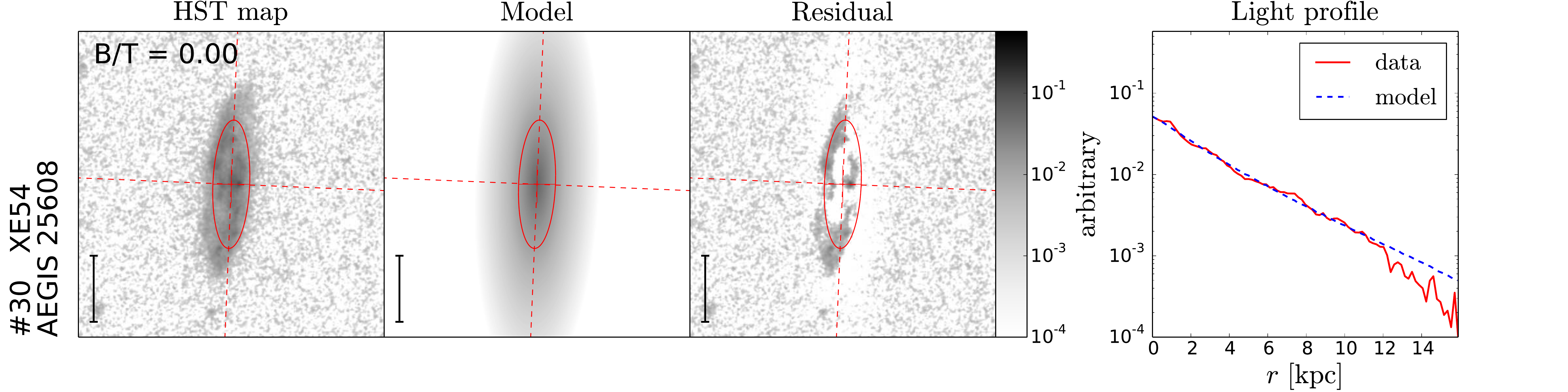}
        \hfill\includegraphics[height=0.1623\textwidth,clip,trim=0 0 32cm 1.1cm]{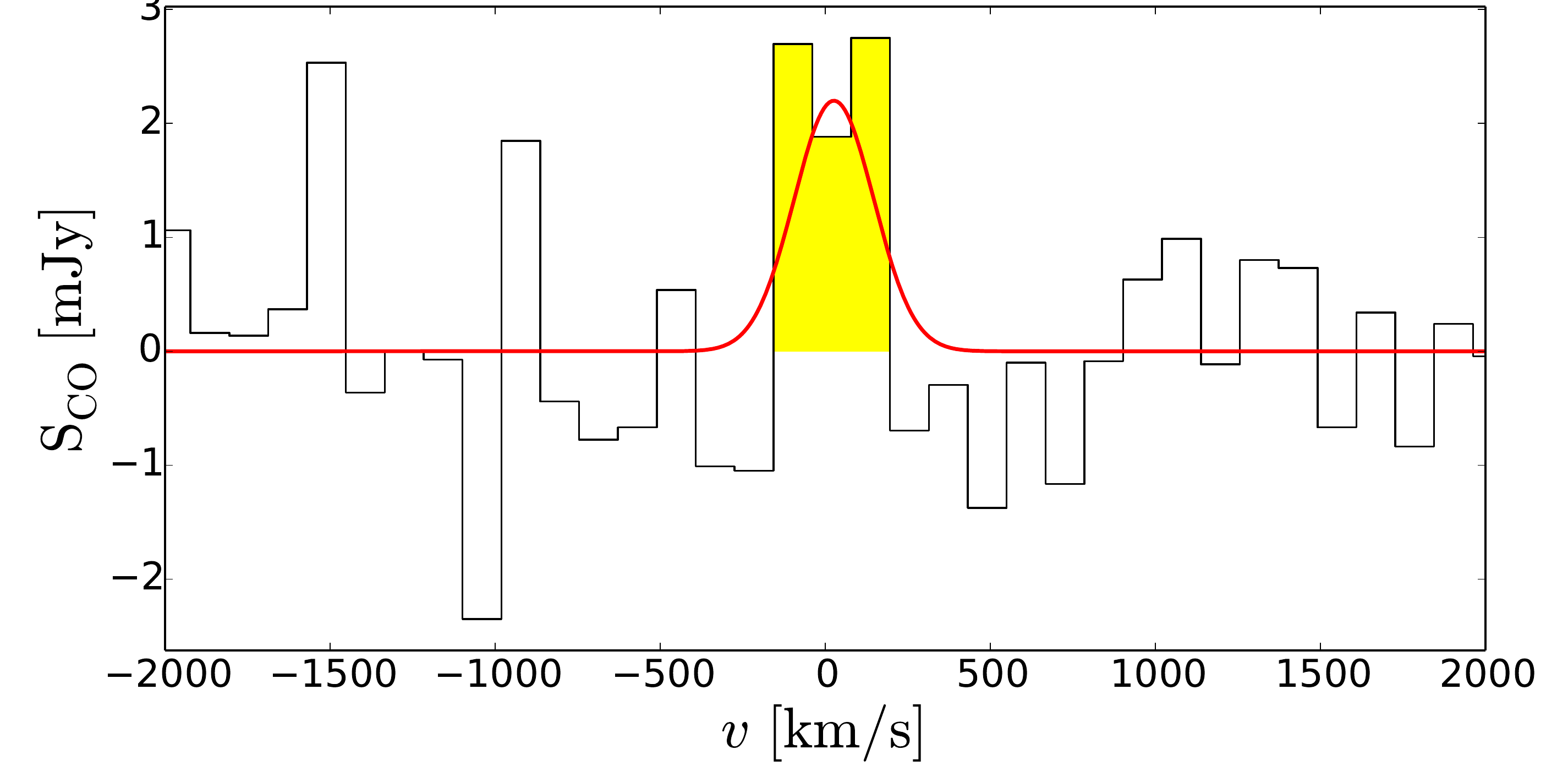}
        \\
        \includegraphics[height=0.1623\textwidth,clip,clip,trim=0 0 4cm 1.1cm]{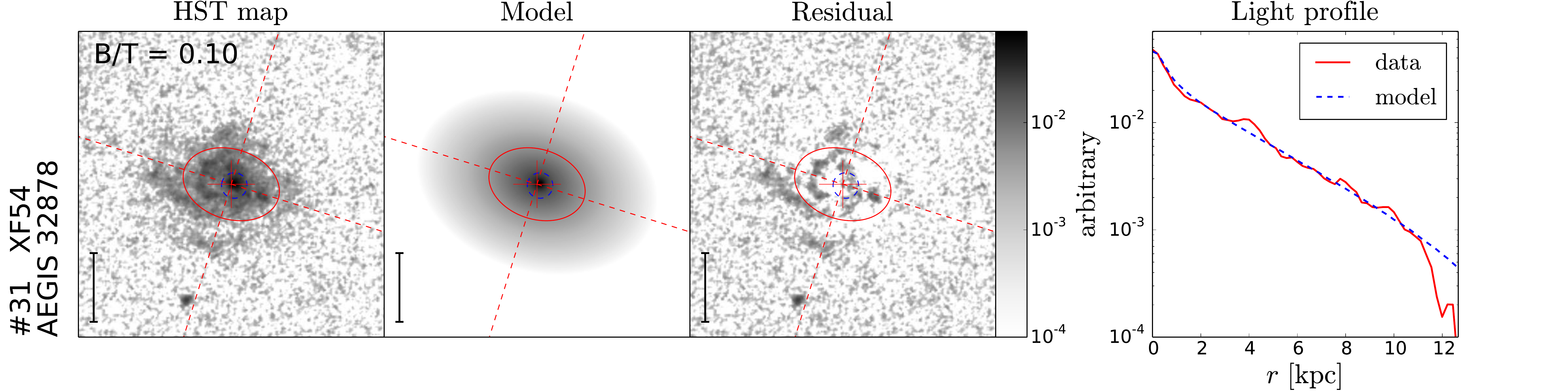}
        \hfill\includegraphics[height=0.1623\textwidth,clip,trim=0 0 32cm 1.1cm]{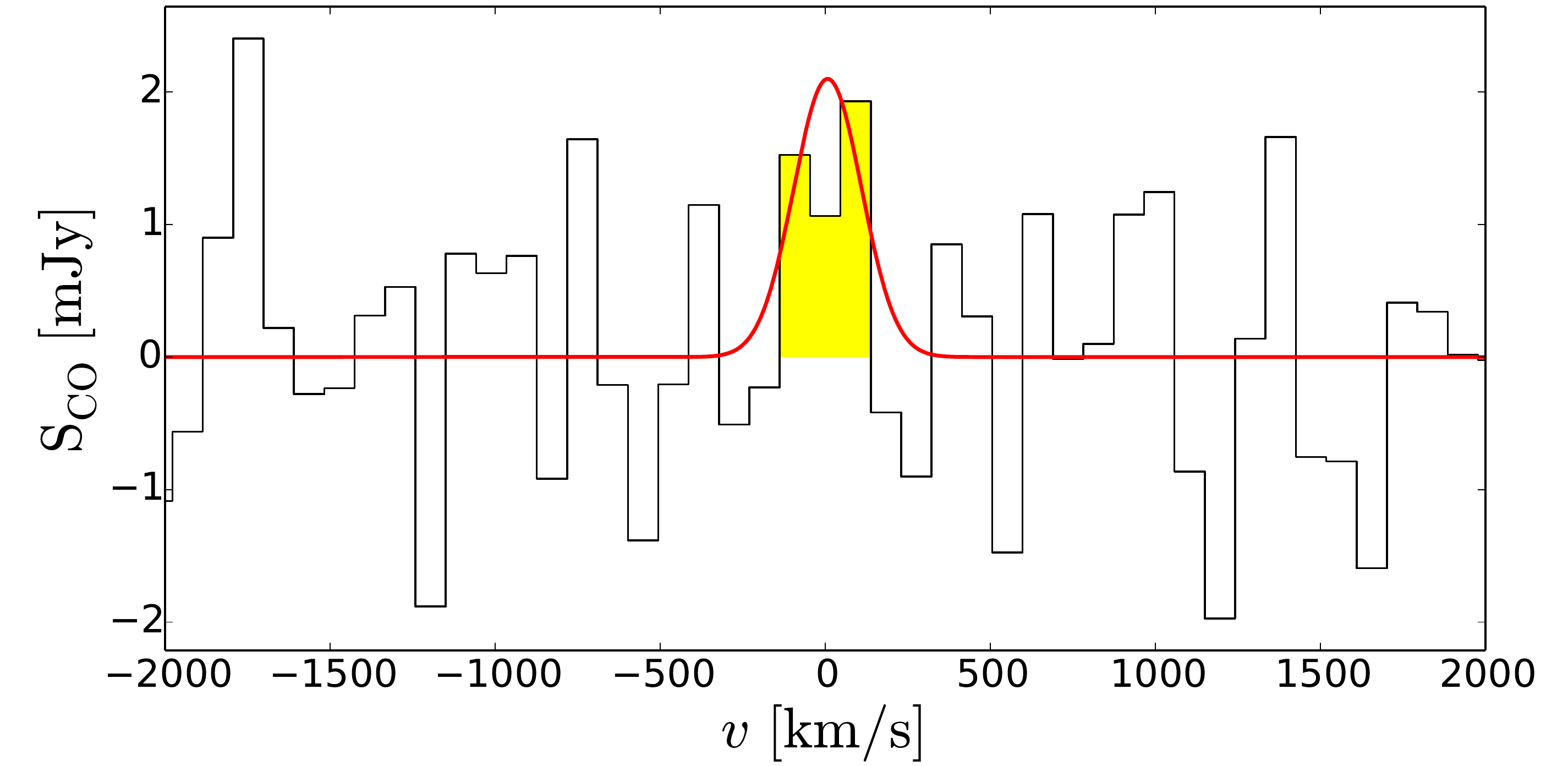}
        \\
        \includegraphics[height=0.1623\textwidth,clip,clip,trim=0 0 4cm 1.1cm]{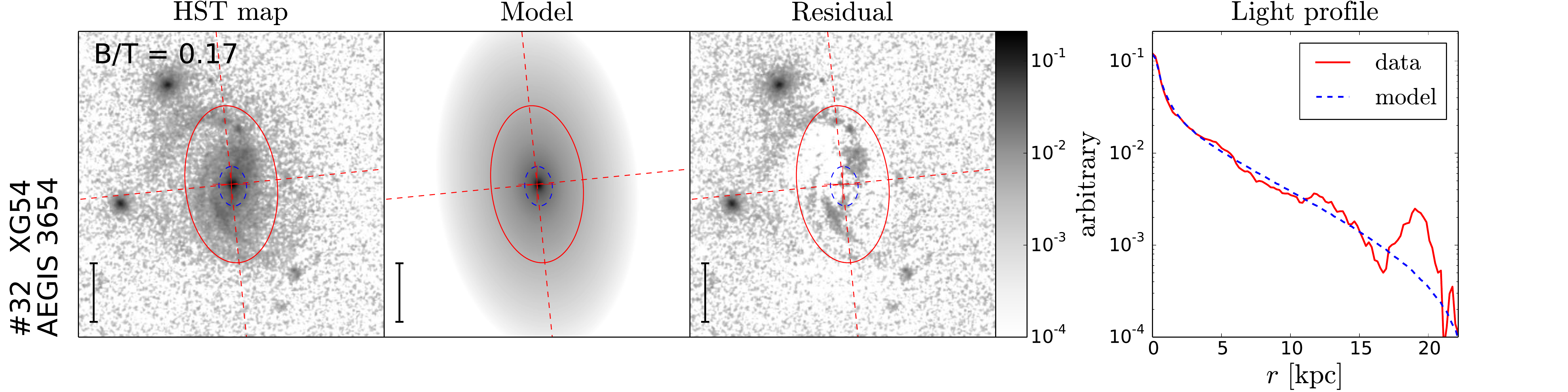}
        \hfill\includegraphics[height=0.1623\textwidth,clip,trim=0 0 32cm 1.1cm]{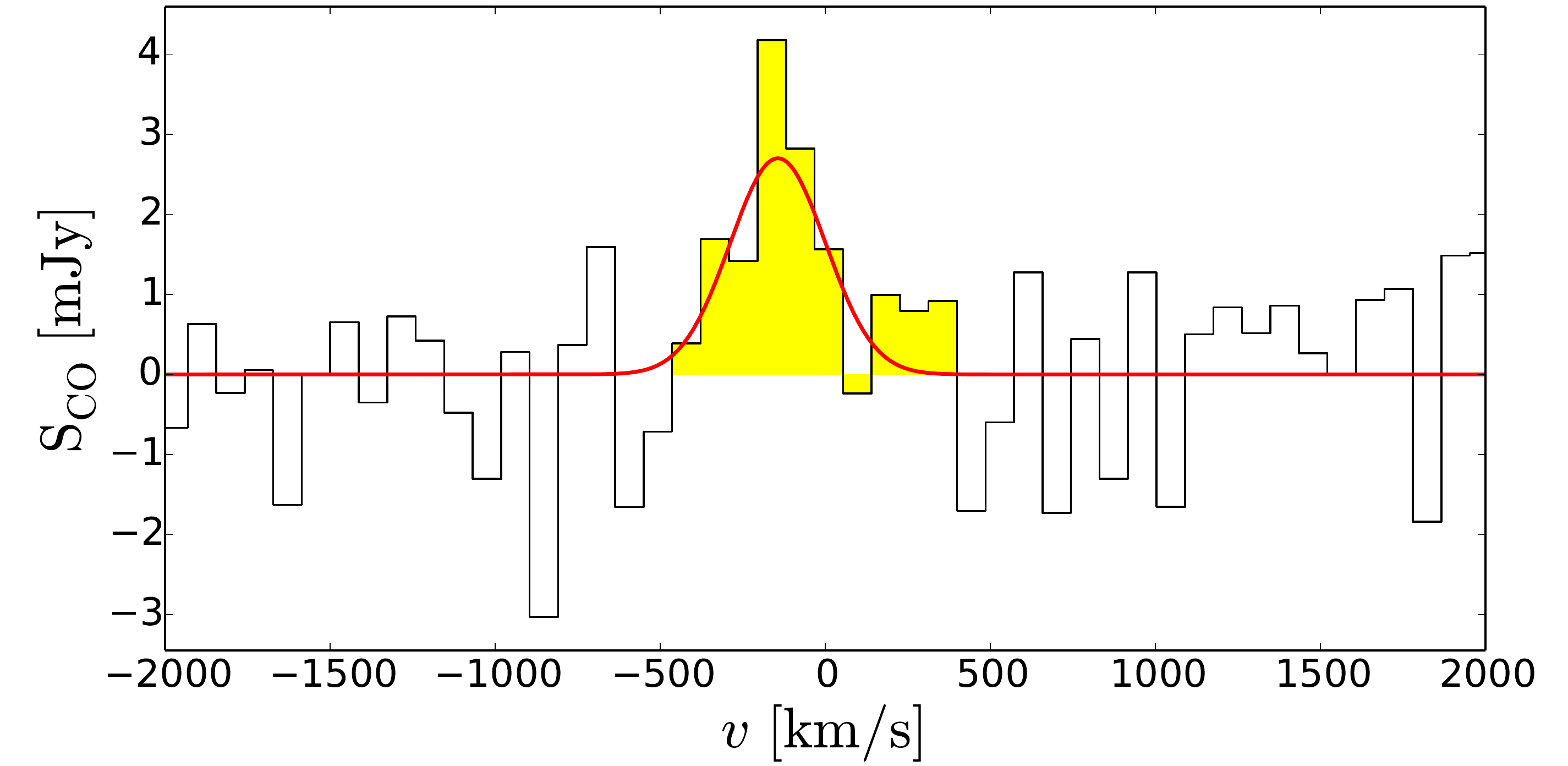}
        \\
        \includegraphics[height=0.1623\textwidth,clip,clip,trim=0 0 4cm 1.1cm]{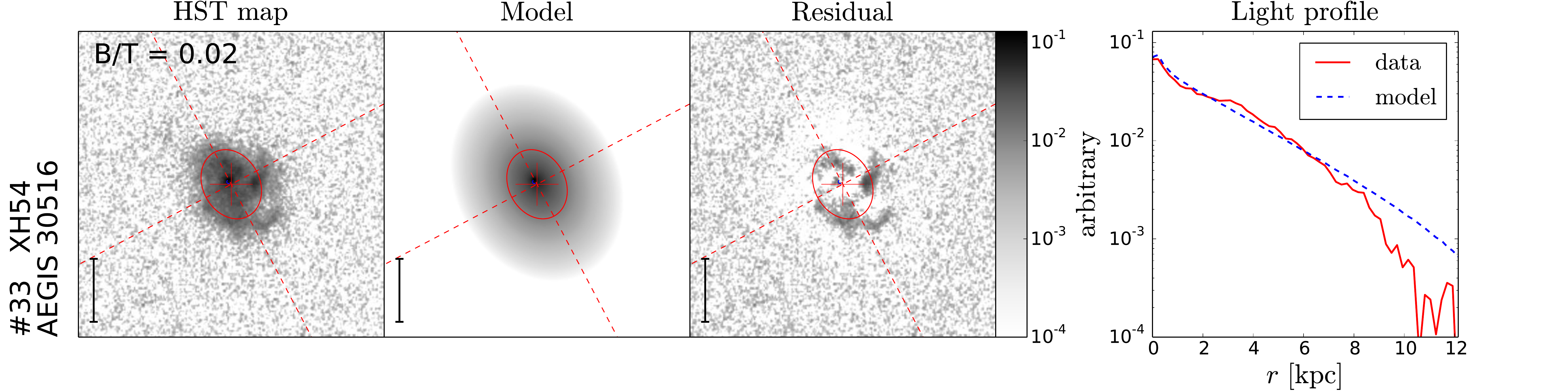}
        \hfill\includegraphics[height=0.1623\textwidth,clip,trim=0 0 32cm 1.1cm]{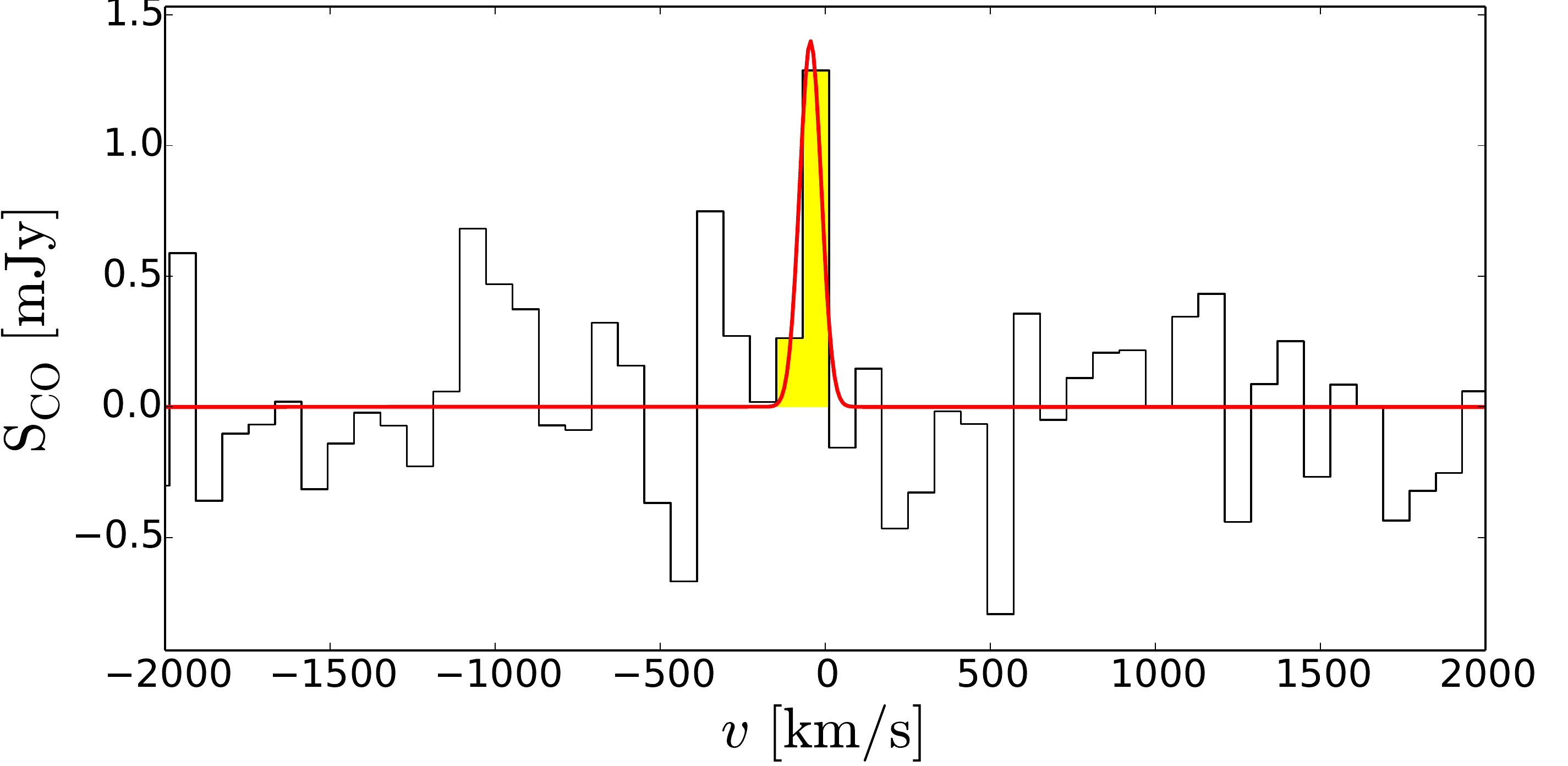}
        \\
        \includegraphics[height=0.1623\textwidth,clip,clip,trim=0 0 4cm 1.1cm]{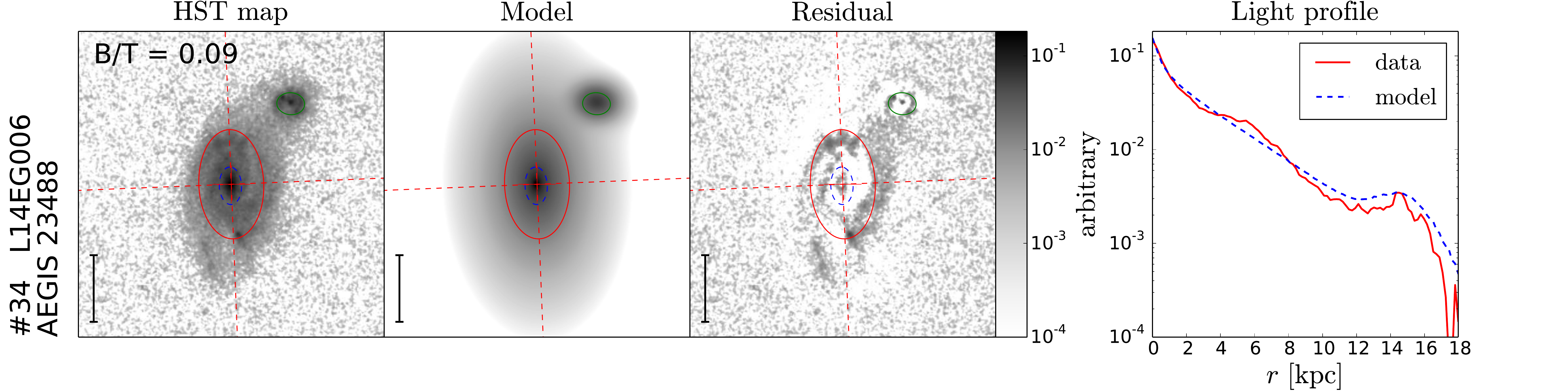}
        \hfill\includegraphics[height=0.1623\textwidth,clip,trim=0 0 32cm 1.1cm]{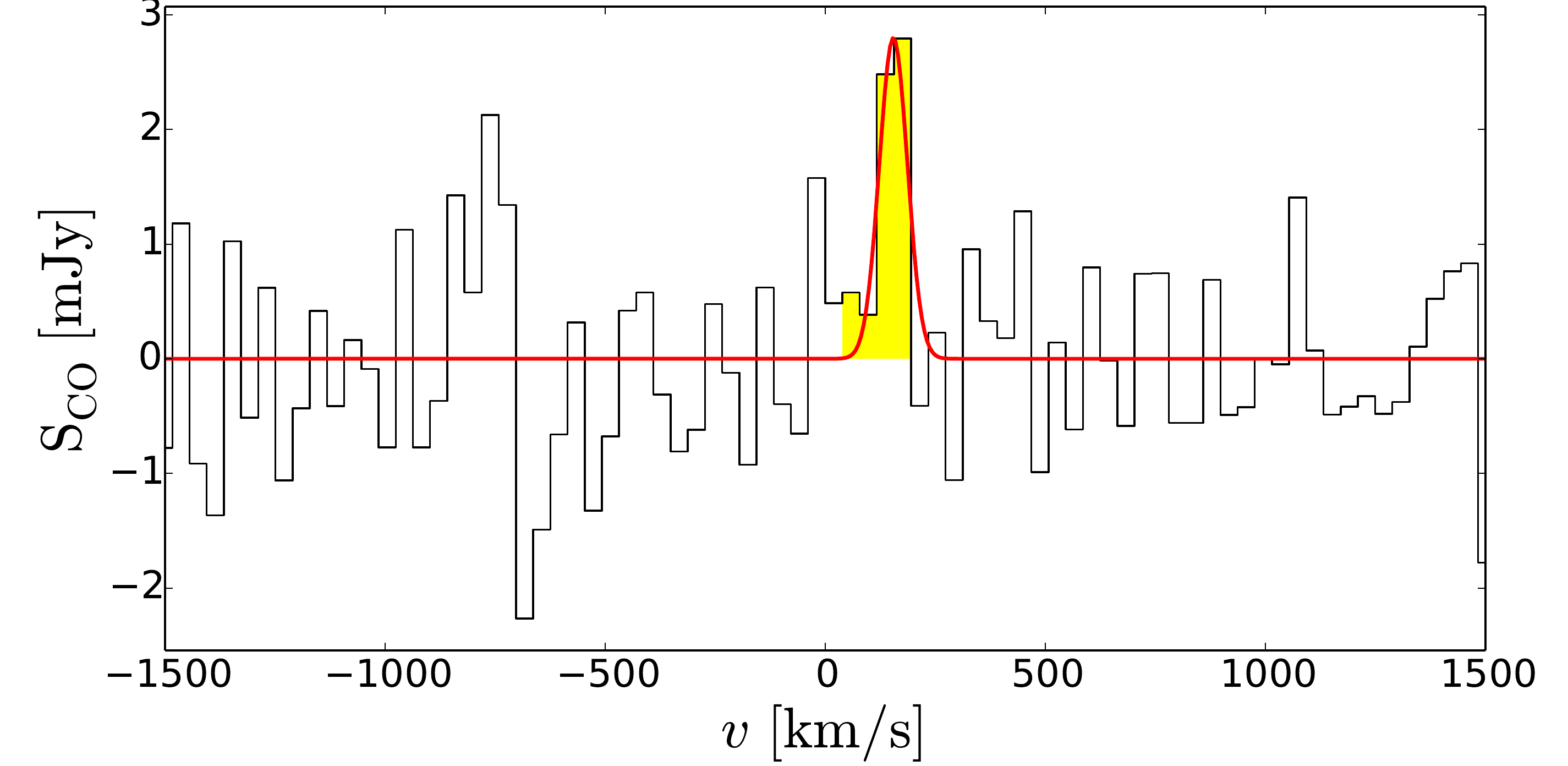}
        \\
        \includegraphics[height=0.1623\textwidth,clip,clip,trim=0 0 4cm 1.1cm]{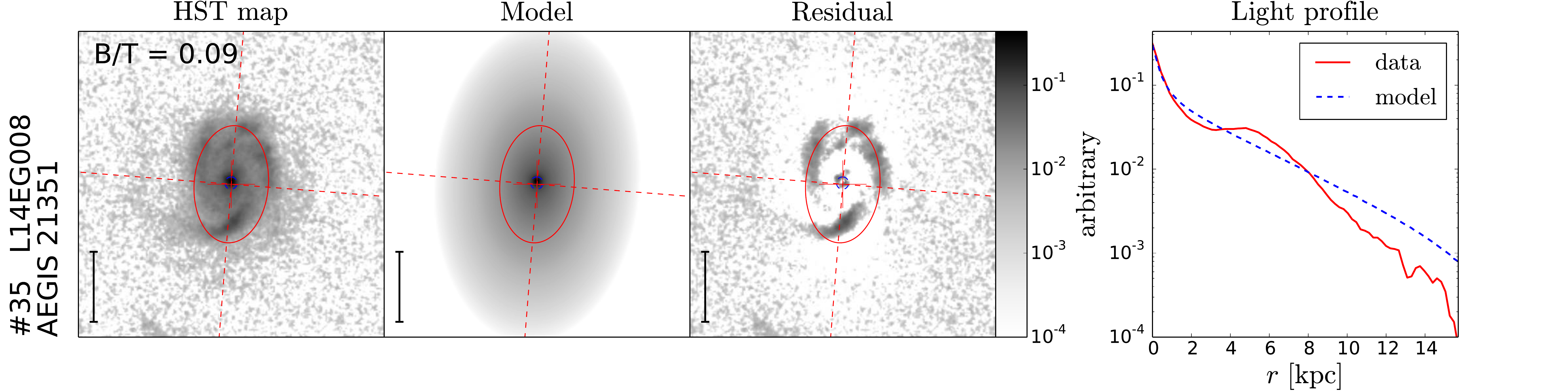}
        \hfill\includegraphics[height=0.1623\textwidth,clip,trim=0 0 32cm 1.1cm]{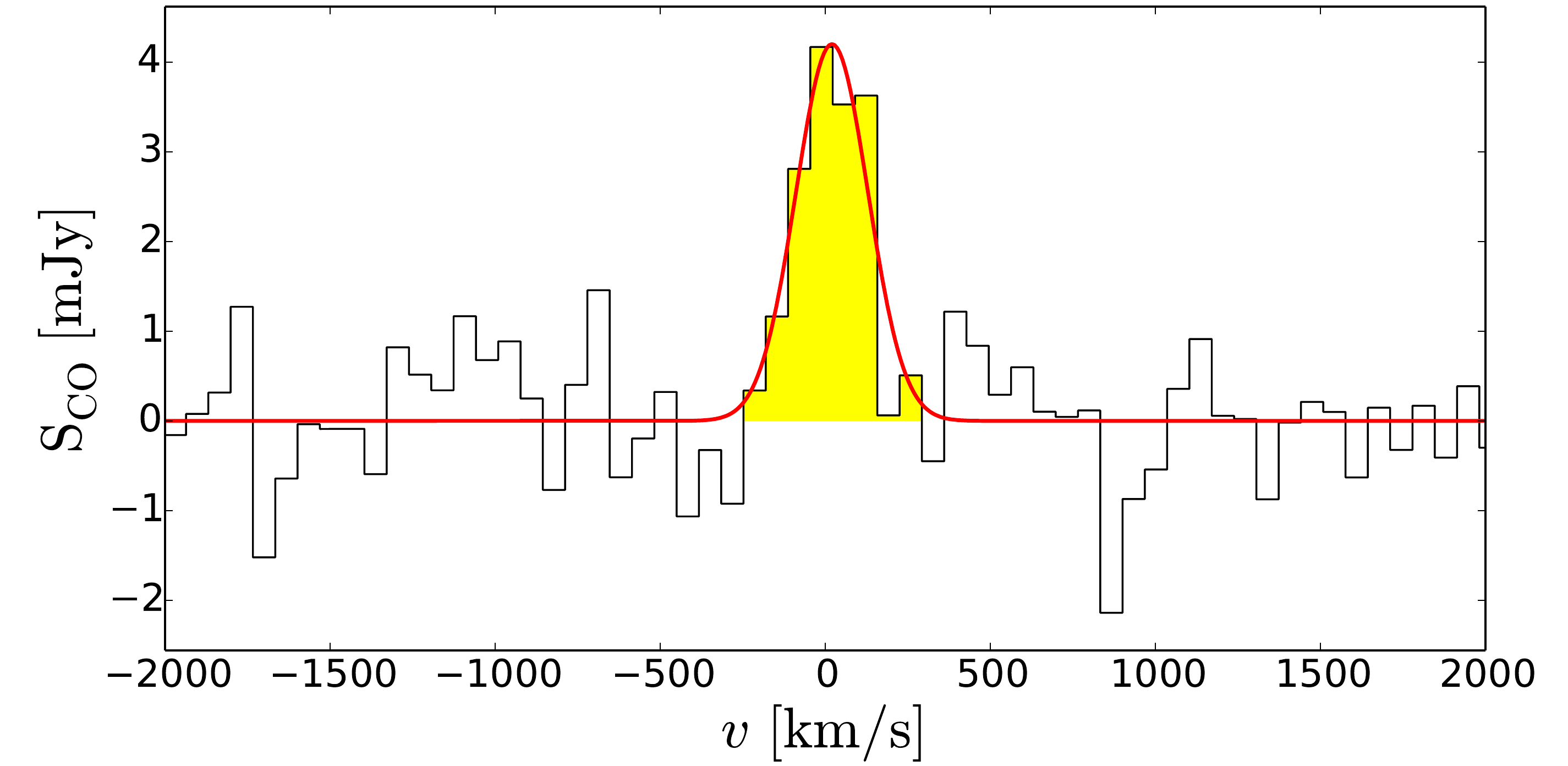}
        \\
\end{figure*}

\begin{figure*}[h!]
        \ContinuedFloat
        \flushleft
        \includegraphics[height=0.175\textwidth,clip,trim=0 0 4cm 0]{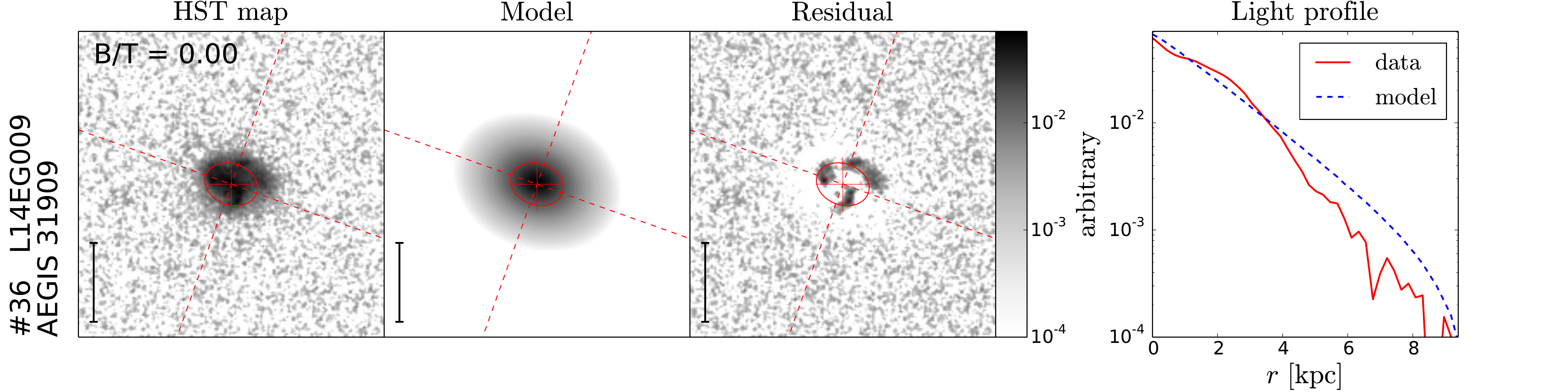}
        \hfill\includegraphics[height=0.175\textwidth,clip,trim=0 0 32cm 0]{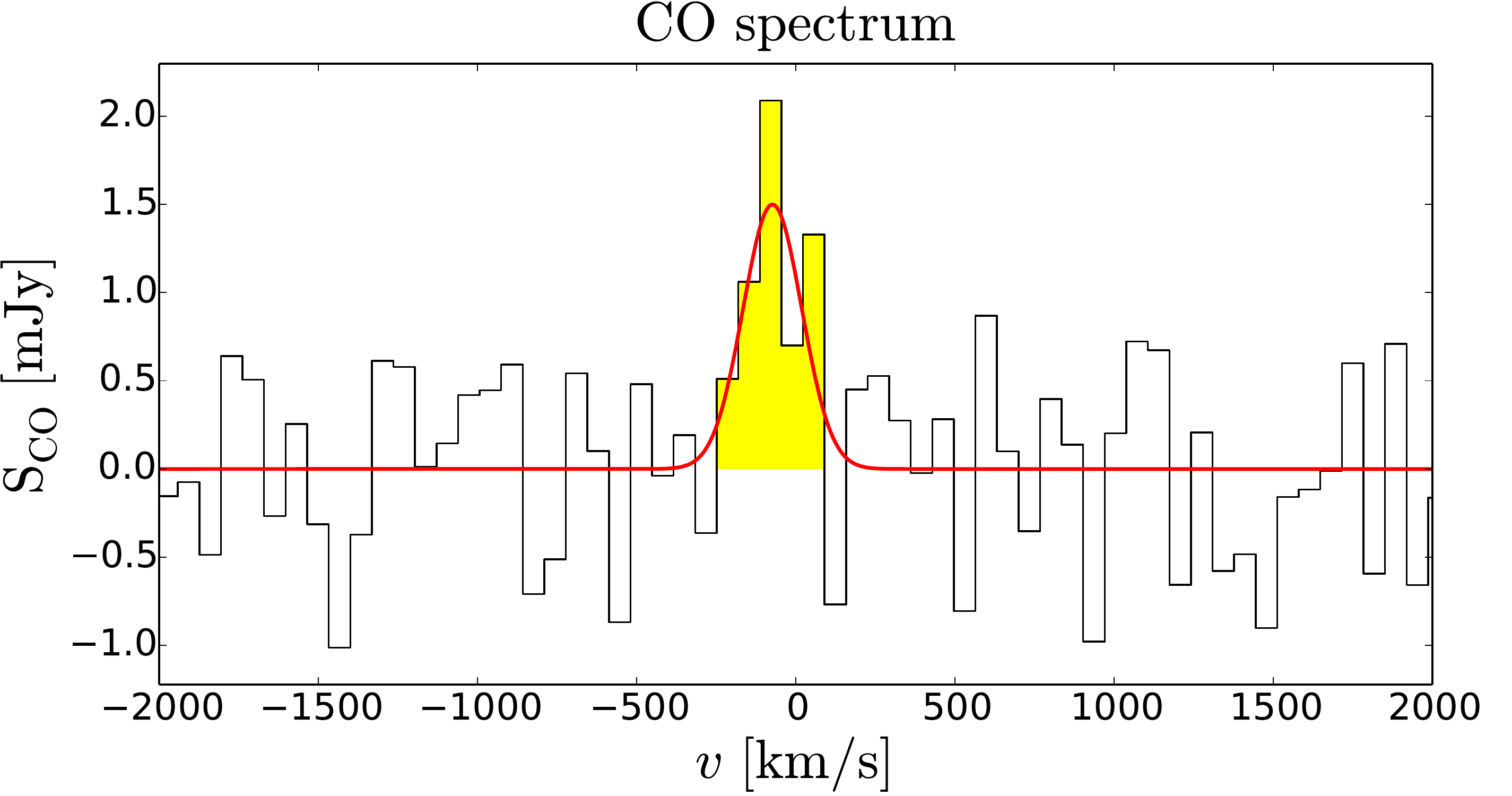}
        \\
        \includegraphics[height=0.1623\textwidth,clip,clip,trim=0 0 4cm 1.1cm]{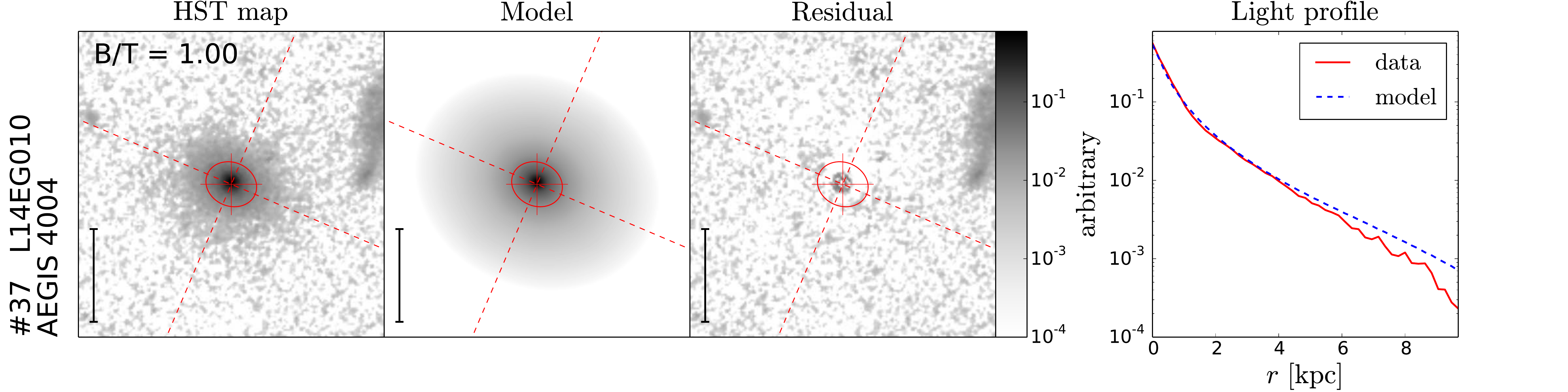}
        \hfill\includegraphics[height=0.1623\textwidth,clip,trim=0 0 32cm 1.1cm]{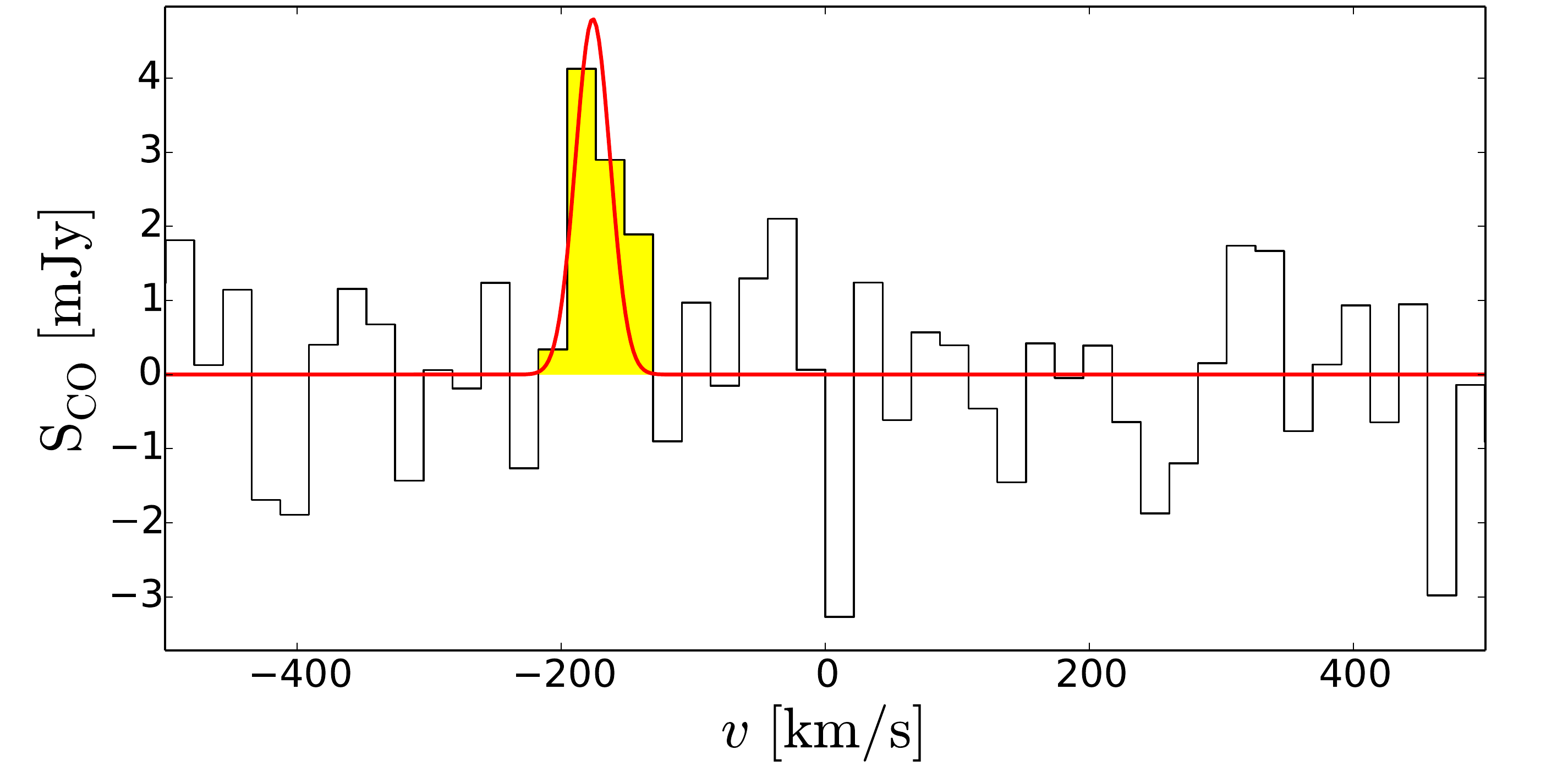}
        \\
        \includegraphics[height=0.1623\textwidth,clip,clip,trim=0 0 4cm 1.1cm]{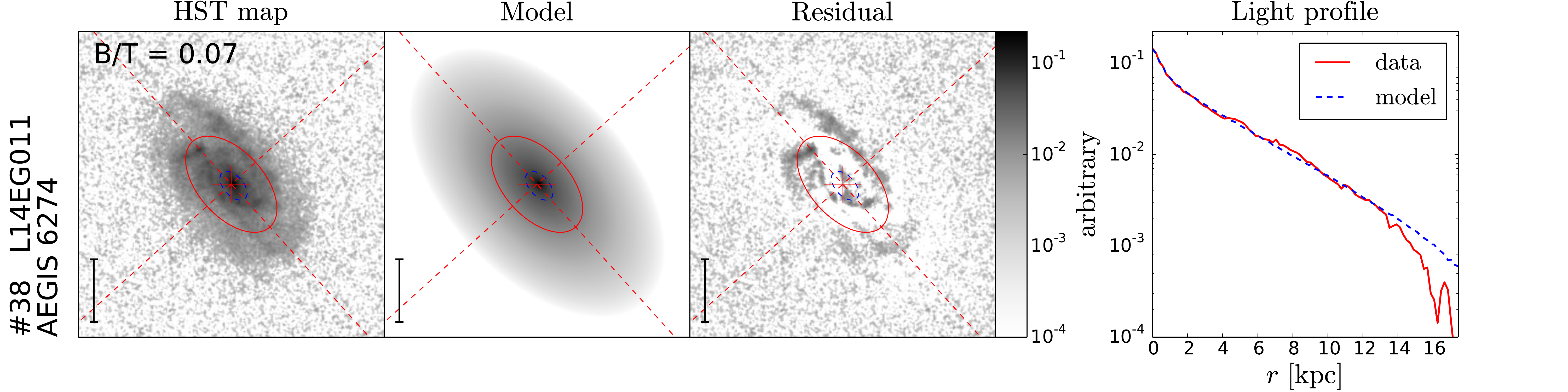}
        \hfill\includegraphics[height=0.1623\textwidth,clip,trim=0 0 32cm 1.1cm]{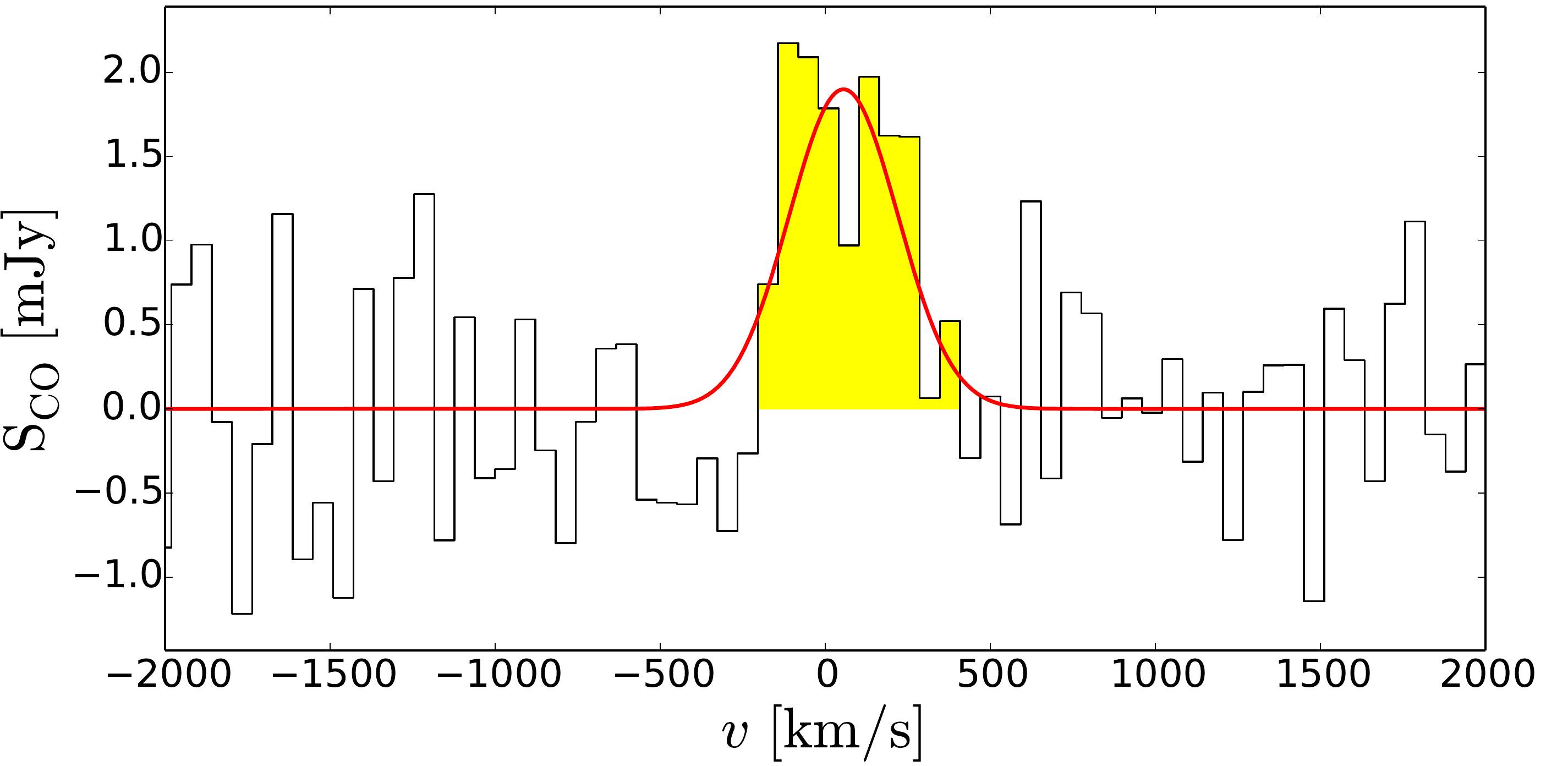}
        \\
        \includegraphics[height=0.1623\textwidth,clip,clip,trim=0 0 4cm 1.1cm]{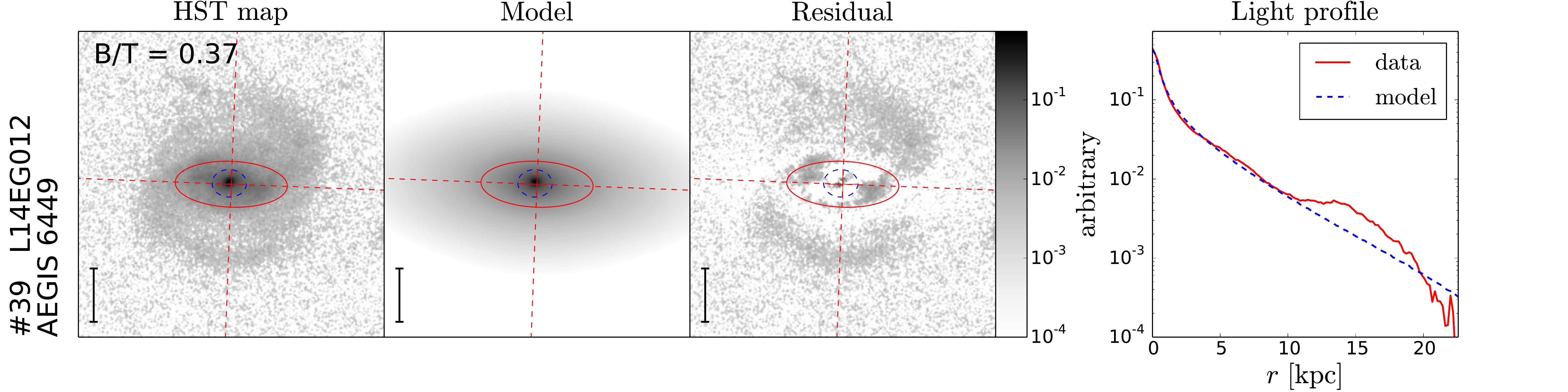}
        \hfill\includegraphics[height=0.1623\textwidth,clip,trim=0 0 32cm 1.1cm]{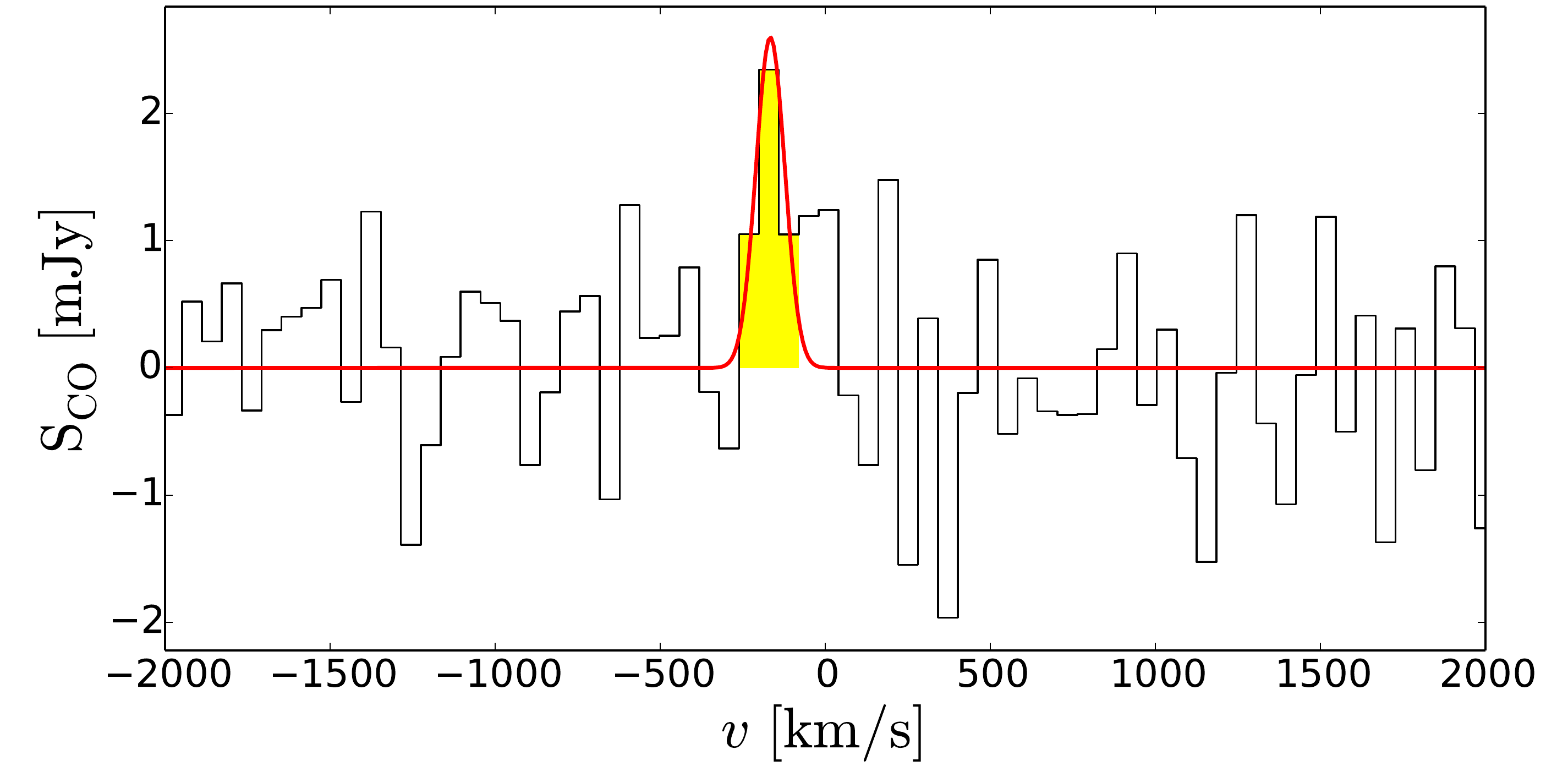}
        \\
        \includegraphics[height=0.1623\textwidth,clip,clip,trim=0 0 4cm 1.1cm]{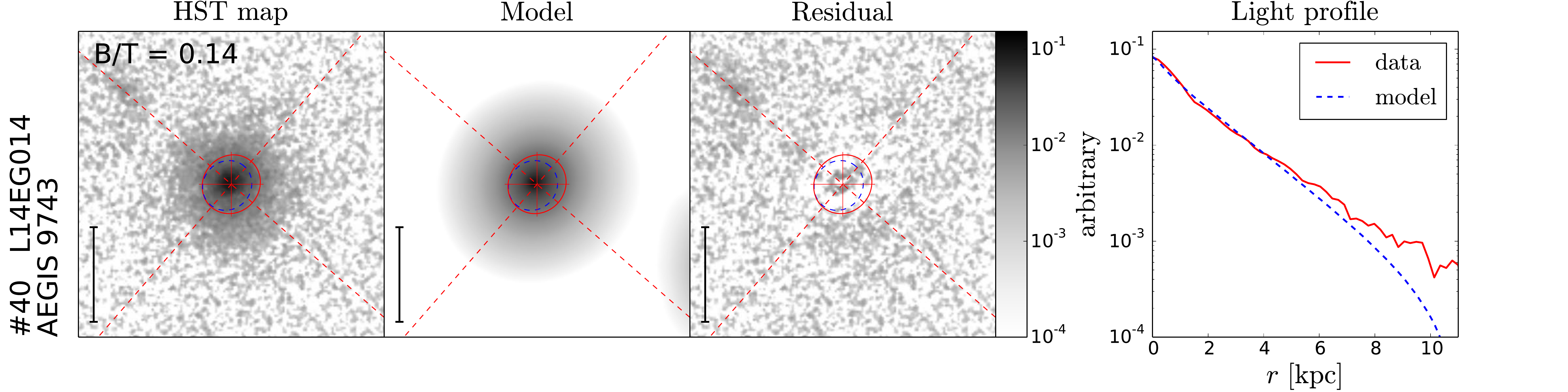}
        \hfill\includegraphics[height=0.1623\textwidth,clip,trim=0 0 32cm 1.1cm]{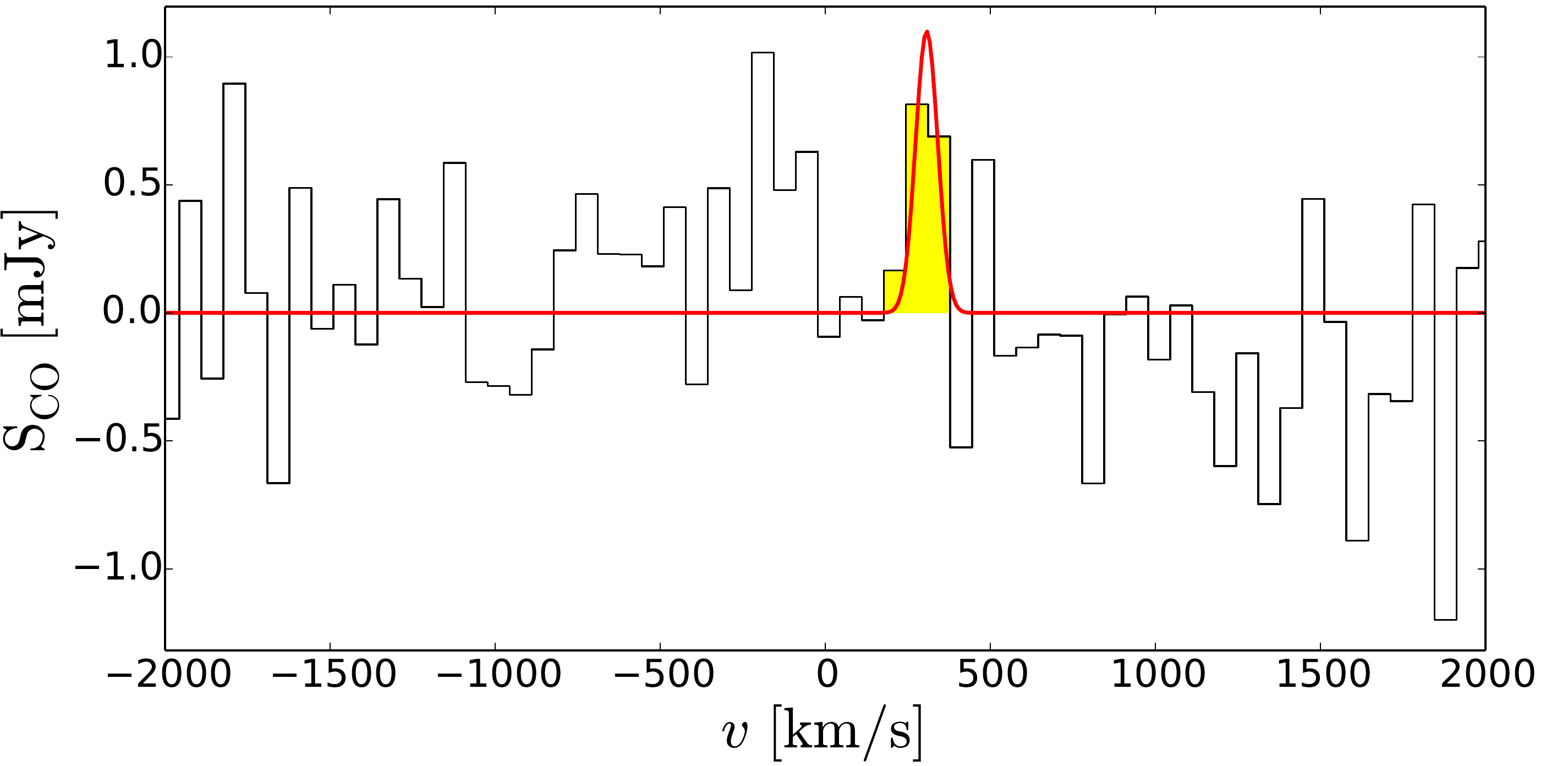}
        \\
        \includegraphics[height=0.1623\textwidth,clip,clip,trim=0 0 4cm 1.1cm]{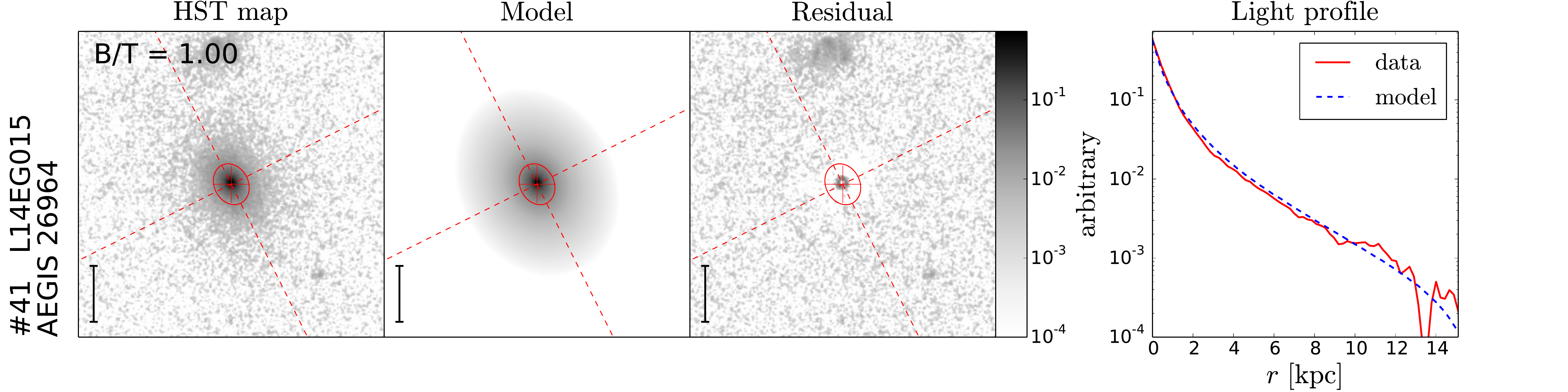}
        \hfill\includegraphics[height=0.1623\textwidth,clip,trim=0 0 32cm 1.1cm]{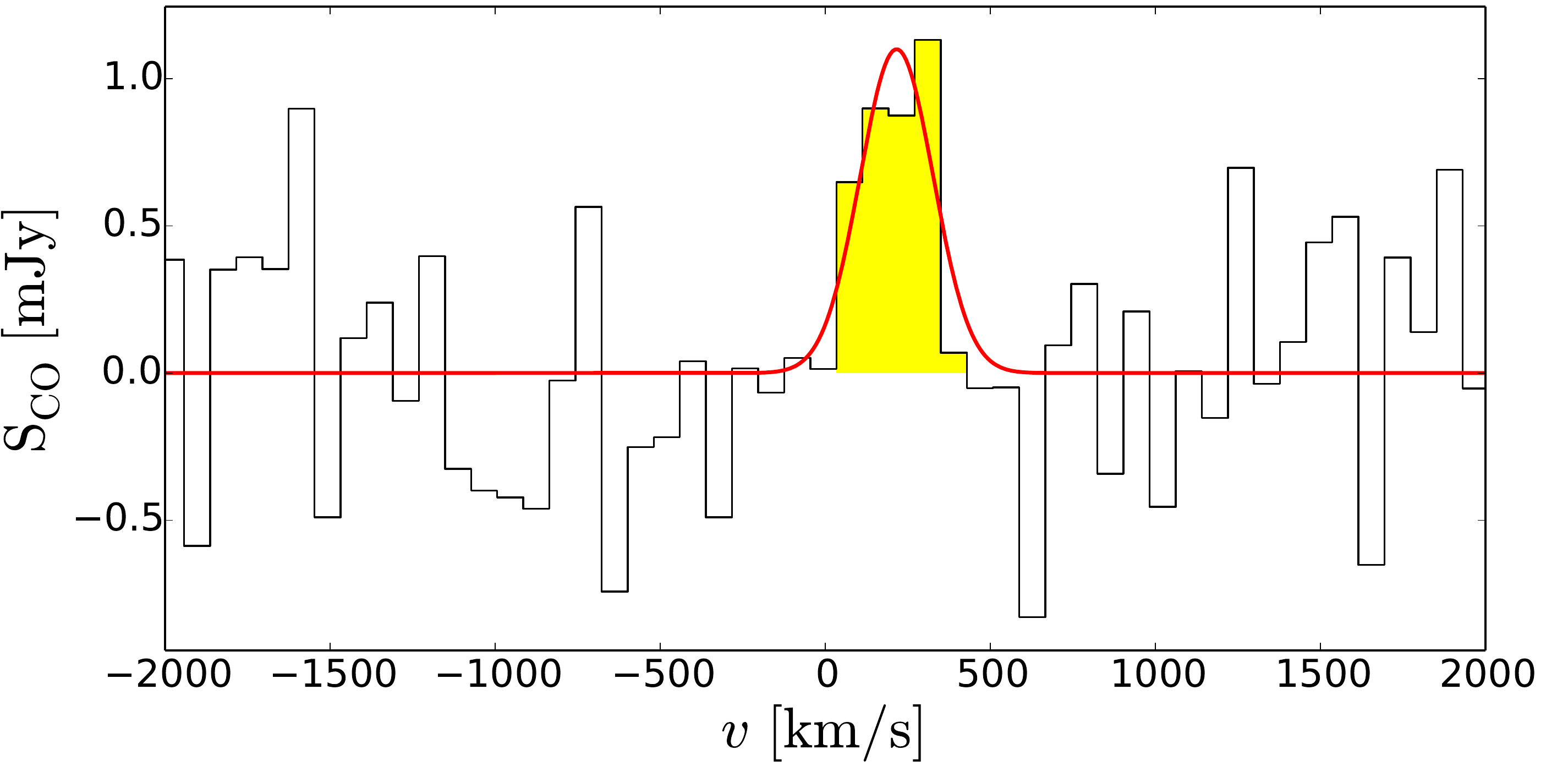}
        \\
        \includegraphics[height=0.1623\textwidth,clip,clip,trim=0 0 4cm 1.1cm]{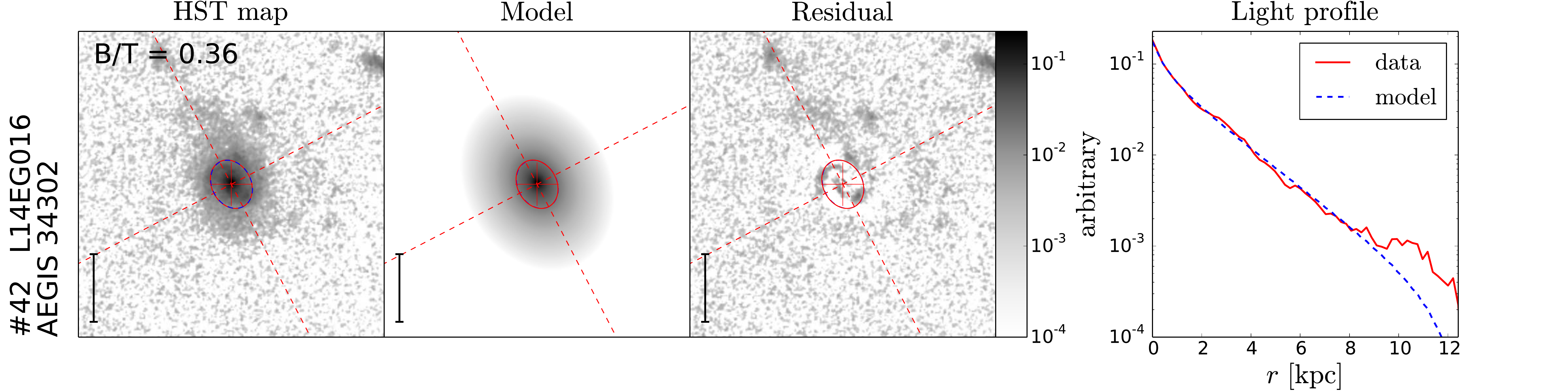}
        \hfill\includegraphics[height=0.1623\textwidth,clip,trim=0 0 32cm 1.1cm]{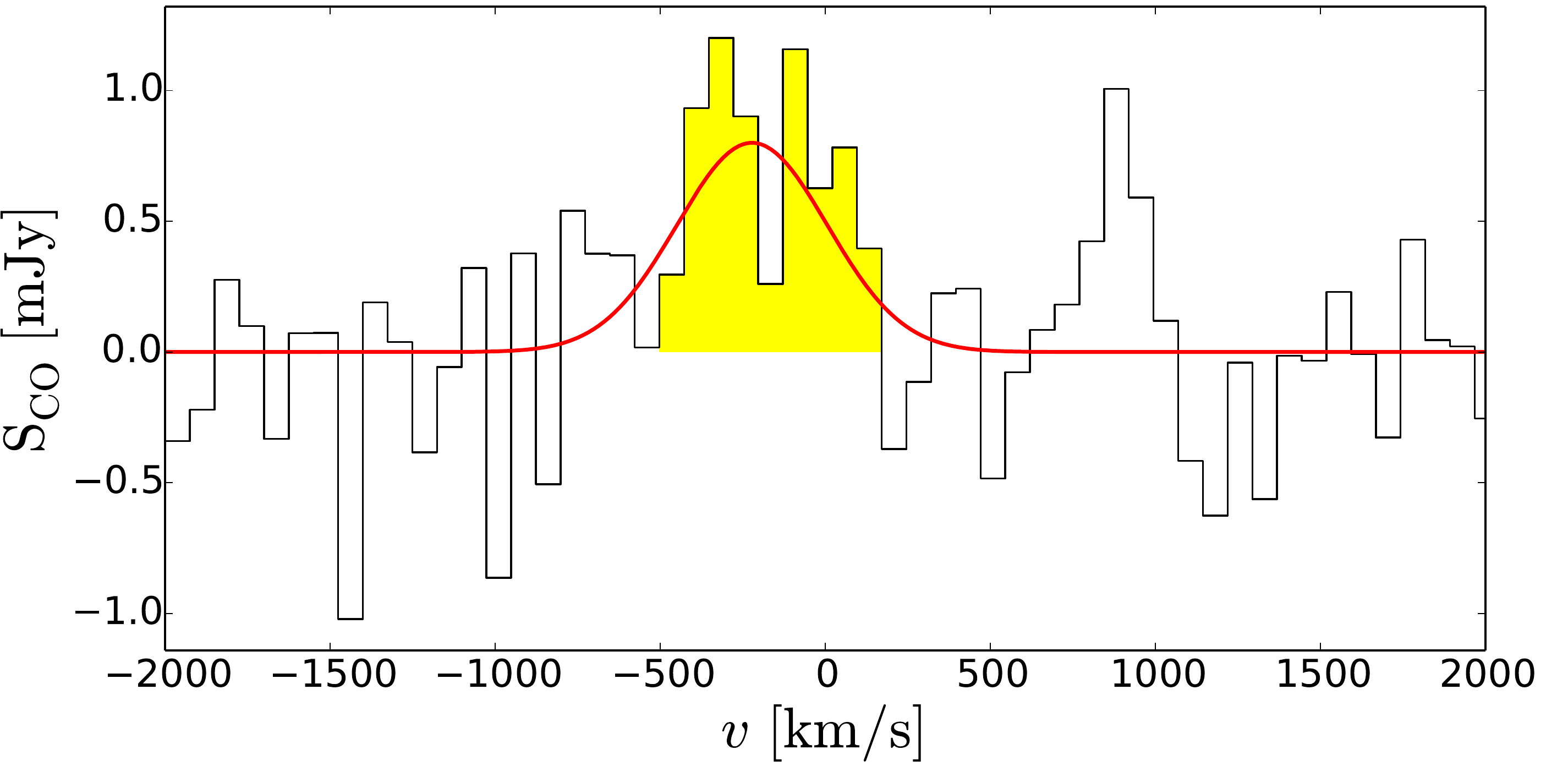}
        \\
\end{figure*}

\begin{figure*}[h!]
        \ContinuedFloat
        \flushleft
        \includegraphics[height=0.175\textwidth,clip,trim=0 0 4cm 0]{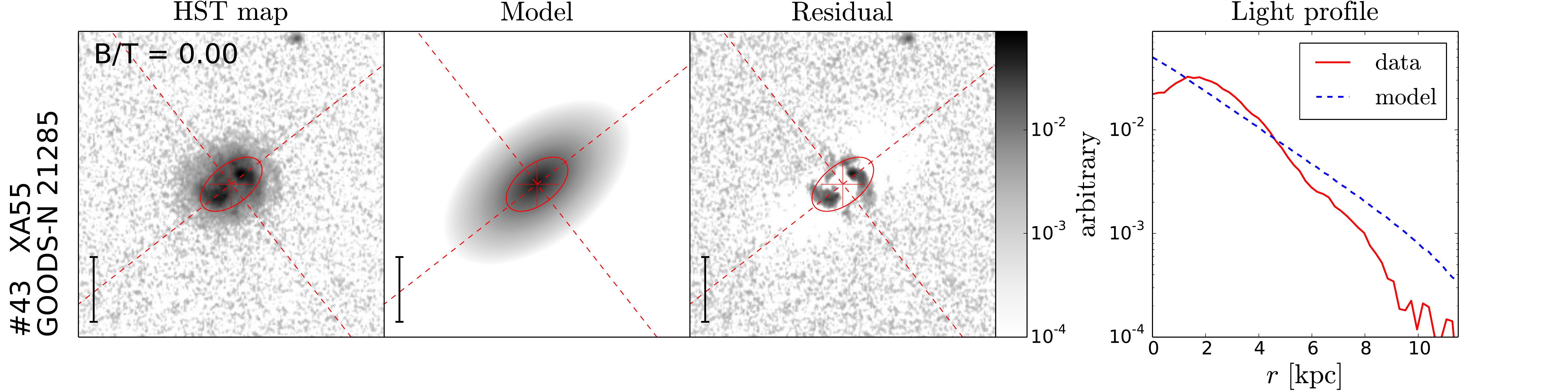}
        \hfill\includegraphics[height=0.175\textwidth,clip,trim=0 0 32cm 0]{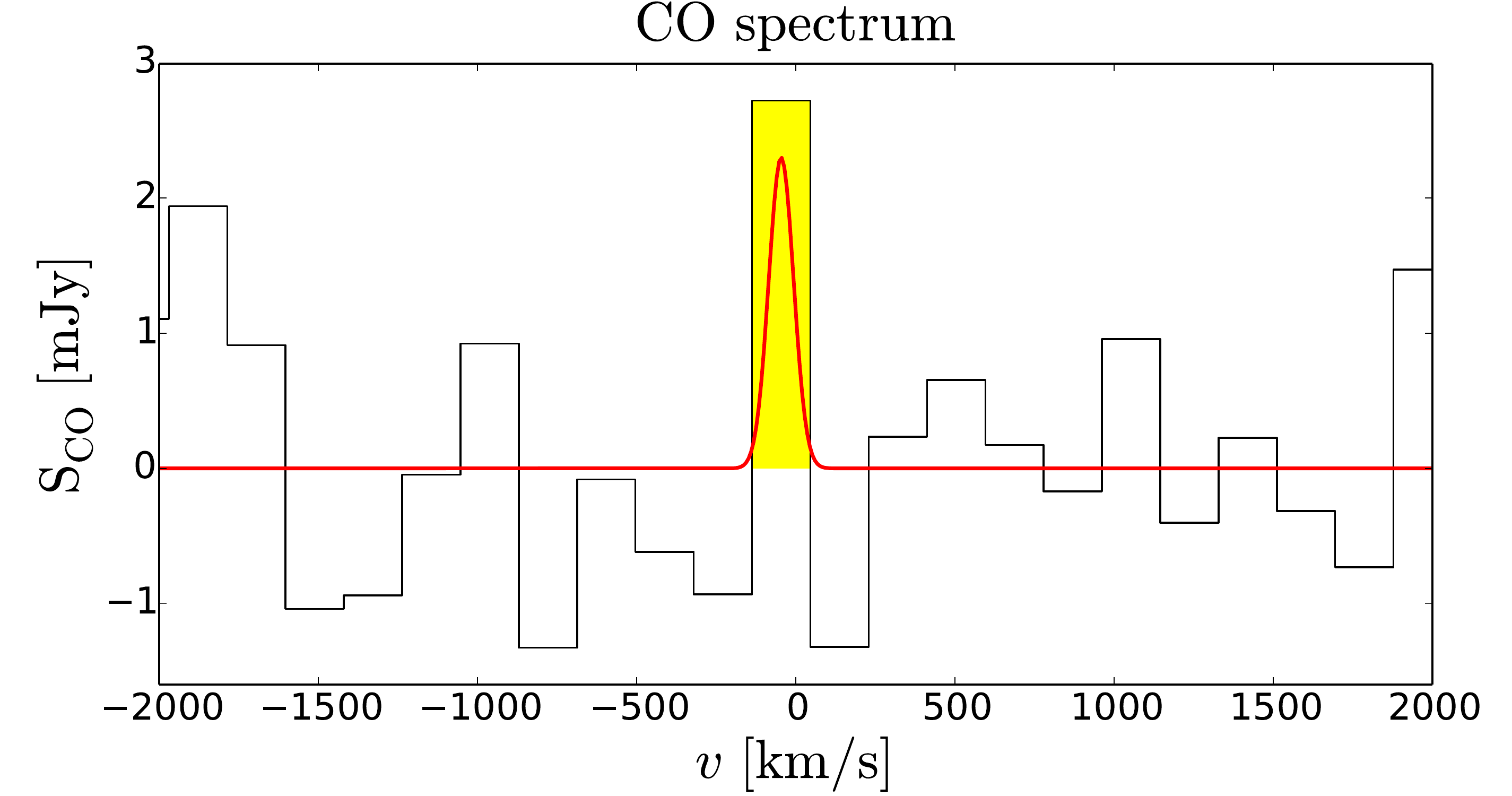}
        \\
        \includegraphics[height=0.1623\textwidth,clip,clip,trim=0 0 4cm 1.1cm]{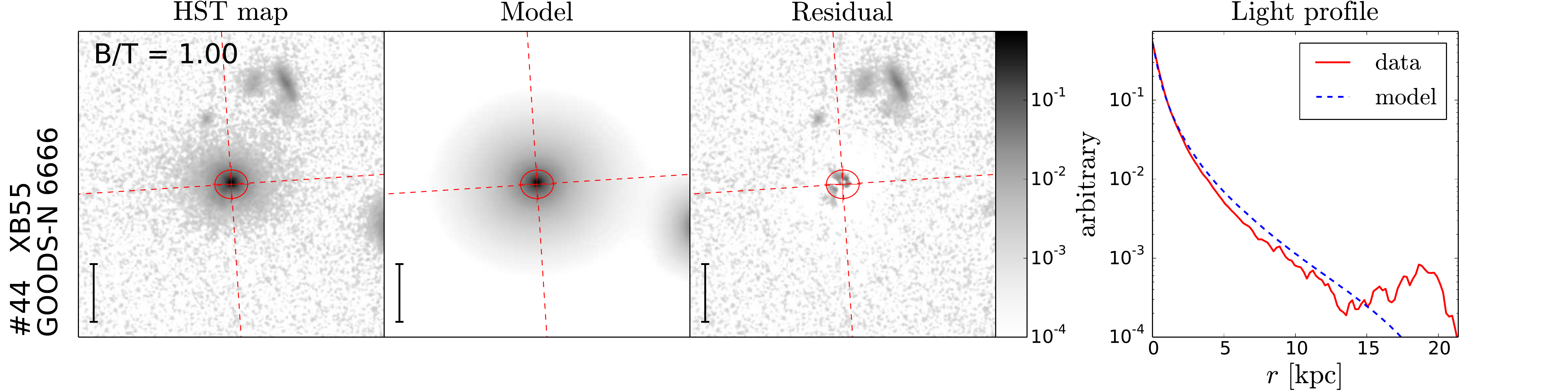}
        \hfill\includegraphics[height=0.1623\textwidth,clip,trim=0 0 32cm 1.1cm]{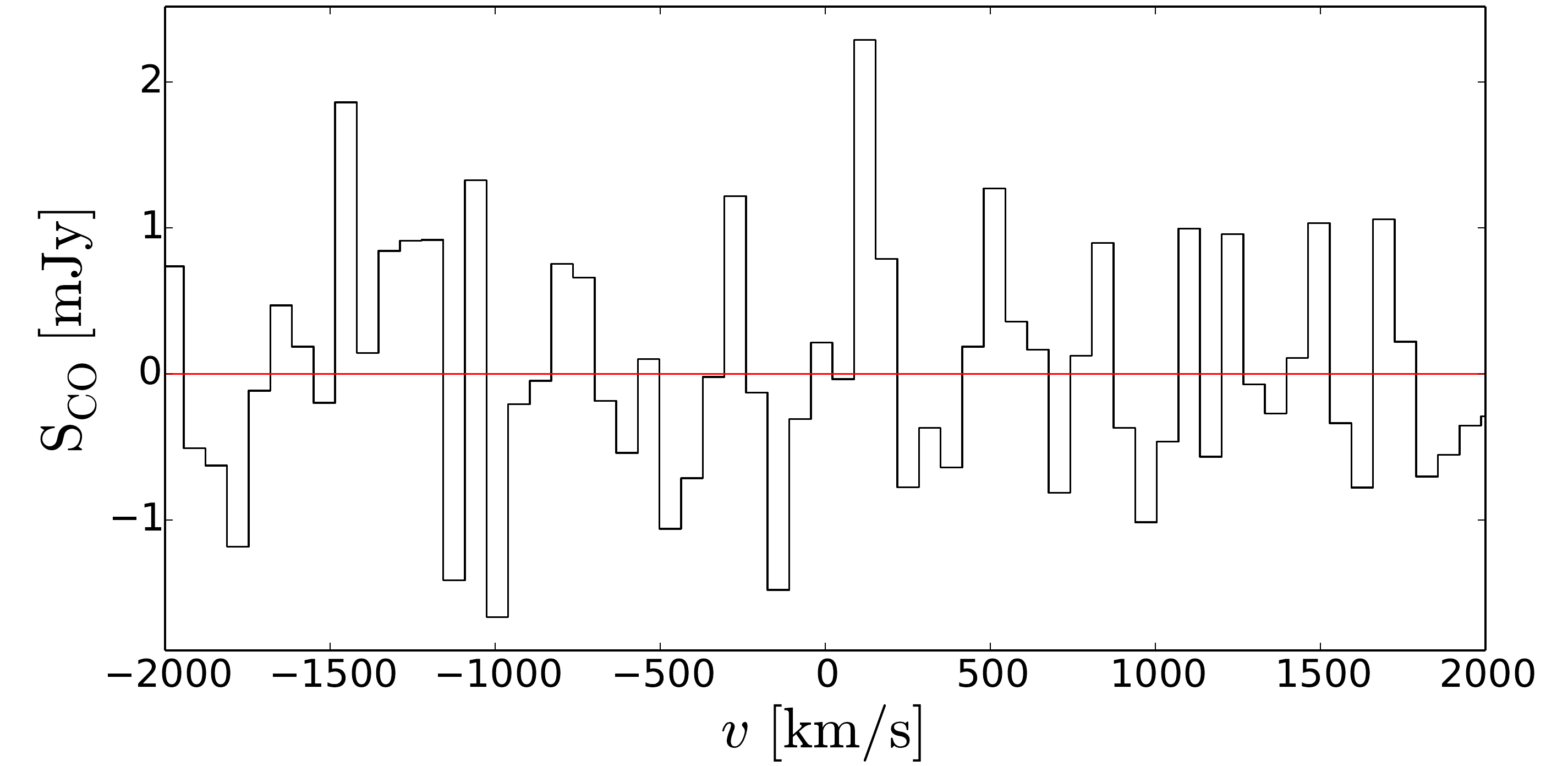}
        \\
        \includegraphics[height=0.1623\textwidth,clip,clip,trim=0 0 4cm 1.1cm]{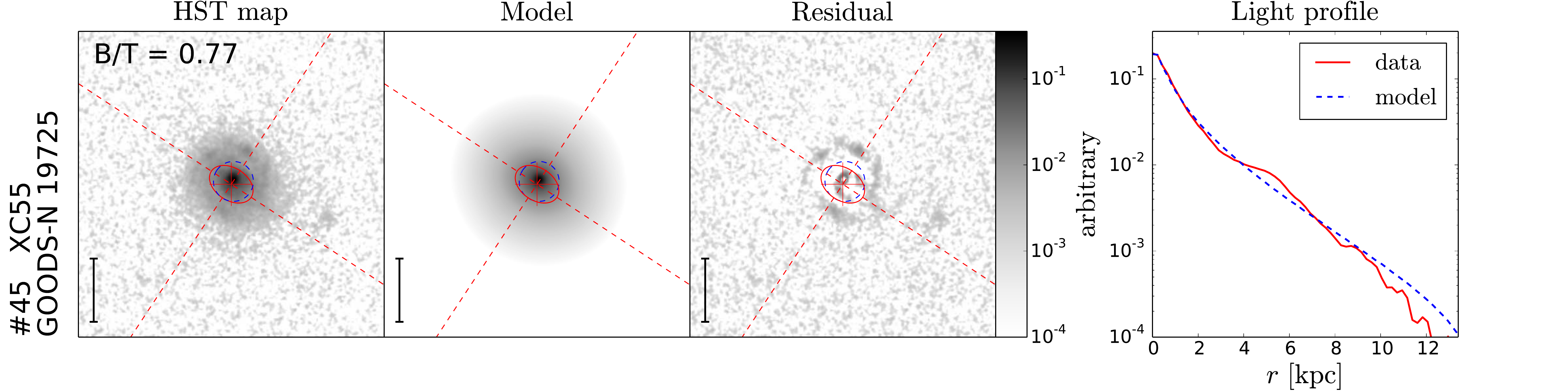}
        \hfill\includegraphics[height=0.1623\textwidth,clip,trim=0 0 32cm 1.1cm]{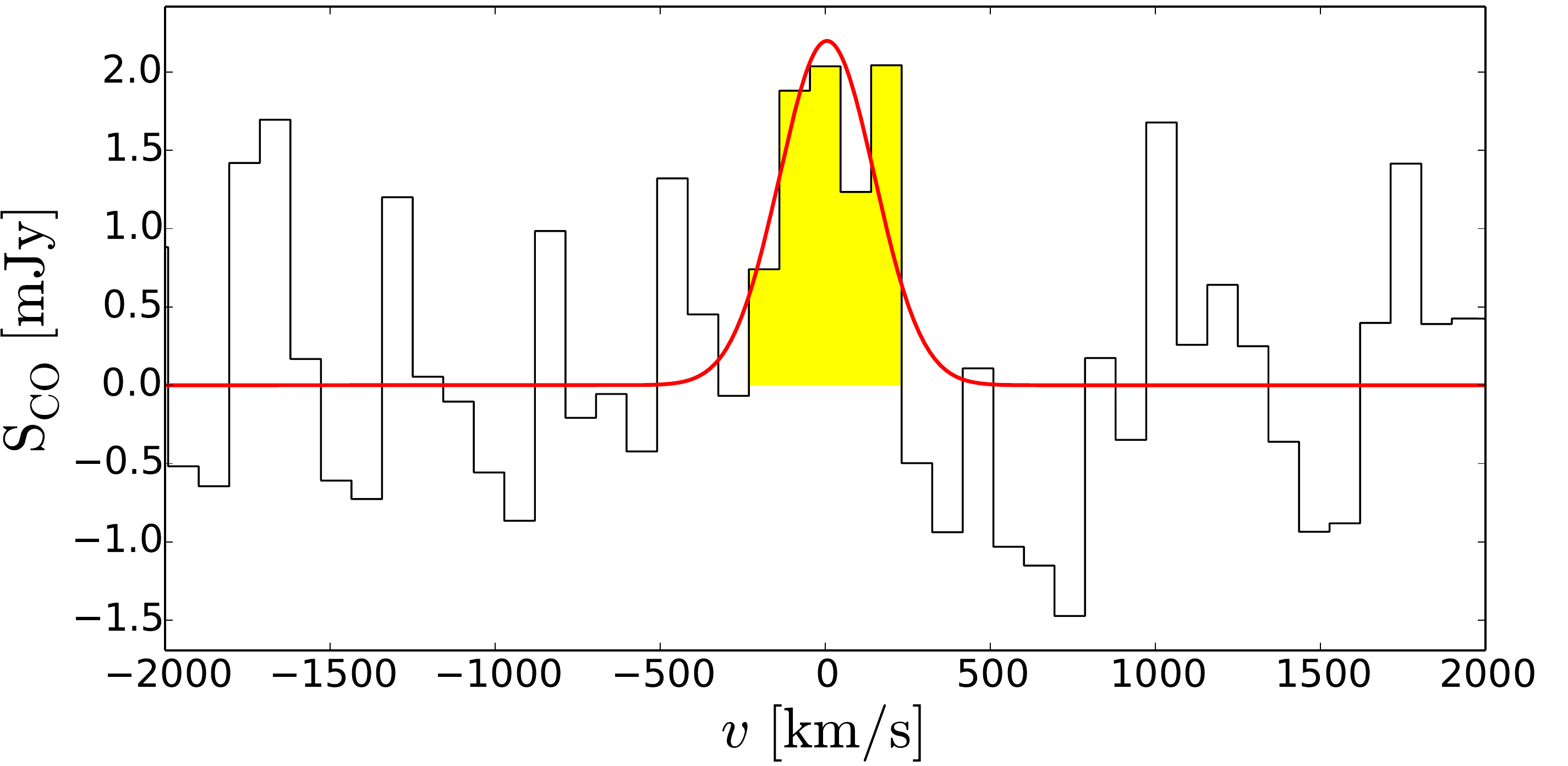}
        \\
        \includegraphics[height=0.1623\textwidth,clip,clip,trim=0 0 4cm 1.1cm]{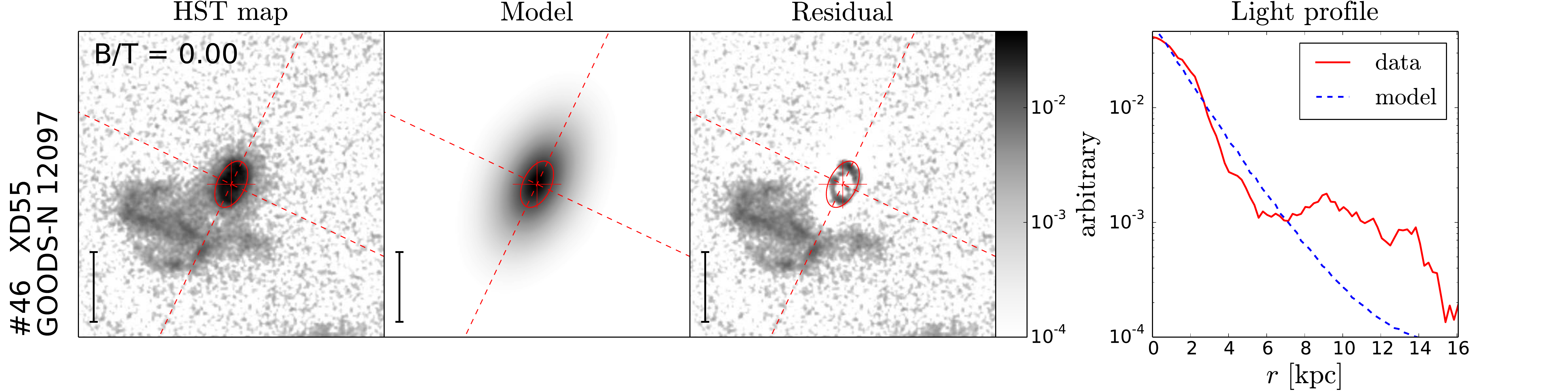}
        \hfill\includegraphics[height=0.1623\textwidth,clip,trim=0 0 32cm 1.1cm]{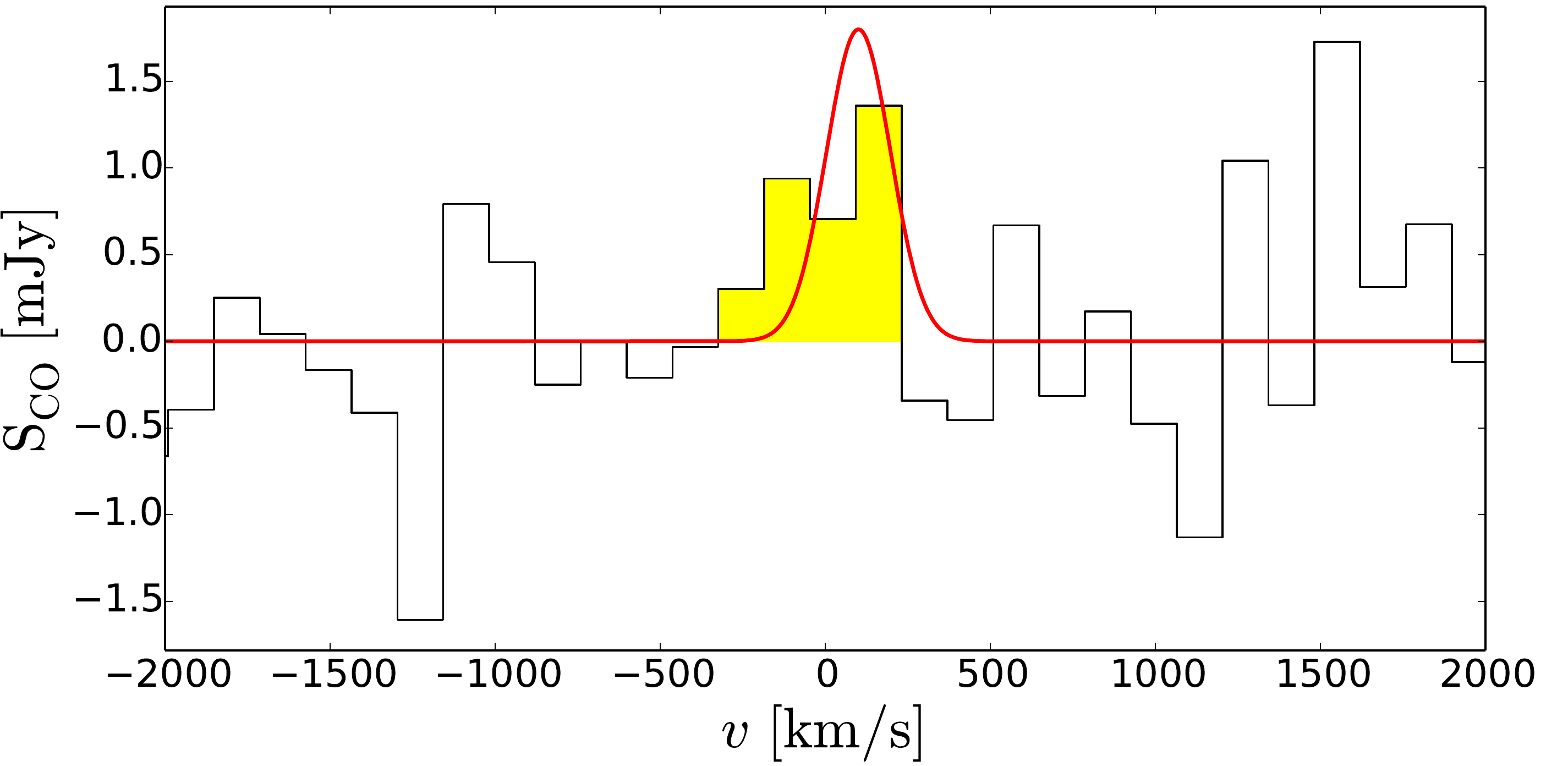}
        \\
        \includegraphics[height=0.1623\textwidth,clip,clip,trim=0 0 4cm 1.1cm]{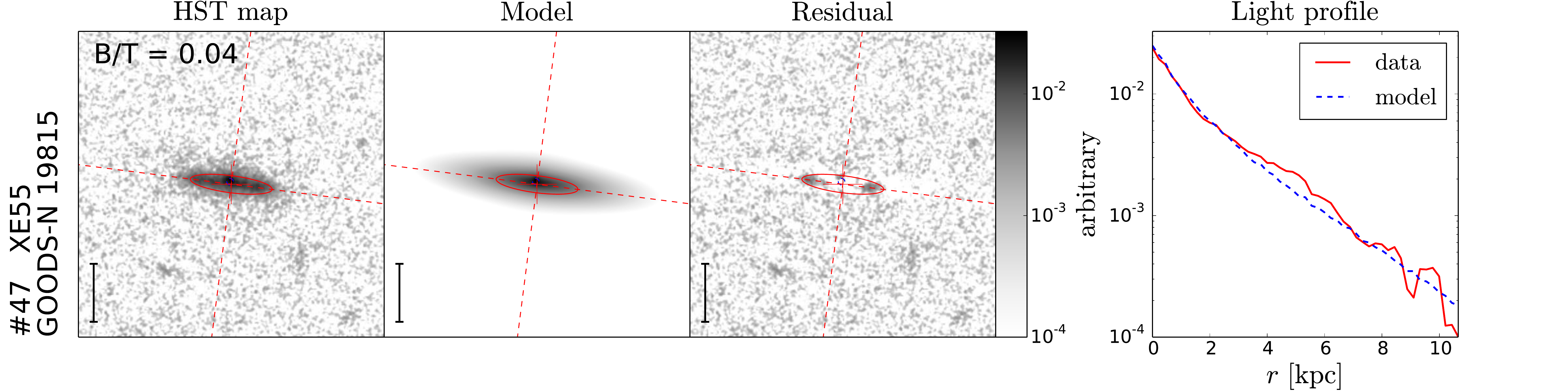}
        \hfill\includegraphics[height=0.1623\textwidth,clip,trim=0 0 32cm 1.1cm]{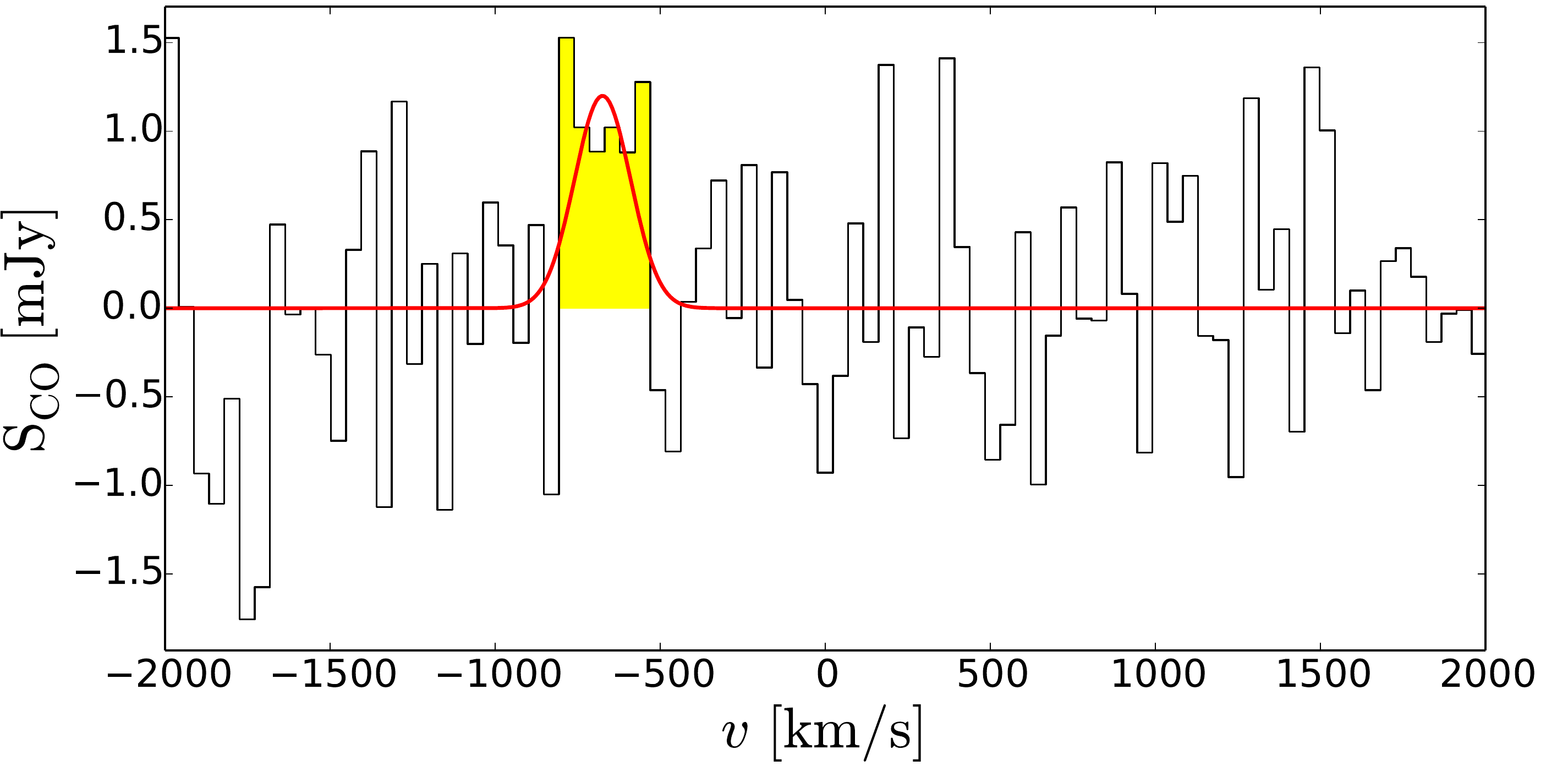}
        \\
        \includegraphics[height=0.1623\textwidth,clip,clip,trim=0 0 4cm 1.1cm]{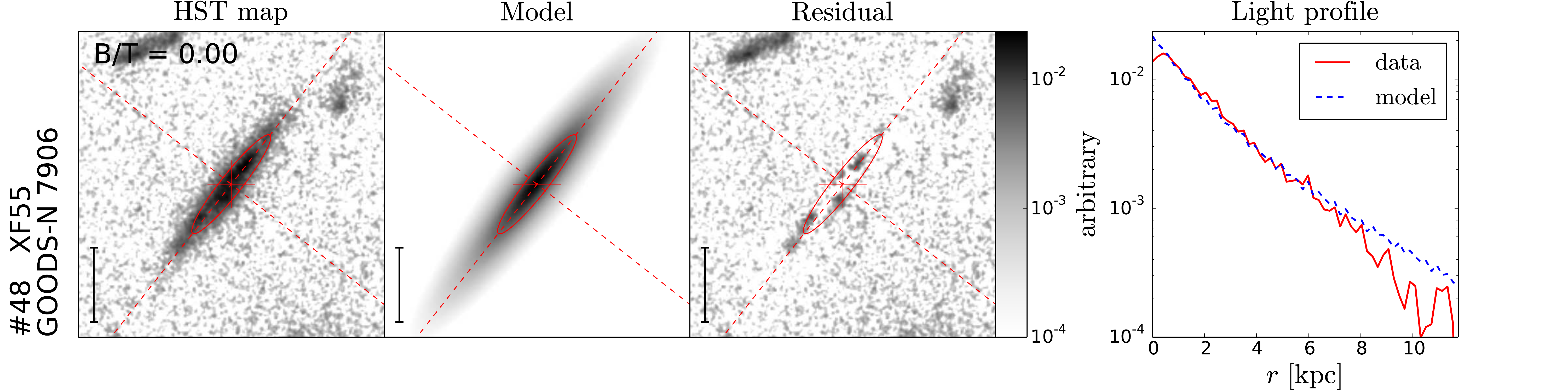}
        \hfill\includegraphics[height=0.1623\textwidth,clip,trim=0 0 32cm 1.1cm]{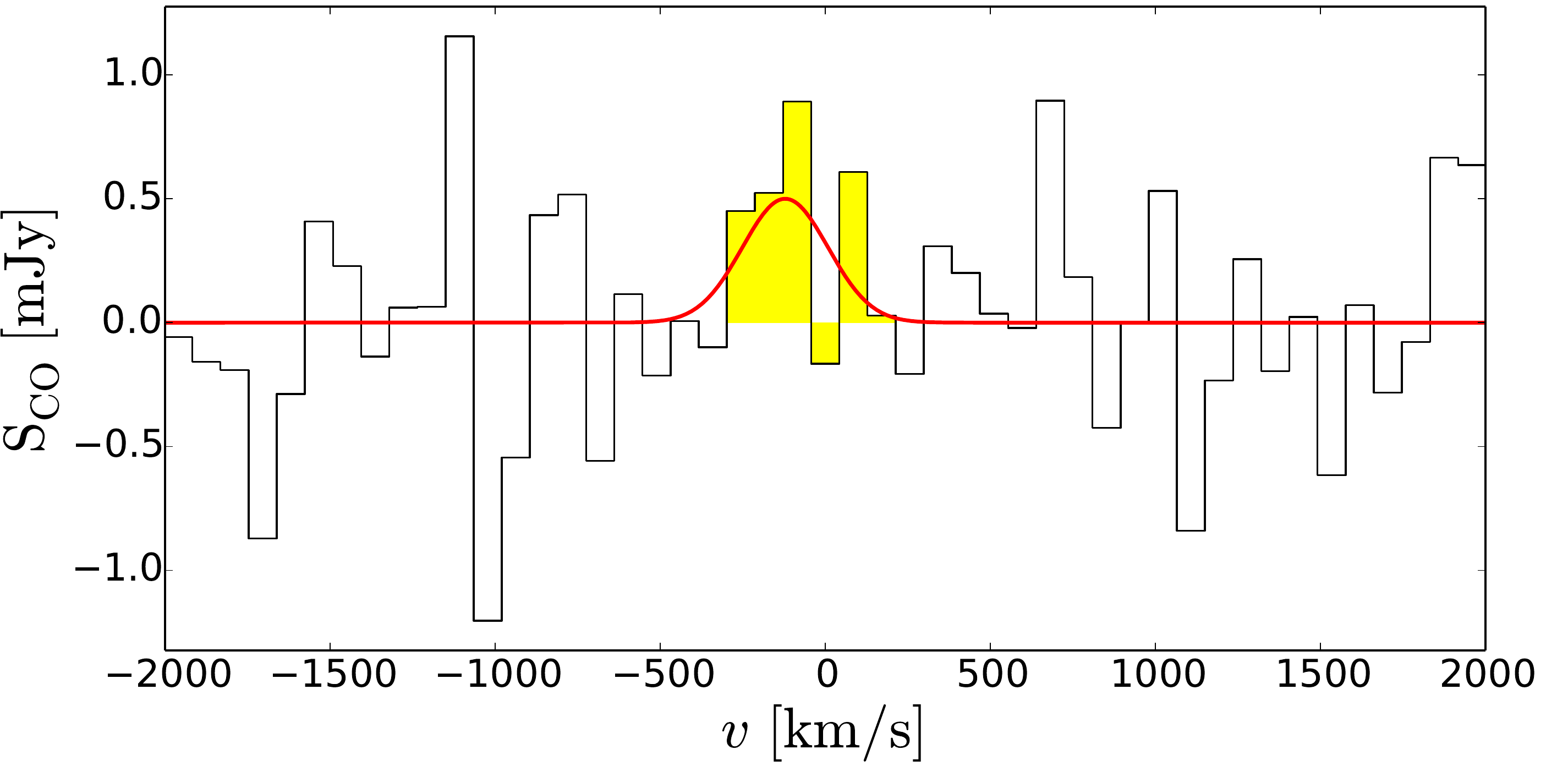}
        \\
        \includegraphics[height=0.1623\textwidth,clip,clip,trim=0 0 4cm 1.1cm]{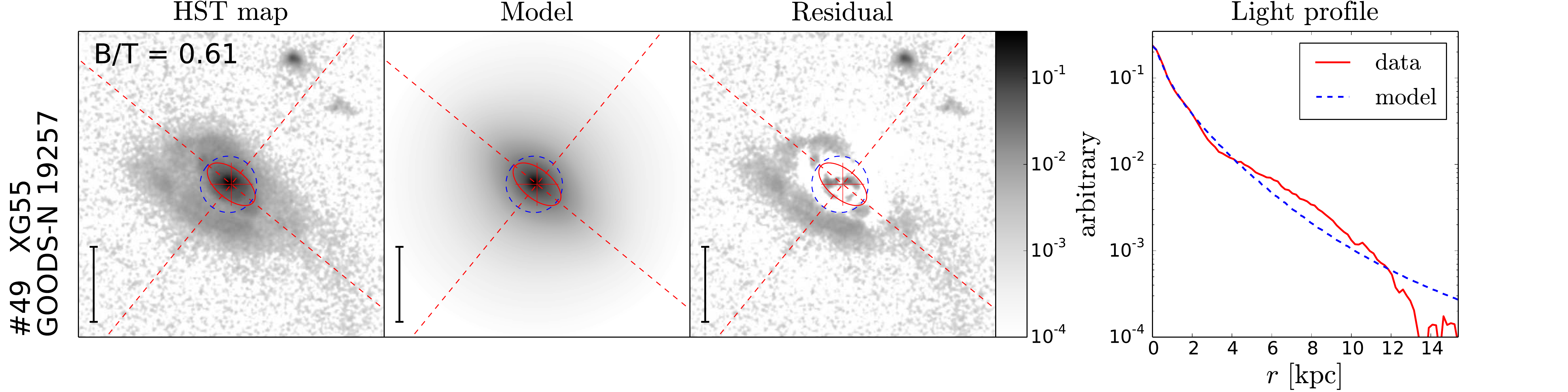}
        \hfill\includegraphics[height=0.1623\textwidth,clip,trim=0 0 32cm 1.1cm]{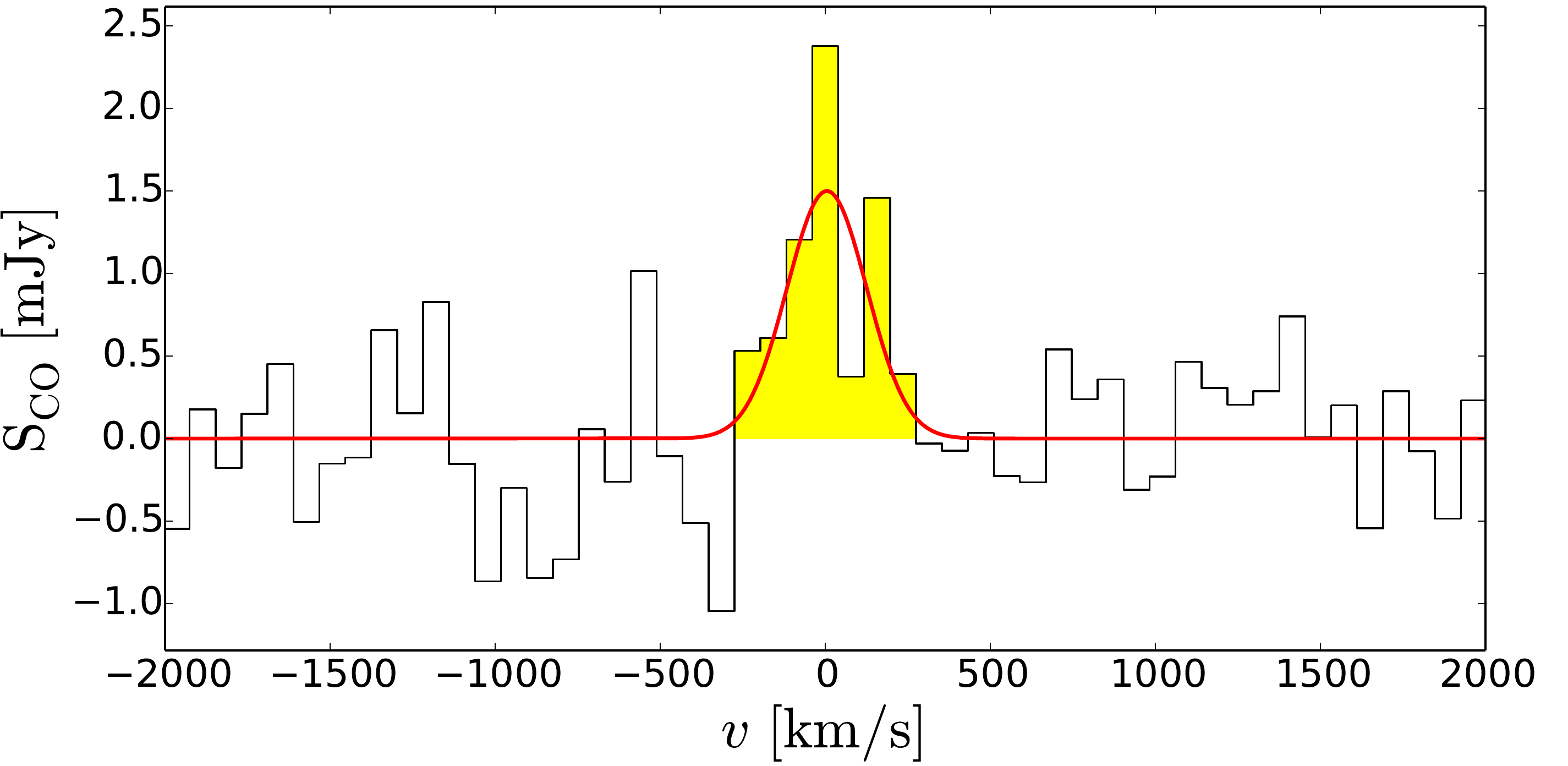}
        \\
\end{figure*}

\begin{figure*}[h!]
        \ContinuedFloat
        \flushleft
        \includegraphics[height=0.175\textwidth,clip,trim=0 0 4cm 0]{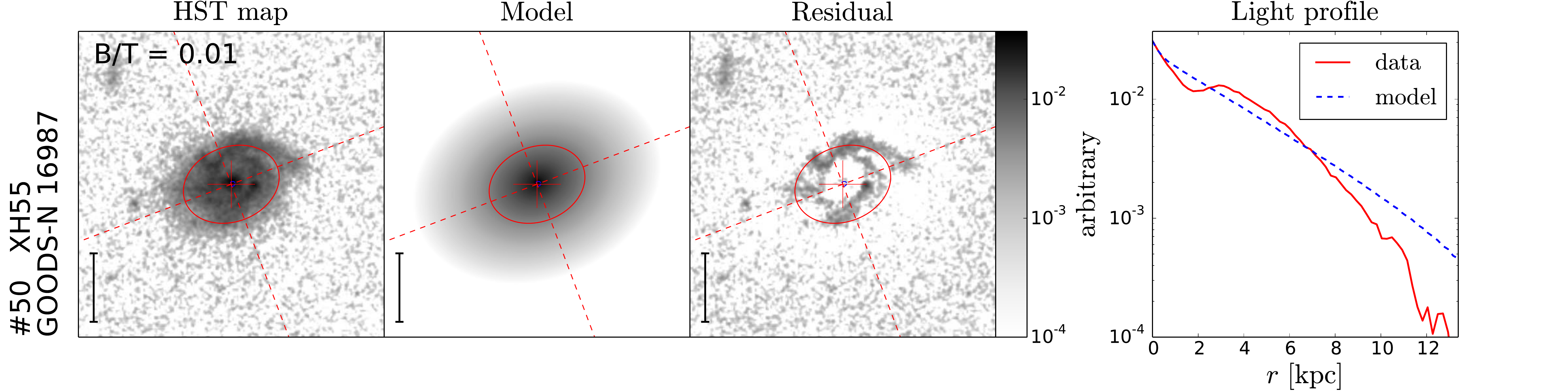}
        \hfill\includegraphics[height=0.175\textwidth,clip,trim=0 0 32cm 0]{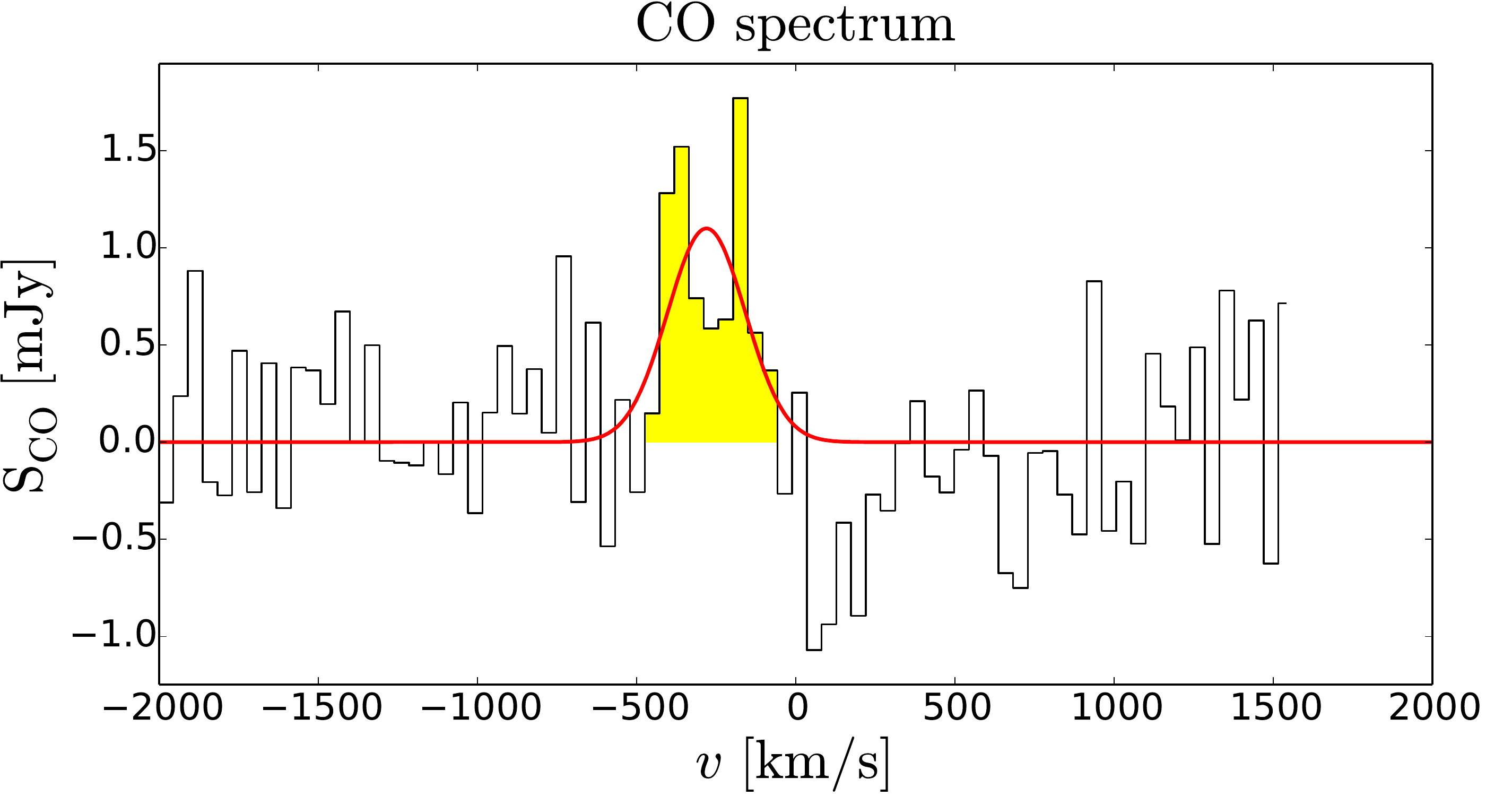}
        \\
        \includegraphics[height=0.1623\textwidth,clip,clip,trim=0 0 4cm 1.1cm]{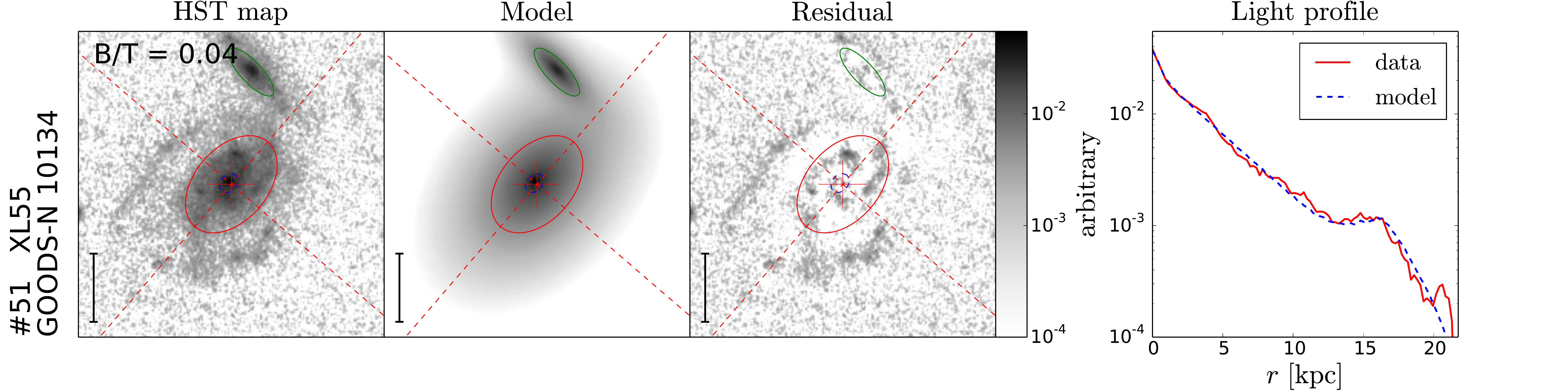}
        \hfill\includegraphics[height=0.1623\textwidth,clip,trim=0 0 32cm 1.1cm]{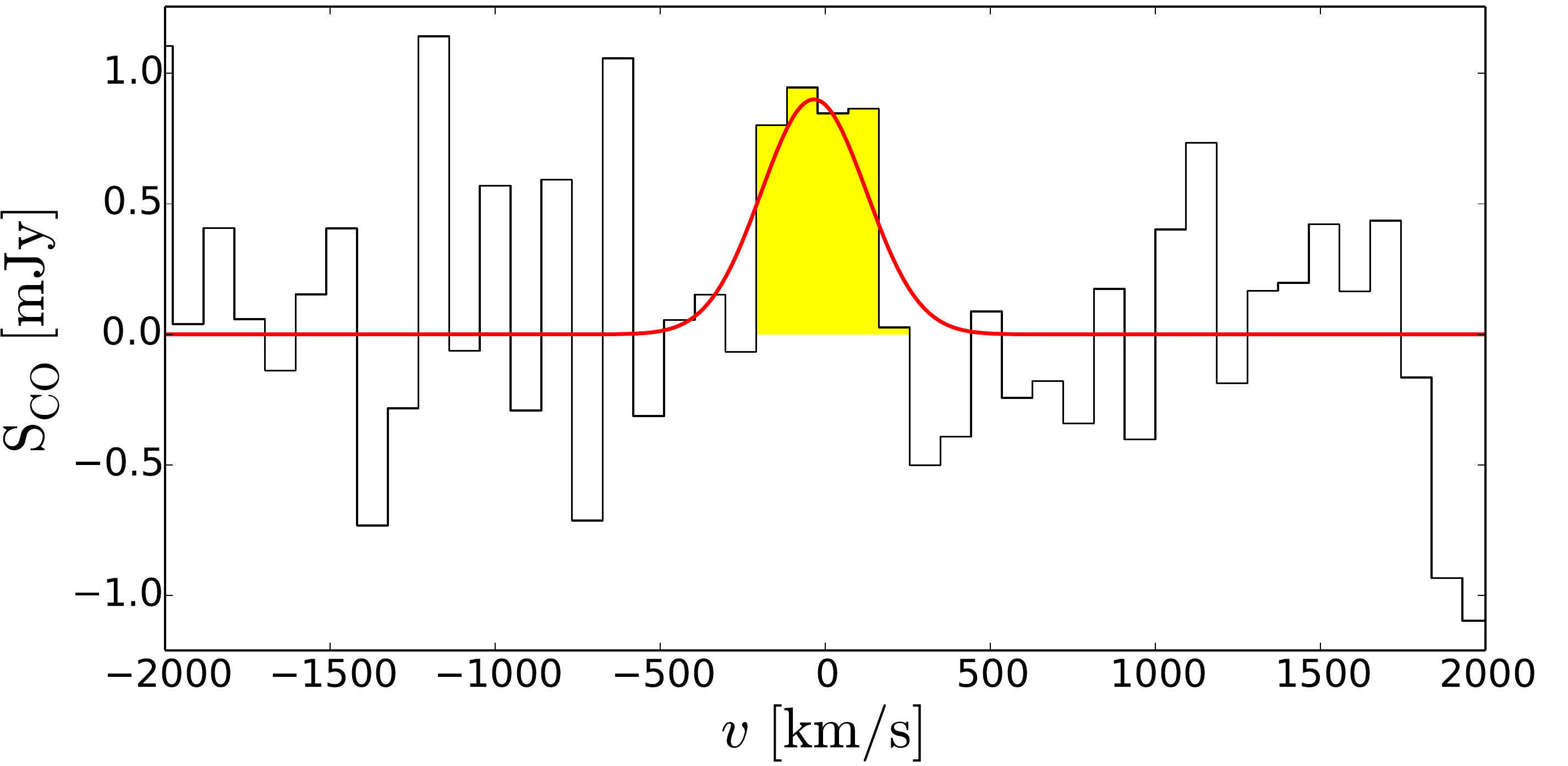}
        \\
        \includegraphics[height=0.1623\textwidth,clip,clip,trim=0 0 4cm 1.1cm]{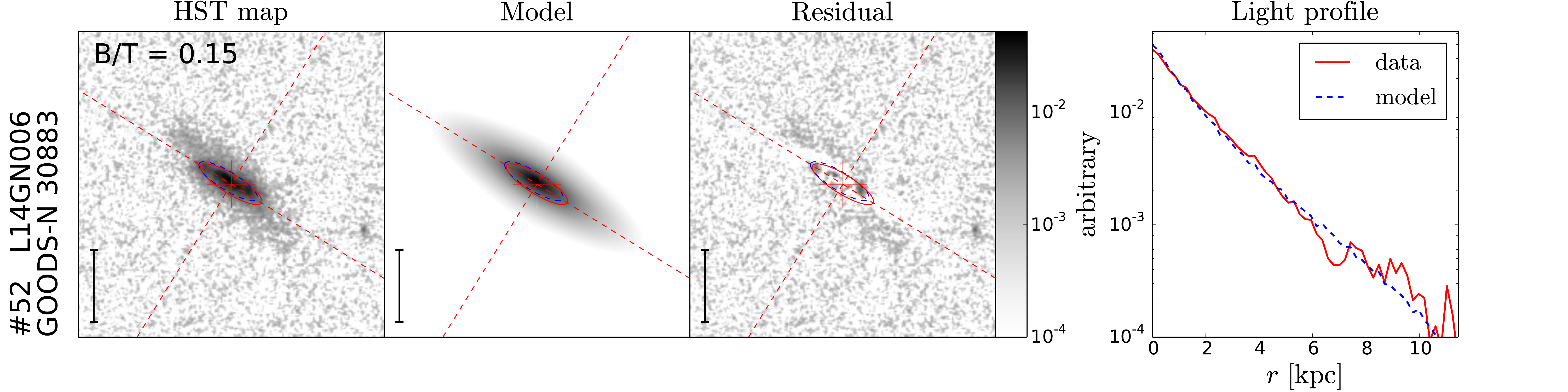}
        \hfill\includegraphics[height=0.1623\textwidth,clip,trim=0 0 32cm 1.1cm]{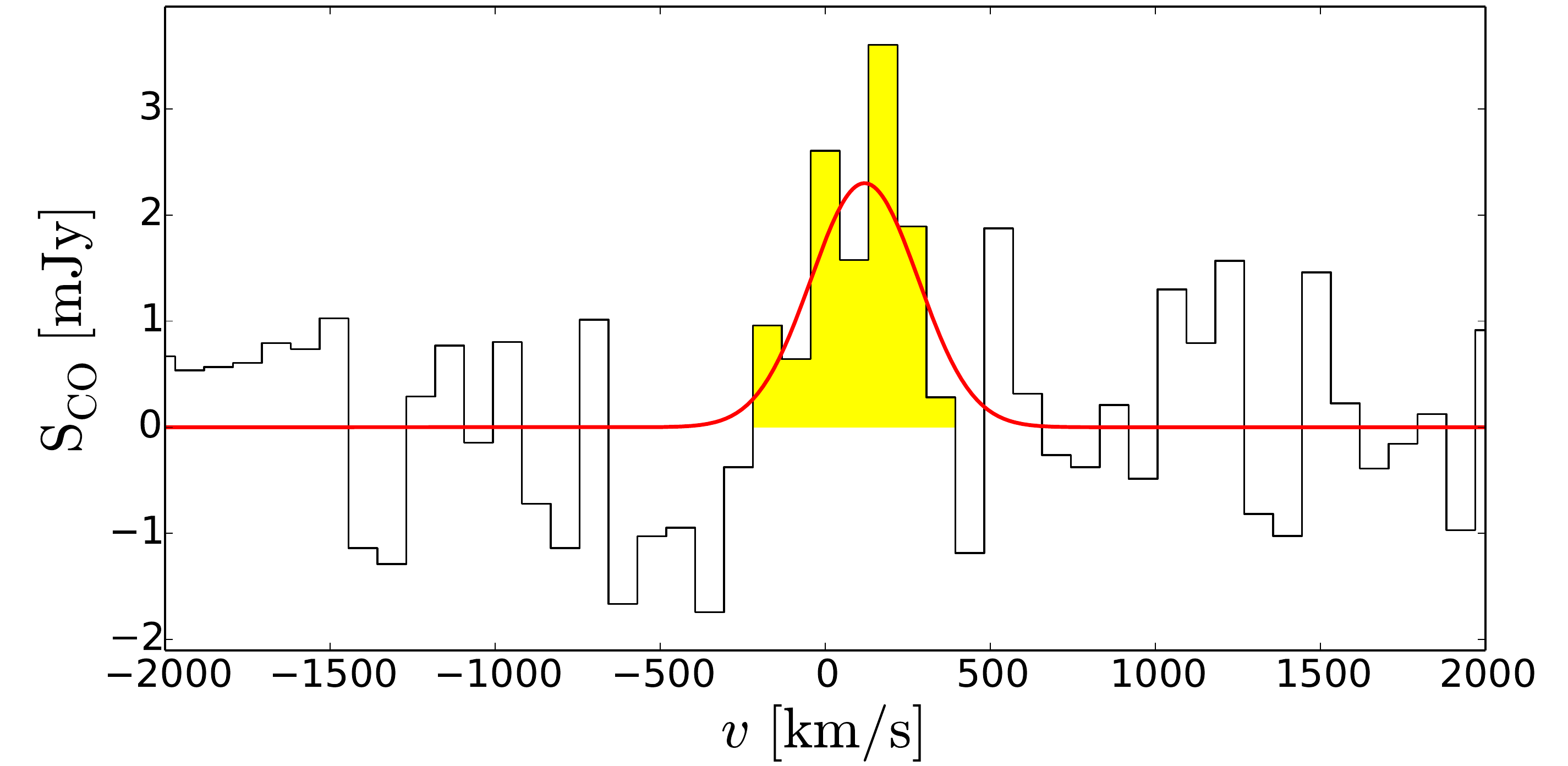}
        \\
        \includegraphics[height=0.1623\textwidth,clip,clip,trim=0 0 4cm 1.1cm]{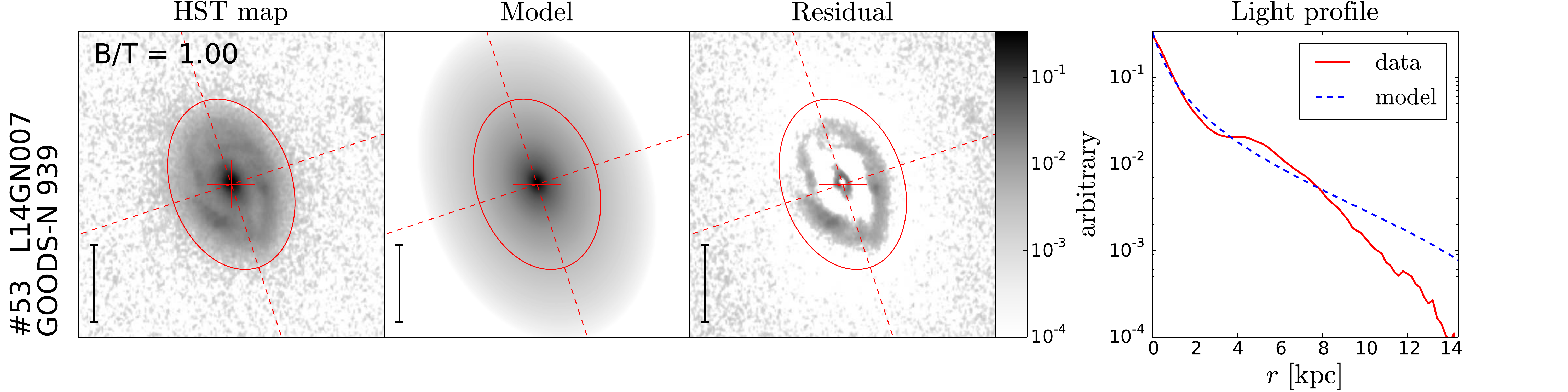}
        \hfill\includegraphics[height=0.1623\textwidth,clip,trim=0 0 32cm 1.1cm]{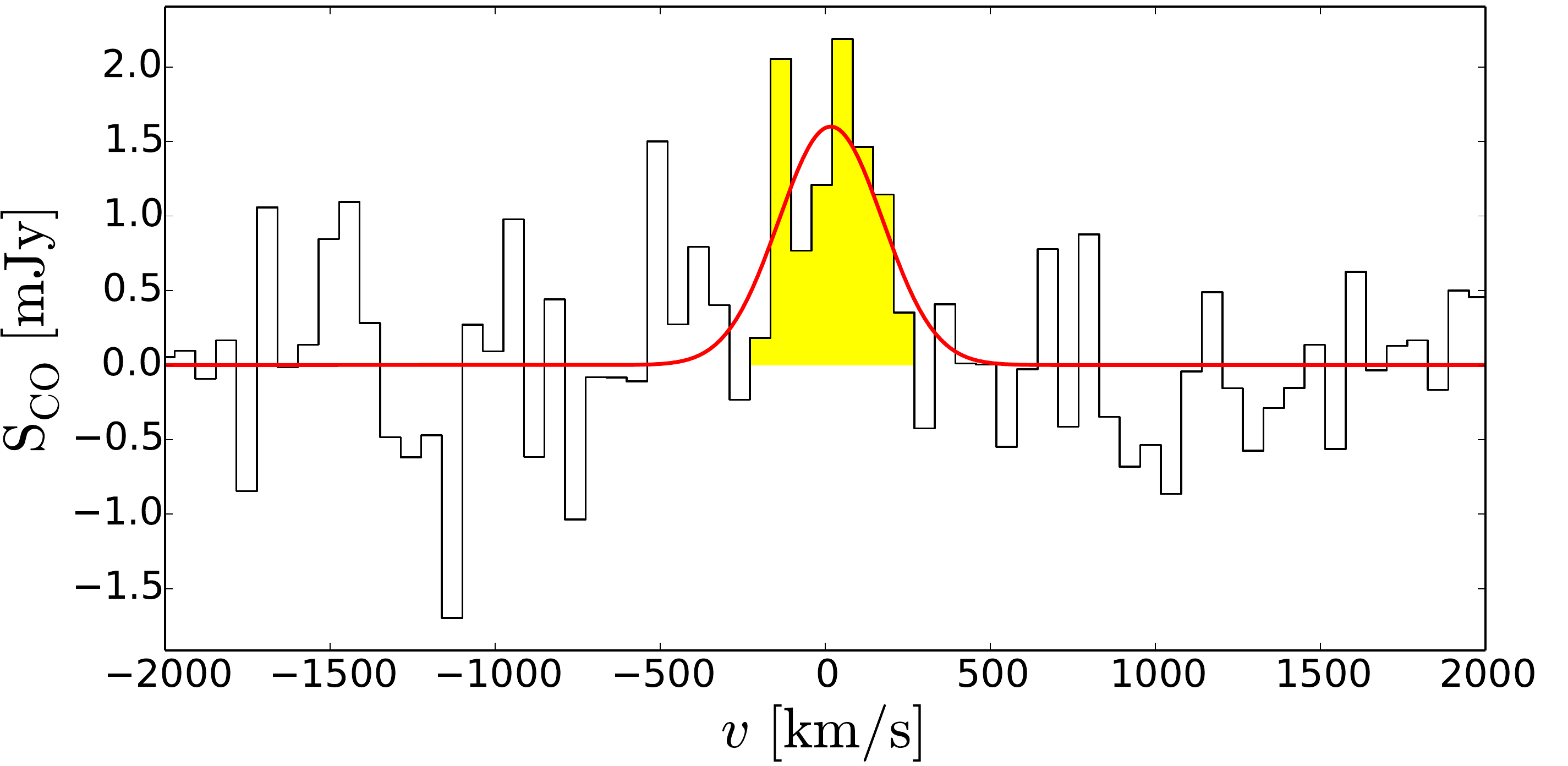}
        \\
        \includegraphics[height=0.1623\textwidth,clip,clip,trim=0 0 4cm 1.1cm]{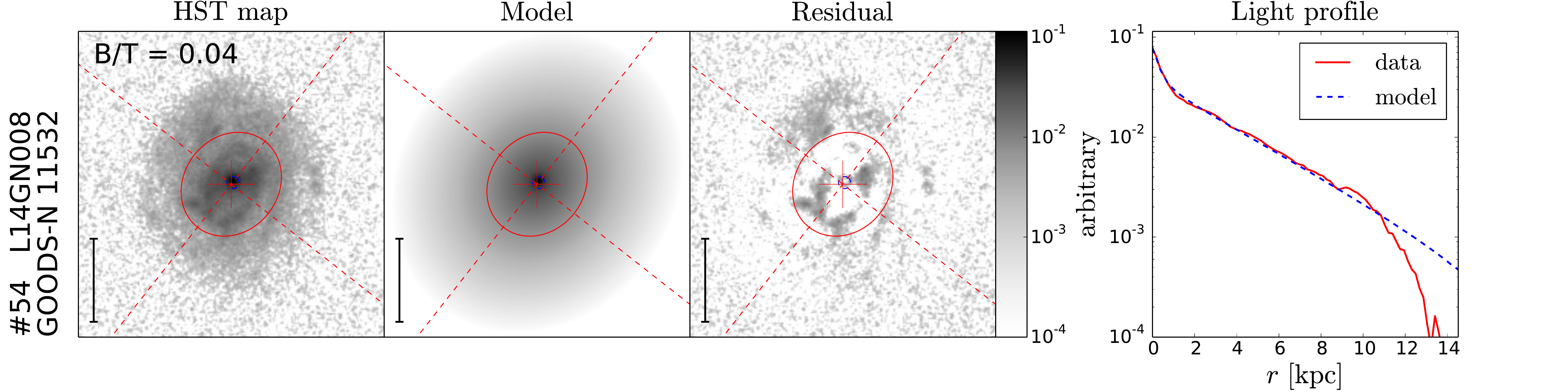}
        \hfill\includegraphics[height=0.1623\textwidth,clip,trim=0 0 32cm 1.1cm]{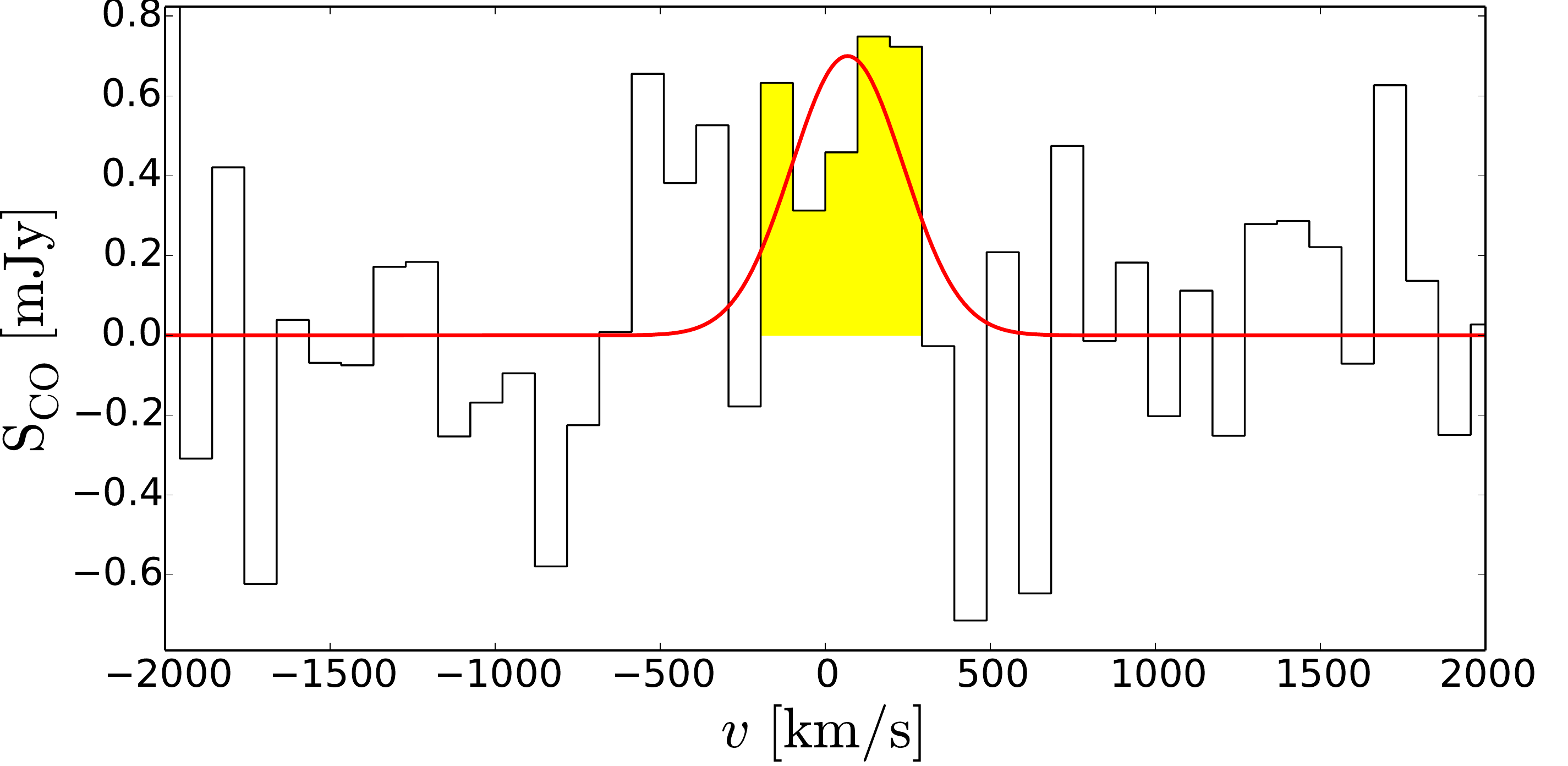}
        \\
        \includegraphics[height=0.1623\textwidth,clip,clip,trim=0 0 4cm 1.1cm]{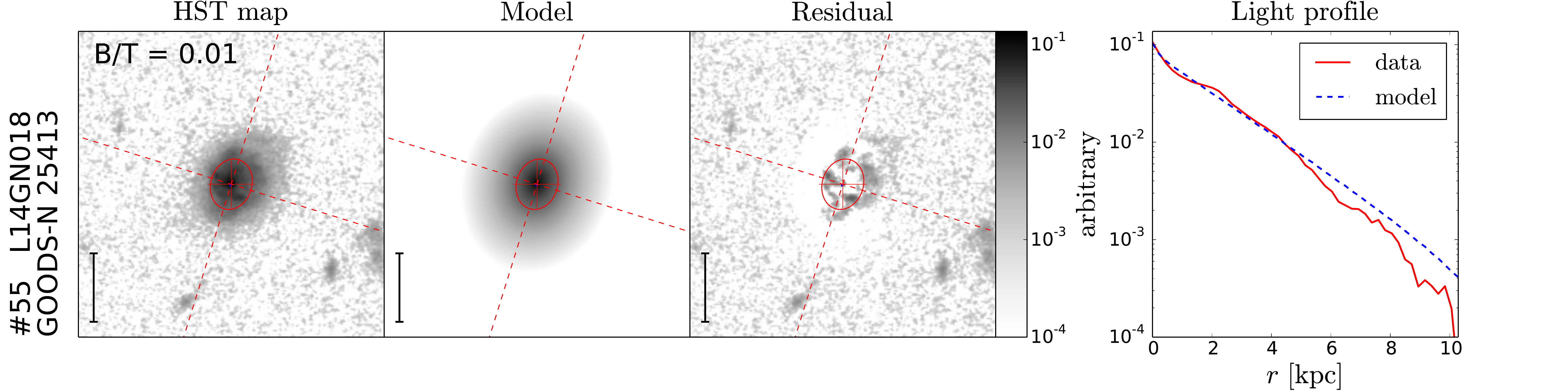}
        \hfill\includegraphics[height=0.1623\textwidth,clip,trim=0 0 32cm 1.1cm]{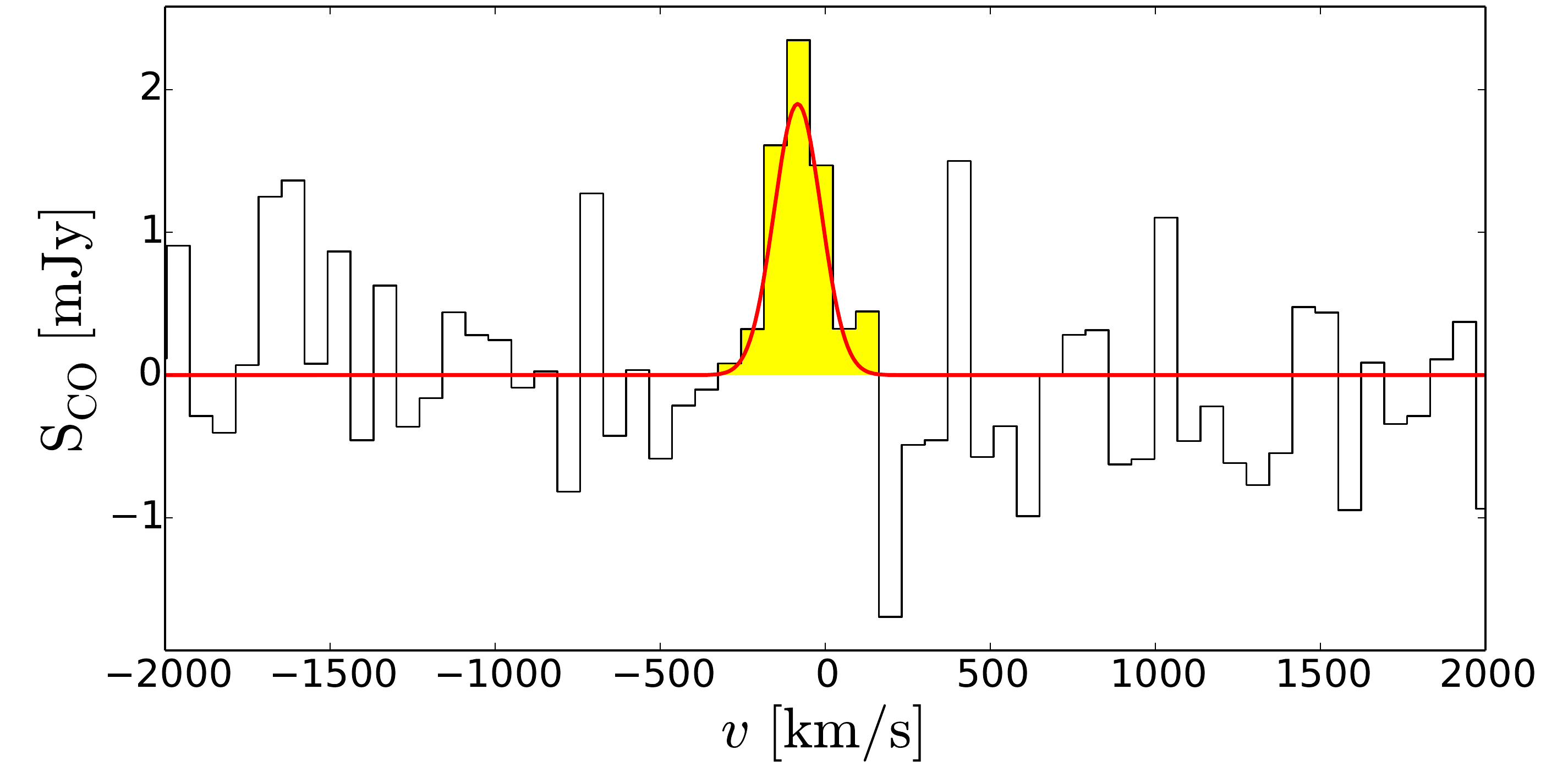}
        \\
        \includegraphics[height=0.1623\textwidth,clip,clip,trim=0 0 4cm 1.1cm]{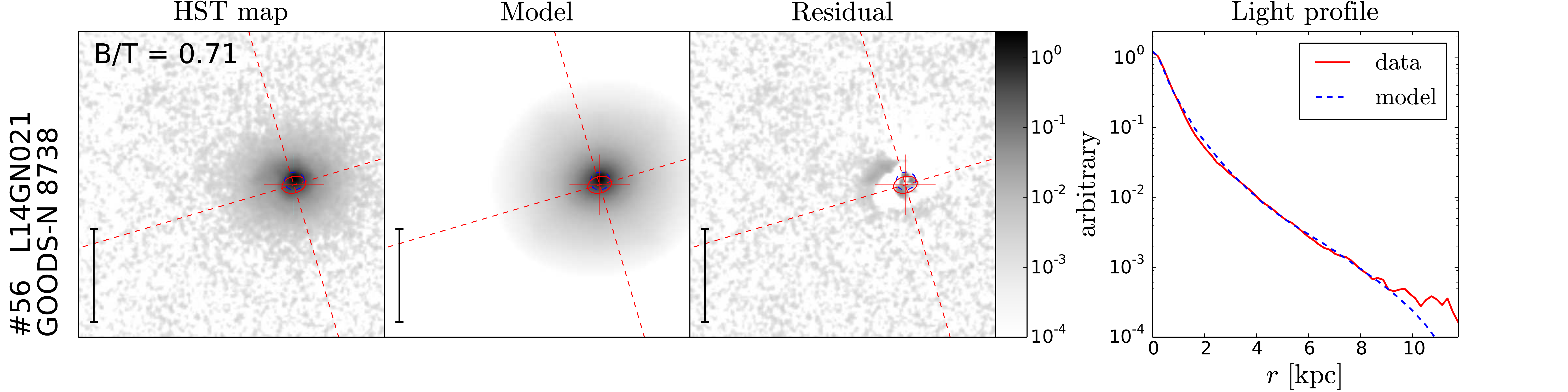}
        \hfill\includegraphics[height=0.1623\textwidth,clip,trim=0 0 32cm 1.1cm]{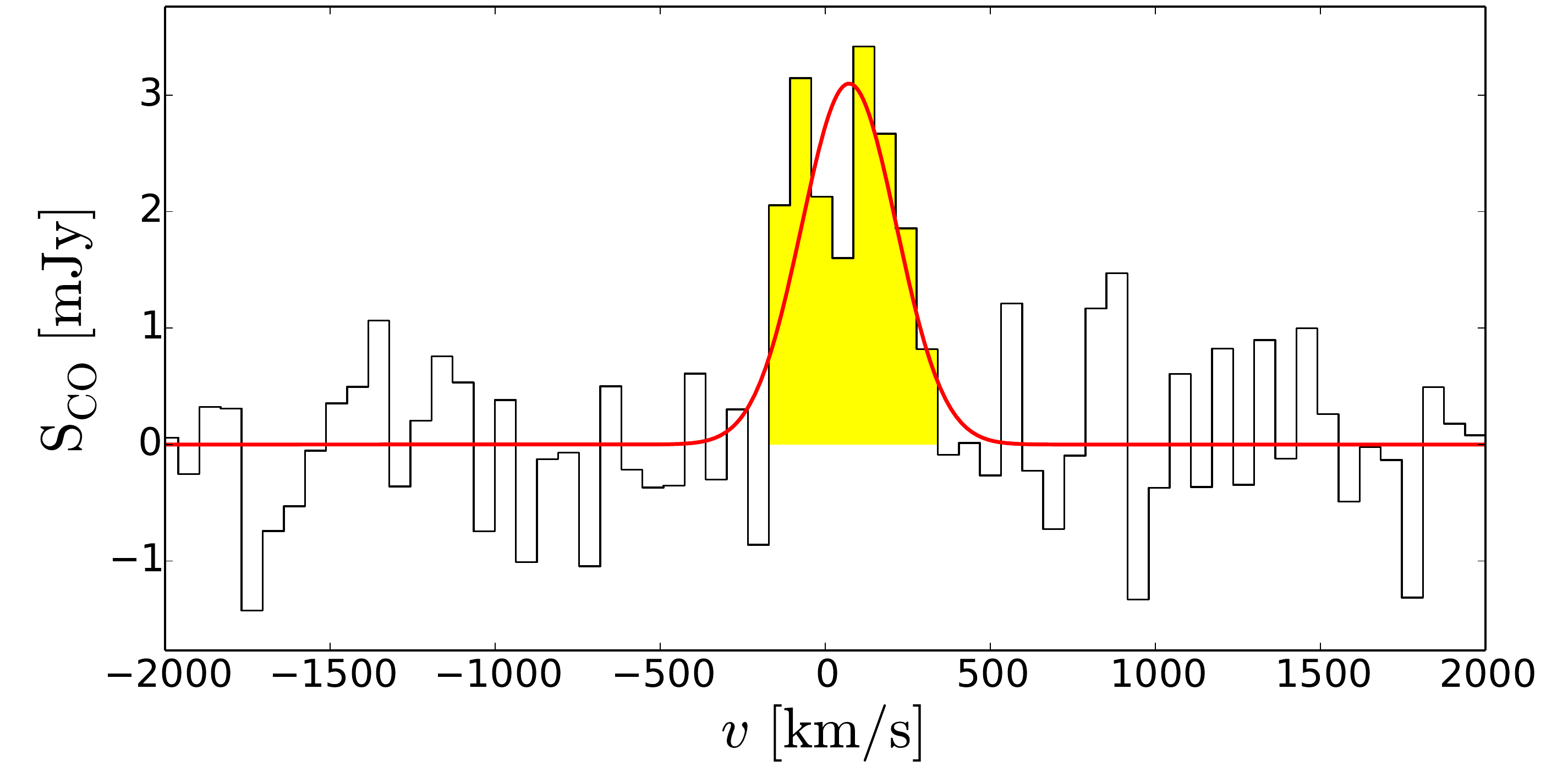}
        \\
\end{figure*}

\begin{figure*}[h!]
        \ContinuedFloat
        \flushleft
        \includegraphics[height=0.175\textwidth,clip,trim=0 0 4cm 0]{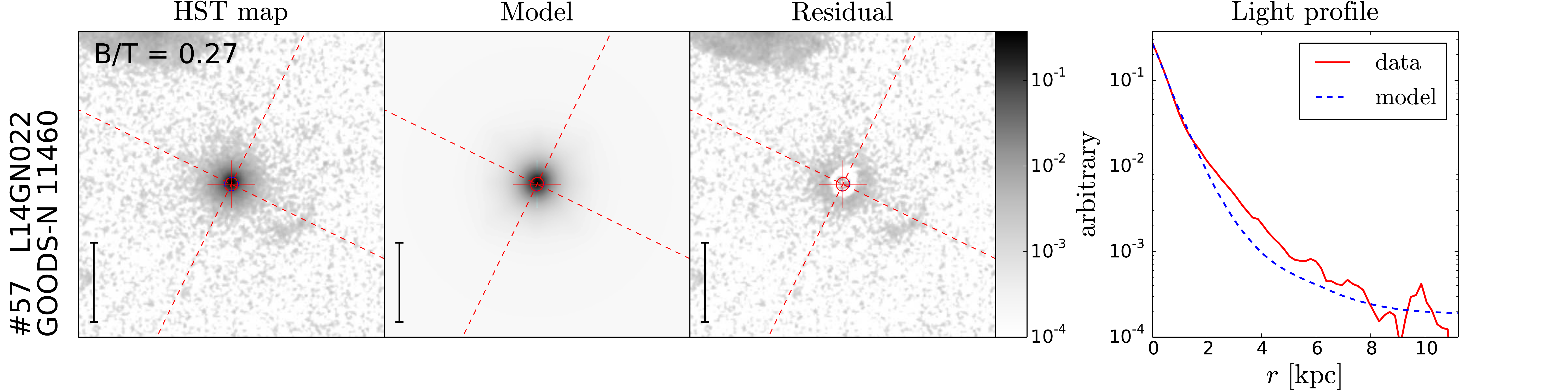}
        \hfill\includegraphics[height=0.175\textwidth,clip,trim=0 0 32cm 0]{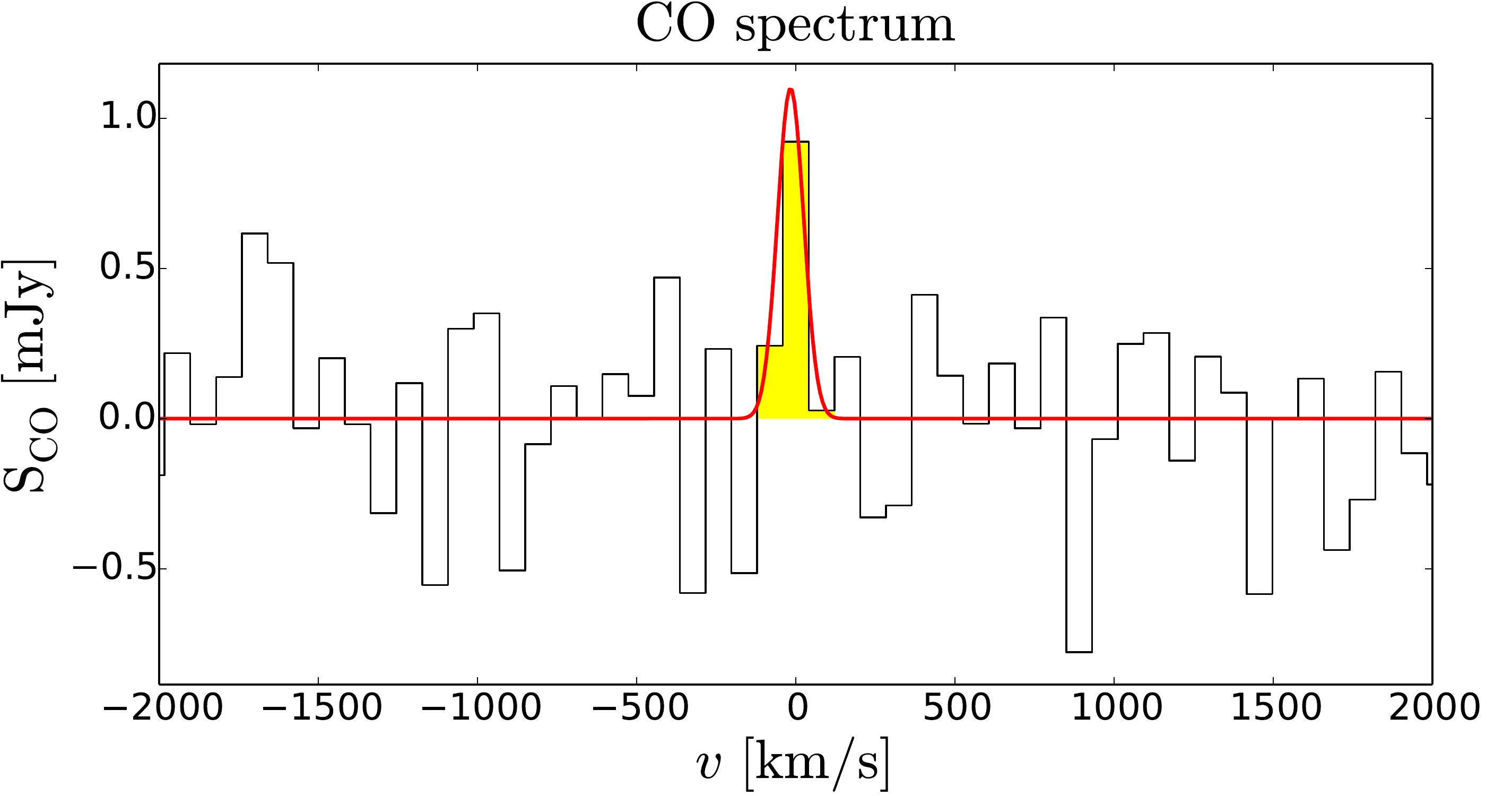}
        \\
        \includegraphics[height=0.1623\textwidth,clip,clip,trim=0 0 4cm 1.1cm]{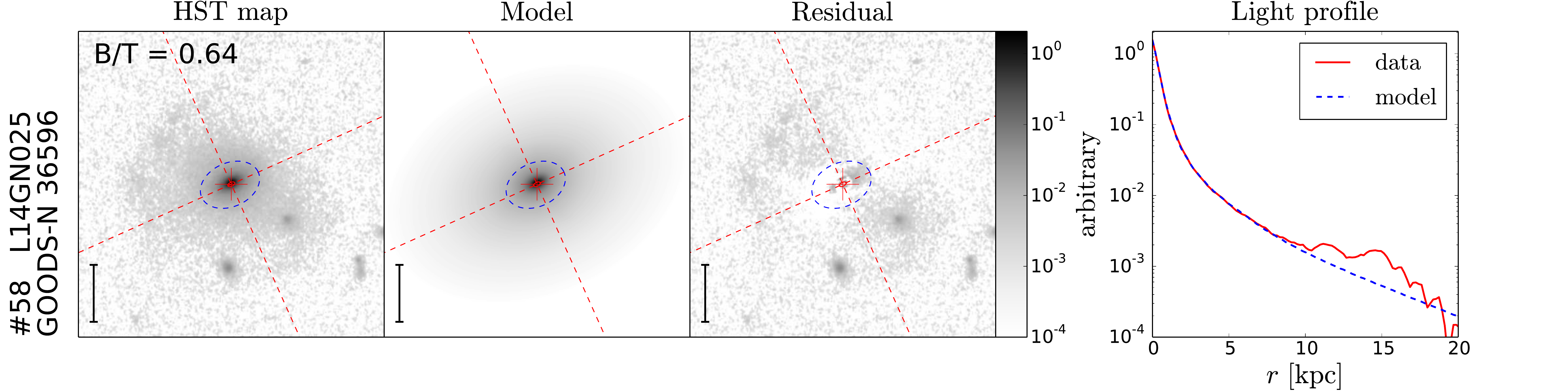}
        \hfill\includegraphics[height=0.1623\textwidth,clip,trim=0 0 32cm 1.1cm]{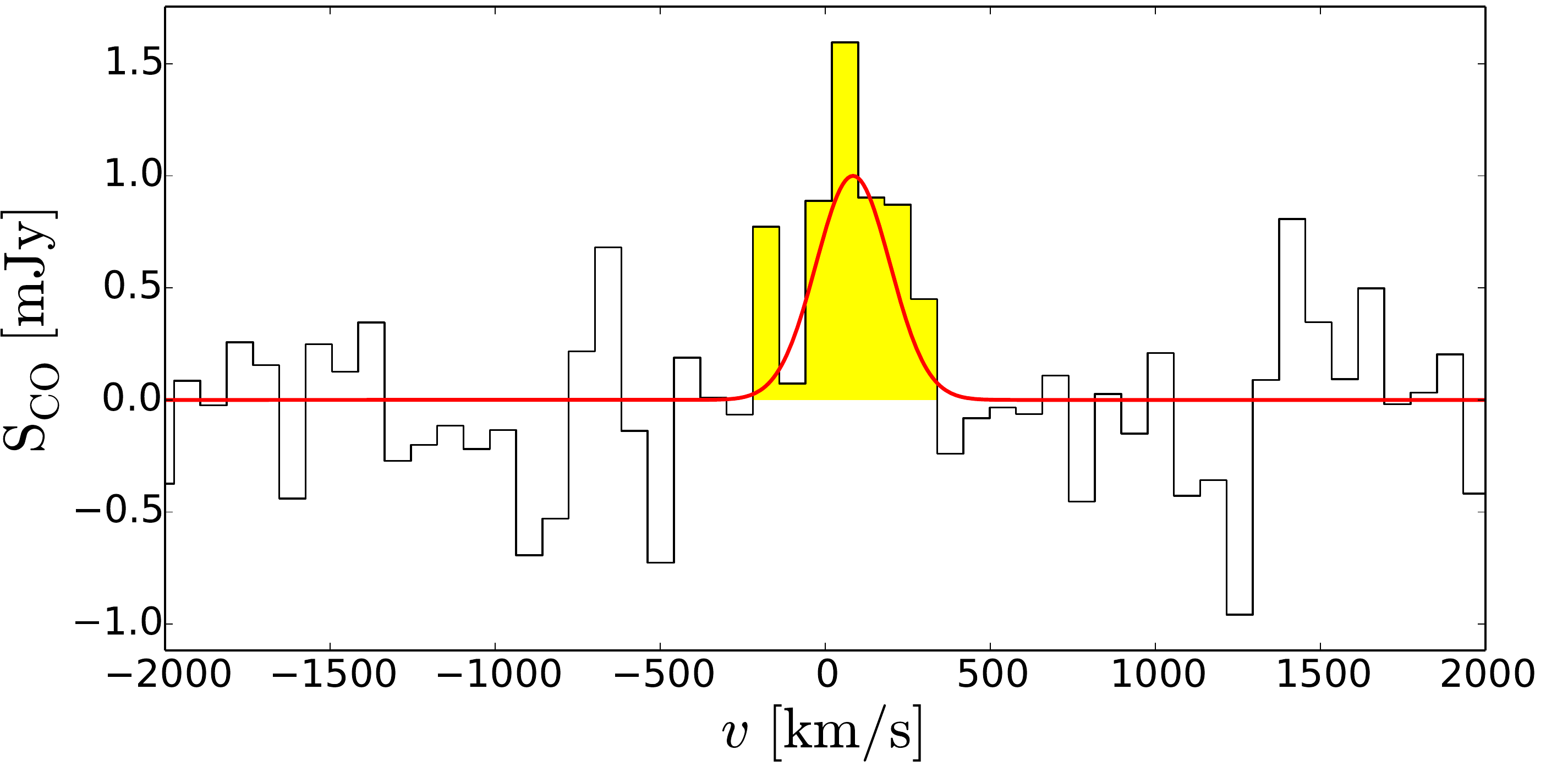}
        \\
        \includegraphics[height=0.1623\textwidth,clip,clip,trim=0 0 4cm 1.1cm]{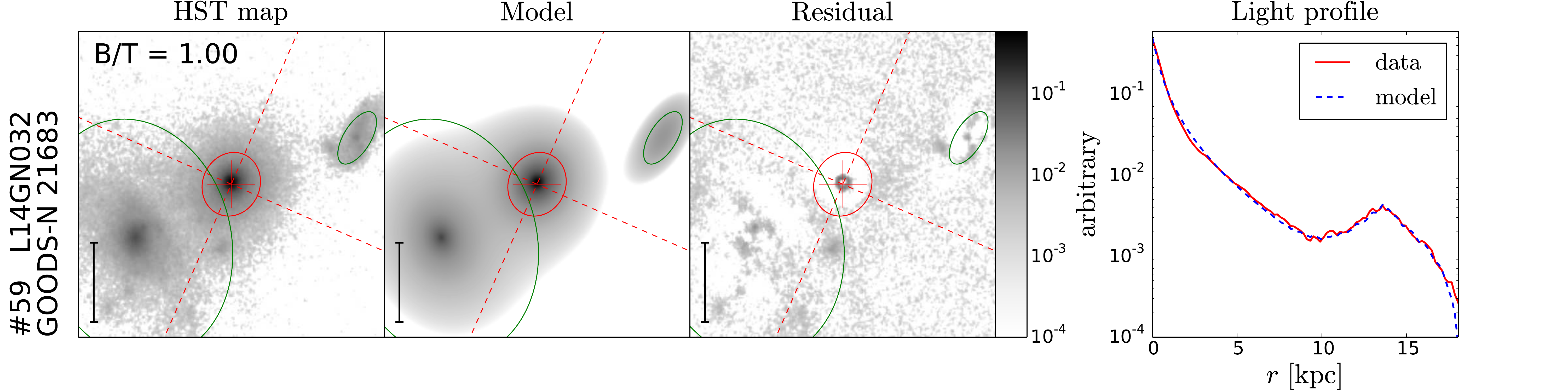}
        \hfill\includegraphics[height=0.1623\textwidth,clip,trim=0 0 32cm 1.1cm]{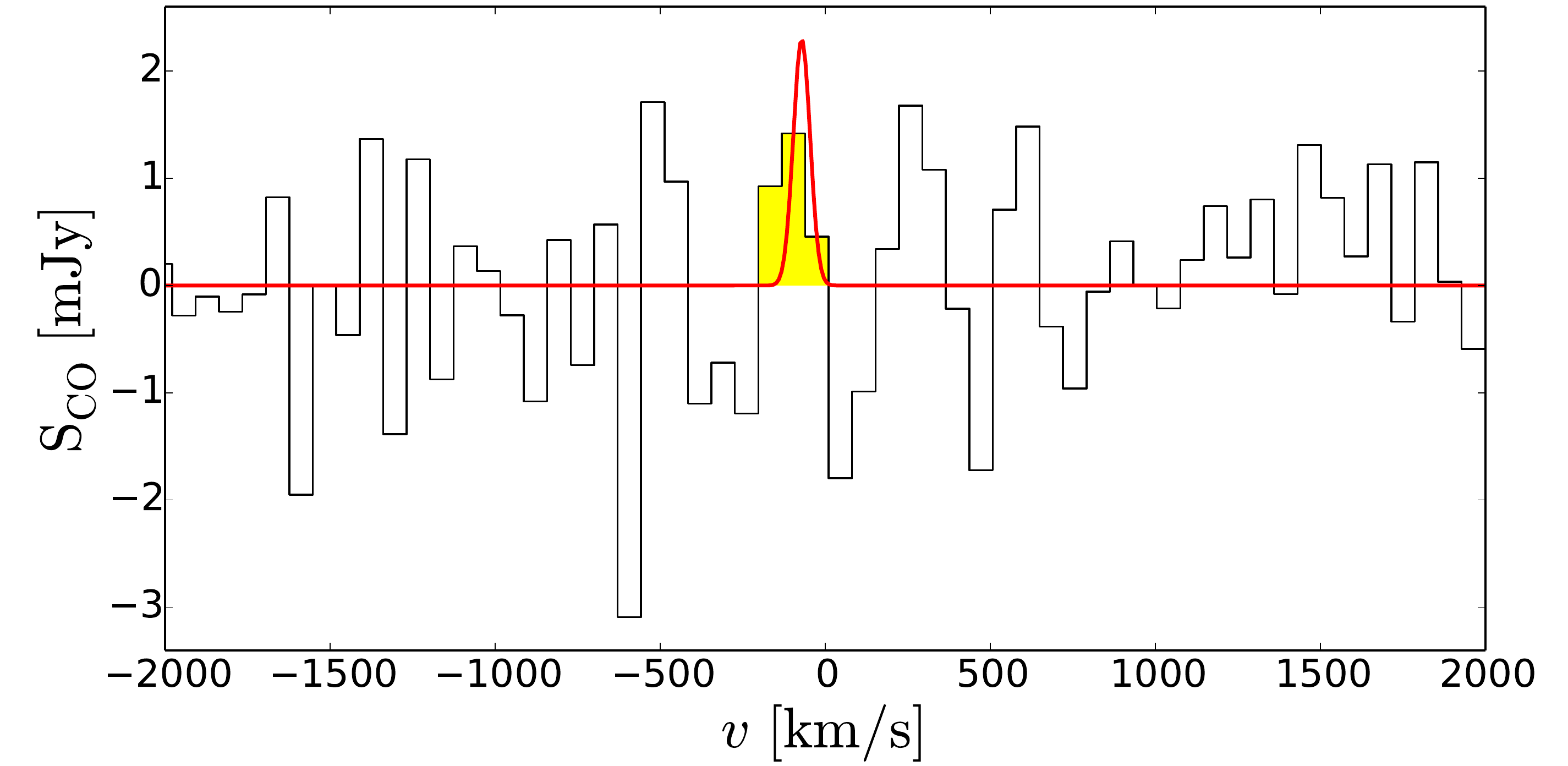}
        \\
        \includegraphics[height=0.1623\textwidth,clip,clip,trim=0 0 4cm 1.1cm]{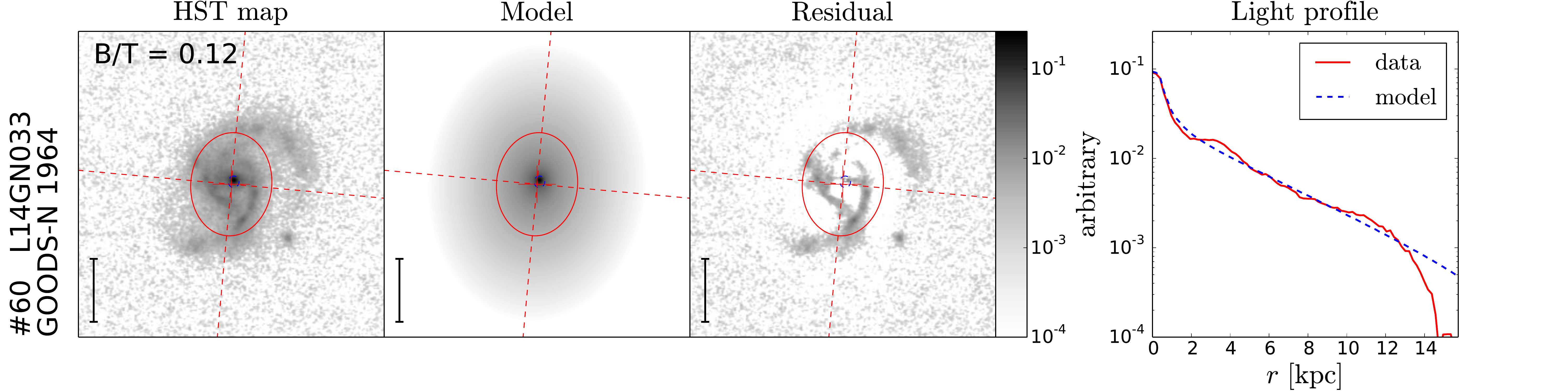}
        \hfill\includegraphics[height=0.1623\textwidth,clip,trim=0 0 32cm 1.1cm]{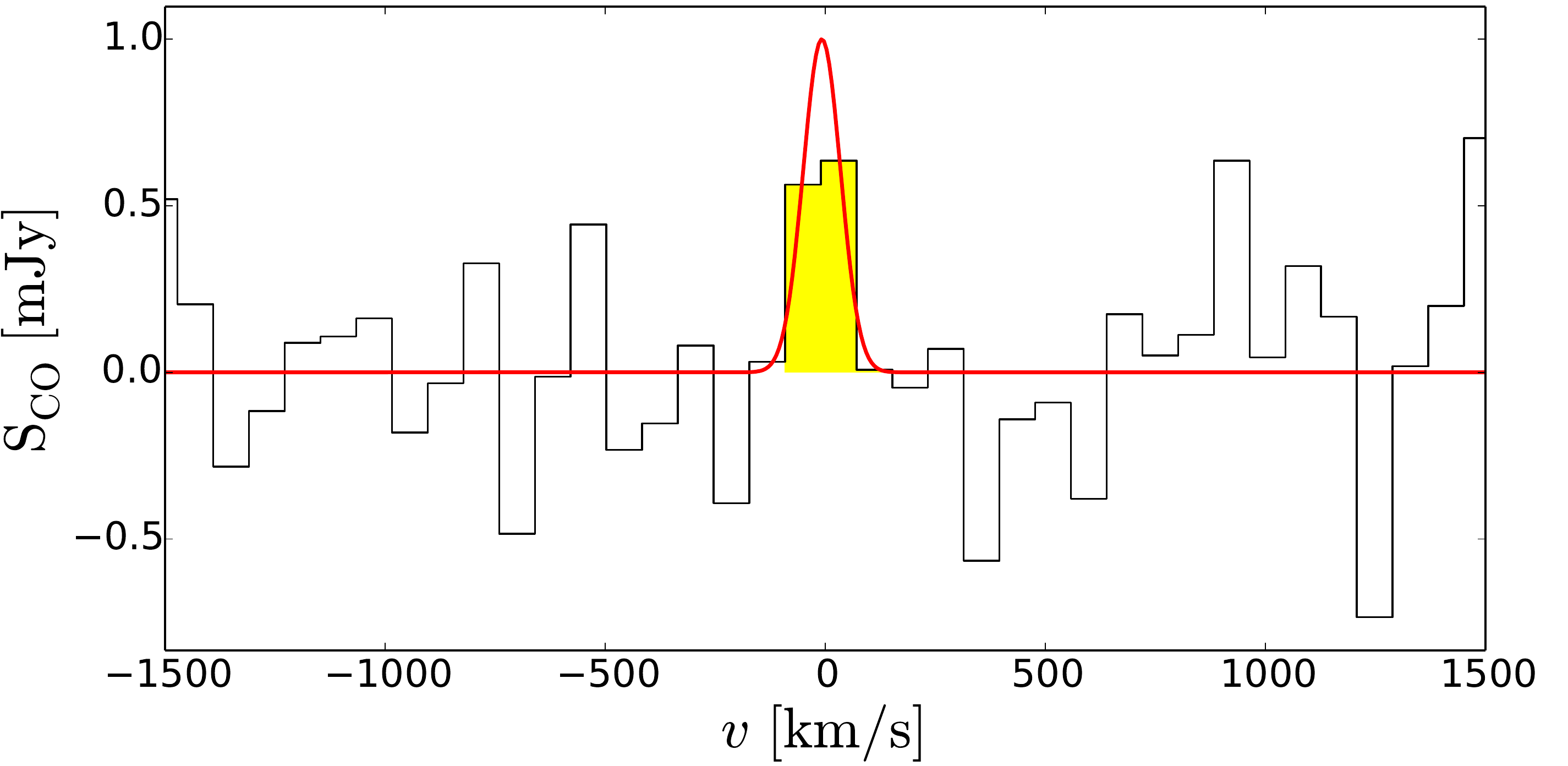}
        \\
        \includegraphics[height=0.1623\textwidth,clip,clip,trim=0 0 4cm 1.1cm]{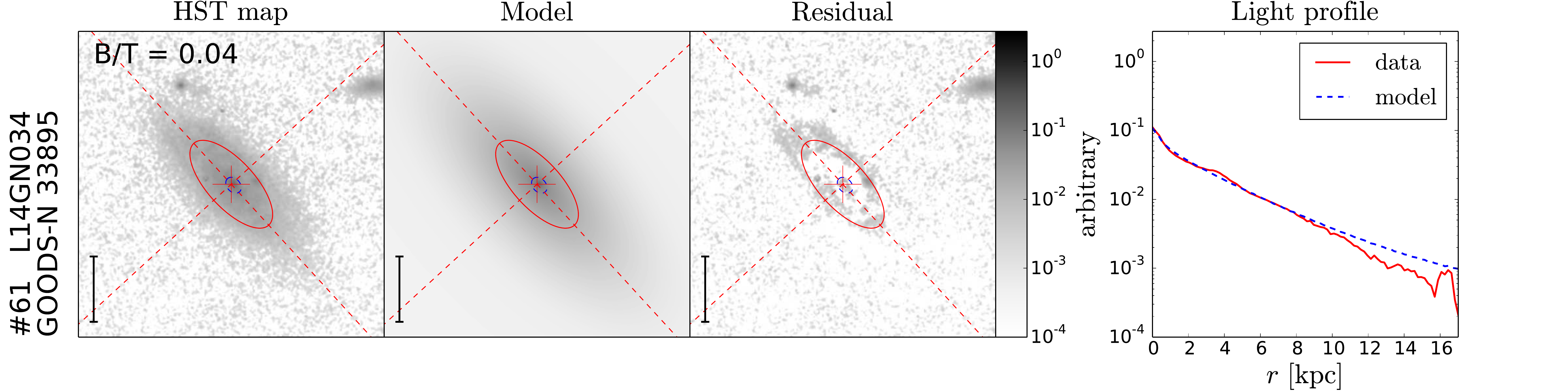}
        \hfill\includegraphics[height=0.1623\textwidth,clip,trim=0 0 32cm 1.1cm]{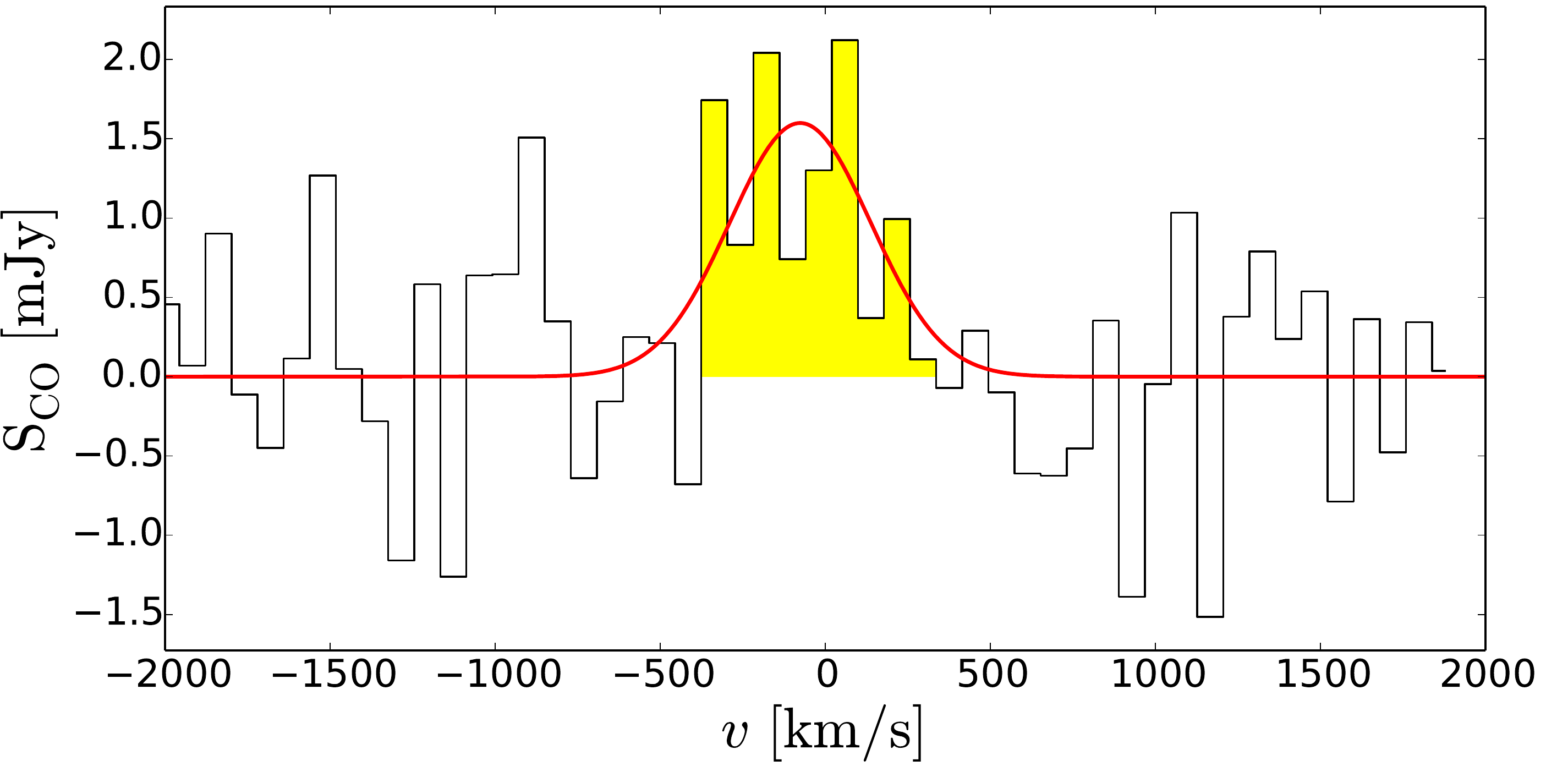}
        \\
\end{figure*}

\end{appendix}

\end{document}